\documentclass[aap]{imsart}

\newcounter{xxx}
\RequirePackage{amsthm,amsmath,amsfonts,amssymb}
\RequirePackage[numbers]{natbib}
\usepackage[autonum, blackhypersetup]{shortex} 
\usepackage{xcolor}
\usepackage{bbm}
\usepackage{algorithm}
\usepackage{algpseudocode}
\usepackage{nicematrix}

\startlocaldefs

\def\TV{\mathrm{TV}}
\endlocaldefs

\begin{document}

\begin{frontmatter}
\title{Uniform ergodicity of parallel tempering \\ with efficient local exploration}
\runtitle{Geometric ergodicity of parallel tempering}

\begin{aug}
\author[A]{\fnms{Nikola}~\snm{Surjanovic}\ead[label=e1]{nikola.surjanovic@stat.ubc.ca}},
\author[B]{\fnms{Saifuddin}~\snm{Syed}\ead[label=e2]{saifuddin.syed@stats.ox.ac.uk}},
\author[A]{\fnms{Alexandre}~\snm{Bouchard-Côté}\ead[label=e3]{bouchard@stat.ubc.ca}}
\and
\author[A]{\fnms{Trevor}~\snm{Campbell}\ead[label=e4]{trevor@stat.ubc.ca}}
\address[A]{Department of Statistics, University of British Columbia\printead[presep={,\ }]{e1,e3,e4}}
\address[B]{Department of Statistics, University of Oxford\printead[presep={,\ }]{e2}}
\end{aug}

\begin{abstract}
    Non-reversible parallel tempering (NRPT) is an effective algorithm for sampling 
    from target distributions with complex geometry, 
    such as those arising from posterior distributions 
    of weakly identifiable and high-dimensional Bayesian models.
    In this work we establish the uniform (geometric) ergodicity of NRPT under a model of  
    \emph{efficient local exploration}. 
    The uniform ergodicity log rates are inversely proportional to an easily-estimable divergence, 
    the global communication barrier (GCB), which was recently introduced in the literature. 
    We obtain analogous ergodicity results for classical reversible parallel tempering, 
    providing new evidence that NRPT dominates its reversible counterpart.
    Our results are based on an analysis of the hitting time of a continuous-time 
    persistent random walk, which is also of independent interest.
    The rates that we obtain reflect real experiments well for distributions where global 
    exploration is not possible without parallel tempering.
\end{abstract}



\end{frontmatter}

\section{Introduction}
\label{sec:intro} 

Sampling from probability distributions, such as a posterior in Bayesian
statistics or a Gibbs distribution in statistical mechanics, is a fundamentally
important task for modern statistical inference. As statistical models for
real-world phenomena become increasingly flexible,  the target distributions
become complex, making direct inference challenging. Markov Chain Monte Carlo
(MCMC) algorithms such as Metropolis--Hastings, Langevin dynamics, and Hamiltonian
Monte Carlo (HMC) are commonly used to approximate samples from complex target
distributions, $\pi_1$, when the density $\pi_1(x)$ can be evaluated up to a
normalizing constant.  MCMC algorithms construct a $\pi_1$-stationary Markov
chain $\cbra{X_t}_{t \geq 0}$ that approximately samples from $\pi_1$ when $t$ is sufficiently
large.  In practice, MCMC methods can struggle to pass through regions of low
density in the target distribution, navigate thin shells, and pass through
entropic barriers \cite{gelman2013bayesian}, and consequently fail to
converge reliably within the given computational budget. 

Parallel tempering (PT) \cite{geyer1991markov,hukushima1996exchange,swendsen1986replica} is a powerful technique often used to improve the mixing of MCMC algorithms. 
PT constructs a Markov chain on an expanded state space targeting $N+1$ distributions on an
\emph{annealing path} $(\pi_\beta)_{\beta \in [0,1]}$ that continuously interpolates
between a \textit{reference} distribution $\pi_0$ where direct inference is
possible, and the \textit{target} distribution $\pi_1$ where inference is
challenging. 
An iteration of a PT algorithm consists of: a \emph{local exploration} move, where
a problem-specific MCMC move is applied in parallel, independently targeting the
annealing distributions; and a \emph{communication} move that proposes
swaps between adjacent states and moves samples along the path. 

PT's empirical performance depends on how frequently the reference and target
communicate along the path. Intuitively, when samples from the reference
successfully reach the target, the target chain undergoes a \emph{restart},
potentially discovering new regions of the state space.  The total number of
restarts empirically correlates with more conventional measures such as mixing
times and effective sample size; however, the analysis of PT's convergence in
terms of these quantities and its ergodic properties is poorly understood.

\begin{figure}[t]
  \centering 
  \begin{subfigure}{0.49\textwidth}
      \centering
      \includegraphics[width=\textwidth]{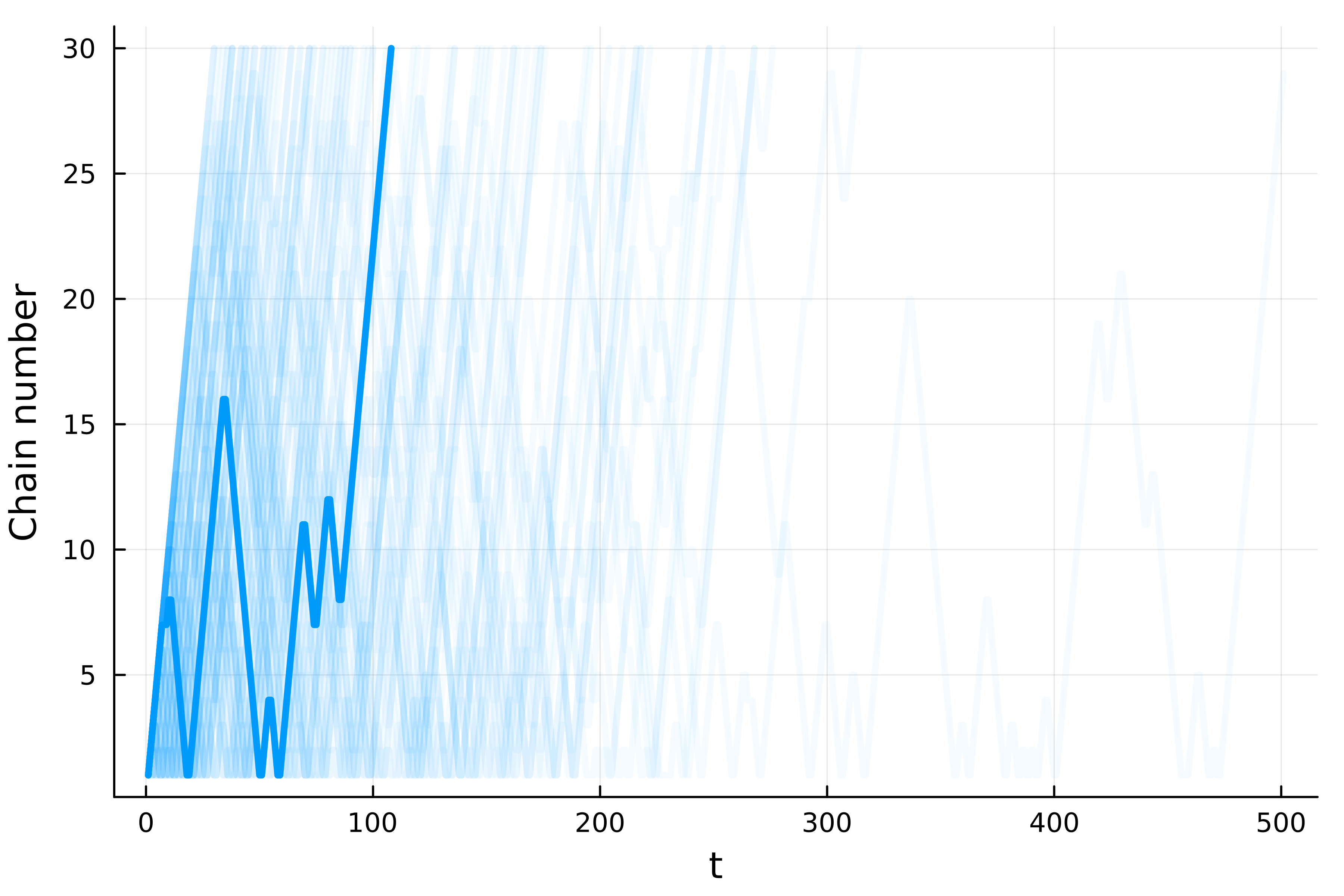}
  \end{subfigure}
  \begin{subfigure}{0.49\textwidth}
      \centering
      \includegraphics[width=\textwidth]{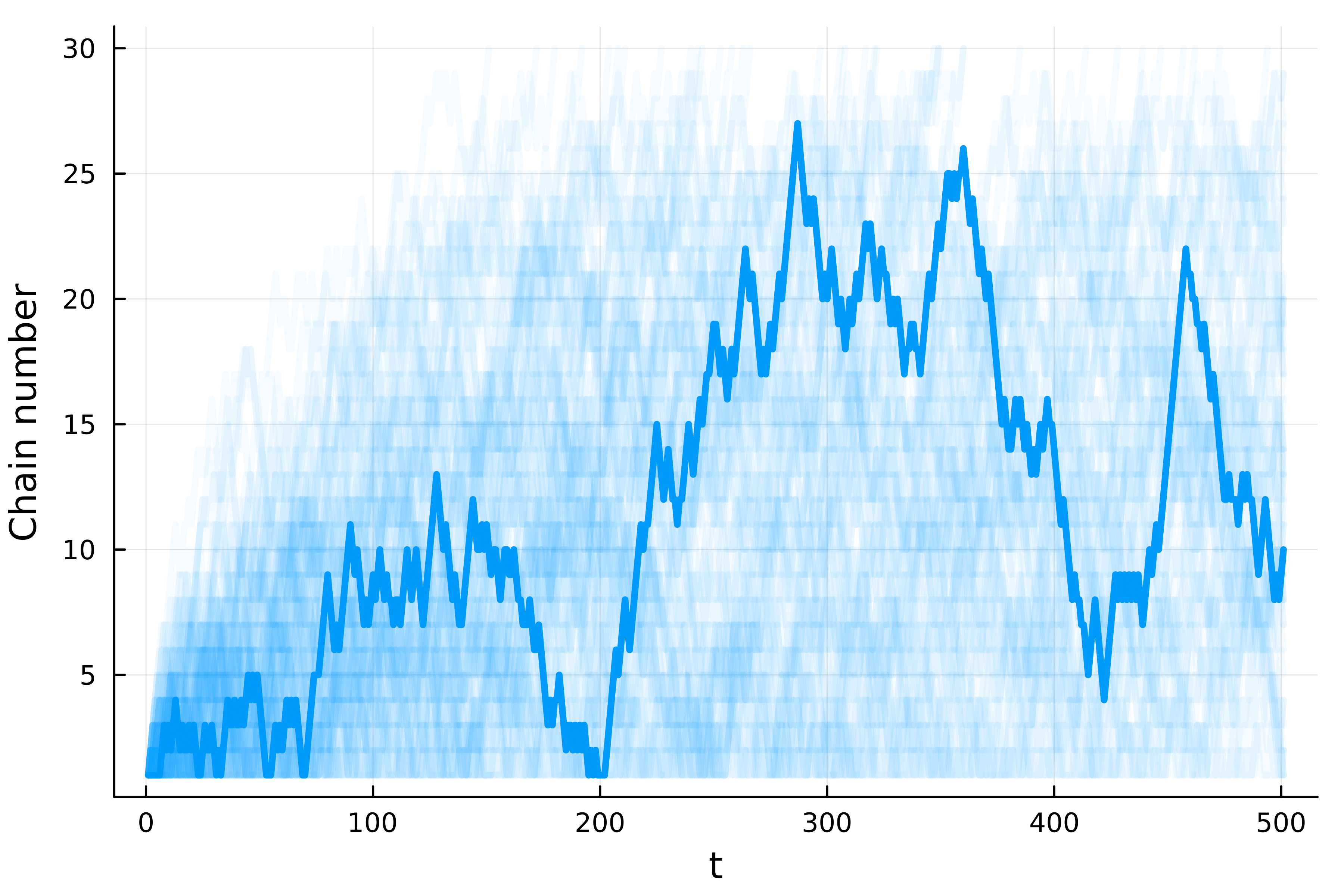}
  \end{subfigure}
  \caption{100 random index process trajectories starting at the reference chain 
  and terminated once they reach the target or at time $t=500$. We show that the rate of geometric ergodicity 
  of PT is 
  governed by the rate at which samples from the reference (bottom) travel to the target (top).
  \textbf{Left:} Communication scheme for non-reversible parallel tempering (NRPT). 
  \textbf{Right:} Communication scheme for reversible parallel tempering (RPT). 
  Both schemes use 30 chains with a rejection rate of $r = 0.1$. 
  The first random trajectory is highlighted.}
  \label{fig:hitting_times}
\end{figure}

Since PT is a ``meta-algorithm,'' which is only fully specified by the
problem-specific local exploration kernels, it results in several ``flavours''
of PT, such as random-walk Metropolis PT, Langevin PT, HMC PT, etc. This poses a
significant challenge when analysing the convergence of PT since it is hard to
disentangle the influence of the problem-specific local exploration move, and
the communication scheme which dramatically impacts the number of empirically observed
restarts.  Notably, \cite{syed2021nrpt} showed that non-reversible
communication (NRPT) \cite{okabe2001replica} dramatically impacts PT's
scalability and efficiency, independent of the exploration move, compared to
traditional reversible communication (RPT): NRPT non-asymptotically dominates
RPT in terms of the number of restarts for a finite $N$. Additionally, as 
$N\to\infty$, the total number of restarts increases for NRPT and decreases to zero 
for RPT. This gap in communication can also be observed in \cref{fig:hitting_times}.

\bigskip \noindent \textbf{Our contributions.}
In this work, we study the uniform ergodicity of PT algorithms under 
efficient local exploration (ELE, defined in \cref{assump:base_erg_collection}). 
To address the above challenges, we 
construct an exploration-agnostic model for PT
algorithms under the ELE assumption, motivated
by \cite{syed2021nrpt}. The ELE allows us to disentangle our analysis from the
problem-specific local exploration moves and analyze the communication moves' role in
the mixing of PT algorithms. This algorithmic model is not expected to hold exactly in practice
but leads to rich and interpretable theoretical results consistent with
experiments. 

In \cref{sec:preliminaries} we introduce PT, related quantities, and our
theoretical assumptions. After this overview, in \cref{sec:marginal_tv} we
show that the target component of the PT chain satisfies a central limit
theorem assuming uniform ergodicity on this component.
We then study the marginal geometric ergodicity of PT in the limit as the
number of parallel chains (annealing path distributions) $N$ tends to infinity. In particular, we obtain 
bounds on the mixing time for the target chain of $O(N\Lambda)$ and $O(N^2)$
for NRPT and RPT, respectively. Here, $\Lambda\geq 0$ is the global
communication barrier (GCB), a PT-specific divergence between $\pi_0$ and
$\pi_1$ introduced in \cite{syed2021nrpt}, which quantifies the intrinsic
difficulty of communicating along the annealing path. Unlike commonly used
quantities like spectral gaps, the GCB is easily estimable by running PT and
relates to optimal tuning guidelines and informal MCMC diagnostics
\cite{syed2021nrpt,syed2021optimized,surjanovic2022VPT}. We then study the
ergodicity of PT when $N$ is finite in \cref{sec:erg_finite}. Various
experiments are included in \cref{sec:experiments}. We end with a discussion in
\cref{sec:discussion}, in which we offer a more detailed discussion of the
ELE assumption and some practical guidance for practitioners.

\bigskip \noindent \textbf{Related work.}
Several papers on the ergodicity and mixing of (reversible) PT exist in the
literature
\cite{miasojedow2013adaptive,woodard2009torpid,woodard2009rapid,lee2023mixing}.
However, these works generally assume that the number of parallel PT chains, $N+1$, is finite and
that the communication scheme is reversible. Additionally, imposing structure on the
target distribution, state space, or exploration kernel is often required. Moreover,
the convergence of PT is usually expressed in terms of computationally intractable
quantities, such as spectral gaps and conductance. 
In contrast, our work analyzes how the ergodicity of PT scales with the 
number of annealing distributions and how it is affected by the chosen 
communication scheme without making any additional
structural assumptions on the target or state space. 

The ergodic properties of the global PT chain with a fixed number of $N+1$ chains and a
reversible communication scheme are analyzed in past work \cite{woodard2009torpid,woodard2009rapid}.
Specifically, sufficient conditions for the rapid mixing
(polynomial decrease of the spectral gap in target distribution dimension $d$) of PT 
\cite{woodard2009rapid} and torpid mixing (exponential decrease of the spectral gap in $d$) of 
PT \cite{woodard2009torpid} have been identified. 
The theoretical analysis of the entire PT chain in previous work \cite{woodard2009rapid,woodard2009torpid}
(which targets $\pi_{\beta_0} \times \cdots \times \pi_{\beta_N}$) may be too strict;
often we are interested in the final component of the PT chain targeting $\pi_1$ and not the
global PT chain itself, such as when computing expectations with respect to the
target. We attempt to close this apparent gap between theory and experiment by
analysing the ergodicity of only the marginal of target states under an efficient
local exploration assumption. Unlike previous work
\cite{woodard2009torpid,woodard2009rapid,lee2023mixing}, our analysis can
also identify the influence of the communication scheme on the
mixing time, as well as the influence of the number of chains. 

A previous analysis of PT under the restricted setting where random-walk Metropolis
local exploration kernels are used concludes that PT is geometrically ergodic
under appropriate conditions \cite{miasojedow2013adaptive}. An adaptive version
of PT has also been considered and convergence properties of the algorithm have been 
established \cite{miasojedow2013adaptive}.
Our work is more general in the sense that no restriction is placed on the
type of local exploration kernel being used, aside from an assumption of 
efficient local exploration. Finally, approaches to estimating
bounds on total variation and Wasserstein distance using $L$-lag couplings exist 
in the literature \cite{biswas2019lagcoupling}, including experiments with PT.
One drawback of such methods is that the practitioner is required to select a value 
of $L$, with performance of the method dependent on whether an appropriate value was chosen. 

\section{Preliminaries} 
\label{sec:preliminaries}

We begin with an overview of annealing in \cref{sec:annealing}. 
We then introduce PT algorithms and two communication variants, NRPT and RPT, in
\cref{sec:PT}. We conclude by defining the global communication
barrier (GCB) divergence in \cref{sec:GCB}.

\subsection{Annealing}
\label{sec:annealing} 
Parallel tempering begins with a specification of a \emph{target 
distribution} $\pi_1$ with full support on a measurable state space $(\mcX, B(\mcX))$, 
where $B(\mcX)$ is a standard Borel $\sigma$-algebra.
In Bayesian settings, the target distribution is often 
the posterior, with density $\pi_1(x)$ only known up to a normalizing constant. 
That is, for $x \in \mcX$, with respect to some dominating measure $\dee x$, 
we have that the density $\pi_1(x)$ of the target distribution is
\[
  \pi_1(x) = \frac{\gamma_1(x)}{Z},
\]
where $\gamma_1$ can be evaluated pointwise and $Z=\int_\mcX \gamma_1(x)\dee x$ is a normalizing constant. 
We abuse notation by using the same symbol to denote a distribution and its corresponding
density with respect to a dominating measure; it will be clear from context whether 
we are referring to the distribution or the density.

We also assume the existence of a \emph{reference distribution} $\pi_0$ that is
mutually absolutely continuous with respect to the target $\pi_1$ with density
$\pi_0(x)$. Unlike the target, we assume it is possible to tractably obtain \iid
samples from $\pi_0$ and evaluate the normalized density $\pi_0(x)$ for all
$x\in \mcX$. In the Bayesian framework, $\pi_0$ is usually the prior, although
this is not necessary and can be suboptimal. For example, it may also be chosen to belong
to a variational family and selected to be one that minimizes a divergence 
between $\pi_0$ and $\pi_1$, resulting in samples from the reference 
travelling to the target more frequently \cite{surjanovic2022VPT}. 

Given a reference and target, an \emph{annealing path},
$(\pi_\beta)_{\beta\in[0,1]}$ is a continuum of probability distributions where
the density $\pi_\beta(x)$ continuously interpolates between $\pi_0(x)$ and
$\pi_1(x)$ for all $x\in\mcX$ as the \emph{annealing parameter}, $\beta$,
increases from zero to one. In this paper, we assume for $\beta\in[0,1]$
that $\pi_\beta$ is on the \emph{linear path} with density
\[
 \label{eq:linearpath}
 \pi_\beta(x) \propto \pi_0(x)\exp(-\beta V(x)), \qquad 
  V(x)=\log \pi_0(x) - \log\pi_1(x).
\]
In the above expression, $V(x)$ is referred to as the \emph{energy}. 
Although the linear path is standard in the literature, much of the work in this
paper extends without significant modification to more general annealing paths
(e.g., see \cite{syed2021optimized}). 
To ensure the linear path is sufficiently regular, we will assume the energy has
third moments with respect to the reference and target. 

\bassump
\label{assump:third_moment}
$\pi_0(|V|^3) < \infty$ and $\pi_1(|V|^3) < \infty$.
\ebassump

Finally, an \emph{annealing schedule} $\mcB_N = (\beta_n)_{n=0}^N$ 
of size $N+1$ is a sequence of annealing parameters 
$0 = \beta_0 < \beta_1 < \ldots < \beta_{N-1} < \beta_N = 1$. The annealing schedule
induces a sequence of \emph{annealing distributions} $\rbra{\pi_{\beta_n}}_{n=0}^N$,
discretizing the annealing path $(\pi_\beta)_{\beta \in [0,1]}$.

\subsection{Parallel tempering}
\label{sec:PT}
Given an annealing schedule of size $N+1$, the PT algorithm constructs a Markov
chain $\{\bar X_t\}_{t \geq 0}$ on the expanded state space $\bar{\mcX} \equiv
\mcX^{N+1}$, invariant with respect to the joint density of the annealing
distributions \[
  \label{eq:joint_target}
  \bar{\pi}(\bar x) = \prod_{n=0}^N \pi_{\beta_n}(x^n), \qquad 
  \bar x = (x^0, x^1, \ldots, x^N)\in \bar{\mcX}.
\]
Given $\bar{X}_{t-1}$, PT generates $\bar{X}_{t}$ by performing
a \emph{local exploration} move followed by a \textit{communication} move. 
The local exploration move generates a state $\bar{X}=(X^0,\dots,X^N)\in\bar
\mcX$ by updating the $n^\text{th}$ component of
$\bar{X}_{t-1}=(X^0_{t-1},\ldots,X^N_{t-1})$ independently according to a
$\pi_{\beta_n}$-invariant Markov Kernel $K^{(\beta_n)}$: 
\[ 
  X^n | X^n_{t-1} \sim K^{(\beta_n)}(X_{t-1}^n,\cdot), \qquad   
  n = 0,\ldots,N.
\] For $\beta\in (0,1]$,
$K^{(\beta)}$ corresponds to an MCMC algorithm such as Metropolis--Hastings or
Hamiltonian Monte Carlo. When $\beta=0$, we assume $K^{(0)}$ generates an
\iid sample from the reference distribution.

The communication move in PT involves
proposing swaps between adjacent components of $\bar{X}$ according to a
Metropolis--Hastings move targeting $\bar{\pi}$. At each iteration $t$, the
communication move for PT is specified by a collection of swap proposals
$S_t\subseteq \{0,\dots, N-1\}$. For $n\in S_t$, components $n$ and $n+1$ of
$\bar x$ are swapped with probability $\alpha^{(n,n+1)}(\bar{x})$, where 
\[
  \label{eq:acceptance_prob} 
  \alpha^{(n, n+1)}(\bar x) 
  &= \exp\rbra{\min\cbra{0, (\beta_{n+1} - \beta_n) \cdot (V(x^{n+1}) - V(x^n))}}.  
\] 
Let $S_\text{even}$ and $S_\text{odd}$ be the even and odd subsets of
$\{0,\dots,N-1\}$, respectively. We say a communication at iteration $t$ is
even (odd) if $S_t=S_\text{even}$ ($S_t=S_\text{odd}$). 
The even and odd communication moves trigger the largest
set of nearest-neighbour swaps in parallel without interference. We say the
communication scheme corresponds to \emph{reversible} PT (RPT) when 
$S_t$ is chosen between $S_\text{even}$ and $S_\text{odd}$
with equal probability. The \emph{non-reversible} PT (NRPT) communication scheme 
is obtained when $S_t$ deterministically alternates between 
$S_\text{even}$ and $S_\text{odd}$.
Notably, each iteration of RPT can potentially reverse any accepted swap from
the previous iteration, whereas NRPT cannot. This subtle difference between
NRPT and RPT, introduced by Okabe et al.~\cite{okabe2001replica} and subsequently studied by
Syed et al.~\cite{syed2021nrpt}, results in a more efficient PT communication procedure.
We summarize the NRPT algorithm in \cref{alg:NRPT} and note that
RPT can be obtained with the comment included above line \ref{line:DEO}.

In general, the PT Markov chain $\bar{X}_t$ is $\bar\pi$-invariant but not time-homogeneous, 
due to the possible time dependency of the proposed swaps $S_t$ during the communication step.
For $t\in\nats$, we define $\bar K_t$ to denote the marginal $t$-step Markov transition
kernel for the PT chain,
\[
  \bar X_t | \bar X_0 \sim \bar K_t(\bar X_0, \cdot).  
\]
We use $\bar K^\text{NRPT}_t$ and $\bar K^\text{RPT}_t$ to indicate the
$t$-step transition kernel for NRPT and RPT, respectively. It is worth noting
that we can also describe the NRPT and RPT algorithms in terms of a time-homogeneous
Markov chain $\tilde X_t = (\bar{X}_t, S_t)$ if we further extend the state-space with the
set of swap proposals $S_t \in \{S_\text{even},S_\text{odd}\}$.
We refer readers to \cref{sec:PT_kernels} for a formal description of the PT Markov kernel.

\begin{algorithm}[t]
	\footnotesize
\caption{Non-reversible parallel tempering (NRPT)}
\label{alg:NRPT}
\begin{algorithmic}[1]
\Require Initial state $\bar x_0$, annealing schedule $\mcB_N = \rbra{\beta_n}_{n=0}^N$, 
  exploration kernels $\rbra{K^{(\beta_n)}}_{n=0}^N$, \# iterations $T$
\For{$t=1,2,\ldots,T$} 
  \For{$n = 0,1,\ldots,N$}
    \textcolor{gray}{\Comment{Local exploration move}}
    \State $x^n | x_{t-1}^n \sim K^{(\beta_n)}(x_{t-1}^n, \cdot)$
  \EndFor
  \textcolor{gray}{\LineComment{For RPT, replace the line below with: \quad $U \gets \text{Unif}(0,1)$; {\bf if} {$U \leq 0.5$} {\bf then} 
    {$S_t \gets S_\text{even}$} {\bf else} {$S_t \gets S_\text{odd}$}}}
  \State {\bf if} {$t-1$ is even} {\bf then} {$S_t \gets S_\text{even}$} 
    {\bf else} {$S_t \gets S_\text{odd}$} \label{line:DEO}
  \For{$n \in S_t$}
    \textcolor{gray}{\Comment{Communication move}}
    \State $U \gets \text{Unif}(0,1)$
    \State {\bf if} {$U \leq \alpha^{(n,n+1)}(\bar x)$} {\bf then}  
      $(x^{n+1}, \, x^n) \gets ( x^{n}, \, x^{n+1})$ 
      \textcolor{gray}{\Comment{Swap with probability in \cref{eq:acceptance_prob}}}
  \EndFor
  \State $\bar x_t\gets \bar x$
\EndFor
\State \textbf{Return:} $(\bar x_t)_{t=0}^T$
\end{algorithmic}
\end{algorithm}

\subsection{Global communication barrier}
\label{sec:GCB}

The \emph{global communication barrier} (GCB) divergence, denoted $\Lambda$, 
is a quantity that characterizes the difficulty of 
moving draws along a path of distributions from the reference $\pi_0$ to the target $\pi_1$ using NRPT. 
The GCB is a property of only the path itself and can be thought of intuitively 
as a notion of ``total resistance'' that the path offers to draws attempting to traverse
the path from the reference to the target. In this work, since we use the linear path in 
\cref{eq:linearpath}---which is determined entirely by its endpoints $\pi_0$ and
$\pi_1$---the GCB can be seen as a divergence between $\pi_0$ and $\pi_1$, defined as
\[
  \label{eq:GCB}
  \Lambda := \frac{1}{2} \int_0^1 
    \EE\sbra{\abs{V(X_\beta) - V(X_\beta')}} \, \dee \beta,
  \qquad X_\beta, X_\beta' \stackrel{iid}{\sim} \pi_\beta.
\]
The GCB is finite under \cref{assump:third_moment}. We refer readers to \cref{sec:GCB_properties} for
further basic properties of the GCB, such as its $O(d^{1/2})$ scaling with
dimension for certain reference/target pairs (\cref{prop:GCB_submanifold}),
invariance under diffeomorphic pushforwards (\cref{prop:GCB_invariant_transformation}), 
and bounds via the total variation distance (\cref{prop:GCB_TV_bound}).

Note that the annealing schedule $\mcB_N=(\beta_n)_{n=0}^N$ does not appear in \cref{eq:GCB}; 
as stated previously, the GCB is a property of the path itself
rather than the discretization to $N+1$ distributions. Corollary 2 in \cite{syed2021nrpt} links the GCB to 
the behaviour of PT when $N$ is sufficiently large, finding that 
$\Lambda \approx \sum_{n=0}^{N-1} r_{n,N}$.
Here, $r_{n,N}$ is the average rejection rate for swaps between chains $n$ and $n+1$ 
of a $\bar\pi$-stationary PT chain,
\[
  r_{n,N}
  &= 1- \EE[\alpha^{(n,n+1)}(\bar{X})],\quad \bar{X}\sim \bar\pi.
\]

Previously, the GCB has been used to capture various aspects of the behaviour of
PT via this link to rejection rates \cite{syed2021nrpt}. For example, the GCB
is inversely proportional to the large-$N$ asymptotic \emph{restart rate}, which is the rate at
which fresh samples from the reference travel to the target distribution under
appropriate conditions (see Theorem 3 and Assumptions (A1)-(A3) in
\cite{syed2021nrpt}). This identity has been exploited to suggest ways to tune
the reference $\pi_0$ to maximize the restart rate \cite{surjanovic2022VPT}. 
The restart rate and the GCB also appear to be related to the \emph{effective
sample size} of PT samples from the target chain
\cite{syed2021nrpt,surjanovic2022VPT}, although previous conclusions were only
based on empirical results with no formal analysis of this relationship. One aim
of our work is to formalize such relationships.

\section{Marginal ergodicity and a PT central limit theorem} 
\label{sec:marginal_tv}

In this section, we introduce the notion of marginal uniform ergodicity for
PT and show that it guarantees that the target states, $X^N_t$, satisfy a
central limit theorem.

\subsection{Uniform ergodicity}

Let $\bar X_t$ be a $\bar\pi$-invariant PT Markov chain with initial
distribution $\bar X_0\sim \bar\mu$ at time $t$. We say $\bar{X}_t$ is
\emph{uniformly ergodic} with rate $0 < \rho < 1$ if 
there exists $0 \leq C< \infty$ such that 
for all $t \geq 0$ and any initial distribution $\bar\mu$ we have
\[
  \label{eq:unif_geom_erg}
  \mathrm{TV}(\bar\mu\bar K_t, \bar \pi) \leq C \rho^t.
\]
Here, $\bar\mu\bar K_t(\dee \bar x')=\int_{\bar\mcX} \bar \mu(\dee \bar
x)\bar K_t(\bar x,\dee x')$ is the law of $\bar X_t$ at time $t$, and
$\TV(\bar\mu,\bar\mu')$ is the \emph{total variation distance} between
probability measures $\bar{\mu},\bar{\mu}'$ on the expanded state space
$\bar\mcX$, 
\[
  \mathrm{TV}(\bar\mu,\bar\mu') 
  = \frac{1}{2} \sup_{\bar f \in \bar \mcF} \abs{\bar \mu(\bar f) - \bar\mu'(\bar f)}.
\]
In the expression above, $\bar \mcF$ denotes the set of measurable functions $\bar f:\bar \mcX \to [-1,1]$, 
and $\bar \mu(\bar f)$ is the expectation of $\bar f$ with respect to $\bar \mu$.
The uniform ergodicity condition then guarantees that draws from the chain 
satisfy a central limit theorem (CLT) for a broad class of functions of the 
states \cite[Chapter 21]{douc2018markov}.

Uniform ergodicity is also classically used to analyse the mixing properties of MCMC
algorithms. 
The \emph{mixing time}, $t_\text{mix}(\eps)$, is the time required for $\bar X_t$ to approximate a
sample from $\bar \pi$ within a tolerance of $\epsilon > 0$:  
\[
  t_\text{mix}(\eps) 
  = \min \cbra{t \geq 0 : \sup_{\bar\mu} \mathrm{TV}( \bar\mu \bar K_t, \bar \pi) \leq \eps}.
\]
Notably, uniform ergodicity immediately implies that
for all $\eps>0$, $t_\text{mix}(\eps)$ is bounded, with
\[
  t_\text{mix}(\eps) \leq \max \cbra{0, \frac{\log \eps - \log C}{\log \rho}}.
\]

\subsection{Marginal uniform ergodicity}
Uniform ergodicity is a strong condition since convergence in total variation
requires all $N+1$ components of the PT chain to progress towards the stationary
distribution simultaneously. 
In the context of PT, we are usually not interested in the entire $(N+1)$-component 
PT chain $\bar{X}\in\bar\mcX$, but rather only in the last marginal state
$X^N_t\in \mcX$ corresponding to the target samples. Since the marginal states
$X^N_t$ are not generally Markovian, to analyse their convergence we need to weaken the
uniform ergodicity condition and quantify the distance between measures 
in terms of the total variation distance of one of their marginals. 

We define the \emph{marginal total variation distance}, 
\[
  \label{eq:TV_N}
  \mathrm{TV}_N(\bar{\mu}, \bar{\mu}') 
  = \frac{1}{2} \sup_{\bar f \in \bar \mcF_N} \abs{\bar{\mu}(\bar f) - \bar{\mu}'(\bar f)},  
\]
where $\bar\mcF_N$ denotes the set of measurable functions 
$\bar f:\bar \mcX \to [-1,1]$ 
such that $\bar f(\bar x)=f(x^N)$ for some measurable $f:\mcX\mapsto\reals$. 

Equipped with the marginal TV distance, we say the PT chain is
\emph{marginally uniformly ergodic} with rate $0 < \rho < 1$ if 
there exists $0 \leq C < \infty$ such that
for all $t \geq 0$ and any initial distribution $\bar \mu$ we have 
\[
  \label{eq:marginal_unif_geom_erg}
  \mathrm{TV}_N(\bar\mu \bar K_t, \bar\pi) \leq  C \rho^t.
\] 
Note that marginal uniform ergodicity is weaker than assuming uniform ergodicity
since $\TV_N(\bar\mu, \bar\mu')\leq \TV(\bar\mu, \bar\mu')$. 

The \emph{marginal mixing time}, $t^N_\text{mix}(\eps)$, is the time required 
for $X^N_t$ to approximate a sample from the target $\pi_1$ within a tolerance of $\epsilon>0$:
\[
  t^N_\text{mix}(\eps) 
  = \min \cbra{t \geq 0 : \sup_{\bar\mu} \mathrm{TV}_N( \bar\mu \bar K_t, \bar \pi) \leq \eps}.
\]
For all $\eps>0$, marginal uniform ergodicity ensures that $t^N_\text{mix}(\eps)$ is bounded, with
\[
  t^N_\text{mix}(\eps) \leq \max \cbra{0, \frac{\log \eps - \log C}{\log \rho}}.
\]

\cref{lem:CLT}, which is based on a modification of CLT results for uniformly ergodic Markov chains
\cite[Chapter 21]{douc2018markov},
shows that marginal uniform ergodicity of PT can imply that functions of the
target states $X^N_t$ satisfy a CLT even if the marginal is not Markovian.
Following the discussion in \cref{sec:PT}, let $\tilde{X}_t=(\bar X_t,S_t)$ be 
the time-homogeneous PT Markov chain with initial condition 
$(\bar{X}_0, S_\text{even})$ and stationary distribution 
$\tilde{\pi}=\bar{\pi} \times \mathrm{Uniform}(\{S_\text{even},S_\text{odd}\})$.
The definitions of marginal uniform ergodicity and the marginal mixing time are 
defined analogously for the time-homogenous PT Markov chain $\tilde X_t$ on the 
extended space $\bar \mcX \times \cbra{S_\text{even}, S_\text{odd}}$.

\bthm
\label{lem:CLT}
Suppose the PT chain $\tilde{X}_t$ is a marginally uniformly ergodic Markov chain
that admits a unique invariant probability measure $\tilde\pi$ with $\tilde{X}_0 \sim \tilde \pi$.
If $f:\mcX\mapsto\reals$ is a measurable function such that there exists a $\delta > 0$ with
$\pi_1(|f|^{2+\delta})<\infty$ and $\pi_1(f)=0$, then
\[
  \label{eq:CLT_thm}
  \frac{1}{\sqrt{T}} \sum_{t=0}^T f(X_t^N) 
  \xRightarrow{T \to \infty} \distNorm(0, \sigma^2(f)),
\]
where the asymptotic variance $\sigma^2(f)$ is given by 
\[
  \sigma^2(f) = \EE[f(X_0^N)^2] + 2 \sum_{t=1}^\infty \EE[f(X_0^N)f(X_t^N)].
\]
\ethm 

We remark that for both NRPT and RPT, the major assumption in \cref{lem:CLT} is  
that of marginal uniform ergodicity, which we establish in the following sections. 
For the remainder of our theoretical analysis in
\cref{sec:erg_infinite,sec:erg_finite}, we establish the marginal uniform ergodicity of PT 
for both a finite number of chains and as $N \to \infty$, under an assumption of efficient 
local exploration.

\section{Uniform ergodicity of parallel tempering: infinite number of chains}
\label{sec:erg_infinite}

In this section, we analyze the uniform ergodicity of PT in the large-$N$
asymptotic limit. We obtain the following easily interpretable results: 
(1) the rate of uniform ergodicity of NRPT can be characterized via the GCB $\Lambda$ alone; and 
(2) the rate of uniform ergodicity of RPT can be characterized in a problem-independent manner 
with an appropriate scaling of time. 
In particular, assuming efficient local exploration (ELE), 
NRPT is geometrically ergodic with log rate approximately bounded by  
$1/((\Lambda + 2) N)$ (\cref{thm:PDMP_hitting_time}), and 
RPT is geometrically ergodic with log rate approximately bounded by 
$\pi^2/(8N^2)$ (\cref{thm:Brownian_hitting_time}). 
This implies that our bound on the mixing time of NRPT scales linearly with $N$ and 
is proportional to $\Lambda$, while our bound on the mixing time of RPT scales 
quadratically with $N$ and is independent of $\Lambda$. 

To establish our large-$N$ uniform ergodicity results, we  derive a bound on the hitting time for a 
special piecewise-deterministic Markov process (PDMP), which can be viewed as 
a continuous-time version of a persistent random walk with two boundaries. 
This result may be of separate interest to 
readers in the mathematical and physics communities; we are not aware of 
such simple closed-form bounds on the cumulative distribution function of the hitting time 
in the existing literature, despite extensive work in the area on Telegrapher's equations 
and persistent/correlated random walks \cite{rossetto2018persistent,masoliver1992solutions,
masoliver1993maximum,foong1994properties,goldstein1951diffusion,renshaw1981correlated,masoliver1993solution}.

\subsection{Notation and assumptions}

We introduce a set of assumptions sufficient for various results on ergodicity. 
For a given finite number of chains in PT, we write $X_t^n$ for the state in 
chain $n$ at time $t$ for $t \geq 0$ 
and $V_* \pi_{\beta_n} (\dee v)$ for the push-forward of $\pi_{\beta_n}$ by $V$, 
i.e., $V_* \pi_{\beta_n}(\dee v) = \pi_{\beta_n}(V^{-1}(\dee v))$.

\bassump
\label{assump:base_erg_collection}
We have 
\begin{enumerate}
    \item \emph{Independent samples from the reference:} $K^{(\beta_0)}(x^0, \cdot) = \pi_0(\cdot)$.
    \label{assump:iid_reference}

    \item \emph{(Strong) efficient local exploration (ELE):} 
    For each $n=0, 1, \ldots, N$ and any initial distribution $\mu$, we have that 
    if $X \sim \mu$ and $X'|X \sim K^{(\beta_n)}(X, \cdot)$, then 
    $V(X')$ is independent of $V(X)$ and $V(X') \sim V_* \pi_{\beta_n}$.
    \label{assump:ELE}
\end{enumerate}
\ebassump

\cref{assump:base_erg_collection}.\ref{assump:iid_reference} 
can be easily satisfied if the user chooses the reference $\pi_0$ so that it is possible to obtain 
\iid samples from the distribution, which is what we strongly recommend in practice.
Generally, prior distributions in the Bayesian setting are straightforward to obtain \iid samples from, 
but may not always perform well if the prior is different from the posterior. 
One can instead use variational PT \cite{surjanovic2022VPT}
to obtain a reference distribution from which \iid sampling is possible and that lies close 
to the target distribution with respect to a given divergence.
\cref{assump:base_erg_collection}.\ref{assump:ELE} is related to 
the ELE assumption introduced in 
\cite{syed2021nrpt}, except that we require the efficient exploration to hold 
for all initial distributions $\mu$ and not just $\mu = \pi_{\beta_n}$ 
(i.e., not just at stationarity). We abuse terminology and use the terminology ELE for 
\cref{assump:base_erg_collection}.\ref{assump:ELE}. 
This assumption allows us to study a \emph{model of PT}, agnostic to the specific choices 
of local explorers in each chain.
In \cref{sec:discussion} we discuss the assumption in greater detail. 
The interpretation of the ELE is that the univariate energy values in a given chain 
are i.i.d., which is in general much weaker than an assumption on the states themselves. 
As a simple example, an explorer that efficiently
obtains samples from one mode of a bimodal distribution but has difficulty crossing 
between modes can satisfy this assumption. 

We also assume that the annealing schedule is chosen to be 
asymptotically optimal with respect to the restart rate of samples from the reference 
\cite[Section 5.1]{syed2021nrpt}. 
In particular, a simple tuning procedure can be defined that approximately 
achieves this \cite{syed2021nrpt}. 
Let $X_{\beta'}, X'_{\beta'} \stackrel{iid}{\sim} \pi_{\beta'}$ and define 
\[
  \label{eq:schedule_generator}
  \Lambda(\beta) = \frac{1}{2} \int_0^\beta \EE\sbra{\abs{V(X_{\beta'}) - V(X'_{\beta'})}} \, \dee \beta', \qquad 
  \Lambda \equiv \Lambda(1).
\]
The schedule obtained with $\beta_n = \Lambda^{-1}(\frac{\Lambda n}{N})$ 
will be close to the optimal choice of temperatures based on equi-acceptance 
communication probabilities for a fixed $N$ \cite{syed2021nrpt}.

\bassump
\label{assump:schedule_generator}
For a given $\pi_0$, $\pi_1$, and $N \geq 1$, the annealing schedule 
$\mcB_N = \rbra{\beta_n}_{n=0}^N$ satisfies $\beta_n = \Lambda^{-1}(\frac{\Lambda n}{N})$.
\ebassump

\subsection{Non-reversible parallel tempering}
\label{sec:erg_NRPT_infinite}

We present uniform ergodicity results for NRPT and relate the rate of 
convergence to the hitting time of a continuous-time process. 
Let $Z_\Lambda(t) = (W_\Lambda(t), \eps_\Lambda(t))$ where $W_\Lambda(t) \in [0,1]$ and 
$\eps_\Lambda(t) \in \{-1, 1\}$.
To distinguish between statements made for NRPT and RPT, we introduce 
two different probability measures, $\Pr_\text{NRPT}$ and $\Pr_\text{RPT}$.
Under $\Pr_\text{NRPT}$, a piecewise-deterministic Markov process (PDMP) 
is initialized at $Z_\Lambda(0) = (0, 1)$ and the process $W_\Lambda(t)$ moves with a 
velocity of $\eps_\Lambda(t)$. 
Sign changes in $\eps_\Lambda(t)$ occur with reflections at the boundaries and otherwise occur at a
constant rate of $\Lambda \geq 0$ according to a Poisson process. 
For a formal definition of the infinitesimal generator of $Z_\Lambda(t)$ we refer readers 
to \cite{syed2021nrpt}.
With $Z_\Lambda(t)$ defined, under $\Pr_\text{NRPT}$ we introduce the 
hitting time
\[
  \tau_\infty = \inf \cbra{t \geq 0 : Z_\Lambda(t) = (1,1)}. 
\]
An illustration of the hitting times of the discretized index process 
is provided in \cref{fig:hitting_times}. 
We begin by relating the hitting time of the PDMP to ergodicity of NRPT 
with an infinite number of chains.

\blem
\label{thm:erg_NRPT}
Suppose \cref{assump:third_moment,assump:base_erg_collection,assump:schedule_generator} hold for all $N$.
Let $\bar{\pi}$ be the joint target given by 
\cref{eq:joint_target} for $N+1$ chains. Then, for all $t \geq 1$, we have
\[
  \limsup_{N \to \infty} \sup_{\bar \mu} 
  \mathrm{TV}_N(\bar{\mu} \bar{K}_{tN}^\text{NRPT}, \bar{\pi}) 
  \leq \Pr_\text{NRPT}(\tau_\infty > t-1),
\]
where the inner supremum is taken over all distributions $\bar \mu$ on $\bar \mcX$. 
\elem

We remark that the preceding result holds for all sequences of initial distributions 
$\bar\mu_1, \bar\mu_2, \ldots$, where each $\bar\mu_N$ is a probability distribution 
over $\mcX^{N+1}$. The efficient local exploration assumption  
gives us this strong result for our exploration-agnostic model of PT.
In practice, we do not expect the initial distribution to vary arbitrarily as the
number of chains increase; it is more usual that the initialization is a product
of $N$ copies of the reference distribution or prior. 

The previous result also establishes that if time is scaled by a factor of $N$, the TV 
distance of marginal PT samples targeting $\pi_1$ is bounded by the survival function of 
a PDMP hitting time.
It is straightforward to establish that the survival function of the hitting time 
$\tau_\infty$ decays at least exponentially with a simple 
dependence on $\Lambda$, establishing uniform ergodicity.
The following is a coarse result that does not require careful analysis to
obtain and technically demonstrates uniform ergodicity, but is too loose for practical purposes.

\bprop 
\label{prop:PDMP_hitting_time_loose}
For any $t,\Lambda \geq 0$, we have 
$\Pr_\text{NRPT}(\tau_\infty > t) \leq \left(1-e^{-2\Lambda}\right)^{\floor{t/2}}$. 
Hence, under the assumptions of \cref{thm:erg_NRPT}, for any $t \geq 1$ we have 
\[
  \limsup_{N \to \infty} \sup_{\bar\mu} 
  \mathrm{TV}_N(\bar{\mu} \bar{K}_{tN}^\text{NRPT}, \bar{\pi}) 
  &\leq \left(1-e^{-2\Lambda}\right)^{\floor{(t-1)/2}}.
\]
\eprop

This bound does not involve an intricate analysis; it can be obtained by means of a 
geometric argument for the PDMP, but is very coarse due to the nature of 
its construction. Based on properties of the PDMP, we know that there is a positive 
probability $e^{-2\Lambda}$ of reaching the target chain within two units of time, irrespective 
of the starting position. This results in $\floor{t/2}$ time blocks, where we assume 
that we initialize at the worst-case starting point at the beginning of each time block.
Because it is calculated using such a worst-case hitting time bound, it is most 
useful for small $\Lambda$ (e.g., $0 \leq \Lambda \leq 1$). 
Note that in practice, for a collection of realistic problems, 
$\Lambda$ lies within the range $[0, 90]$ \cite{syed2021nrpt}. Even for a value of $\Lambda = 10$, the bound 
above results in a rate of about $1 - 2\cdot 10^{-9}$, which is unrealistic 
and too slow for practical purposes.

\begin{figure}[t]
  \centering
  \includegraphics[width=0.5\textwidth]{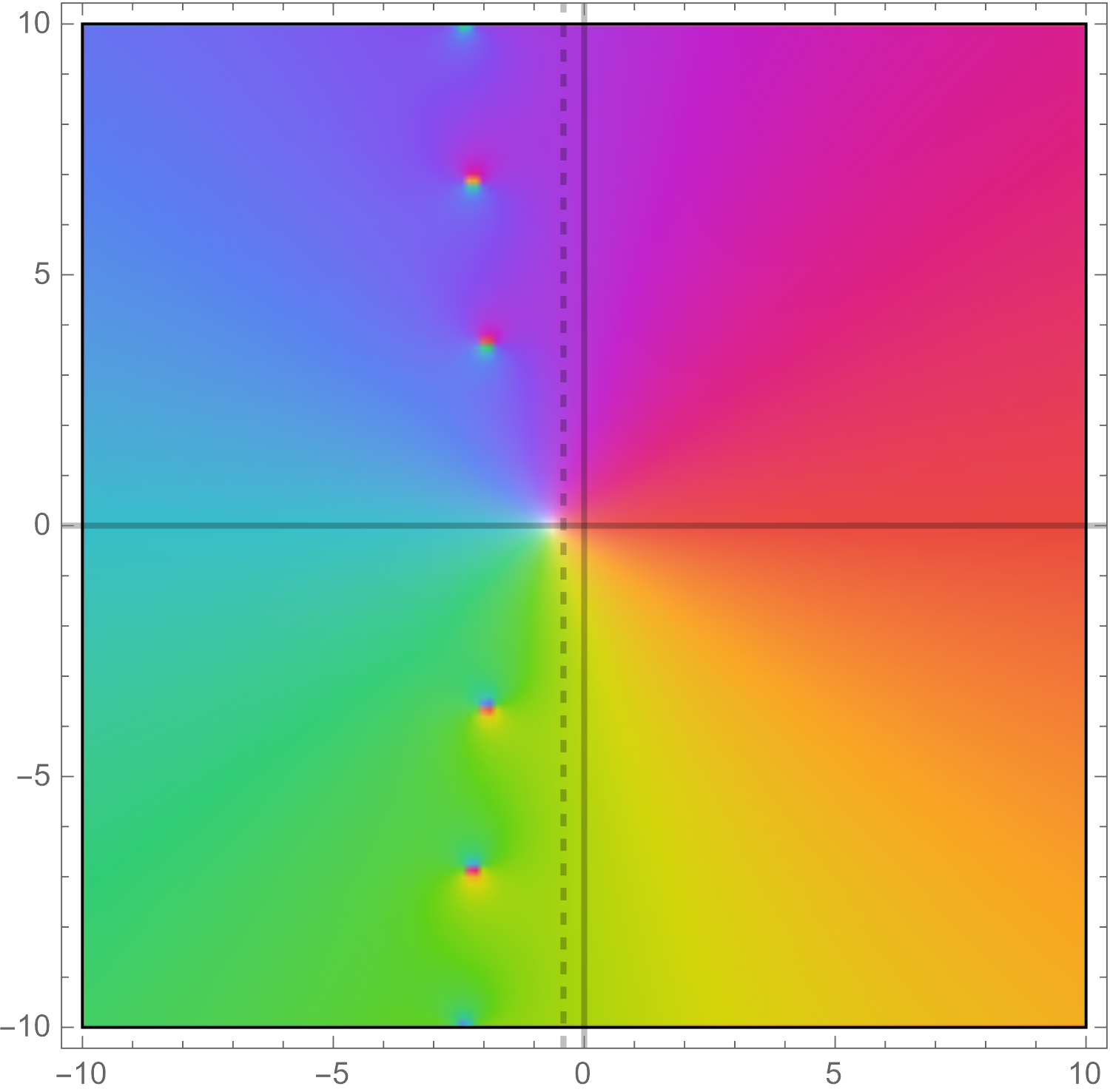}
  \caption{Plot of $F(z)$ with the real part of $z$ on the abscissa and the imaginary part on the 
  	ordinate. Brightness indicates the magnitude of $\abs{F(z)}$ 
    and colour indicates the argument of $F(z)$. 
    The dashed vertical line represents our bound on the real component of the largest pole, 
    whose real value is less than or equal to $-1/(\Lambda+\sqrt{2})$. 
    In this figure, $\Lambda = 1$ and the plotting range is $\cbra{z = x+iy : -10 \leq x, y \leq 10}$.}
  \label{fig:F_plot}
\end{figure}

To tighten the preceding bound, we perform a more careful analysis of the PDMP hitting time 
by bounding the locations of poles of the Laplace transform of $t \mapsto \Pr_\text{NRPT}(\tau_\infty > t+1)$. 
For all $z \in \comps$ such that the following integral is defined and finite, 
the Laplace transform is
\[
  F(z) = \int_0^\infty \Pr_\text{NRPT}(\tau_\infty > t+1) e^{-zt} \, \dee t.
\]
$F$ is defined for at least $\text{Re}(z) \geq 0$, 
as $\int_0^\infty \Pr_\text{NRPT}(\tau_\infty > t+1) \, \dee t \leq \EE_\text{NRPT}[\tau_\infty] < \infty$ 
\cite[Thm. 1]{syed2021nrpt}.
The locations of poles of $F$, as shown in \cref{fig:F_plot}, dictate the rate of decay of 
$t \mapsto \Pr_\text{NRPT}(\tau_\infty > t)$: 
the more negative the real value of the right-most pole, the faster the rate of convergence of NRPT.
For $\Lambda \geq 1$, on at least 
$\cbra{z \in \comps : z \neq 0, \, \text{Re}(z) > -1/(\Lambda + \sqrt{2})}$ we have that
an analytic continuation of $F$ is given by 
\[
  \begin{aligned}
  \label{eq:Laplace_transform}
  F(z) &= \frac{D(z) - e^z}{z \cdot D(z)}, \quad
  D(z) = \cosh(r(z)) + \frac{z}{r(z)} \sinh(r(z)), \label{eq:D} \quad
  r(z) = \sqrt{(z+\Lambda)^2 - \Lambda^2},
  \end{aligned}
\]
where $\sqrt{z}$ is the principal square root of $z$.
By placing bounds on the Bromwich integral (Laplace inversion), we obtain our convergence rate for NRPT.

\bthm
\label{thm:PDMP_hitting_time}
For $\Lambda \geq 1$ and any $t > 1$,
\[
  \label{eq:PDMP_hitting_time_1}
  \Pr_\text{NRPT}(\tau_\infty > t) \leq C(\Lambda) \cdot \exp\rbra{ \frac{-(t-1)}{\Lambda+2} },
\]
with
\[
  \label{eq:C_Lambda} 
  C(\Lambda) 
  &= \sup_{t \geq 0} \lim_{R \to \infty} 
    \frac{1}{\pi} \int_0^R \text{Re}\rbra{e^{ixt} F\rbra{\frac{-1}{\Lambda+2} + ix}} \, \dee x \\
  &\leq 106.
\]
Under the additional assumptions of \cref{thm:erg_NRPT}, for any $t > 2$,
\[
  \limsup_{N \to \infty} \sup_{\bar\mu} \mathrm{TV}_N(\bar{\mu} \bar{K}_{tN}^\text{NRPT}, \bar{\pi}) 
    &\leq C(\Lambda) \cdot \exp\rbra{ \frac{-(t-2)}{\Lambda+2} }.
\]
\ethm

We make some comments on \cref{thm:PDMP_hitting_time}.
First, the results suggest that for NRPT, $t_\text{mix}^N(\eps) = O(N \Lambda \log(1/\eps))$.
Second, note that we establish $C(\Lambda) = O(1)$.
The constant $C(\Lambda)$ should be less than 1.2 based on experiments presented in \cref{sec:experiments}.
We also remark that we expect the log rate $1/(\Lambda+2)$ to be tight for large values of $\Lambda$; 
simulations in \cref{sec:experiments} confirm that as $\Lambda \to \infty$ 
the appropriate log rate is around $1/\Lambda$.
Finally, the above result presents a substantial improvement compared to \cref{prop:PDMP_hitting_time_loose}. 
For instance, considering again $\Lambda = 10$, we see that the rate in this improved  
bound is now approximately $0.920$, a considerable improvement compared 
to the rate of $1 - 2\cdot 10^{-9}$ coming from the loose bound in  \cref{prop:PDMP_hitting_time_loose}.

\subsection{Reversible parallel tempering}
\label{sec:erg_reversible_infinite}

We present uniform ergodicity results for RPT
and relate the rate of convergence to a hitting time of 
Brownian motion with reflections. 
Under $\Pr_\text{RPT}$ let $W(t)$ denote Brownian motion with reflective 
boundaries on $[0,1]$ for $t \geq 0$ and initialized at $W(0) = 0$, unless stated otherwise. 
Under $\Pr_\text{RPT}$, set
\[
  \tau_\infty = \inf \cbra{t \geq 0 : W(t) = 1}.
\]

\blem
\label{thm:erg_reversible}
Suppose \cref{assump:third_moment,assump:base_erg_collection,assump:schedule_generator} hold for all $N$.
Let $\bar{\pi}$ be the joint target given by 
\cref{eq:joint_target} for $N+1$ chains. Then, for all $t \geq 1$ we have
\[
  \limsup_{N \to \infty} \sup_{\bar\mu}
  \mathrm{TV}_N(\bar{\mu} \bar{K}_{tN^2}^\text{RPT}, \bar{\pi}) 
  \leq \Pr_\text{RPT}(\tau_\infty > t - 1).
\]
\elem

We draw the reader's attention to the $O(N^2)$ scaling of time in \cref{thm:erg_reversible} for RPT 
in contrast to the $O(N)$ scaling for NRPT. 
A bound on the hitting time of Brownian motion with a reflection is presented below, 
which is independent of $\Lambda$. 
Therefore, as more chains are added to RPT, its performance in terms of ergodicity 
is asymptotically independent of the given $\pi_0$ and $\pi_1$. However, this 
comes at the cost of requiring $O(N^2)$ iterations to reach stationarity 
compared to $O(\Lambda N)$ for NRPT.

\bthm 
\label{thm:Brownian_hitting_time}
For $t \geq 0$, we have 
\[ 
  \Pr_\text{RPT}(\tau_\infty > t) 
  = \sum_{k=1}^\infty \frac{2}{k \pi} (1-(-1)^k) \sin\rbra{\frac{k \pi}{2} }
    \exp\rbra{-\frac{k^2 \pi^2 t}{8}}.
\] 
For $t \geq 1$ we conclude that 
\[ 
  \Pr_\text{RPT}(\tau_\infty > t) \leq 2 \exp\rbra{-\frac{\pi^2 t}{8}}.
\]
Hence, under the assumptions of \cref{thm:erg_reversible}, for all $t \geq 2$, 
\[
  \limsup_{N \to \infty} \sup_{\bar\mu} \mathrm{TV}_N(\bar{\mu} \bar{K}_{tN^2}^\text{RPT}, \bar{\pi}) 
    &\leq 2 \exp\rbra{-\frac{\pi^2 (t-1)}{8}}.
\]
\ethm

\cref{thm:Brownian_hitting_time} suggests that for RPT, $t_\text{mix}^N(\eps) = O(N^2 \log(1/\eps))$.
Comparing the bounds on the rates in \cref{thm:PDMP_hitting_time,thm:Brownian_hitting_time}, we see that 
when $1/((\Lambda+2) N) > \pi^2/(8 N^2)$ we should expect the total variation distance 
bound to converge to zero more quickly for NRPT than it does for RPT. 
This holds when $N \gtrsim \Lambda$, suggesting that NRPT is more robust to 
the selection of too many chains, which is in line with previous results comparing 
NRPT to RPT \cite{syed2021nrpt}.

\section{Uniform ergodicity of parallel tempering: finite number of chains}
\label{sec:erg_finite}

Having performed an asymptotic analysis of PT in \cref{sec:erg_infinite}, which established 
simple expressions for the rate of uniform ergodicity in terms of the GCB, $\Lambda$, 
we now turn to the case used in practice with finite $N$.
In this section we provide results that are less interpretable but apply precisely for 
a given fixed $N$.
We establish that NRPT and RPT with a finite number of chains are uniformly 
ergodic under our exploration-agnostic ELE model of parallel tempering. 
Except for a few special cases, our bounds on the \emph{rates} in the finite-chain case for NRPT 
and RPT do not have simple closed forms.
Instead, we present convenient computational methods to compute total variation 
bounds exactly at any given \emph{fixed time point} 
(\cref{prop:persistent_walk_exact,prop:random_walk_exact}).

\subsection{The index process and coupling strategies}

To formalize our study of the ergodicity of PT with a finite number of chains, we define the \textit{index process},
which captures information about the trajectories of samples initialized at each 
of the $N+1$ chains. Define the index process 
$(\bar{I}_t, \bar{\eps}_t)$ for $t \geq 0$, where 
$\bar{I}_t = (I_t^0, I_t^1, \ldots, I_t^N)$ and 
$\bar{\eps}_t = (\eps_t^0, \eps_t^1, \ldots, \eps_t^N)$. 
The process $I_t^n$, as a function of $t$, is 
the trajectory of the sample initialized at the $n^\text{th}$ chain. For instance, 
the trajectories in \cref{fig:hitting_times} correspond to samples initialized at 
the reference chain. 
The $\eps_t^n$ indicate the proposed direction (up or down) of the trajectory 
for the next time step.
The states are initialized as $\bar{I}_0 = (0, 1, 2, \ldots, N)$ and 
$\bar{\eps}_0 = (1, -1, 1, \ldots, (-1)^N)$.
On occasion, we drop the superscript and write $I_t$ for an arbitrary index process initialized 
at a point specified according to $I_0$. 

\begin{figure}[t]
    \includegraphics[width=0.55\textwidth]{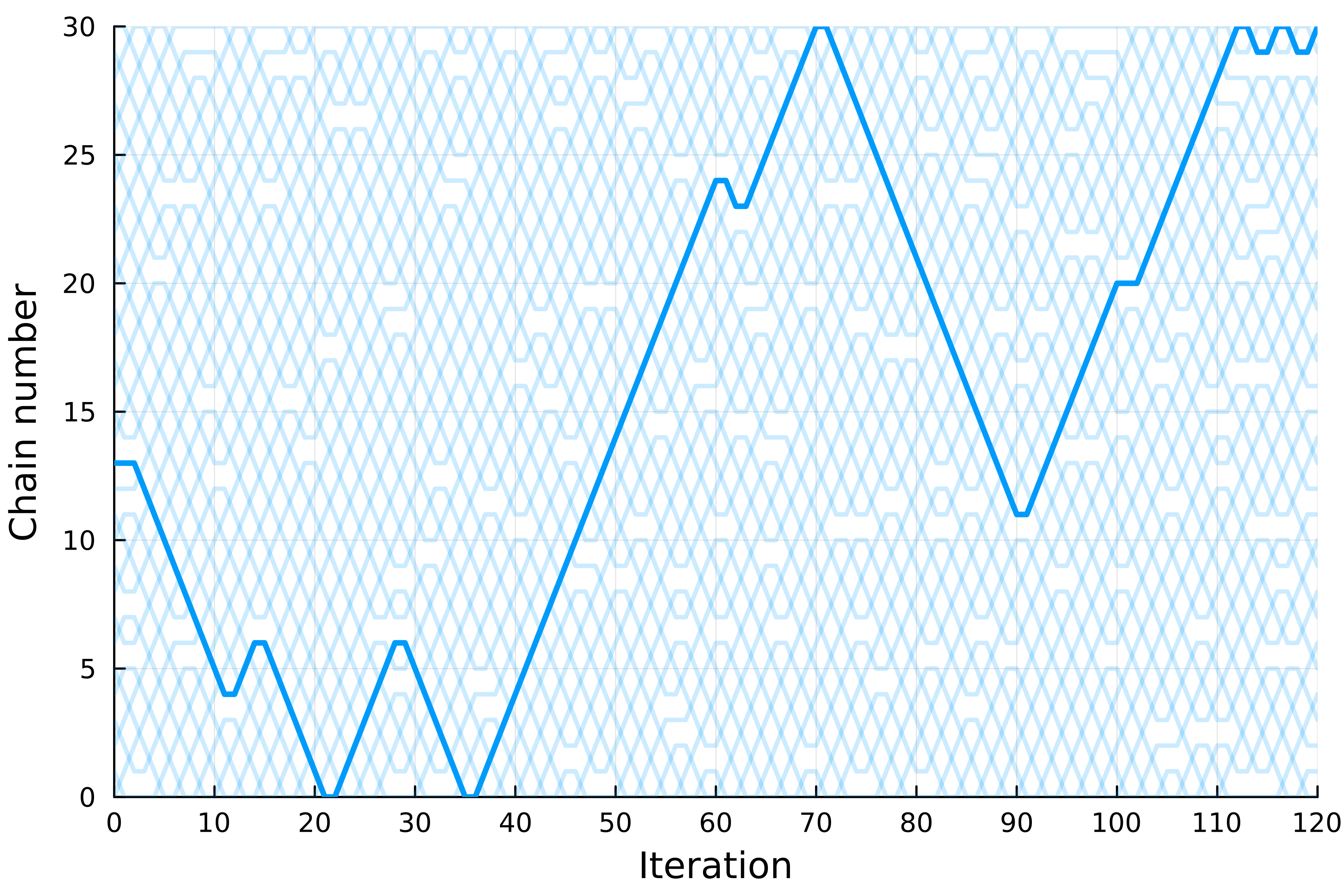}
    \caption{An illustration of one realization of an ancestral process (bold) at time $t=120$
    for $N=30$ along with other index processes. For this realization, the ancestral 
    process hits the reference and the states for the process are coupled after the 
    first time such an event occurs.}
    \label{fig:ancestral_process}
\end{figure}

Our proof of uniform ergodicity for a finite number of chains is based on coupling techniques 
(the infinite-chain results of \cref{sec:erg_infinite} are obtained by taking appropriate limits 
of the following results).
For our coupling we create two instances of PT: one with states initialized 
according to $\bar{\mu}$ and another initialized according to the stationary target $\bar{\pi}$.
Because the reference distribution is selected so that i.i.d.~sampling is possible
(\cref{assump:base_erg_collection}.\ref{assump:iid_reference})
and because the energies are i.i.d.~(\cref{assump:base_erg_collection}.\ref{assump:ELE}), 
we are able to partially couple the two instances of PT
by coupling both the \textit{index processes} and the \textit{samples from the reference chain}.
(Note, however, that we do not couple the states in $\bar\mcX$ other than those that 
had visited the reference chain at some point in the past of their index process.)
The main proof technique we use is to show that the distribution of 
$X_t^N$ at any given time $t$
is a weighted mixture between $\pi_1$ and some other distribution.
Geometric ergodicity is later established by showing that the weight on the latter 
distribution decays geometrically.
Whether or not $X_t^N$ is distributed according to $\pi_1$ is determined by the 
history of an appropriate index process, which we refer to as the \textit{ancestral process}. 
Intuitively, if the path of the index process associated with $X_t^N$ had hit the 
reference at some time $0 \leq s < t$, then $X_t^N$ is distributed according to $\pi_1$
because it can be coupled with the instance of PT initialized at stationarity.
Therefore, the rate of uniform ergodicity is related to the hitting time of a 
(persistent) random walk via the partial coupling.
An example of an ancestral process is illustrated in \cref{fig:ancestral_process}.

Existing simple schedule tuning methods very reliably achieve a near-equal communication acceptance 
rate in practice \cite{syed2021nrpt}, 
and we therefore often assume that for a given value of $N$, an annealing schedule that maximizes 
the restart rate is used. 
For NRPT, the optimal annealing schedule is the one that sets the average communication 
acceptance rate between adjacent chains to be equal \cite[Section 5.1]{syed2021nrpt}. 
The following assumption is the exact finite-chain analogue of 
\cref{assump:schedule_generator}.
Under \cref{assump:equi-acceptance}, the rejection rate is approximately $\Lambda/N$ 
\cite{syed2021nrpt}.

\bassump
\label{assump:equi-acceptance}
The average communication rejection rates between adjacent chains 
from $\pi_0$ to $\pi_1$ are all equal to $r_N$ for some $0 \leq r_N < 1$.
That is, for all $n \in \{0, 1, \ldots, N-1\}$,
\[
  \label{eq:equi-acceptance}
  r_{0,N} = r_{1,N} = \ldots = r_{N-1,N} = r_N.
\]
\ebassump

\subsection{Non-reversible parallel tempering}
The rate of geometric ergodicity for NRPT is determined by the hitting time 
(defined under $\Pr_\text{NRPT}$)
\[
  \tau_N = \inf \cbra{ t \geq 0 : (I_t, \eps_t) = (0,-1) }, 
\] 
where $I_t$ is in this case initialized with $I_0 = (N, -1)$. 
Under $\Pr_\text{NRPT}$ the process $\{I_t\}_{t \geq 0}$ forms a 
persistent random walk with reflections at the boundaries $\cbra{0, N}$.
Further, under equi-acceptance \cref{assump:equi-acceptance}, 
the persistent random walk at time $t$ 
moves in the direction $\eps_t$ with constant probability $1-r_N$, so that 
$I_{t+1} = I_t + \eps_t$ and $\eps_{t+1} = \eps_t$.
If the move in direction $\eps_t$ is rejected (with probability $r_N$), then 
$I_{t+1} = I_t$ and $\eps_{t+1} = -\eps_t$.
Under \cref{assump:equi-acceptance}, the hitting time can also be viewed as the time to hit $(N,1)$ 
starting from $(0,1)$, which is in line with the definition of $\tau_\infty$.  

The following is the finite-dimensional analogue of \cref{thm:erg_NRPT}.

\blem
\label{thm:erg_NRPT_finite}
Under \cref{assump:base_erg_collection},
for any $t, N \geq 1$, 
\[
  \mathrm{TV}_N(\bar{\mu} \bar{K}_t^\text{NRPT}, \bar{\pi}) 
  \leq \Pr_\text{NRPT}(\tau_N > t-1).
\]
\elem

With the hitting time result of \cref{thm:erg_NRPT_finite}, we establish the 
uniform ergodicity of NRPT under ELE for any value of $N$. 
The following is the finite-dimensional analogue of \cref{prop:PDMP_hitting_time_loose} 
and is again a coarse result; it does not require a careful analysis to
obtain and establishes uniform ergodicity, but is too loose for practical purposes.

\bprop
\label{prop:persistent_walk_bound}
Suppose that 
\cref{assump:base_erg_collection,assump:equi-acceptance} hold. For any $t \geq 0$, $N \geq 1$, 
the hitting time $\tau_N$ satisfies
\[
  \Pr_\text{NRPT}(\tau_N > t) 
  \leq [1 - (1-r_N)^{2N}]^{\floor{t/(2N+1)}}.
\]
Hence, for $t \geq 1$ and $N \geq 1$ we have 
\[
  \mathrm{TV}_N(\bar{\mu} \bar{K}_t^\text{NRPT}, \bar{\pi}) 
  \leq [1 - (1-r_N)^{2N}]^{\floor{(t-1)/(2N+1)}}.
\]
\eprop

The previous result is most useful for situations in which $r_N N$ is small; we therefore provide 
an alternative way to calculate $\Pr_\text{NRPT}(\tau_N > t)$ \textit{exactly} for any
$N, t \geq 1$ and $0 \leq r_N < 1$, under the assumption of equi-acceptance. 
Because estimates of $r_N$ may be easily obtained 
from PT, each of these values can be assumed to be known by the practitioner.
The following result may be used to compute bounds 
on the total variation distance for samples from the target chain.

Define the $(2N+2) \times (2N+2)$ NRPT transition matrices
\[
  (A_\text{DEO}(N,r_N))_{ij} = 
  \begin{cases}
    1, & i = j = 1 \\
    r_N, & i = 2k \text{ for some } 1 \leq k \leq N, j=2k-1 \\
    r_N, & i = 2k+1 \text{ for some } 1 \leq k \leq N, j = 2k+2 \\
    1-r_N, & i = 2k \text{ for some } 1 \leq k \leq N, j=2k+2 \\
    1-r_N, & i = 2k+1 \text{ for some } 1 \leq k \leq N, j=2k-1 \\
    1 & i = 2N+2, j = 2N+1 \\
    0 & \text{otherwise}.
  \end{cases}
\]
For instance, 
\[
  A_\text{DEO}(2, r_2) = 
  \begin{pNiceMatrix}[columns-width=auto] 
    1 & 0 & 0 & 0 & 0 & 0 \\
    r_2 & 0 & 0 & 1-r_2 & 0 & 0 \\
    1-r_2 & 0 & 0 & r_2 & 0 & 0 \\
    0 & 0 & r_2 & 0 & 0 & 1-r_2 \\
    0 & 0 & 1-r_2 & 0 & 0 & r_2 \\
    0 & 0 & 0 & 0 & 1 & 0 
  \end{pNiceMatrix}.  
\]

\bthm
\label{prop:persistent_walk_exact}
Fix $t, N \geq 1$ and $0 \leq r_N < 1$. 
Under the assumptions of \cref{prop:persistent_walk_bound}, 
\[
  \label{eq:persistent_walk_exact}
  \Pr_\text{NRPT}(\tau_N > t) 
  = 1 - (0,  \ldots, 0, 1, 0)^\top A_\text{DEO}(N,r_N)^{t} (1, 0, \ldots, 0).
\]
Hence, for $t \geq 2$ and $N \geq 1$ we have 
\[
  \mathrm{TV}_N(\bar{\mu} \bar{K}_t^\text{NRPT}, \bar{\pi}) 
  \leq 1 - (0,  \ldots, 0, 1, 0)^\top A_\text{DEO}(N,r_N)^{t-1} (1, 0, \ldots, 0).
\]
\ethm

We make some comments on \cref{prop:persistent_walk_exact}.
For a given fixed $t$, the exact TV distance bound for NRPT can be obtained by 
taking a matrix exponent of a sparse matrix.
Again, the entries of the matrix can assumed to be known by practitioners from the output 
of a trial run of PT.
The matrix $A_\text{DEO}$ is of size $(2N+2) \times (2N+2)$ to account for the position in 
$\cbra{0,1,\ldots,N}$ and momentum in $\cbra{-1, 1}$ of the index process. 
The second-to-last row of $A_\text{DEO}^t$ corresponds to trajectories 
starting from $(0, 1)$ and the first column corresponds to those ending at 
$(N,1)$.

\subsection{Reversible parallel tempering}
We begin by showing that the rate of uniform ergodicity for RPT 
is determined by the hitting time of a random walk with reflections.  
For an index process initialized at $I_0 = N$, under $\Pr_\text{RPT}$ define 
\[
  \tau_N = \inf \cbra{ t \geq 0 : I_t = 0 }. 
\]
The following results are finite-dimensional analogues of \cref{thm:erg_reversible,thm:Brownian_hitting_time}.

\blem
\label{thm:erg_reversible_finite}
Under \cref{assump:base_erg_collection}, for any $t, N \geq 1$,
\[
  \mathrm{TV}_N(\bar{\mu} \bar{K}_t^\text{RPT}, \bar{\pi}) 
  \leq \Pr_\text{RPT}(\tau_N > t-1).
\]
\elem

To prove that RPT is uniformly ergodic for any $N$, we bound the hitting time 
of a random walk with reflections. 

\bprop
\label{prop:random_walk_bound}
Suppose 
\cref{assump:base_erg_collection,assump:equi-acceptance} hold. For 
$t \geq 0$ and $N \geq 1$, 
\[
  \Pr_\text{RPT}(\tau_N > t) 
  \leq \left[1 - \left(\frac{1-r_N}{2}\right)^N\right]^{\floor{t/N}}.
\]
Hence, for any $t \geq 1$ and $N \geq 1$, we have 
\[
  \mathrm{TV}_N(\bar{\mu} \bar{K}_t^\text{RPT}, \bar{\pi}) 
  \leq \left[1 - \left(\frac{1-r_N}{2}\right)^N\right]^{\floor{(t-1)/N}}.
\]
\eprop 

The bound in \cref{prop:random_walk_bound} is not very tight and we therefore
provide an alternative way to calculate $\Pr_\text{RPT}(\tau_N > t)$ 
\textit{exactly}. For the following result we define 
the $(N+1) \times (N+1)$ matrix $A_\text{SEO}(N, r_N)$ with entries
\[
  (A_\text{SEO}(N, r_N))_{ij} = 
  \begin{cases}
    1, & i=j=1 \\
    r_N, & 2\leq i=j \leq N \\
    \frac{1}{2}(1-r_N), & |i-j| = 1, 2 \leq i \leq N+1 \\
    \frac{1}{2}(1+r_N), & i = j = N+1, \\
    0, & \text{ otherwise}. 
  \end{cases}
\]
For instance, 
\[
  A_\text{SEO}(3, r_3) = 
  \begin{pNiceMatrix}
    1 & 0 & 0 & 0 \\
    \frac{1}{2}(1-r_3) & r_3 & \frac{1}{2}(1-r_3) & 0 \\
    0 & \frac{1}{2}(1-r_3) & r_3 & \frac{1}{2}(1-r_3) \\
    0 & 0 & \frac{1}{2}(1-r_3) & \frac{1}{2}(1+r_3) 
  \end{pNiceMatrix}.
\]

\bthm
\label{prop:random_walk_exact}
Under the assumptions of \cref{prop:random_walk_bound}, 
for $t, N \geq 1$,
\[
  \label{eq:random_walk_exact}
  \Pr_\text{RPT}(\tau_N > t) 
  = 1 - (0,  \ldots, 0, 1)^\top A_\text{SEO}(N,r_N)^t (1, 0, \ldots, 0).
\]
Hence, for any $t \geq 2$ and $N \geq 1$, we have 
\[
  \mathrm{TV}_N(\bar{\mu} \bar{K}_t^\text{RPT}, \bar{\pi}) 
  \leq 1 - (0,  \ldots, 0, 1)^\top A_\text{SEO}(N,r_N)^{t-1} (1, 0, \ldots, 0).
\]
\ethm

\section{Experiments}
\label{sec:experiments}
We conduct several experiments to assess our assumptions and theoretical results. 
We start with a comparison of our theoretical bounds on TV distance with experimental 
results obtained on an Ising model in \cref{sec:ising}.
We then demonstrate how the infinite-chain bounds can be obtained from a limit 
of the finite-chain results in \cref{sec:finite_to_infinite}, along with other 
experiments confirming our theory.  
In \cref{sec:ELE} we consider the ELE model and assess whether it is 
approximately satisfied in practice. 
For our simulations, we use the Julia Pigeons package \cite{surjanovic2023pigeons}, 
which has parallel and distributed implementations of NRPT.

\subsection{Ising model: empirical and theoretical TV bounds}
\label{sec:ising} 

We compare empirical estimates of the total variation distance between samples 
from the target chain and the true target distribution to our theoretical bounds.
To facilitate estimation of the total variation distance, we consider a discrete target,
which is in this case a simplified Ising model on a $4 \times 4$ grid. 
The target distribution is a discrete distribution with all possible parameter 
values supported on a finite set of size $2^{4 \times 4} = 65,536$.
Each one of $4 \times 4 = 16$ parameters takes on a value $X_i \in \cbra{-1,1}$, $i=1,2,\ldots,16$, 
and the probability mass function is given by 
\[
  \pi(x_1, \ldots, x_{16}) \propto \exp\left(\sum_{x_i \sim x_j} x_i x_j\right), 
\]
where $x_i \sim x_j$ for $i \neq j$ if they are connected on the lattice. In our 
example, we embed the lattice on a torus (e.g., the top-left vertex of the grid 
is connected to the top-right and bottom-left vertices). 

We run 100,000 instances of PT with $6$ chains for 25 time steps. 
From a preliminary run, we estimate $\hat{\Lambda} = 2.2$.
The local explorer in this case is a Gibbs sampler that attempts a sign flip at 
each vertex on the $4 \times 4$ grid three times per iteration. 
At every time step for $1 \leq t \leq 24$ we compare the empirical distribution 
of the 100,000 samples from the target chain $X_t^N$ to the known target distribution 
$\pi_1$. 

\begin{figure}[t]
  \begin{subfigure}{0.48\textwidth}
      \centering
      \includegraphics[width=\textwidth]{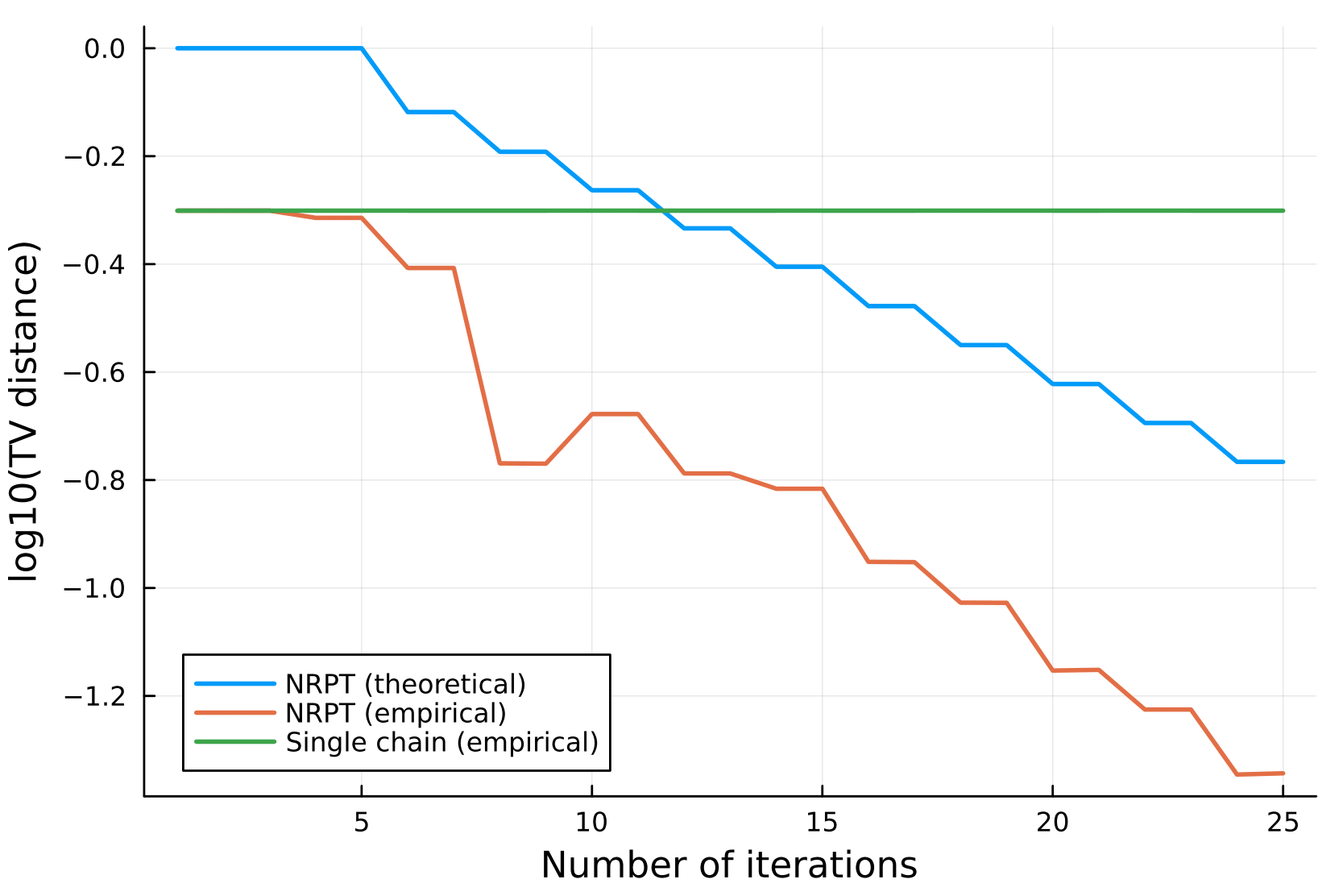}
  \end{subfigure}
  \begin{subfigure}{0.48\textwidth}
      \centering
      \includegraphics[width=\textwidth]{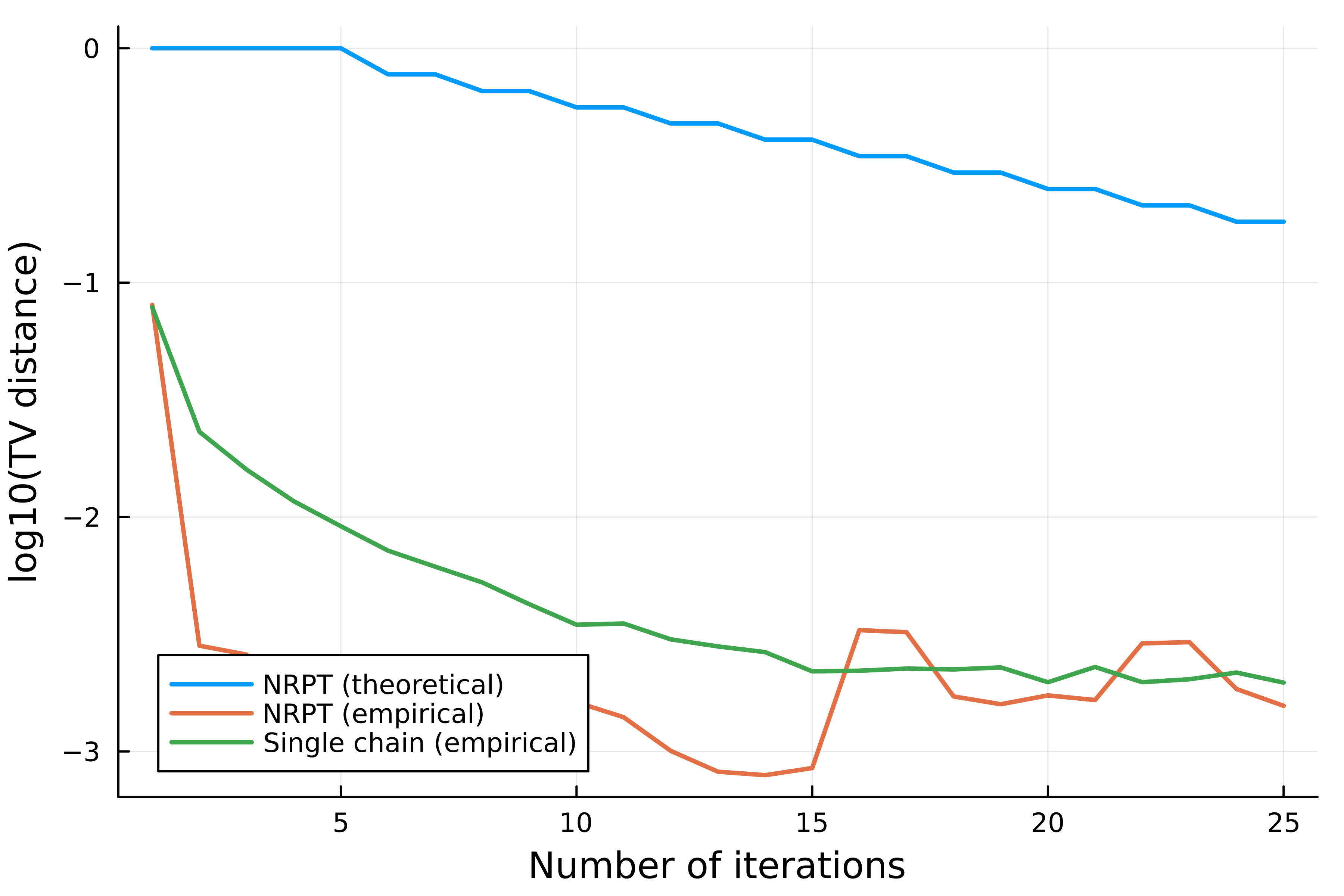}
  \end{subfigure}
  \caption{Theoretical finite-chain TV distance bounds for NRPT compared to empirical TV estimates 
    obtained from PT with several chains.
    \textbf{Left:} PT and the single-chain sampler initalized at the Ising state 
    with all entries equal to $-1$. 
    \textbf{Right:} PT and the single-chain sampler initialized at a random state uniform 
    over $\{-1, 1\}^{4 \times 4}$.}
  \label{fig:TV_distance_empirical}
\end{figure}

The plot of the empirical TV estimates and our bound is displayed in 
\cref{fig:TV_distance_empirical}. 
In the left panel we expect the single-chain Gibbs sampler to struggle because of the initialization 
at $\cbra{-1, -1, \ldots, -1}$, which is a state with very high probability and hence moving 
away from that state with a Gibbs sampler is unlikely. 
Consequently, almost all of the exploration that can be achieved is only through PT 
restarts.
From the left panel we observe that our theoretical bound 
is relatively close to the empirical TV distance in terms of their rates of decay. 
Due to the nature of the Ising model and the chosen initialization, 
it is difficult for the Gibbs sampler alone to explore the entire target distribution. 
In the right panel, we consider a random initialization of both PT and the Gibbs sampler 
from a uniform distribution on $\cbra{-1,1}^{4 \times 4}$. 
In this case, the Gibbs sampler more adequately explores the state space 
and we expect our theoretical bounds on the TV distance to be more conservative than 
the empirical estimates. This is because the TV distance may decrease due to 
the contribution of the local exploration kernels, and not just from PT restarts. 
We note that in the right panel, as the TV distance for PT drops below $10^{-3}$, 
the estimate of TV distance is dominated by noise.

\subsection{Further confirmation of theoretical results} 
\label{sec:finite_to_infinite} 

\begin{figure}[t]
    \centering
    \begin{subfigure}{0.49\textwidth}
      \centering
      \includegraphics[width=\textwidth]{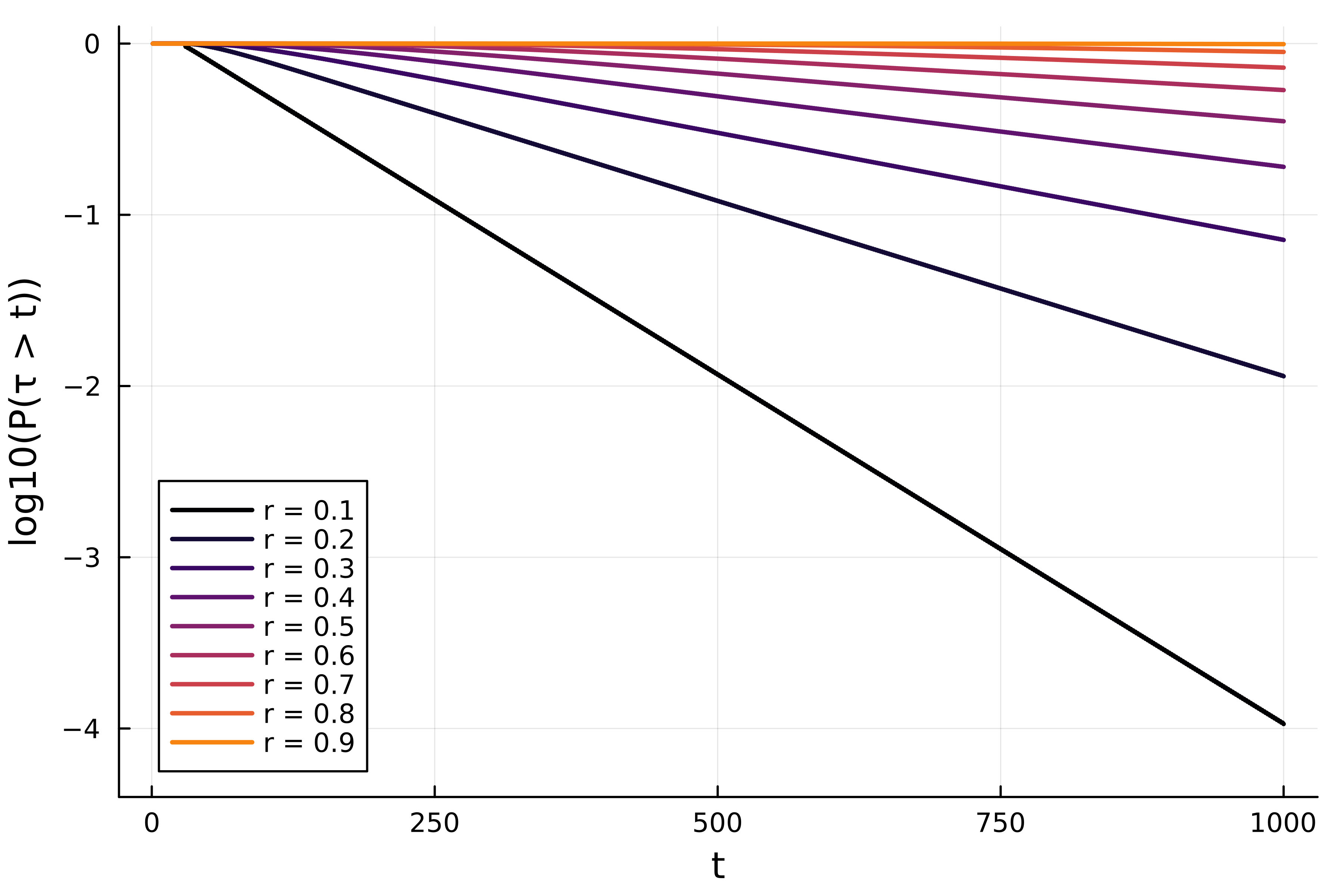}
    \end{subfigure}
    \begin{subfigure}{0.49\textwidth}
      \centering
      \includegraphics[width=\textwidth]{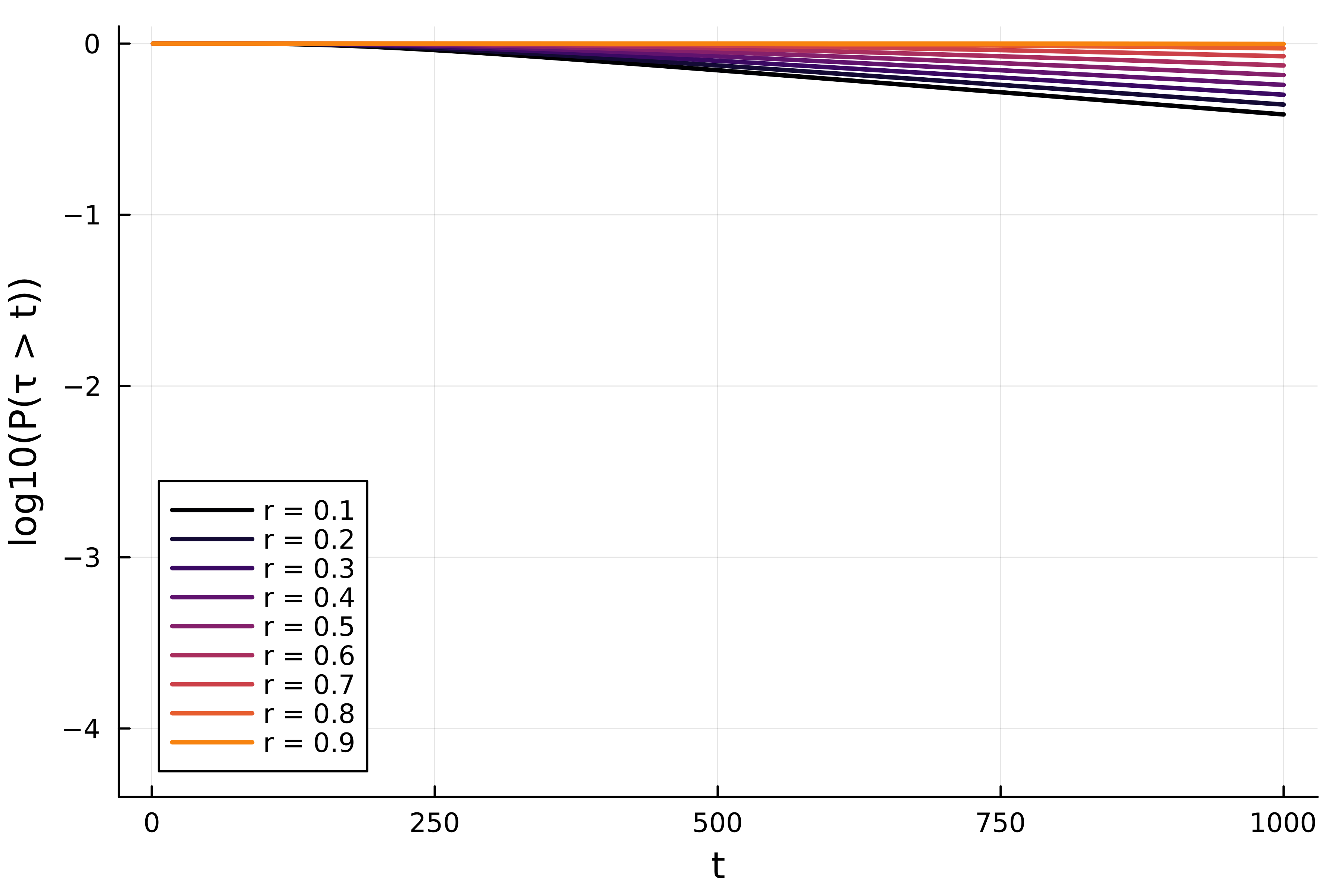}
    \end{subfigure}
    \caption{Survival curves, $\Pr(\tau > t)$, for $N=30$ and
    $r \in \cbra{0.1, 0.2, \ldots, 0.9}$.
    \textbf{Left:} NRPT communication scheme.
    \textbf{Right:} RPT communication scheme.}
    \label{fig:survival_curves}
\end{figure}

We estimate the survival curves of the finite-chain hitting times for given values of 
$0 \leq r < 1$ and $t$ for both NRPT and RPT in \cref{fig:survival_curves}.
In these examples it is clear that NRPT dominates RPT in terms of the time 
it takes for a sample from the reference to reach the target, which has also been 
supported by a previous analysis of the expected hitting time for both 
communication schemes \cite{syed2021nrpt}.

\begin{figure}[t]
  \centering 
  \begin{subfigure}{0.48\textwidth}
      \centering
      \includegraphics[width=\textwidth]{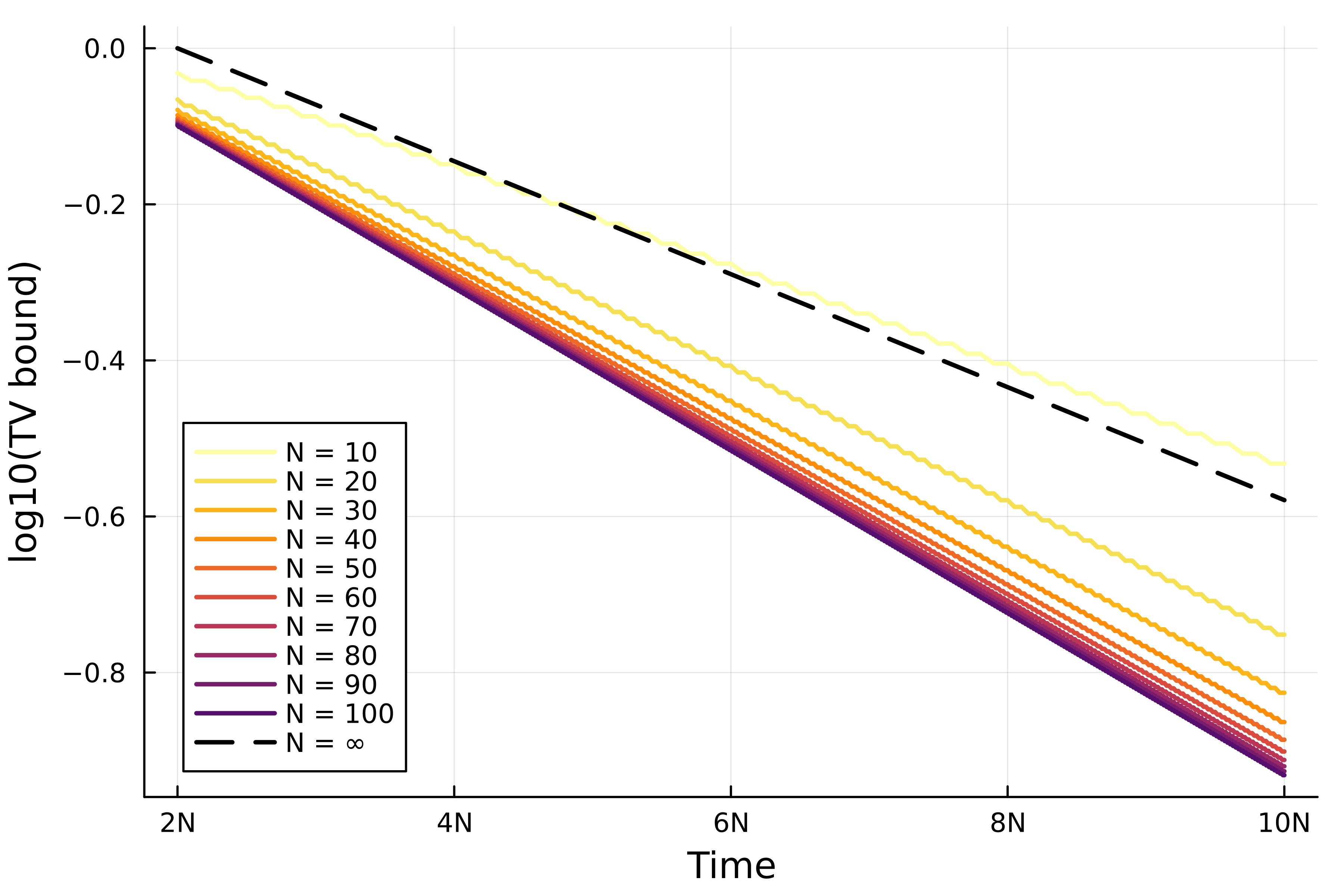}
  \end{subfigure}
  \begin{subfigure}{0.48\textwidth}
      \centering
      \includegraphics[width=\textwidth]{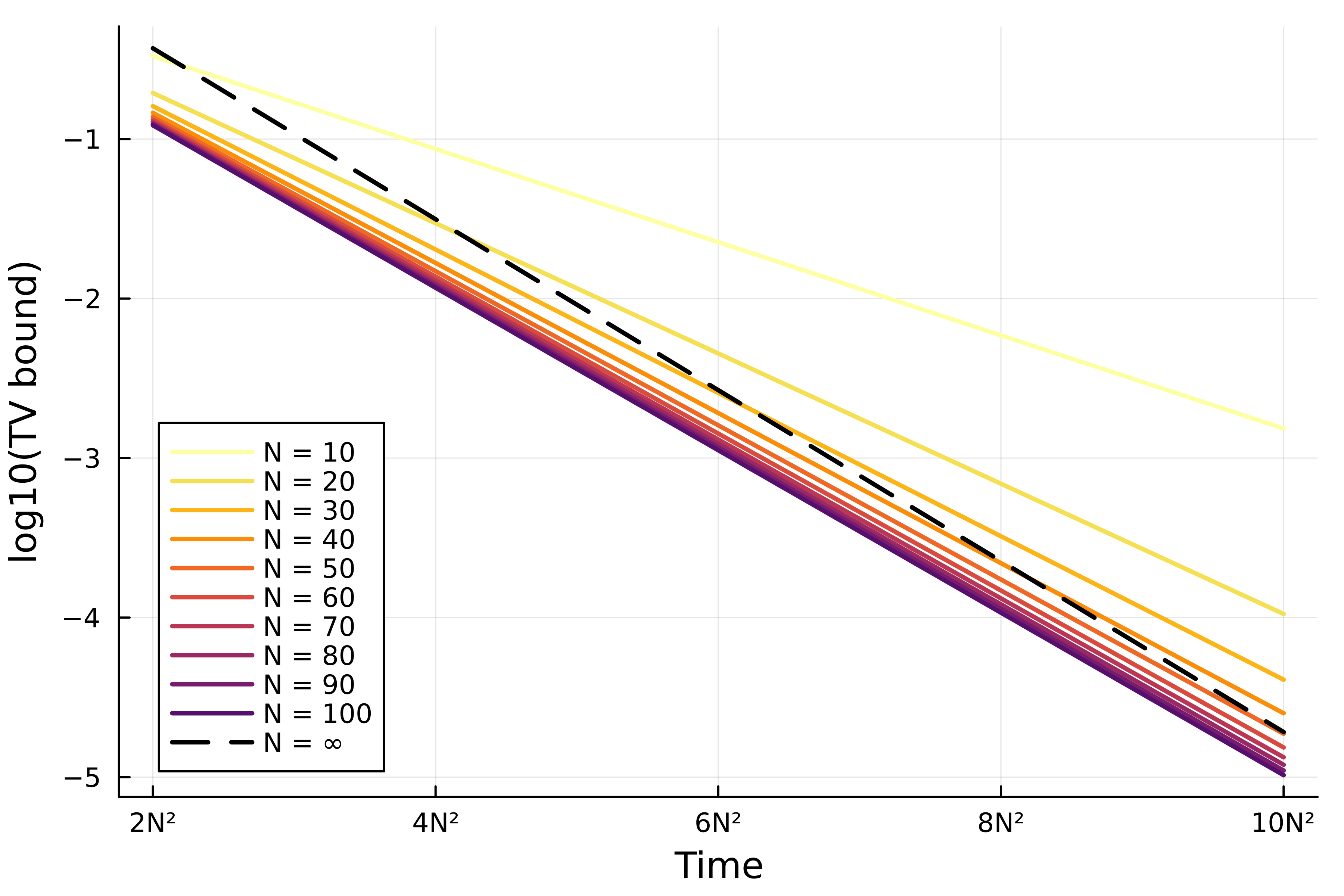}
  \end{subfigure}
  \caption{Finite-chain TV distance bounds for varying $N$ and $\Lambda = 4$, 
    along with corresponding theoretical \emph{rates} as $N \to \infty$.
    Within each figure, $N$ varies in the set $N \in \cbra{10, 20, \ldots, 100}$.
    \emph{The horizontal (time) axis is scaled by a factor of $N$ and $N^2$ for NRPT and RPT, respectively.}
    Several additional plots for different vales of $\Lambda$ are presented in 
    \cref{fig:additional_finite_infinite_chain}.
    \textbf{Left:} NRPT communication scheme. 
    \textbf{Right:} RPT communication scheme.}
  \label{fig:finite_versus_infinite_chain}
\end{figure}

We also compare our finite-chain bounds to the ones obtained as the number of 
chains tends to infinity. We do this by fixing a value of $\Lambda$ (i.e., the 
difficulty of the sampling problem) and setting $r_N = \Lambda/N$ 
(see \cite{syed2021nrpt}). We then increase $N$ and scale time to compare to the 
infinite-chain limit.
In \cref{fig:finite_versus_infinite_chain} we see a 
comparison between the finite- and infinite-chain bounds.  
As $N$ increases, we find that the asymptotic bounds on the \emph{rates} for NRPT and RPT that we 
obtain in \cref{sec:erg_infinite} begin to coincide with the bounds obtained in \cref{sec:erg_finite}.

\begin{figure}[t]
  \centering 
  \begin{subfigure}{0.49\textwidth}
      \centering
      \includegraphics[width=\textwidth]{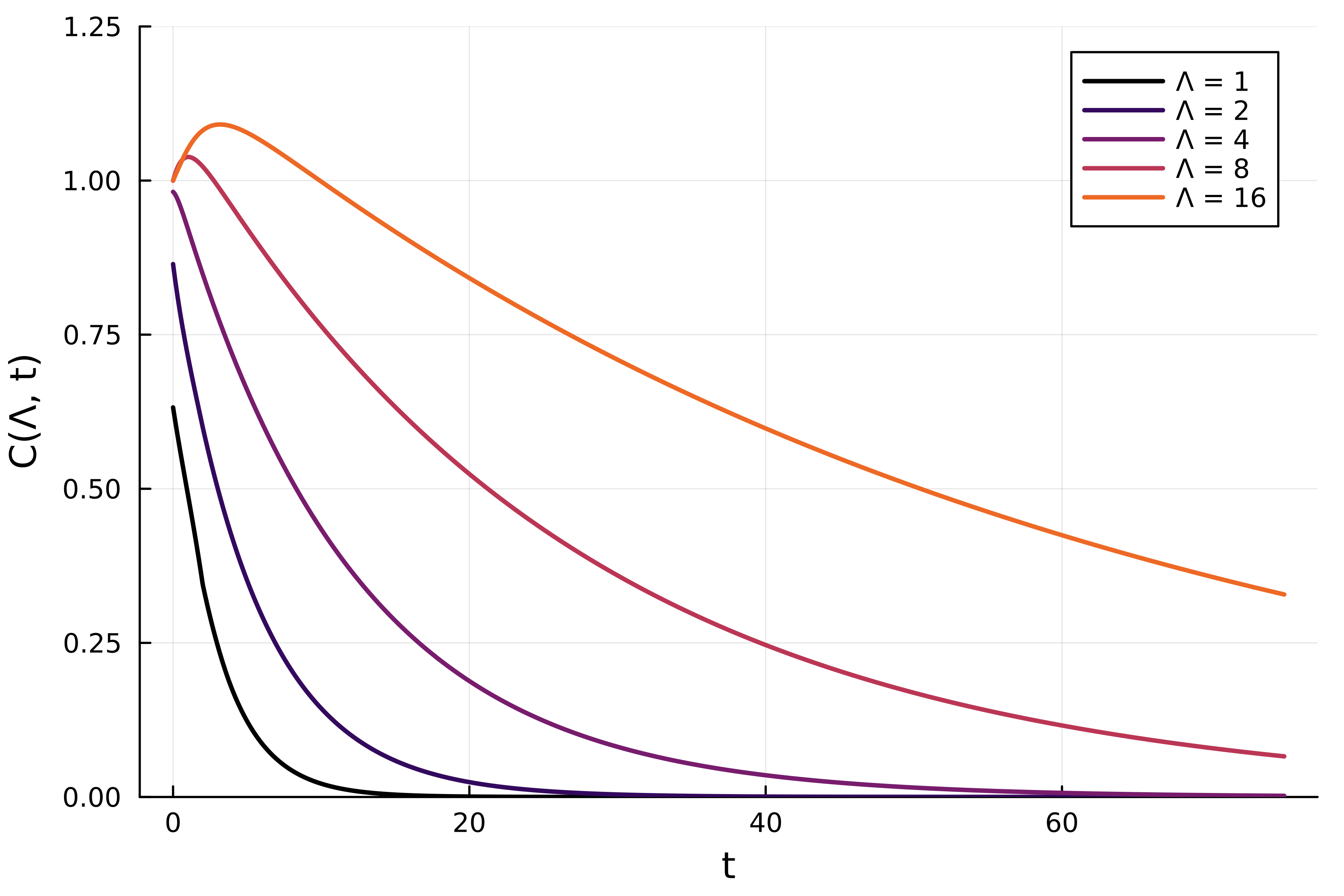}
    \end{subfigure}
    \begin{subfigure}{0.49\textwidth}
      \centering
      \includegraphics[width=\textwidth]{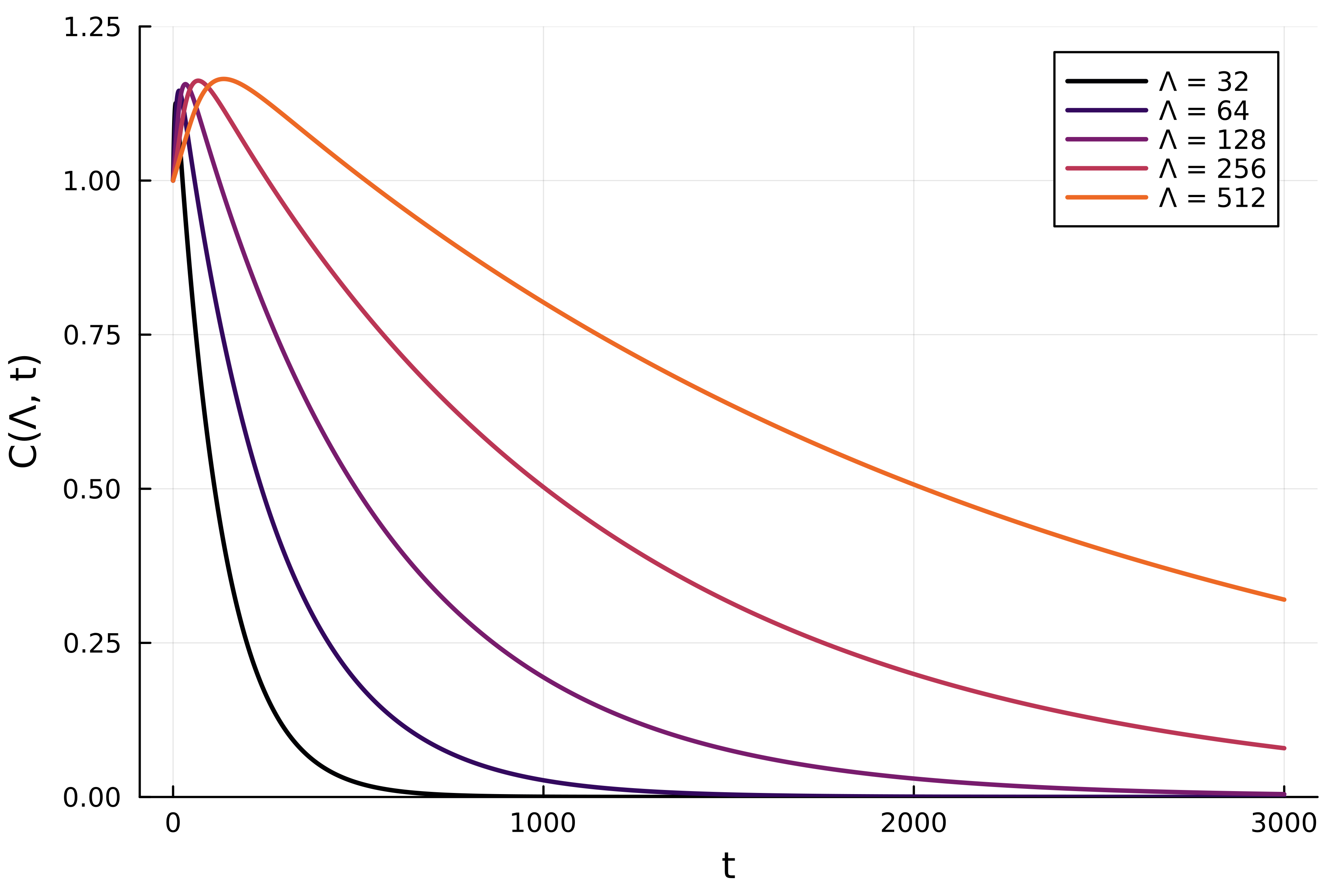}
    \end{subfigure}
  \caption{Curves for $C(\Lambda, t)$ as a function of $t$ at  
  for $\Lambda = 2^j$, $j=0,1,\ldots, 9$. For the range of values considered, 
  $C(\Lambda, t)$ is approximately less than or equal to 1.2 and is decreasing 
  for sufficiently large values of $t$.
  The grid of $t$ values considered consists of 10,000 evenly-spaced points on 
  $[0.001, 75.0]$ (left) and $[0.001, 3000.0]$ (right). 
  \textbf{Left:} $\Lambda \in \cbra{1, 2, 4, 8, 16}$. 
  \textbf{Right:} $\Lambda \in \cbra{32, 64, 128, 256, 512}$.}
  \label{fig:C_Lambda_t}
\end{figure}

Our bound on the total variation distance for samples obtained from NRPT, given 
by \cref{thm:PDMP_hitting_time}, contains a constant $C(\Lambda) = O(1)$.
Note that $C(\Lambda) = \sup_{t \geq 0} C(\Lambda, t)$, where 
\[
 C(\Lambda, t) 
 = \lim_{R \to \infty} \frac{1}{\pi} \int_0^R \text{Re}\rbra{e^{ixt} 
    F\rbra{-\frac{1}{\Lambda+2} + ix}} \, \dee x.
\]
To complement the universal bound on $C(\Lambda)$ provided in \cref{thm:PDMP_hitting_time}, here 
we also compute $C(\Lambda, t)$ numerically for various values of 
$\Lambda$ and $t$.
The results are presented in \cref{fig:C_Lambda_t} and seem to suggest that for a given value of 
$\Lambda$, we have that $t \mapsto C(\Lambda, t)$ is eventually a decreasing function for large values of $t$. 
In \cref{fig:C_Lambda_t} we consider $\Lambda = 2^j$ for $j=0,1,\ldots,9$. 
We present a table of the estimated supremum over the considered values of $t$
and $\Lambda$ in \cref{tab:C_Lambda_t_supremum}.

\begin{table}[t]
  \centering
  \begin{tabular}{r|r|r|r|r|r|r|r|r|r}
  $\Lambda = 1$ & $\Lambda = 2$ & $\Lambda = 4$ & $\Lambda = 8$ & $\Lambda = 16$ & 
    $\Lambda = 32$ & $\Lambda = 64$ & $\Lambda = 128$ & $\Lambda = 256$ & $\Lambda = 512$ \\ 
  \hline
  0.632 & 0.864 & 0.982 & 1.038 & 1.091 & 1.126 & 1.146 & 1.156 & 1.162 & 1.165 \\ 
  \end{tabular}
  \caption{Estimates of $C(\Lambda)$ based on the supremum of $\sup_t C(\Lambda, t)$ 
  over the range of values of $t$ considered in \cref{fig:C_Lambda_t}.}
  \label{tab:C_Lambda_t_supremum}
\end{table}

Finally, in \cref{sec:appendix_experiments} we also provide a collection of plots of $F(z)$ 
to allow for visual inspection of its poles, which determine the rate of convergence for NRPT.
In \cref{fig:F_plot} we provide one of these plots.

\subsection{Efficient local exploration}
\label{sec:ELE}

Our theoretical results on the uniform ergodicity of NRPT and RPT assume that
the ELE model (\cref{assump:base_erg_collection}.\ref{assump:ELE}) holds. 
This assumption cannot be expected to hold exactly in practice and 
we assess to what extent the assumption is satisfied.

We consider simulations on the same models as in Figure 4 of \cite{syed2021nrpt}: 
a Bayesian mixture model, a model involving 
estimation of parameters in an ODE, an Ising model, and a copy number inference problem. 
The target distributions for 
each of the four models are multimodal and we would not 
expect mixing to occur efficiently within the state space $\mcX$ without PT. 

\begin{figure}[t]
  \centering
  \begin{subfigure}{0.245\textwidth}
      \centering
      \includegraphics[width=\textwidth]{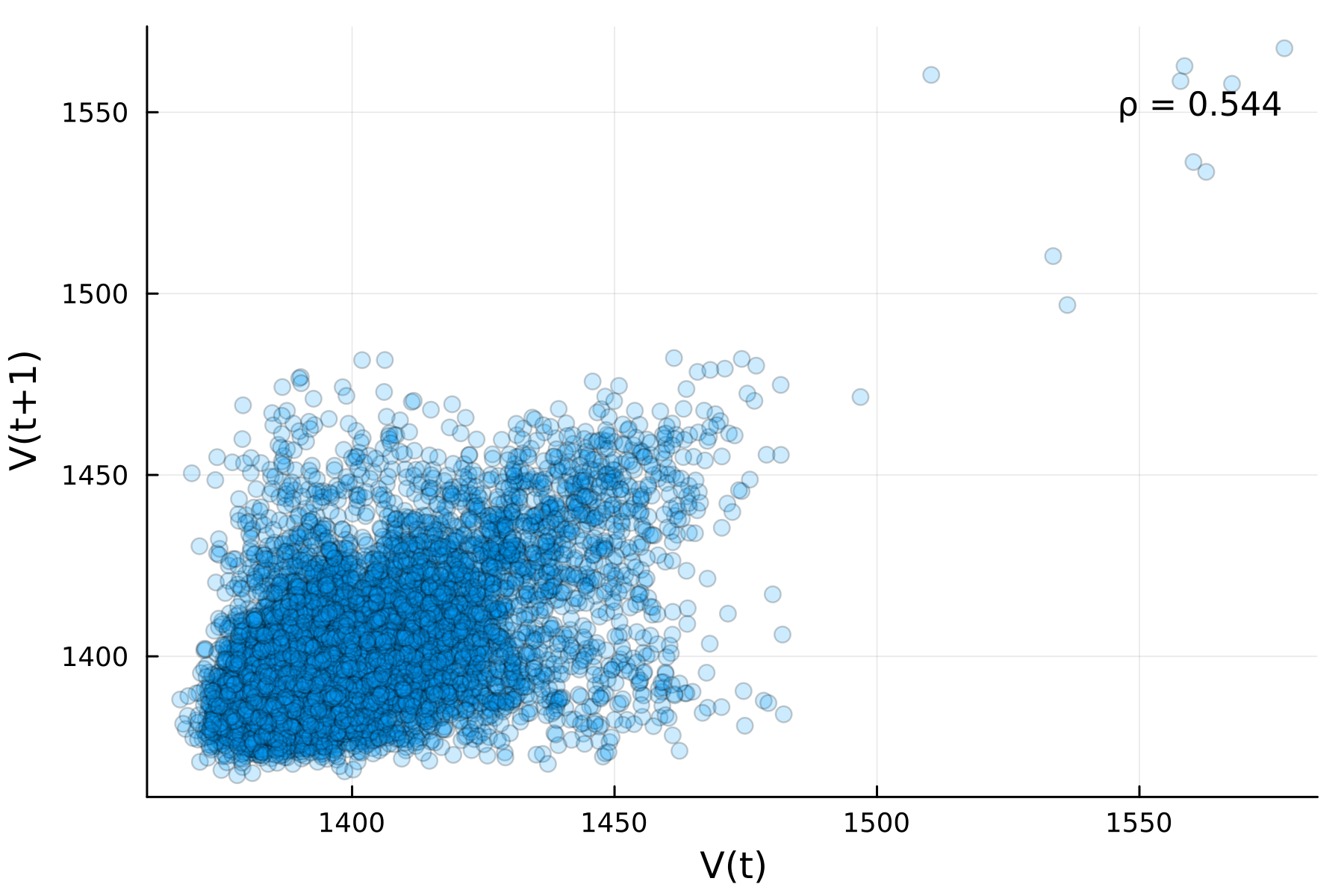}
  \end{subfigure}
  \begin{subfigure}{0.245\textwidth}
      \centering
      \includegraphics[width=\textwidth]{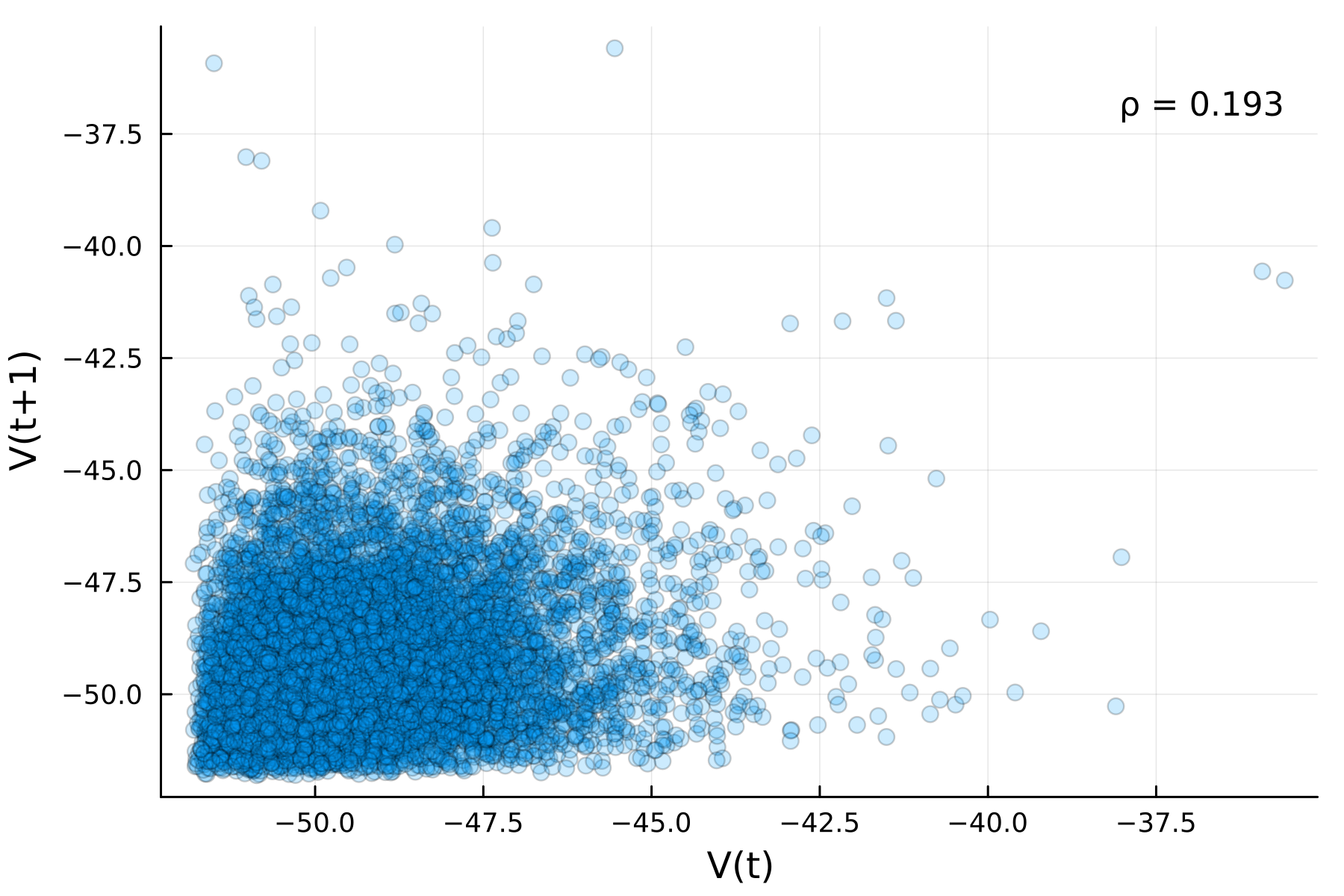}
  \end{subfigure}
  \begin{subfigure}{0.245\textwidth}
      \centering
      \includegraphics[width=\textwidth]{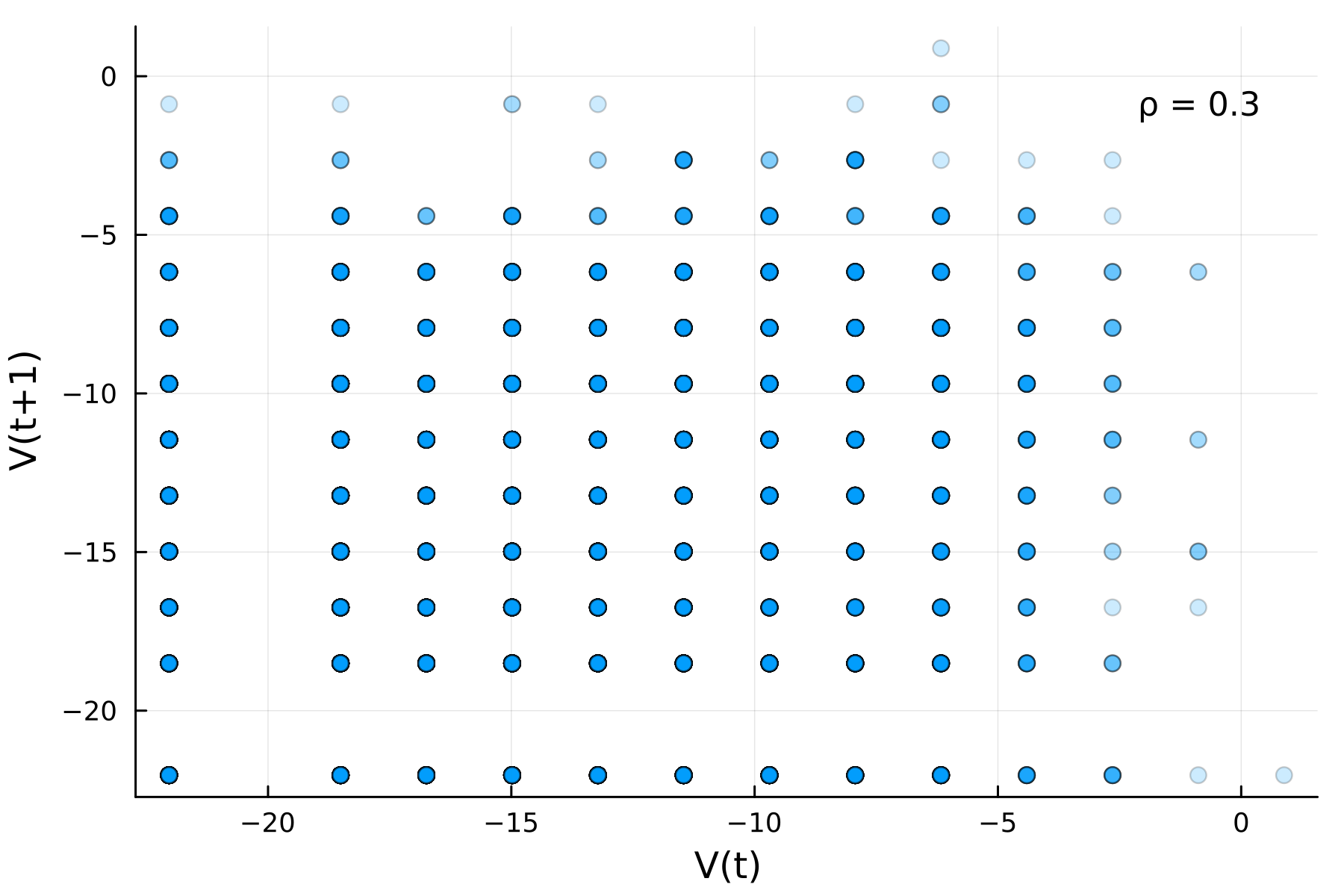}
  \end{subfigure}
  \begin{subfigure}{0.245\textwidth}
      \centering
      \includegraphics[width=\textwidth]{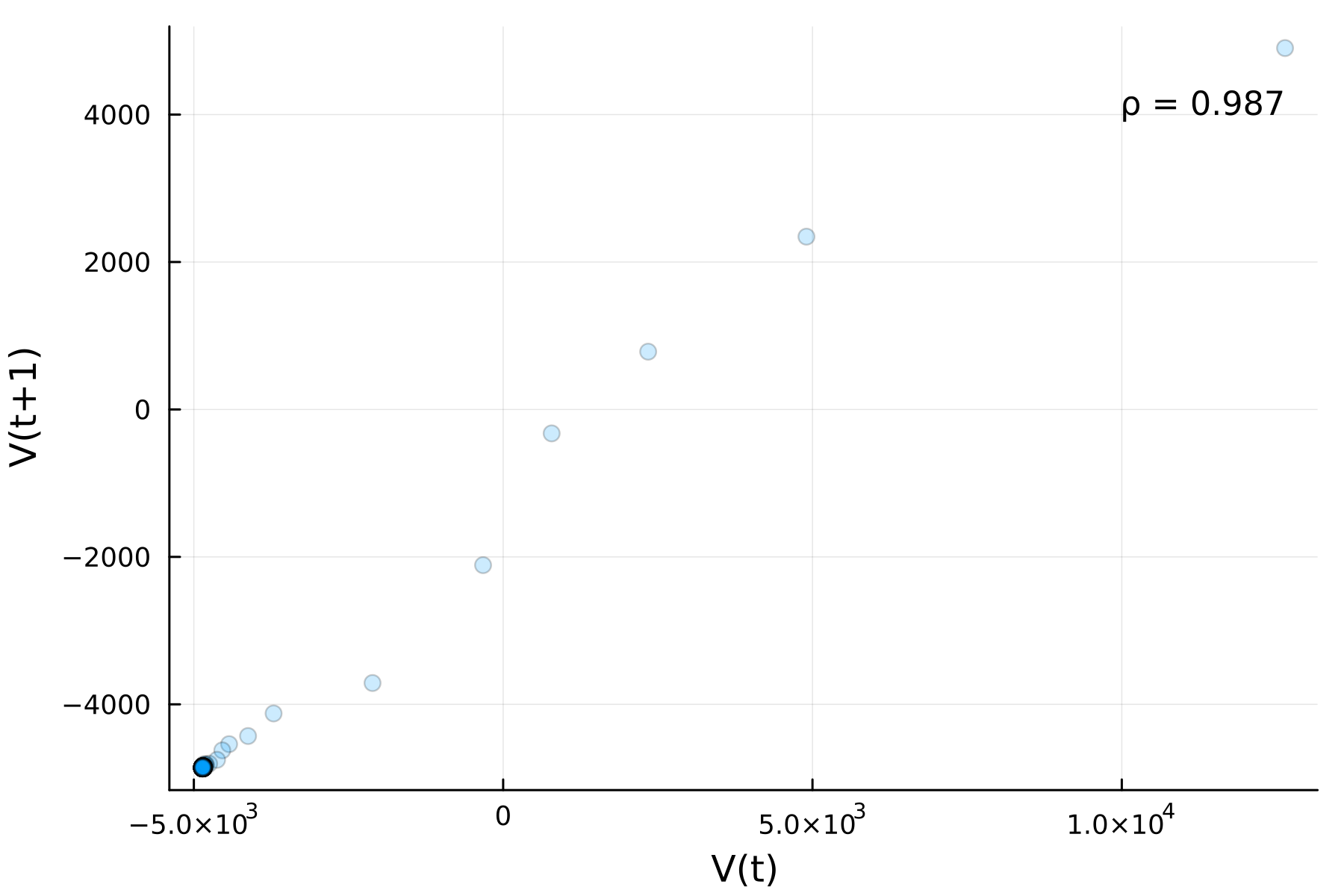}
  \end{subfigure}
  \caption{Pairwise correlation plots for the energy pairs $\{(V_t, V_{t+1})\}_{t=0}^{T-1}$
  from the target chain where $T=10,000$.
  \textbf{Left to right:} Bayesian mixture model, ODE parameter estimation, Ising model, 
  copy number inference. Additional pairwise correlation plots and traceplots of the energies 
  for several chains along the annealing path are presented in \cref{sec:appendix_experiments}.}
  \label{fig:pairwise_correlation}
\end{figure}

We begin by examining whether energy samples in a given chain $n$ are approximately 
independent. In a given chain we obtain the energy values $V_1, \ldots, V_T$ 
up to time $T$, where $V_t = V(X_t)$. 
We present plots of the pairs $\{(V_t, V_{t+1})\}_{t=1}^{T-1}$ and estimate the 
correlation between the pairs, obtaining an estimate of the lag-one autocorrelation
of the energies.
In \cref{fig:pairwise_correlation} we see the pairwise correlation plots for the energies
in the target chain for each of the four models. 
We see that the estimated correlation values for the energy pairs are relatively small in 
magnitude for the first three models, suggesting that the correlation between consecutive energies in the target chain 
is relatively low and the ELE assumption is at least approximately satisfied in the target chain.
This does not hold, however, for the fourth model, showing that is possible for the 
assumption to be seriously violated. Fortunately, this diagnostic can be produced at 
no extra cost from the PT output. 

\begin{figure}[t]
	\centering
		\includegraphics[width=\textwidth]{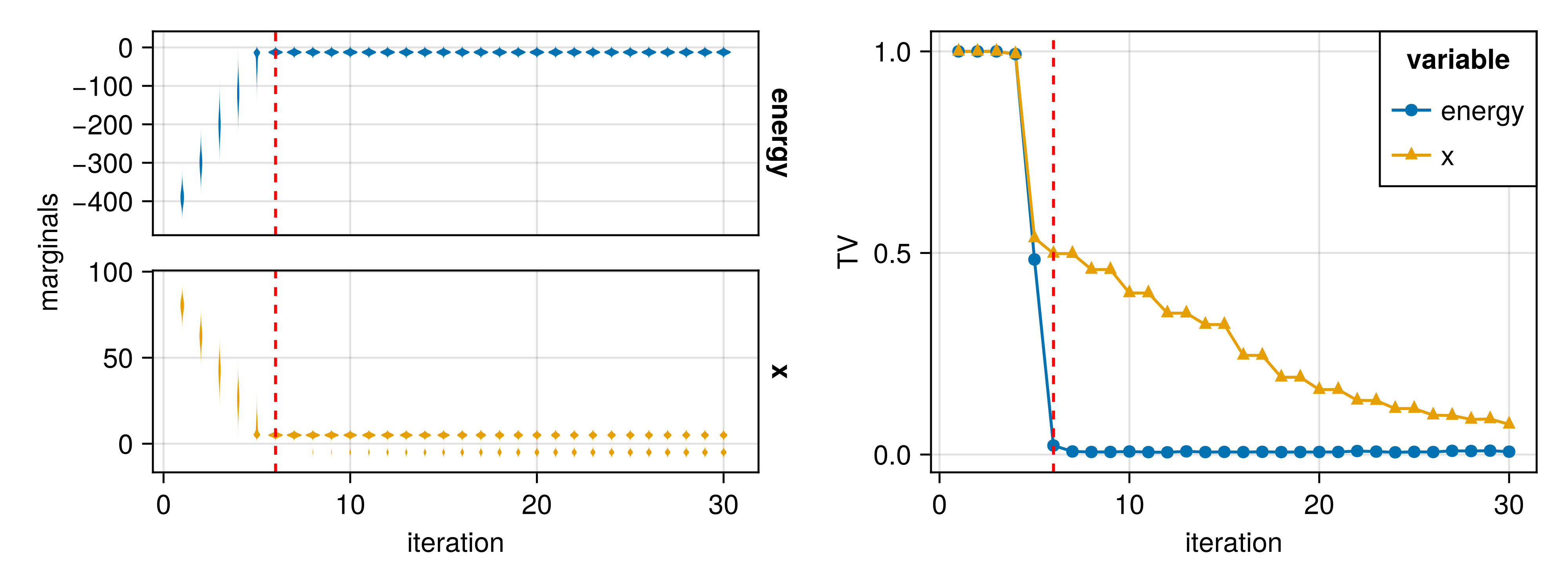}
	\caption{Practical implication of our theory, illustrated on a bimodal, univariate target 
		distribution ($\Lambda  \approx 2.5$, $N = 10$): in the context of PT algorithms, monitoring the energy traces 
		is an effective summary for convergence diagnostic purposes. Left panel: marginal 
		distribution of 50,000 independent NRPT algorithms initialized at $x = 100$. 
		Here our key assumption, the ELE (\cref{assump:base_erg_collection}.\ref{assump:ELE}), is well 
		approximated \emph{once the distribution of the energy  has small total variation to the true energy distribution} 
		(refer to right panel: this is achieved around iteration 6, denoted as a vertical dashed line in both panels). 
		Our theory makes predictions on the number of additional iterations past the vertical line required to decrease the 
		target chain TV distance under any specified threshold. For example, 
		for a TV distance of $\eps = 0.07$ (achieved empirically $30-6=24$ iterations after the dashed line), 
		our theory gives a bound of $150$ (taking $C=1>0.982$ based on \cref{tab:C_Lambda_t_supremum}).
		Only a single energy trace $V_t = V(X_t^N)$ is shown for simplicity, but the other chains 
    converge at the same rate in this example for $n > 0$, and instantaneously at 
    $n = 0$ with \iid reference sampling.}
	\label{fig:marginals}
\end{figure}

Having considered the ``independent'' part of the ELE's \iid condition, we now 
move to the ``identically distributed'' part. 
To assess the distributional assumption under the ELE model, we note 
that this condition can be interpreted as stating that the stochastic 
process tracking the evolution of the energy $V_t$ has reached stationarity. 
In real-world scenarios, a burn-in period is typically needed before we have 
``convergence of the energies,'' i.e., before the 
marginal distribution of the energy $V(X_t^n)$ is close to $V_* \pi_{\beta_n}$ 
(see vertical dashed line in \cref{fig:marginals}). The number of iterations needed to burn-in 
the energies is problem- and initialization-dependent and therefore out of reach of 
our explorer-agnostic setup. Once this initial period is over, however,
our theory predicts the number of additional iterations needed for PT to translate 
convergence of univariate energies into convergence of any statistic computed on the 
target chain. An example is presented in \cref{fig:marginals}.

\section{Practical guidance and discussion}
\label{sec:discussion}

Combining our theoretical results with existing work in the PT literature leads to useful guidelines for practitioners: 
\begin{enumerate}
  \item When the target distribution exhibits a complex geometry, NRPT is preferable 
    over single-chain MCMC methods and RPT. \label{guideline:use_PT}

  \item In the context of PT algorithms, monitoring the energy traces 
  is an effective summary for convergence diagnostic purposes. 
  This feature is especially attractive when PT is used to perform 
  inference over non-Euclidean spaces (discrete graphs, 
  phylogenetic trees, etc.), where a trace plot on the state space might be non-obvious to 
  visualize, whereas 
  the energies are always real-valued and hence easy to visualize. 
  \label{guideline:traceplots}

  \item After the energies have burned in, and subject to an energy correlation diagnostic illustrated in 
  \cref{fig:pairwise_correlation}, we expect our ergodicity results 
    to hold approximately. When the target distribution is challenging, mixing 
    can only be achieved by performing restarts from the reference chain for 
    which \iid sampling is possible. For these challenging problems, we
    expect our TV distance bounds to be fairly tight. 
    For distributions where the local explorer is sufficient for some $\beta > 0$, 
    our bounds should be conservative. \label{guideline:ergodicity}
\end{enumerate}

In the remainder of this section we offer some additional comments on the ELE
model, as well as the points above. 

\subsection{Discussion on the ELE model}
Parallel tempering is a \emph{meta-algorithm}: the annealing and 
communication procedures can be defined independently of the local explorers. 
This leads to many different ``flavours'' of PT, including: 
Hamiltonian Monte Carlo \cite{neal2011hmc} PT, 
slice sampler \cite{neal_slice_2003} PT, 
Metropolis-adjusted Langevin algorithm \cite{rossky1978brownian} PT, 
among many others. 
As a consequence, 
any formal study of PT as an algorithm requires one to either fix a specific type of 
local explorer or introduce a hypothesis that allows for an \emph{explorer-agnostic} 
analysis. In this work we took the latter approach and worked under the ELE 
model. Although this assumption is somewhat strong, it allowed us to study 
PT independently of the type of local explorer used. 

One may pose two valid questions about the ELE assumption: 
Does ELE only ever hold approximately? 
Does ELE automatically imply uniform ergodicity of even single-chain explorers?
Importantly, the answer to both questions is ``no.''
We acknowledge that the ELE model can only be approximately satisfied in general. 
However, we present simple instructional examples in which the condition is 
exactly met in the target chain.
For instance, the explorer's Markov transition kernel in the following example 
is not irreducible (and therefore certainly not ergodic). The kernel is only efficient at \emph{locally}
exploring its state space and is a very simple example of where PT could aid 
in mixing and achieving geometric ergodicity across the global space. 
The example is characteristic of certain multimodal target distributions.

\bexa 
\label{ex:disjoint_modes}
Let $R_1 = [-15, -5]$, $R_2 = [5, 15]$ and $R = R_1 \cup R_2$. 
Consider the target 
\[
  \pi_1(x) &\propto 0.5 \cdot \distNorm(x; -10, 1) \cdot \mathbbm{1}(x \in R_1) + 
    0.5 \cdot \distNorm(x; 10, 1) \cdot \mathbbm{1}(x \in R_2).
\]
The Markov transition kernel $K$ defined by 
\[
  K(x, A) = 2 \pi_1(A \cap R_1) \cdot \mathbbm{1}(x \in R_1) 
    + 2 \pi_1(A \cap R_2) \cdot \mathbbm{1}(x \in R_2)
\] 
obtains ``i.i.d.'' samples from the mode in which it is currently located.
If $\mu$ is any distribution supported on $R$,
$X_0 \sim \mu$, and $X_t|X_{t-1} \sim K(X_{t-1}, \cdot)$ for $t \geq 1$, then 
$-\log \pi_1(X_t)$ is distributed according to $V_* \pi_1$ and is independent of $X_{t-1}$.
However, the Markov kernel $K$ is unable to travel between the two modes and 
hence is not irreducible and cannot be ergodic. 
\eexa

Yet another example of an exploration kernel that satisfies the ELE model 
is a Gibbs sampler for a target distribution concentrated on a thin region. 
Bayesian posterior inverse problems can have this kind of ``thin shell'' structure, 
such as when estimating the parameters of an ODE for an mRNA transfection model \cite{ballnus2017comprehensive}.
We find that the following sampler satisfies ELE in the target chain, 
a condition on \emph{local} exploration, but mixes arbitrarily poorly on a \emph{global} scale. 

\bexa 
\label{ex:concentration} 
Fix $a > 0$ and consider the target 
\[
  \pi_1(x_1, x_2) = \distNorm(x_2; a x_1, 1) \mathbbm{1}(x_1 \in [0,1]).
\] 
We can define a Gibbs sampler kernel $K$ by noting that  
\[
  \pi_1(x_2 | x_1) &= \distNorm(x_2; a x_1, 1), \qquad
  \pi_1(x_1 | x_2) \propto \distNorm(x_1; x_2/a, 1/a^2) \mathbbm{1}(x_1 \in [0,1]).
\]
The Markov kernel $K$ based on the Gibbs sampler has
$-\log \pi_1(X_t)$ distributed according to $V_* \pi_1$ and independent of $X_{t-1}$ for any $a > 0$.
However, as $a$ increases, its mixing performance on a \emph{global} scale deteriorates. 
\eexa

\subsection{Discussion on guidelines}

For guideline (\ref{guideline:use_PT}), to illustrate the advantage of PT over traditional 
single-chain MCMC methods when the target distribution has a complex geometric structure,
we first consider a bimodal example where HMC exhibits slow mixing.

\bexa 
\label{ex:HMC_slow_mixing}
Consider the target 
\[
  \pi_1(x) = 0.5 \cdot \distNorm(-100, 1)(x) + 0.5 \cdot \distNorm(100, 1)(x).  
\]
Suppose we initialize $X_0 = -100$ and draw a momentum $p_0 \sim \distNorm(0,1)$.
We then simulate Hamiltonian dynamics exactly for \emph{any length of time} 
and obtain $X_1$. We iteratively refresh the momentum and simulate the dynamics 
for any length of time, obtaining $X_1, X_2, \ldots$.
We have 
\[
  \Pr(X_1, \ldots, X_t \in (-\infty, 0)) 
  \geq \Pr\rbra{\chi^2_t < 2 \Delta },
\]
where 
$\Delta = -\log \pi_1(0) + \log \pi_1(-100) \geq 0.5 (100)^2 + \log(0.5) \approx 4999.31$, 
and 
$\chi^2_t$ is a chi-squared random variable with $t$ degrees of freedom.
For instance, even if we obtain 9500 samples $X_1, \ldots, X_{9500}$ by 
simulating Hamiltonian dynamics exactly, we conclude that 
\[
  \Pr(X_1, \ldots, X_{9500} \in (-\infty, 0)) \geq 0.9997,
\]
and the probability of visiting the second mode within 9500 iterations is $\approx 0$.
\eexa

With this example in mind, we provide some intuition as to why we should 
expect NRPT to perform well on this target distribution, as well as why 
RPT is an improvement over HMC but still does not perform as well as NRPT.

\bexa 
\label{ex:PT_fast_mixing}
Consider the target $\pi_1$ from \cref{ex:HMC_slow_mixing} and set the reference  
\[
  \pi_0(x) = \distNorm(x; 0, 100^2+1).  
\]
From simulations we estimate $\hat\Lambda = 2.75$ and therefore 
based on informal recommendations in \cite{syed2021nrpt}, 
we set $N = 6 \approx 2\hat\Lambda$.
Assuming $r \approx \Lambda/N \approx 0.46$ (see \cite{syed2021nrpt}), and based on the 
hitting time bounds for finite-chain PT given by 
\cref{prop:persistent_walk_exact,prop:random_walk_exact},
\[
  \Pr_\text{NRPT}(\tau_6 \leq 20) \approx 0.323, \qquad 
  \Pr_\text{RPT}(\tau_6 \leq 20) \approx 0.104.
\]
Informally, there is a 50\% probability that a sample 
from the reference is from the second mode in $(0, \infty)$ and a 32.3\% (10.4\%) 
chance that we see a restart within the first 20 NRPT (RPT) iterations. We therefore expect 
to obtain a sample from the second mode of the distribution much more quickly with PT
than with HMC. We also see that NRPT should be expected to perform better than RPT. 
\eexa

For guideline (\ref{guideline:traceplots}) on assessing burn-in with univariate 
energy traceplots from the target chain, we note that 
it is common for a certain proportion of the initial MCMC samples to be discarded 
until the samples have ``burned in.'' Typically, traceplots are observed to aid in 
this process. However, if the samples are multivariate or do not live in  
Euclidean space, it can be tedious or impossible to monitor the 
traceplots of each of the components of the samples. 
For instance, if the target distribution is over a combinatorial space of phylogenetic 
trees, it is not always clear how to monitor the ``traceplots'' of such trees. 
A practical approach in PT is to monitor the real-valued univariate energies of 
the samples from the target chain.
Regardless of the structure of the underlying state space, the energies are always 
univariate and real-valued and can therefore be easily monitored. The justification for monitoring 
the energies in PT follows from the need to monitor the univariate ELE assumption.

Finally, for guideline (\ref{guideline:ergodicity}), we would like to know whether 
our theoretical ergodicity results hold in practice once the energies have burned in.
Such results can give practitioners a practical guideline for a minimum amount of time  
to run the PT Markov chain before obtaining samples representative of the target 
distribution. We see that this is the case for the Ising model example in 
\cref{sec:ising}. In general, when the target distribution is sufficiently complex 
that global exploration is only possible with PT, we expect our TV distance bounds 
to be appropriate. Our bounds are based on an analysis where the only 
improvement in TV distance can come from an \iid sample from the reference travelling 
to the target and do not take into account the mixing of local explorers. 
Even though the ELE model is introduced, it is only used to perform a coupling 
of energy statistics in the proof. Therefore, our obtained bounds should be more conservative 
in cases where global exploration is possible even without PT.

\subsection{Summary and future work} 

In this work we presented new results on the uniform ergodicity of PT under 
efficient local exploration 
and related its rate of convergence to the global communication barrier (GCB). 
This quantity is a divergence specific to parallel tempering that quantifies the difficulty of the 
sampling problem given a reference distribution $\pi_0$ and a target $\pi_1$. 
Under an efficient \textit{local} exploration model of PT---whereby local 
samplers explore the univariate energies efficiently---we 
established that NRPT and RPT are both uniformly 
ergodic. This result holds with both a finite number of chains and in the limit 
as the number of chains tends to infinity with a proper scaling of time. 

There are several possible directions for future work and we believe that our proof 
technique based on coupling the energies of PT can be extended to 
prove other properties.
One direction is to weaken the assumption of efficient local exploration 
kernels and instead consider only samplers that are ``locally ergodic,'' substituting 
exact local independence with approximate local independence. 
Finally, 
although our analysis of the rate of ergodicity is fairly tight, we speculate that 
using geometric ergodicity to infer a central limit theorem for the samples 
from the target is suboptimal in the case of PT. 
This is because the rate of ergodicity is determined by the time of the \emph{first restart}
where a sample from the reference travels to the target, which cannot happen before 
time $N$. In contrast, in the long run, restarts can occur as often as at 50\% of 
PT iterations and we suspect that the \emph{long-run restart rate} should control the asymptotic 
variance of samples. Modifications of our proof technique could therefore possibly yield 
new central limit theorems for PT with tighter bounds on the asymptotic variance of samples 
from the target chain.

\clearpage


\begin{appendix}
\section{Properties of the GCB}
\label{sec:GCB_properties}

In this section we consider the scaling properties of the GCB, extending previous 
work \cite{syed2021nrpt}, and connect it to well-known divergences. 
We first consider a tractable example to examine the scaling 
limit of the GCB in high dimensions in a case where the target does \emph{not} have 
independent components (\cref{prop:GCB_submanifold}), connecting to previously obtained high-dimensional 
scaling results of the GCB under independence assumptions \cite{syed2021nrpt}.
We then show that the GCB divergence is invariant with respect to one-to-one
and differentiable transformations of the reference and target 
distributions (\cref{prop:GCB_invariant_transformation}), further reinforcing that the 
GCB scales roughly as $O(d^{1/2})$ with respect to dimension for a considerably wide 
range of reference $\pi_0$ and target $\pi_1$ pairs.
Next, we prove a sub-additivity property of the GCB when $\pi_0$ and $\pi_1$ have 
independent components (\cref{prop:GCB_independent_components}), 
showing that we can obtain a coarse \emph{non-asymptotic} 
dimensional scaling result with a rate of $O(d)$ instead of $O(d^{1/2})$.
Finally, we relate the GCB to well-known quantities such as the 
total variation distance and the inclusive/exclusive KL divergences 
(\cref{prop:GCB_TV_bound} and \cref{prop:GCB_KL_bounds}).

One might might be concerned that performance of NRPT potentially deteriorates when the reference $\pi_0$ 
is bad compared to the target $\pi_1$, especially for high-dimensional targets.
However, in the following example we show that the GCB between multivariate Gaussians 
is $O(d^{1/2})$ with a mismatched covariance and sufficiently close means.
Although the following result concerns only Gaussian distributions, 
it is a useful model for understanding the behaviour of NRPT and the GCB in
high-dimensional and approximately unimodal settings with possible concentration on a submanifold.

\bprop
\label{prop:GCB_submanifold}
Fix $\rho \in [0,1)$. Let $\pi_{0,d} \sim \distNorm(0, \mathbb{I}_d)$ and 
$\pi_{1,d} \sim \distNorm(\mu_d, \Sigma_d)$
where $(\Sigma_d)_{ij} = \rho$ for $i \neq j$ and $(\Sigma_d)_{ii} = 1$ for $1 \leq i \leq d$.
Suppose $\limsup_{d\to \infty} \|\mu_d\|^2/d \leq m$. Then,
\[
  \limsup_{d \to \infty} d^{-1/2} \cdot \Lambda(\pi_{0,d}, \pi_{1,d}) 
  \leq
  \left(-\frac{1}{2} \log(1-\rho) + \frac{1}{2} \rbra{\frac{1}{1-\rho}(1+m) + 1}\right)^{1/2}.
\]
\eprop

Next, we establish that the GCB is invariant with respect to appropriate 
transformations (diffeomorphisms) of the reference and target distributions.
For instance, one can then immediately apply \cref{prop:GCB_submanifold} to 
conclude that the GCB scales as $O(d^{1/2})$ with respect to a wide range of 
reference $\pi_0$ and target $\pi_1$ pairs.
For a measurable function $h:\mcX \to \mcX$ we write $h_* \pi$ 
for the push-forward measures of $\pi$.

\bprop
\label{prop:GCB_invariant_transformation}
Let $h:\mcX \to \mcX$ be a diffeomorphism. Suppose
$\pi_0, \pi_1, h_*(\pi_0),$ and $h_*(\pi_1)$ have densities with respect to a common
dominating measure and that $\pi_0, \pi_1$ and $h_*(\pi_0), h_*(\pi_1)$ are 
mutually absolutely continuous. Then,
\[
  \Lambda(h_*(\pi_0), h_*(\pi_1)) = \Lambda(\pi_0, \pi_1).
\]
\eprop

When $\pi_0$ and $\pi_1$ are $d$-dimensional products of their marginal distributions, 
\cite{syed2021nrpt} established that $\Lambda(\pi_0, \pi_1) \sim d^{1/2}$ as 
$d \to \infty$. We present an analogous result except with a non-asymptotic bound that holds 
for all $d$, establishing that in such a case the GCB is upper bounded by the 
sum of the GCBs between each of the individual marginals.

\bprop
\label{prop:GCB_independent_components}
Suppose $\pi_0 = \pi_0^{(1)} \times \cdots \times \pi_0^{(d)}$ and 
$\pi_1 = \pi_1^{(1)} \times \cdots \times \pi_1^{(d)}$.
If each pair $\pi_0^{(j)}$ and $\pi_1^{(j)}$ for $j=1,2,\ldots,d$ is mutually absolutely 
continuous and $\pi_0^{(j)}, \pi_1^{(j)}$ have densities with respect to a common dominating measure, then
\[ 
  \Lambda(\pi_0, \pi_1) \leq \sum_{j=1}^d \Lambda(\pi_0^{(j)}, \pi_1^{(j)}).
\]
\eprop

Finally, it is useful to bound the GCB in terms of other well-known divergences 
and distances. In \cref{prop:GCB_TV_bound} we present a bound in terms of the 
TV distance. 

\bprop
\label{prop:GCB_TV_bound}
For any annealing schedule $0 = \beta_0 < \beta_1 < \ldots < \beta_N = 1$, we have
\[
  \Lambda(\pi_0, \pi_1) 
  \leq 2 \sbra{ \sum_{n=0}^{N-1} \frac{
    \mathrm{TV}(\pi_{\beta_n},\pi_{\beta_{n+1}})}{1-\mathrm{TV}(\pi_{\beta_n},\pi_{\beta_{n+1}})} }.
\]
\eprop

Note that setting $N=1$ in the preceding result yields a bound on the GCB in terms 
of $\mathrm{TV}(\pi_0,\pi_1)$. 
Additionally, by applying Lemma 2.6 of \cite{tsybakov2009introduction} we obtain the 
following upper bound on the GCB in terms of the inclusive and exclusive KL divergences.
The KL divergence $\kl{\pi_1}{\pi_0}$ between two probability 
distributions such that $\pi_1 \ll \pi_0$ is defined as 
\[
  \kl{\pi_1}{\pi_0}  
  = \int \log\frac{\dee \pi_1}{\dee \pi_0} \dee \pi_1.
\]
The KL divergence $\kl{\pi_0}{\pi_1}$ is defined analogously provided that 
$\pi_0 \ll \pi_1$.
For the following bound we apply the Bretagnolle--Huber inequality to obtain non-vacuous 
bounds in terms of the KL divergences when $\kl{\pi_0}{\pi_1}$ and $\kl{\pi_1}{\pi_0}$ 
are large. However, a similar bound can be obtained using Pinsker's inequality, which 
can be more suitable for small values of the KL divergences.

\bprop
\label{prop:GCB_KL_bounds}
The following inequalities hold:
\[
  \Lambda(\pi_0, \pi_1) \leq 2 \cdot \min 
    \cbra{\frac{g(\kl{\pi_1}{\pi_0})}{1-g(\kl{\pi_1}{\pi_0})}, \, 
    \frac{g(\kl{\pi_0}{\pi_1})}{1-g(\kl{\pi_0}{\pi_1})}},
\]
where $g(x) = 1 - \frac{1}{2} \exp(-x)$.
\eprop

We now prove each of the various properties of the GCB presented above.

\subsection{Proof of \cref{prop:GCB_submanifold}}
We start with a preliminary lemma.
\blem
\label{lem:eigen_Sigma}
Fix $\rho \in [0,1)$. Let $\Sigma_d$ be the $d \times d$ matrix where $(\Sigma_d)_{ij} = 
\rho$ for $i \neq j$ and $(\Sigma_d)_{jj} = 1$ for $1 \leq j \leq d$. Then, $\Sigma_d$
has $1-\rho$ as an eigenvalue of geometric multiplicity $d-1$ and $(d-1)\rho + 1$ as
an eigenvalue of geometric multiplicity one.
\elem
\bprfof{\cref{lem:eigen_Sigma}}
It suffices to find $d-1$ linearly independent eigenvectors with eigenvalue $1-\rho$ and
one eigenvector with eigenvalue $(d-1)\rho + 1$. Note that the set
\[ 
  \cbra{(1,-1,0,\ldots,0)^\top, (1,0,-1,0,\ldots,0)^\top, \ldots, (1,0,\ldots,0,-1)^\top}
\]
is an eigenbasis for an eigenspace of dimension $d-1$ corresponding to the eigenvalue $1-\rho$. Finally, $(1, 1, \ldots, 1)^\top$ is an eigenvector with eigenvalue $(d-1)\rho + 1$.
\eprfof

\bprfof{\cref{prop:GCB_submanifold}}
Let $g(x) = \abs{\log \pi_1(x) - \log \pi_0(x)}$. By Theorem 3.5 of 
\cite{surjanovic2022VPT}, 
\[
  \Lambda(\pi_{0,d}, \pi_{1,d}) \leq \sqrt{\frac{1}{2}\rbra{\EE_{\pi_{0,d}}[g] + \EE_{\pi_{1,d}}[g]}}.
\]
Now, 
\[
  g(x)
  &= \abs{-\frac{1}{2} \log\abs{\Sigma_d} - \frac{1}{2} (x-\mu_d)^\top \Sigma_d^{-1} (x-\mu_d) 
    + \frac{1}{2} x^\top x} \\
  &\leq \frac{1}{2}\rbra{\abs{\log|\Sigma_d|} + \abs{(x-\mu_d)^\top \Sigma_d^{-1} (x-\mu_d)} 
    + \abs{x^\top x}}.
\]
We bound each of the terms in the expression above. 
From \cref{lem:eigen_Sigma} we have that
\[ \log|\Sigma_d| = (d-1) \log(1-\rho) + \log((d-1)\rho + 1). \]
For the second term, note that the eigenvalues of $\Sigma_d^{-1}$ are 
$1/(1-\rho)$ and $1/((d-1)\rho+1)$, and that for any $d$ or $\rho$,
$1/((d-1)\rho+1) < 1/(1-\rho)$, and so the largest eigenvalue is $1/(1-\rho)$. Therefore, 
\[
  \abs{(x-\mu_d)^\top \Sigma_d^{-1} (x-\mu_d)} \leq \|x-\mu_d\|^2 \cdot \frac{1}{1-\rho}.
\]
Taking expectations of each component in the bound for $g(x)$, 
we obtain the bound for any $d$ and $\rho$,
\[
  \Lambda(\pi_{0,d}, \pi_{1,d}) \leq 
  \sqrt{\frac{1}{2}\rbra{u_1 + u_2}},
\]
where
\[
  u_1 &= \abs{(d-1) \log(1-\rho) + \log((d-1)\rho + 1)} \\
  u_2 &= \frac{1}{1-\rho} (d + \|\mu_d\|^2) + d
\]
Dividing by $d^{1/2}$ and taking the limit supremum as $d \to \infty$ we obtain the stated result.
\eprfof

\subsection{Proof of \cref{prop:GCB_invariant_transformation}}
\bprfof{\cref{prop:GCB_invariant_transformation}}
Let $X_\beta \sim \pi_\beta \propto \pi_0^{1-\beta} \cdot \pi_1^\beta$ for $\beta \in 
[0,1]$ and set $Y_\beta = h_*(X_\beta)$. We write $\mu_\beta$ for the density of $Y_\beta$,
which has the same dominating measure $\nu$ as $\pi_\beta$ for all $\beta$. 
We first show that $\mu_\beta$ forms a linear annealing path
so that we can apply the standard definition of the global communication barrier. 
That is, we show that $\mu_\beta \propto \mu_0^{1-\beta} \cdot \mu_1^\beta$ 
for $0 \leq \beta \leq 1$. To see this, note that
\[
  \mu_0(y) = \pi_0(h^{-1}(y)) \cdot \abs{J}, \qquad 
  \mu_1(y) = \pi_1(h^{-1}(y)) \cdot \abs{J},
\]
where $J$ is the Jacobian of the inverse transformation $h^{-1}$ and $|J|$ denotes its determinant.
Therefore,
\[
  \mu_\beta(y) 
  &= \pi_\beta(h^{-1}(y)) \cdot \abs{J} \\
  &\propto \pi_0^{1-\beta}(h^{-1}(y)) \cdot \pi_1^\beta(h^{-1}(y)) \cdot \abs{J}^{1-\beta} \cdot \abs{J}^\beta \\
  &= \mu_0^{1-\beta}(y) \cdot \mu_1^\beta(y).
\]

Because $\mu_\beta$ forms a linear path for $0 \leq \beta \leq 1$, we may write
\[
  &2\Lambda(\pi_0, \pi_1) \\
  &= \int_0^1 \int_{\mcX \times \mcX} \abs{\log \frac{\pi_1(x)}{\pi_0(x)}
    - \log \frac{\pi_1(x')}{\pi_0(x')}} \pi_\beta(x) \pi_\beta(x') \, \dee(\nu \times \nu) \, \dee\beta \\
  &= \int_0^1 \hspace{-1.5mm} \int_{\mcY \times \mcY} \abs{\log \frac{\pi_1(h^{-1}(y))}{\pi_0(h^{-1}(y))}
    - \log \frac{\pi_1(h^{-1}(y'))}{\pi_0(h^{-1}(y'))}} \pi_\beta(h^{-1}(y)) \pi_\beta(h^{-1}(y')) \cdot \abs{J}^2 \, \dee(\nu \times \nu) \, \dee\beta \\
  &= \int_0^1 \int_{\mcY \times \mcY} \abs{\log \frac{\pi_1(h^{-1}(y))}{\pi_0(h^{-1}(y))}
    - \log \frac{\pi_1(h^{-1}(y'))}{\pi_0(h^{-1}(y'))}} \mu_\beta(y) \mu_\beta(y') \, \dee(\nu \times \nu) \, \dee\beta \\
  &= \int_0^1 \int_{\mcY \times \mcY} \abs{\log \frac{\mu_1(y)}{\mu_0(y)}
    - \log \frac{\mu_1(y')}{\mu_0(y')}} \mu_\beta(y) \mu_\beta(y') \, \dee(\nu \times \nu) \, \dee\beta \\
  &= 2\Lambda(h_*(\pi_0), h_*(\pi_1)).
\]
\eprfof

\subsection{Proof of \cref{prop:GCB_independent_components}}
\bprfof{\cref{prop:GCB_independent_components}}
Note that for $\beta \in [0,1]$,
\[
  \pi_\beta(x) &\propto \pi_0^{1-\beta}(x) \cdot \pi_1^\beta(x) \\
  &\propto \prod_{j=1}^d \pi_\beta^{(j)}(x^{(j)}).
\]
That is, annealing the entire distribution is the same as 
annealing each marginal distribution.
Let $X_\beta, Y_\beta' \stackrel{iid}{\sim} \pi_\beta$. Then,
\[
  \Lambda(\pi_0, \pi_1)
  &= \frac{1}{2} \int_0^1 
    \EE\sbra{\abs{\sum_{j=1}^d \log \frac{\pi_1^{(j)}(X_\beta^{(j)})}{\pi_0^{(j)}(X_\beta^{(j)})} 
      - \sum_{j=1}^d \log \frac{\pi_1^{(j)}(Y_\beta^{(j)})}{\pi_0^{(j)}(Y_\beta^{(j)})}}} \, \dee\beta \\
  &\leq \sum_{j=1}^d \frac{1}{2} \int_0^1 
    \EE\sbra{\abs{\log \frac{\pi_1^{(j)}(X_\beta^{(j)})}{\pi_0^{(j)}(X_\beta^{(j)})} 
      - \log \frac{\pi_1^{(j)}(Y_\beta^{(j)})}{\pi_0^{(j)}(Y_\beta^{(j)})}}} \, \dee\beta \\
  &= \sum_{j=1}^d \Lambda(\pi_0^{(j)}, \pi_1^{(j)}).
\]
\eprfof

\subsection{Proof of \cref{prop:GCB_TV_bound,prop:GCB_KL_bounds}}
We begin with some intermediate lemmas, first showing that the normalizing constant 
for each of the distributions on the annealing path $\pi_\beta$
is bounded away from zero. Define 
\[
  Z(\beta) = \int_\mcX \pi_0^{1-\beta}(x) \cdot \pi_1^\beta(x) \, \dee \mu, \quad 
    0 \leq \beta \leq 1,  
\]
and note that $Z(0) = Z(1) = 1$.

\blem
\label{lem:Z_continuous}
  $Z(\beta)$ is continuous as a function of $\beta \in [0,1]$ and 
  \[\inf_{\beta \in [0,1]} Z(\beta) > 0.\]
  Additionally, $Z(\beta) \leq 1$ for all $\beta \in [0,1]$.
\elem
\bprfof{\cref{lem:Z_continuous}}
  We show that for any sequence $(\beta_n)_{n \geq 1}$ converging to $\beta$, 
  we have $\lim_{n \to \infty} Z(\beta_n) = Z(\beta)$. Note that for any 
  $\beta \in [0,1]$, by the weighted arithmetic-geometric mean inequality,
  \[
    \pi_0^{1-\beta}(x) \cdot \pi_1^\beta(x) 
    &\leq (1-\beta) \cdot \pi_0(x) + \beta \cdot \pi_1(x) \\
    &\leq \pi_0(x) + \pi_1(x).
  \]
  We apply the dominated convergence theorem and conclude that
  \[
    \lim_{n\to\infty} Z(\beta_n)
    &= \lim_{n\to\infty} \int_{\mcX} \pi_0^{1-\beta_n}(x) \cdot \pi_1^{\beta_n}(x) \, \dee\mu \\
    &= \int_{\mcX} \lim_{n\to\infty} \pi_0^{1-\beta_n}(x) \cdot \pi_1^{\beta_n}(x) \, \dee\mu \\
    &= \int_{\mcX} \pi_0^{1-\beta}(x) \cdot \pi_1^{\beta}(x) \\
    &= Z(\beta).
  \]
  By the above argument we conclude that $Z(\beta)$ is continuous on the compact set $[0,1]$.
  We also know that $Z(0) = Z(1) = 1$ and $Z(\beta) > 0$ for all $\beta \in [0,1]$.
  We therefore have that $\inf_{\beta \in [0,1]} Z(\beta) > 0$.
  From the application of the arithmetic-geometric mean inequality above, we also conclude that
  \[
    Z(\beta) \leq \int_{\mcX} \sbra{(1-\beta) \cdot \pi_0(x) + \beta \cdot \pi_1(x)} \, \dee\mu = 1.
  \]
\eprfof

We show an intermediate result to \cref{prop:GCB_TV_bound}.
\blem
\label{lem:GCB_TV_bound_halfway}
\[
  \Lambda(\pi_0, \pi_1) \leq \frac{2 \cdot \mathrm{TV}(\pi_0,\pi_1)}{\inf_{\beta \in [0,1]} Z(\beta)} < \infty.
\]
\elem

\bprfof{\cref{lem:GCB_TV_bound_halfway}}
  Set $M = 1/\inf_{\beta \in [0,1]} Z(\beta) < \infty$. We have
  \[
    \Lambda(\pi_0, \pi_1)
    &= \frac{1}{2} \int_0^1 \EE\sbra{ \abs{\log \frac{\pi_1(X_\beta)}{\pi_0(X_\beta)} - 
      \log \frac{\pi_1(X_\beta')}{\pi_0(X_\beta')}} } \, \dee\beta \\
    &\leq \int_0^1 \EE\sbra{ \abs{\log \frac{\pi_1(X_\beta)}{\pi_0(X_\beta)}} } \, \dee\beta \\
    &= \int_0^1 \int_{\mcX} \abs{\log \frac{\pi_1(x)}{\pi_0(x)}} \cdot 
      \frac{\pi_0^{1-\beta}(x) \cdot \pi_1^\beta(x)}{Z(\beta)} \, \dee\mu \, \dee\beta \\ 
    &= \int_{\mcX} \abs{\log \frac{\pi_1(x)}{\pi_0(x)}} \cdot \pi_0(x) \cdot 
      \int_0^1 \frac{[\pi_1(x) / \pi_0(x)]^\beta}{Z(\beta)} \, \dee\beta \, \dee\mu \quad \text{(Fubini)} \\
    &\leq M \int_{\mcX} \abs{\log \frac{\pi_1(x)}{\pi_0(x)}} \cdot \pi_0(x) \cdot 
      \int_0^1 \left(\frac{\pi_1(x)}{\pi_0(x)}\right)^\beta \, \dee\beta \, \dee\mu \\
    &= M \int_{\mcX} \abs{\log \frac{\pi_1(x)}{\pi_0(x)}} \cdot \pi_0(x) \cdot 
      \left[\log \frac{\pi_1(x)}{\pi_0(x)}\right]^{-1} 
      \cdot \rbra{\frac{\pi_1(x)}{\pi_0(x)} - 1} \, \dee\mu \\
    &= M \int_{\mcX} \abs{\log \frac{\pi_1(x)}{\pi_0(x)}} \cdot 
      \left|\log \frac{\pi_1(x)}{\pi_0(x)}\right|^{-1} \cdot \pi_0(x) 
      \cdot \abs{\frac{\pi_1(x)}{\pi_0(x)} - 1} \, \dee\mu \\
    &= M \int_{\mcX} \abs{\pi_1(x) - \pi_0(x)} \, \dee\mu \\
    &= 2 M \cdot \mathrm{TV}(\pi_0,\pi_1).
  \]
\eprfof

\blem
\label{lem:Z_lower_bound}
\[
  Z(\beta) \geq 1 - \mathrm{TV}(\pi_0,\pi_1).  
\]
\elem

\bprfof{\cref{lem:Z_lower_bound}}
  Note that
  \[
    Z(\beta) 
    &= \int_{\mcX} \pi_0^{1-\beta}(x) \cdot \pi_1^\beta(x) \, \dee\mu \\
    &\geq \int_{\mcX} \pi_0(x) \wedge \pi_1(x) \, \dee\mu \\
    &= 1 - \mathrm{TV}(\pi_0,\pi_1).
  \]
\eprfof

\bprfof{\cref{prop:GCB_TV_bound}}
By applying \cref{lem:GCB_TV_bound_halfway} and \cref{lem:Z_lower_bound}, we 
obtain the desired result for $\pi_0$ and $\pi_1$.
Noting that for any annealing schedule $\cbra{\beta_n}_{n=0}^N$ we have 
$\Lambda(\pi_0, \pi_1) = \sum_{n=0}^{N-1} \Lambda(\pi_{\beta_n}, \pi_{\beta_{n+1}})$, 
we establish the second part of the result.
\eprfof

\bprfof{\cref{prop:GCB_KL_bounds}}
Because $\Lambda(\pi_0, \pi_1) \leq 2 \cdot \mathrm{TV}(\pi_0,\pi_1)/(1-\mathrm{TV}(\pi_0,\pi_1))$ 
and the function $f(x) = 2x/(1-x)$ is strictly increasing on $[0,1)$,
the result follows by applying the Bretagnolle--Huber inequality 
(see Lemma 2.6 in \cite{tsybakov2009introduction}).
We do not need to consider the case 
where $\mathrm{TV}(\pi_0,\pi_1) = 1$ because the mutual absolute continuity of 
$pi_0$ and $\pi_1$ guarantees that the total variation distance is strictly less than one.
\eprfof

\newpage
\section{Proof of CLT for the marginal of a Markov chain}
\bprfof{\cref{lem:CLT}}
Following the discussion in \cref{sec:PT_kernels_homogeneous}, let
$\tilde{X}_t=(\bar X_t,S_t)$ be the time-homogeneous PT Markov chain with
initial condition $(\bar{X}_0, S_\text{even})$ and transition
kernel $\tilde{K}$. By assumption, $\tilde{X}_t$ has a unique invariant probability 
distribution $\tilde{\pi}=\bar{\pi} \times \mathrm{Uniform}(\{S_\text{even},S_\text{odd}\})$ 
and $\tilde X_0 \sim \tilde \pi$.
Because the marginal of a Markov chain is not necessarily Markovian, 
we operate on the extended state space $\tilde\mcX$ of the time-homogeneous kernel $\tilde K$ but  
restrict the class of measurable functions $\tilde f:\tilde{\mcX}\mapsto\reals$ such that 
$\tilde{\pi}(\tilde{f})=0$,
$\tilde{\pi}(|\tilde f|^{2+\delta}) < \infty$ for some $\delta>0$, and
$\tilde{f}(\bar{x},s)=f(x^N)$ for some measurable $f:\mcX\mapsto \reals$. 

We prove that the PT chain satisfies the CLT for all such functions $\tilde f$.
To establish the desired CLT in terms of $\tilde f$, 
we follow the proof of Theorem 21.4.4 in \cite{douc2018markov} and verify that a CLT holds
under the weakened assumptions of our \cref{lem:CLT}. In particular, Assumption
(21.4.12) in \cite{douc2018markov} is replaced with our marginal uniform
ergodicity assumption given by \cref{eq:marginal_unif_geom_erg}.
The result follows by verifying that the proof of Theorem 21.4.4 in \cite{douc2018markov}
remains unchanged with this substitution of conditions.
To complete the proof, Lemma 21.4.3 of \cite{douc2018markov} must also be verified 
with the same replacement of ergodicity assumptions with our marginal variant.
We conclude the desired CLT result with initial distribution $\tilde \pi$ 
by applying Theorem 21.4.1 and Remark 21.4.2 in \cite{douc2018markov}, 
which do not make assumptions on the uniform ergodicity of the kernel $\tilde K$ 
and hence do not need to be modified.
\eprfof

\newpage
\section{Equivalence of two forms of the ELE assumption}

The ELE assumption can be formulated as: 
\begin{itemize}
  \item[(i)] For each $n=0, 1, \ldots, N$ and any initial distribution $\mu$, we have that 
    if $X \sim \mu$ and $X'|X \sim K^{(\beta_n)}(X, \cdot)$, then 
    $V(X')$ is independent of $X$ and $V(X') \sim V_* \pi_{\beta_n}$.
\end{itemize}
We show that this assumption is equivalent to: 
\begin{itemize}
  \item[(ii)] For each $n=0, 1, \ldots, N$ and any initial distribution $\mu$, we have that 
    if $X \sim \mu$ and $X'|X \sim K^{(\beta_n)}(X, \cdot)$, then 
    $V(X')$ is independent of $V(X)$ and $V(X') \sim V_* \pi_{\beta_n}$.
\end{itemize}

We begin by showing that (i) implies (ii). 
This clearly follows because independence of $V(X')$ and $X$ implies 
independence of $V(X')$ and $V(X)$. 

We now establish that (ii) implies (i). 
Set $V' = V(X')$ and note that $V' | X \sim K_{V'|X}(X, \cdot)$, 
where for any $x \in \mcX$ and measurable $B \subset \reals$, 
\[
  K_{V'|X}(x, B) = K^{(\beta_n)}(x, V^{-1}(B)).  
\]
Our goal is to show that for all measurable sets $A \subset \mcX$ and $B \subset \reals$, 
and any initial distribution $X \sim \mu$, that
\[
  \Pr(V(X') \in B, X \in A) = \Pr(V(X') \in B) \Pr(X \in A).  
\]
Suppose this is not the case. Then, there exist sets $A, B$ and a distribution 
$X \sim \mu$ such that  
\[
  \Pr(V(X') \in B, X \in A) \neq \Pr(V(X') \in B) \Pr(X \in A).  
\]
We claim then that we must have $x_1, x_2 \in \mcX$ such that 
$K_{V'|X}(x_1, B) > K_{V'|X}(x_2, B)$. If this were not the case, then
$K_{V'|X}(x_1, B) = K_{V'|X}(x_2, B)$ for all $x_1, x_2 \in \mcX$. 
Then, necessarily, $K_{V'|X}(\cdot, B) = V_* \pi_{\beta_n}(B)$, and we would have 
\[
  \Pr(V(X') \in B, X \in A) 
  &= \int_{A} \mu(\dee x) K_{V'|X}(x, B) \\
  &= V_* \pi_{\beta_n}(B) \mu(A) \\ 
  &= \Pr(V(X') \in B) \Pr(X \in A),
\]
which we assumed did not hold. Therefore, $K_{V'|X}(x_1, B) > K_{V'|X}(x_2, B)$
for some $x_1, x_2$.
Consider now the initial distribution 
$\mu = \frac{1}{2} \delta_{x_1} + \frac{1}{2} \delta_{x_2}$. Then,
\[
  \Pr(V(X') \in B)
  &= \int_{\mcX} \mu(\dee x) K_{V'|X}(x, B) 
  = 0.5 K_{V'|X}(x_1, B) + 0.5 K_{V'|X}(x_2, B).
\] 
On the other hand, we also have that, for $v = V(x_1)$, 
\[
  \Pr(V(X') \in B | V(X) = v)
  &= \Pr(V(X') \in B | X = x_1)
  = K_{V'|X}(x_1, B) 
  > \Pr(V(X') \in B),
\] 
which is a contradiction to the statement of (ii). We conclude that (i) must hold.

\newpage 
\section{Coupling technique for PT}

\subsection{Definition of PT kernels} 
\label{sec:PT_kernels} 

We define the joint the exploration kernel for $A = A_0 \times \cdots \times A_N$, 
where each $A_n \in \mcF$, as
\[
  \bar K^\text{expl}(\bar x, A) 
  &= \prod_{n=0}^N K^{(\beta_n)}(x^n, A_n).
\]
For the communication kernels we define 
$\bar x^{(n, n+1)}$ to be the state $\bar x$ with positions $n$ and $n+1$ 
swapped for $n=0,1,\ldots, N-1$. 
Whether or not a communication swap is performed depends on the acceptance 
probability For two kernels $K_1$ and $K_2$, define 
$(K_1 K_2)(x_1, A) = \int K_1(x_1, \dee x_2) K_2(x_2, A)$.
The communication kernels for adjacent chains, $\bar K^{(n,n+1)}$, and 
for the entire PT Markov chain, $\bar K^\text{even}$ and $\bar K^\text{odd}$, 
are
\[ 
  \bar K^{(n, n+1)}(\bar x, \cdot) 
    &= (1 - \alpha^{(n,n+1)}(\bar x)) \cdot \delta_{\bar x}(\cdot) + 
    \alpha^{(n,n+1)}(\bar x) \cdot \delta_{\bar x^{(n,n+1)}}(\cdot) \\
  \bar K^\text{even} &= \prod_{\substack{n \text{ even} \\ 0 \leq n \leq N-1}} 
    \bar K^{(n, n+1)} \\
  \bar K^\text{odd} &= \prod_{\substack{n \text{ odd} \\ 0 \leq n \leq N-1}} 
    \bar K^{(n, n+1)}.
\]
We define the deterministic even-odd (DEO) communication kernel 
corresponding to NRPT, introduced by \cite{okabe2001replica}, as 
\[
  \label{eq:DEO_kernel}
  \bar K^\text{DEO}_t 
  &= \begin{cases}
    \bar K^\text{even}, & t \text{ even}, \\
    \bar K^\text{odd}, & t \text{ odd},
  \end{cases}
  \qquad t \in \nats,
\]
which alternates deterministically between even and odd communication swaps.
The stochastic even-odd (SEO) communication kernel corresponding to RPT is 
\[
  \label{eq:SEO_kernel}
  \bar K^\text{SEO}_t 
  = \frac{1}{2} \bar K^\text{even} + \frac{1}{2} \bar K^\text{odd}, 
  \qquad t \in \nats,
\]
which chooses an even or odd communication swap with equal probability 
for any given time step $t$.

Recall that the PT Markov chain $\{\bar X_t\}_{t \geq 0}$ is created 
by initializing $\bar X_0$ according to some distribution $\bar{\mu}$ 
and drawing 
\[
  \bar X_t | \bar X_{t-1} 
  \sim \bar{K}^\text{expl} \cdot \bar{K}_{t-1}^\bullet(\bar X_{t-1}, \cdot)
\]
for $\bullet \in \cbra{\text{DEO}, \text{SEO}}$ and $t \geq 1$. 
We define a sequence of PT 
kernels that perform both local exploration and communication, 
$\bar{K}^\bullet_t$ for $\bullet \in \cbra{\text{NRPT},\text{RPT}}$. 
These kernels yield the step $t$ marginals of the PT Markov chain, so that 
\[
  \bar X_t | \bar X_0 \sim \bar{K}^\bullet_t(\bar X_0, \cdot).  
\]
For NRPT, these kernels are recursively defined by 
\[
  \bar{K}_1^\text{NRPT}  &= \bar{K}^\text{expl} \cdot \bar{K}_0^\text{DEO} \\
  \bar{K}_t^\text{NRPT} &= \bar{K}_{t-1}^\text{NRPT} \cdot \bar{K}^\text{expl} \cdot 
    \bar{K}_{t-1}^\text{DEO}, \qquad t \geq 2. 
\]
Similarly, the kernels for RPT are obtained by replacing the DEO communication kernels 
with SEO communication,
\[
  \bar{K}_1^\text{RPT}  &= \bar{K}^\text{expl} \cdot \bar{K}_0^\text{SEO} \\
  \bar{K}_t^\text{RPT} &= \bar{K}_{t-1}^\text{RPT} \cdot \bar{K}^\text{expl} \cdot 
    \bar{K}_{t-1}^\text{SEO}, \qquad t \geq 2. 
\]

\subsection{Time-homogeneous representation of PT kernels} 
\label{sec:PT_kernels_homogeneous} 
It is often convenient to work with the time-inhomogeneous
Markov chain $\bar{X}_t\in \bar\mcX$ for NRPT and RPT. 
However, we can also express NRPT and RPT as time-homogenous Markov chains
$(\bar{X}_t, S_t) \in \tilde{\mcX}$, initialised at
$(\bar{X}_0,S_{\text{even}})\in\tilde{\mcX}$, where
$\tilde{\mcX}=\bar{\mcX}\times \{S_\text{even},S_\text{odd}\}$ is the extended
state space augmented with the parity of the swap proposal.

Formally, the Markov transition kernel $\tilde{K}^\text{NRPT}$ on $\tilde{\mcX}$
for NRPT generates $(\bar{X}_{t+1},S_{t+1})$ given $(\bar{X}_t,S_t)$ as follows:
\[
  \bar{X}_{t+1}|\bar{X}_t,S_t &\sim 
  \begin{cases}
  \bar{K}^\text{expl}K^\text{even}(\bar{X}_t,\cdot),& S_t = S_\text{even}\\
  \bar{K}^\text{expl}K^\text{odd}(\bar{X}_t,\cdot),& S_t = S_\text{odd}
  \end{cases},\\
  S_{t+1}|\bar{X}_t,S_t &= 
  \begin{cases}
  S_\text{odd}, & S_t = S_\text{even}\\
  S_\text{even}, & S_t = S_\text{odd}
  \end{cases}.
\]
Similarly, the Markov transition kernel $\tilde{K}^\text{RPT}$ on
$\tilde{\mcX}$ for RPT generates $(\bar{X}_{t+1},S_{t+1})$ given
$(\bar{X}_t,S_t)$ as follows: 
\[
  \bar{X}_{t+1}|\bar{X}_t,S_t &\sim 
  \begin{cases}
  \bar{K}^\text{expl}K^\text{even}(\bar{X}_t,\cdot),& S_t = S_\text{even}\\
  \bar{K}^\text{expl}K^\text{odd}(\bar{X}_t,\cdot),& S_t = S_\text{odd}
  \end{cases},\\
  S_{t+1}|\bar{X}_t,S_t &\sim 
  \mathrm{Uniform}(\{S_\text{even},S_\text{odd}\}).
\]
For $\bullet\in\{\text{NRPT},\text{RPT}\}$, it follows that if
$\bar{K}_t^\bullet$ is a $\bar{\pi}$-invariant time-inhomogeneous kernel on
$\bar{\mcX}$ at time $t$, then $\tilde{K}^\bullet$ is a $\tilde{\pi}$-invariant
time-homogeneous kernel on $\tilde{\mcX}$, where the stationary distribution is 
$\tilde{\pi}=\bar{\pi}\times \mathrm{Uniform}(\{S_\text{even},S_\text{odd}\})$.

\subsection{Definition of coupled kernels}

Consider a Markov chain defined on the
expanded state space $\bar{\mcX}^2$ on which we will perform our coupling. 
By construction, the first $N+1$ marginals of this chain will
also form a Markov chain, representing PT initialized according to $\bar{\mu}$, 
while another copy of PT will be initialized according to $\bar{\pi}$.
We denote the states of the coupled copy of PT with $\bar{y}$ 
(corresponding to $\bar{\pi}$)
instead of $\bar{x}$ (corresponding to $\bar{\mu}$).
Let the initial distribution of the joint Markov chain be $\nu$, where
\[
  \nu(\dee(\bar{x},\bar{y}))
  = \pi_0(\dee x^0) \delta_{x^0}(\dee y^0) 
    \bar\mu_{1:N}(\dee x^{1:N}) \prod_{n=1}^N \pi_{\beta_n}(\dee y^n).
\]
Note that in the \textit{reference} chain the states are already coupled. 

The Markov transition kernels for the coupled chain are defined below. 
To distinguish between kernels defined on the original versus the expanded state space, 
we use the letter $P$ for \textit{paired} kernels associated with the full, coupled Markov chain.
The original \emph{kernels} are denoted with a $K$.

We begin by describing the exploration kernels.
Note that for $x \in \mcX$ and $A \in \mcF$, 
under \cref{assump:base_erg_collection}.\ref{assump:ELE} and the fact that 
$\mcF$ is a standard Borel $\sigma$-algebra, we can decompose 
\[
  K^{(\beta_n)}(x, A) 
  &= \int_{V(\mcX)} V_*(\pi_{\beta_n})(\dee v) \cdot K^{(\beta_n | V)}(x, v; A),  
\]
where $K^{(\beta_n | V)}(x, v; \cdot)$ is a kernel.
(See Theorem 3.4 and Corollary 3.6 on disintegration in \cite{kallenberg2021foundations}.)
For $A, A' \in \mcF$ and $B \in \mcF^2$ we define the following 
coupled kernels and give an explanation below. 
For $x, y \in \mcX$ let
\[
  \tilde{P}^{(\beta_n)}(x, y; B) &= \int_{B} K^{(\beta_n)}(x, \dee x') 
    \delta_{x'}(\dee y'), \qquad 0 \leq n \leq N \\
  \hat{P}^{(\beta_n)}(x, y; A \times A') &= \int_{V(\mcX)} V_*(\pi_{\beta_n})(\dee v) \cdot 
    K^{(\beta_n | V)}(x, v; A) \cdot K^{(\beta_n | V)}(y, v; A'), \,\, 1 \leq n \leq N \\
  P^{(\beta_0)}(x, y; B) &= \tilde{P}^{(\beta_0)}(x, y; B) \\
  P^{(\beta_n)}(x, y; B) &= \tilde{P}^{(\beta_n)}(x, y; B) \mathbbm{1}(x = y) +
    \hat{P}^{(\beta_n)}(x, y; B) \mathbbm{1}(x \neq y),
    \qquad 1 \leq n \leq N.
\]
In words, as we see from the kernels $P^{(\beta_n)}$ for $0 \leq n \leq N$, 
the states from the reference chain are always coupled so that almost surely 
for all $t \geq 0$ we have $X_t^0 = Y_t^0$. Furthermore,
the intermediate energy values across \textit{all chains} are also coupled so that 
almost surely for all $t \geq 0$ and any $0 \leq n \leq N$ we have 
$V(\tilde X_t^n) = V(\tilde Y_t^n)$ where $\tilde X$ and $\tilde Y$ are the intermediate 
states after local exploration and before communication.
However, note that for chains $1 \leq n \leq N$ the states 
$X^n$ and $Y^n$ may differ even though the energy values are the same.
For chains $1 \leq n \leq N$, if the 
states are the same across the two copies of PT, then their states also remain coupled
until the next communication move and not just their energies. 
For instance, if chain $n > 0$ is coupled in both instances of PT and after communication 
this state is passed on to chain $n+1$, then in the next exploration phase 
both instances of PT have chain $n+1$ coupled.
Finally, set 
\[
  \bar{P}^{(\beta_n)}((\bar{x}, \bar{y}); 
    A_0 \times \cdots \times A_N \times A_0' \times \cdots \times A_N')
    = P^{(\beta_n)}(x^n, y^n, A_n \times A_n'),
\]
and define
\[
  \bar{P}^\text{expl}(\bar{x}, \bar{x}'; \cdot)
    = \prod_{n=0}^N \bar{P}^{(\beta_n)}(\bar{x}, \bar{x}'; \cdot).
\]
This concludes the specification of the exploration kernels for the coupled Markov chain.

We now define the communication kernels. To emphasize that the communication 
acceptance probability depends \emph{only on the energy levels} and not on the 
states themselves, we redefine 
\[
  \alpha^{(n, n+1)}(\bar v)
  &= \exp\rbra{\min\cbra{0, (\beta_{n+1} - \beta_n) \cdot (v^{n+1} - v^n)}},
\]
and set $\bar v = (v^0, v^1, \dots, v^N)^\top$ and $V(\bar{x}) = (V(x^0), \ldots V(x^N))^\top$. 
For the paired communication kernels, set
\[
  \bar{P}^{(n, n+1)}(\bar{x}, \bar{y}; \cdot) 
    &= (1 - \alpha^{(n,n+1)}(V(\bar{x}))) \delta_{(\bar{x}, \bar{y})}(\cdot) + 
    \alpha^{(n,n+1)}(V(\bar{x})) \delta_{(\bar{x}^{(n,n+1)}, \bar{y}^{(n,n+1)})}(\cdot) \\
  \bar{P}^\text{even} 
    &= \prod_{\substack{n \text{ even} \\ 0 \leq n \leq N-1}} \bar{P}^{(n,n+1)} \\
  \bar{P}^\text{odd} 
    &= \prod_{\substack{n \text{ odd} \\ 0 \leq n \leq N-1}} \bar{P}^{(n,n+1)} \\
  \bar{P}^\text{DEO}_t 
    &= \begin{cases}
    \bar{P}^\text{even}, & t \text{ even} \\
    \bar{P}^\text{odd}, & t \text{ odd}
    \end{cases} \\
  \bar{P}^\text{SEO}_t 
    &= \frac{1}{2} \bar{P}^\text{even} + \frac{1}{2} \bar{P}^\text{odd},
\]
and 
\[
  \bar{P}_1^\text{NRPT} 
    &= \bar{P}^\text{expl} \cdot \bar{P}_0^\text{DEO} \\
  \bar{P}_t^\text{NRPT} 
    &= \bar{P}_{t-1}^\text{NRPT} \cdot \bar{P}^\text{expl} \cdot
    \bar{P}_{t-1}^\text{DEO}, \qquad t \geq 2, 
\]
with $\bar{P}_t^\text{RPT}$ defined analogously.
Note that by the construction of the exploration kernel we have almost surely for all $t \geq 0$ 
that $\alpha^{(n,n+1)}(V(\tilde{\bar{X}}_t)) = \alpha^{(n,n+1)}(V(\tilde{\bar{Y}}_t))$, 
explaining why only $V(\bar{x})$ appears in the definition of $\bar{P}^{(n,n+1)}$.
This concludes the specification of the Markov kernel for the expanded state space. 

\newpage
\section{Uniform ergodicity with a finite number of chains}
This section is split into three parts. We first prove results that relate 
the rate of ergodicity for a finite number of chains 
to the hitting time of discrete processes. In the subsequent two subsections, 
we establish bounds on the survival functions of the hitting times for NRPT 
and RPT, respectively.

To formalize the notion of an index process introduced in the main text, 
we follow the presentation of \cite{syed2021nrpt}. 
We define the indicator variables $P_t^{(n,n+1)} = P_t^{(n+1,n)} \in \cbra{0,1}$
where $P_t^{(n,n+1)} = 1$ if a swap is \emph{proposed} between chains $n$ and $n+1$ 
at time step $t$ and $P_t^{(n,n+1)} = 0$ otherwise. 
We define the swap transition indicators as 
$S_t^{(n,n+1)} = S_t^{(n+1,n)} = P_t^{(n,n+1)} A_t^{(n,n+1)}$ so that the indicator 
is equal to one if and only if the swap is \emph{proposed and accepted} at time 
$t$ between chains $n$ and $n+1$. From here, we initialize $I_0^n = n$ and set $\eps_0^n = 1$ 
if $P_0^{(n,n+1)} = 1$, otherwise $\eps_0^n = -1$. For time steps $t > 0$, we 
recursively define 
\[
  I_{t+1}^n = 
  \begin{cases}
    I_t^n + \eps_t^n, & \text{if } S_t^{(I_t^n, I_t^n + \eps_t^n)} = 1, \\
    I_t^n, & \text{otherwise},
  \end{cases}  
  \qquad 
  \eps_{t+1}^n = 
  \begin{cases}
    1, & \text{if } P_{t+1}^{(I_{t+1}^n, I_{t+1}^n + 1)} = 1, \\
    -1, & \text{otherwise}.
  \end{cases}
\]
Note that under ELE \cref{assump:base_erg_collection}.\ref{assump:ELE}, 
swap proposal acceptance indicators are independent and identically distributed, 
with success probability equal to $\alpha_{n,N}$ when chain $n$ is communicating with chain 
$n+1$.

\subsection{Hitting times and ergodicity}

\begin{figure}
	\includegraphics[width=0.4\textwidth]{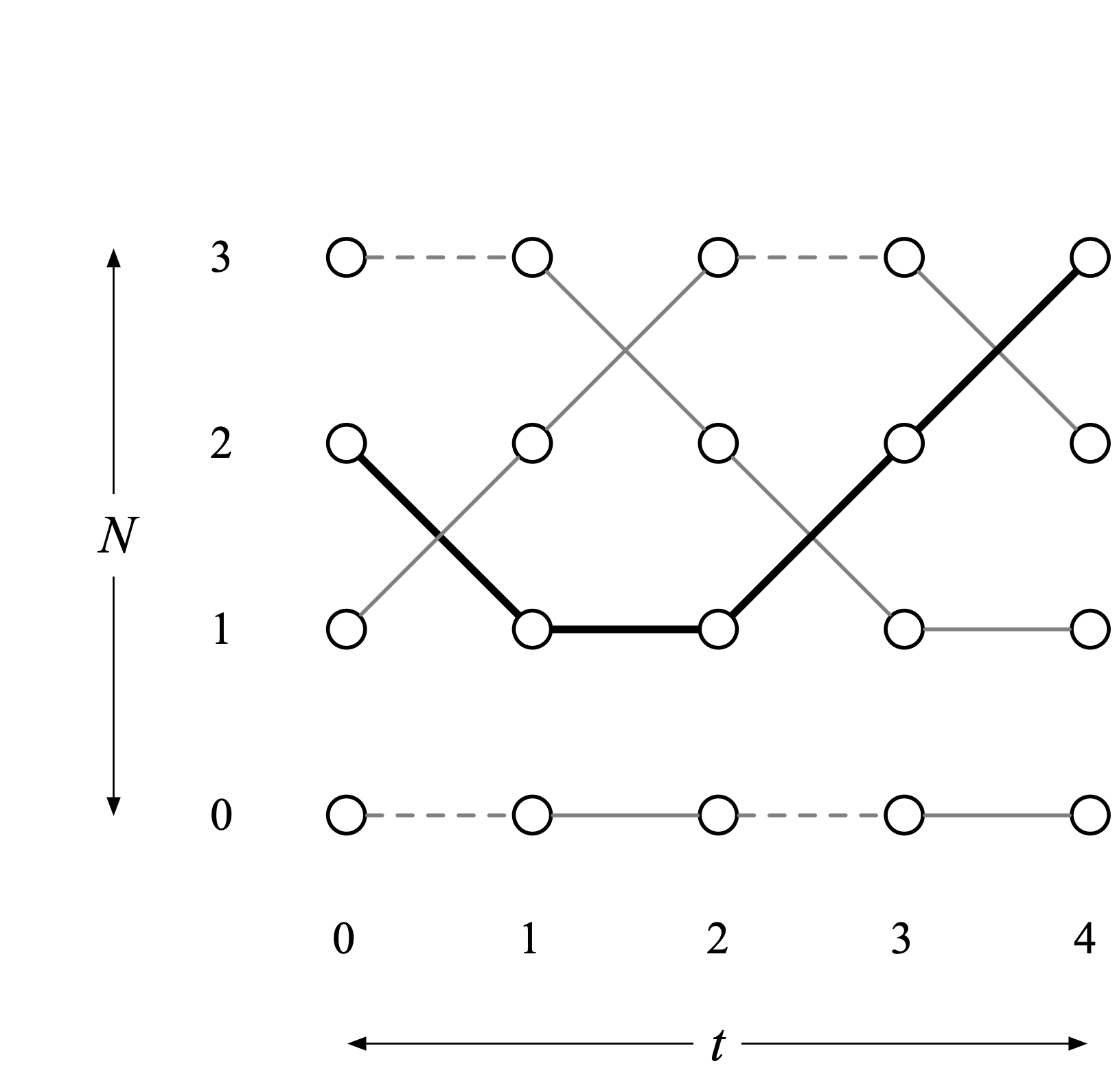}
	\caption{
  Example of an ancestral process (bold line). 
	Dashed lines indicate where no swap is proposed. Solid lines are either accepted 
	or rejected swaps (diagonal and horizontal respectively).  Each index process is a
  connected component in this figure and each index process is identified by its value at 
	the initial iteration. For example, the index process in bold is denoted by  
	$I_\cdot^2 = (2, 1, 1, 2, 3)$, where the superscript 2 indicates it starts at 2. The function 
	$n^\star(t)$ indicates for each iteration $t$ the index process currently occupying the target 
	(top) chain, identifying it again by its initial value. For example $n^\star(4) = 2$ since the bold 
	line is in the top at iteration $t = 4$. These two definitions combined define the ancestral process, 
	$A_t(\cdot) = I_\cdot^{n^*(t)}$, for example $A_t(\cdot) = (2, 1, 1, 2, 3)$ for $t = 4$ in the 
  figure (bold line). Finally, $H_t$ is the event that the ancestral process hits the reference (bottom) chain strictly before $t$. 
	For example, the event $H_4$ does not contain the realization shown in the above figure.}  
	\label{fig:ancestry}
\end{figure}

We begin by introducing the ancestral process, which can be defined for both 
NRPT and RPT (refer to \cref{fig:ancestry} for an example). In the analysis that follows, fix $t \geq 1$. 
For \emph{exactly one} $n^\star(t) \in \{0,1,\ldots,N\}$,
it is true that $I_t^{n^\star(t)} = N$, where $I_t^n$ is the value of the 
index process of the $n$-th machine at time $t$.
We refer to this index process 
on machine $n^\star(t)$ as the \textit{ancestral process} at time $t$
and denote it $A_{s}(t)$ for $0 \leq s \leq t$ (bold line in \cref{fig:ancestral_process}).
Formally, for a fixed $t \geq 1$,
\[
  A_t(s) = I_{s}^{n^\star(t)}, \qquad 0 \leq s \leq t.
\]

We define a collection of events $H_{t,s}$, where for a given $t \geq 1$ and 
$0 \leq s \leq t-1$, we define $H_{t,s}$ to be the set of events where the
ancestral process hits the reference at some point \emph{strictly before} time $t$
with its \emph{last visit} to the reference at time $s$. 
We define $H_t$ to be the event where the ancestral process hits the reference at 
some point strictly before time $t$, which is the union of the $H_{t,s}$ over $s$. 
That is, 
\[
  H_{t,s} &= \cbra{\omega \in \Omega : A_{t}(s)(\omega) = 0, \, 
    A_{t}(s')(\omega) > 0 \text{ for } s < s' \leq t } \\
  H_t &= \bigcup_{0 \leq s \leq t-1} H_{t, s}.
\] 
Now, on the set $H_t$ we have that $X_t^N = Y_t^N$ almost surely. 
This is by construction of the coupled kernel: once the reference chain is hit,
the states in both instances of PT are the same for that index process.
We can therefore write
\[
  X_t^N = Y_t^N \mathbbm{1}_{H_t} + X_t^N \mathbbm{1}_{H_t^c}.
\]
When there is possible ambiguity, we write $\Pr_{\bullet, \bar{\mu}}$ and 
$\EE_{\bullet, \bar{\mu}}$ to denote probability and expectation under the 
scheme $\bullet \in \cbra{\text{NRPT}, \text{RPT}}$, 
for PT initialized at distribution $\bar{\mu}$.
Because $Y_t^N \sim \pi_1$ for all $t$ on the Markov chain (initialized according to 
$\bar{\pi}$) with which we are to couple, 
it follows that for $\bullet \in \cbra{\text{NRPT}, \text{RPT}}$,
\[
  \mathrm{TV}_N(\bar{\mu} \bar{K}_t^\bullet, \bar{\pi}) 
  &= \frac{1}{2} \sup_{f \in \mcF_N} 
    \abs{\bar{\mu} \bar{K}_t^\bullet f - \bar{\pi} f} \\
  &= \frac{1}{2} \sup_{f \in \mcF_N} 
    \abs{\EE_{\bullet, \bar{\mu}}[f(\bar{X}_t) - f(\bar{Y}_t)]} \\
  &= \frac{1}{2} \sup_{f \in \mcF_N} 
    \abs{\EE_{\bullet, \bar{\mu}}[f(\bar{X}_t) - f(\bar{Y}_t) \mid H_t^c]} \cdot 
    \Pr_{\bullet, \bar{\mu}}(H_t^c) \\
  &\leq \Pr_{\bullet, \bar{\mu}}(H_t^c) \\
  &= \Pr_{\bullet}(H_t^c).
\]
The last equality follows from the ELE assumption whereby the behaviour of the 
energies and index process are independent of the initialization of the Markov chain.
Having established the above inequality, it remains to relate this probability 
to the survival time of two different processes for NRPT and RPT.

\subsection{Non-reversible parallel tempering}

\bprfof{\cref{thm:erg_NRPT_finite}}
We show that the event $H_t^c$ can be rewritten in terms of the usual (forward-time)
index process, which is typically easier to work with, instead of the (reverse-time) ancestral process. 
We then conclude that 
$\Pr_\text{NRPT}(H_t^c) \leq \Pr_\text{NRPT}(\tau_N > t-1)$ for any $t \geq 1$.
We use \cref{lem:nonreversible_index_process}, stated after the proof of this theorem, 
which describes properties of the index process in reversed time. 

Our general proof technique is as follows. Recall that $H_{t,s}$ is the event that the ancestral 
process' last visit to the reference is at time $s$. We then partition this event 
into cases where $n^\star(t) = n$ for $n \in \cbra{0, 1, \ldots, N}$. 
After this, we make use of a time reversal (\cref{lem:nonreversible_index_process}) 
to state probabilities in terms of a forward-time index process that is always 
\emph{initialized} at chain $N$ (instead of \emph{terminating} at chain $N$).  
This results in events of the form: the index process starts at chain $N$, hits the reference 
for the first time at time $t-s$, and then ends at chain $n$. 
However, by summing over the partitioned events with respect to the termination point,
we conclude that the probability of $H_{t,s}$ is equal to the probability of 
starting at chain $N$ and then hitting the reference at time $t-s$ for the first time. 
Summing over times $s$, we have that the probability of $H_t$ is the same as 
the probability of ever hitting the reference before time $t$, starting from chain $N$. 

Recall that for the desired result, $t \geq 1$ is fixed.
We note that if the $n^\text{th}$ index process is in initialized according to 
$(I_0^n, \eps_0^n) = (n, (-1)^n)$, then if $I_t^n = N$, we must necessarily have 
that $\eps_t^n = d(t,N)$, where
\[
  d(t, N)
  &=
  \begin{cases}
    1, & \text{$t$ and $N$ are both even or odd} \\
    -1, & \text{otherwise}.
  \end{cases}
\]

Formalizing the above argument, we have
\[
  &\Pr_\text{NRPT}(H_t) \\
  &= \sum_{s=0}^{t-1} \Pr_\text{NRPT}(H_{t,s}) \\
  &= \sum_{s=0}^{t-1} \Pr_\text{NRPT}(I_{s}^{n^\star(t)} = 0, \, 
    I_{s'}^{n^\star(t)} > 0 \text{ for } 0 \leq s < s' \leq t) \\
  &= \sum_{s=0}^{t-1} \sum_{n=0}^N \Pr_\text{NRPT}(I_t^n = N, \, I_{s}^n = 0, \, 
    I_{s'}^n > 0 \text{ for } 0 \leq s < s' \leq t) \qquad \text{(disjoint union)} \\
  &= \sum_{s=0}^{t-1} \sum_{n=0}^N \Pr_\text{NRPT}(I_t = N, \, I_{s} = 0, \, 
    I_{s'} > 0 \text{ for } 0 \leq s < s' \leq t \mid (I_0, \eps_0) = (n, (-1)^n)) \\
  &= \sum_{s=0}^{t-1} \sum_{n=0}^N \Pr_\text{NRPT}((I_t, \eps_t) = (N, d(t,N)), \, I_{s} = 0, \, 
    I_{s'} > 0 \text{ for } 0 \leq s < s' \leq t \mid (I_0, \eps_0) = (n, (-1)^n)) \\
  &= \sum_{s=0}^{t-1} \sum_{n=0}^N \Pr_\text{NRPT}((I_t, \eps_t) = (n, -(-1)^n), \, 
    I_{t - s} = 0, \, I_{t - s'} > 0 \text{ for } 0 \leq s < s' \leq t, 
    \mid (I_0, \eps_0) = (N, -d(t,N))) \\
  &= \sum_{s=0}^{t-1} \Pr_\text{NRPT}(I_{t - s} = 0, \, 
    I_{t - s'} > 0 \text{ for } 0 \leq s < s' \leq t, \mid (I_0, \eps_0) = (N, -d(t,N))) \\
  &= \sum_{s=1}^t \Pr_\text{NRPT}(I_s = 0, \, 
    I_{s'} > 0 \text{ for } 0 \leq s' < s \leq t \mid (I_0, \eps_0) = (N, -d(t,N))) \\
  &= \Pr_\text{NRPT}\rbra{\bigcup_{1 \leq s \leq t} 
    I_{s} = 0, \, I_{s'} > 0 \text{ for } 0 \leq s' < s \leq t \mid (I_0, \eps_0) = (N, -d(t,N))} 
    \quad \text{(disjoint)} \\
  &= \Pr_\text{NRPT}\rbra{\min_{0 \leq s \leq t} I_{s} = 0 \mid (I_0, \eps_0) = (N, -d(t,N))}.
\] 
Therefore, 
\[
  \Pr_\text{NRPT}(H_t^c) 
  &= \Pr_\text{NRPT}\rbra{\min_{0 \leq s \leq t} I_s > 0 \mid (I_0, \eps_0) = (N, -d(t,N))} \\
  &\leq \Pr_\text{NRPT}\rbra{\min_{0 \leq s \leq t-1} I_s > 0 \mid (I_0, \eps_0) = (N, -1)} \\
  &\leq \Pr_\text{NRPT}(\tau_N > t-1).
\]
\eprfof

The proof of \cref{thm:erg_NRPT_finite} relies on a result that describes how to 
work with the ancestral process under DEO communication by viewing it as a 
standard index process in reverse time.

\blem
\label{lem:nonreversible_index_process}
Let 
\[
  p 
  = \Pr_{\text{NRPT}}((I_1, \eps_1) = (i_1, e_1), \ldots, (I_t, \eps_t) = (i_t, e_t) 
    \mid (I_0, \eps_0) = (i_0, e_0)).
\]
Then, under \cref{assump:base_erg_collection}.\ref{assump:ELE},
\[
  p 
  = \Pr_{\text{NRPT}}((I_1, \eps_1) = (i_{t-1}, -e_{t-1}), \ldots, (I_t, \eps_t) = (i_0, -e_0) 
    \mid (I_0, \eps_0) = (i_t, -e_t)).
\]
\elem

\bprfof{\cref{lem:nonreversible_index_process}}
We have, by \cref{assump:base_erg_collection}.\ref{assump:ELE},
\[
  p 
  &= \Pr_{\text{NRPT}}((I_1, \eps_1) = (i_1, e_1), \ldots, (I_t, \eps_t) = (i_t, e_t) 
    \mid (I_0, \eps_0) = (i_0, e_0)) \\
  &= \prod_{s = 1}^t \Pr_\text{NRPT}((I_s, \eps_s) = (i_s, e_s) 
    \mid (I_{s-1}, \eps_{s-1}) = (i_{s-1}, e_{s-1})) 
    \quad \text{(Independence)} \\
  &= \prod_{s = 1}^t \Pr_\text{NRPT}((I_1, \eps_1) = (i_s, e_s) 
    \mid (I_0, \eps_0) = (i_{s-1}, e_{s-1})) 
    \quad \text{(Stationarity)} \\
  &= \prod_{s = 1}^t \Pr_\text{NRPT}((I_1, \eps_1) = (i_{s-1}, -e_{s-1}) 
    \mid (I_0, \eps_0) = (i_s, -e_s)) \\
  &= \prod_{s = 1}^t \Pr_\text{NRPT}((I_{t-s+1}, \eps_{t-s+1}) = (i_{s-1}, -e_{s-1}) 
    \mid (I_{t-s}, \eps_{t-s}) = (i_s, -e_s)) \\
  &= \Pr_{\text{NRPT}}((I_1, \eps_1) = (i_{t-1}, -e_{t-1}), \ldots, (I_t, \eps_t) = (i_0, -e_0) 
    \mid (I_0, \eps_0) = (i_t, -e_t)).
\]
\eprfof

\bprfof{\cref{prop:persistent_walk_bound}}
Fix $t \geq 0$ and $N \geq 1$. Suppose first that $t \geq 2N+1$.
Note that for $t \geq 2N+1$ and any starting position 
$(n, \eps) \in \{0,1,\ldots, N\} \times \{-1, 1\}$, 
we can reach $(N, 1)$ by time $t$ with positive probability by travelling in direction $\eps$ 
and reflecting at boundaries. 
In the worst case, we require $2N+1$ units of time. This occurs if we initialize at $(N,-1)$,
in which we need to reach $(0, -1)$, reflect, and then move back up to $(N, 1)$. 
Taking into account this worst-case scenario, we conclude that 
\[
  \Pr_\text{NRPT}(\exists \, 0 \leq s \leq t : (I_s, \eps_s) = (N, 1) 
    \mid (I_0, \eps_0) = (n, \eps)) \geq (1-r)^{2N}, \qquad t \geq 2N+1.
\]
Hence,
\[
  \Pr_\text{NRPT}(\forall \, 0 \leq s \leq t, \, (I_s, \eps_s) \neq (N, 1) 
    \mid (I_0, \eps_0) = (n, \eps)) \leq 1 - (1-r)^{2N}, \qquad t \geq 2N+1.
\]
Now, for any $t \geq 2N+1$, by repeatedly using the fact that segments of the  
index process trajectories are independent (as a consequence of 
\cref{assump:base_erg_collection}.\ref{assump:ELE}), we can split the 
trajectories into blocks of size $2N+1$ and apply the above bound. 
Then, combining these segments we obtain a geometric failure probability, where 
failure here corresponds to not reaching the target chain.
More formally, we find that  
\[
  \Pr_\text{NRPT}(\tau_N > t)
  &= \Pr(\forall \, 0 \leq s \leq t, \, (I_s, \eps_s) \neq (N, 1) 
    \mid (I_0, \eps_0) = (0, 1)) \\
  &\leq [1 - (1-r)^{2N}]^{\floor{t/(2N+1)}}.
\]
The last inequality is obtained by recursively applying the bound 
on the regions $s \in [0, 2N+1], [2N+1, 4N+2], \ldots$.
Note that if $t < 2N+1$, then the inequality trivially holds.
\eprfof

\bprfof{\cref{prop:persistent_walk_exact}}
The proof follows by constructing the transition matrix of the index process, $A_\text{DEO}(N,r)$,
on the expanded state space including the proposed direction of transition $\eps$.
We set the target chain to be an absorbing state. 
One minus the entry in row $2N+1$ and column 
1 of $A_\text{DEO}(N,r)^t$ yields the desired tail probability of the hitting time. 
\eprfof

\subsection{Reversible parallel tempering}

\bprfof{\cref{thm:erg_reversible_finite}}
The proof for the case of RPT follows that of \cref{thm:erg_NRPT_finite}.
All of the arguments are the same, except that the time reversal does not require 
its own lemma due to the inherit reversibility of the index process under the 
SEO communication scheme. 
We have
\[
  \Pr_\text{RPT}(H_t)
  &= \sum_{s=0}^{t-1} \Pr_\text{RPT}(H_{t,s}) \\
  &= \sum_{s=0}^{t-1} \Pr_\text{RPT}(I_{s}^{n^\star(t)} = 0, \, 
    I_{s'}^{n^\star(t)} > 0 \text{ for } 0 \leq s < s' \leq t) \\
  &= \sum_{s=0}^{t-1} \sum_{n=0}^N \Pr_\text{RPT}(I_t^n = N, \, I_{s}^n = 0, \, 
    I_{s'}^n > 0 \text{ for } 0 \leq s < s' \leq t) \qquad \text{(disjoint union)} \\
  &= \sum_{s=0}^{t-1} \sum_{n=0}^N \Pr_\text{RPT}(I_t = N, \, I_{s} = 0, \, 
    I_{s'} > 0 \text{ for } 0 \leq s < s' \leq t \mid I_0 = n) \\
  &= \sum_{s=0}^{t-1} \sum_{n=0}^N \Pr_\text{RPT}(I_t = n, \, 
    I_{t - s} = 0, \, I_{t - s'} > 0 \text{ for } 0 \leq s < s' \leq t, 
    \mid I_0 = N) \\
  &= \sum_{s=0}^{t-1} \Pr_\text{RPT}(I_{t - s} = 0, \, 
    I_{t - s'} > 0 \text{ for } 0 \leq s < s' \leq t, \mid I_0 = N) \\
  &= \sum_{s=1}^t \Pr_\text{RPT}(I_s = 0, \, 
    I_{s'} > 0 \text{ for } 0 \leq s' < s \leq t \mid I_0 = N) \\
  &= \Pr_\text{RPT}\rbra{\bigcup_{1 \leq s \leq t} 
    I_{s} = 0, \, I_{s'} > 0 \text{ for } 0 \leq s' < s \leq t \mid I_0 = N} 
    \qquad \text{(disjoint union)} \\
  &= \Pr_\text{RPT}\rbra{\min_{0 \leq s \leq t} I_{s} = 0 \mid I_0 = N}.
\] 
Therefore, 
\[
  \Pr_\text{RPT}(H_t^c) 
  &= \Pr_\text{RPT}\rbra{\min_{0 \leq s \leq t} I_s > 0 \mid I_0 = N} \\
  &\leq \Pr_\text{RPT}\rbra{\min_{0 \leq s \leq t-1} I_s > 0 \mid I_0 = N} \\
  &\leq \Pr_\text{RPT}(\tau_N > t-1).
\]
Note that the second-to-last inequality is only introduced for consistency with 
the result under the DEO communication scheme, which has to account for the deterministically 
alternating even-odd communication.
\eprfof

\bprfof{\cref{prop:random_walk_bound}}
Suppose first that $t \geq N$.
Note that for $t \geq N$ and any starting position 
$n \in \{0,1,\ldots, N\}$, we can reach chain $N$ by time $t$ with positive probability. 
In the worst case, we initialize at chain $0$, in which case the minimum path length 
of accepting moves in the direction of chain $N$ is the largest. 
Taking into account this worst-case scenario, we conclude that 
for any $t \geq N$ and any $n \in \{0,1,\ldots, N\}$ we have
\[
  \Pr_\text{RPT}(\exists \, 0 \leq s \leq t : I_s = N 
    \mid I_0 = n) \geq (0.5(1-r))^N, \qquad t \geq N.
\]
This is obtained by considering the worst-case scenario in which $I_0 = 0$.
Therefore,
\[
  \Pr_\text{RPT}(\forall \, 0 \leq s \leq t, \, I_s \neq N 
    \mid I_0 = n) \leq 1 - (0.5(1-r))^N, \qquad t \geq N.
\]
Now, for any $t \geq N$, by repeatedly using the fact that segments of the  
index process trajectories are independent (as a consequence of 
\cref{assump:base_erg_collection}.\ref{assump:ELE}), we can split the 
trajectories into blocks of size $N$ and apply the above bound. 
Combining these segments we obtain a geometric failure probability. 
That is, we find that  
\[
  \Pr_\text{RPT}(\tau_N > t)
  &\leq \Pr(\forall \, 0 \leq s \leq t, \, I_s \neq N \mid I_0 = 0) \\
  &\leq [1 - (0.5(1-r))^{N}]^{\floor{t/N}},
\]
where we recursively apply the bound to the intervals 
$s \in [0, N], [N, 2N], \ldots$.
Note that if $0 \leq t < N$, then the inequality trivially holds.
\eprfof

\bprfof{\cref{prop:random_walk_exact}}
The proof follows by constructing the transition matrix of the index process 
(not including $\eps$ momentum states), $A_\text{SEO}(N,r)$,
where the target chain is an absorbing state. One minus the entry in row $N+1$ and column 
1 of $A_\text{SEO}(N,r)^t$ yields the desired tail probability of the hitting time. 
\eprfof

\newpage
\section{Uniform ergodicity with an infinite number of chains}
We begin by proving results that relate 
the rate of ergodicity for an infinite number of chains to the hitting times of 
continuous-time stochastic processes. In the following two subsections we establish bounds 
on the survival functions of the hitting times that correspond to NRPT and RPT.

We employ a modification of Theorem 4 from \cite{syed2021nrpt}, 
which we state below.
The theorem presents a weak convergence result for the scaled index processes. 
We use such a weak convergence result when relating the supremum of the scaled 
index process to the supremum of an appropriate PDMP or Brownian motion under NRPT 
and RPT, respectively. This supremum can be used to examine the hitting time of 
the corresponding index processes, which controls the rate of uniform ergodicity.
Recall from \cref{sec:erg_NRPT_infinite,sec:erg_reversible_infinite} 
the definitions of the continuous-time processes 
$Z_\Lambda(t) = (W_\Lambda(t), \eps_\Lambda(t))$ and $W(t)$, 
which take on values in $[0,1] \times \{-1, 1\}$ and $[0,1]$, respectively.

Let $\mcP_N = \cbra{\beta_0, \beta_1, \ldots, \beta_N}$ 
for $0 = \beta_0 < \beta_1 < \cdots < \beta_N = 1$, 
which we refer to as an \emph{annealing schedule}. 
As in \cite{syed2021nrpt}, we say that $\mcP_N$ is generated by a non-decreasing 
function $G:[0,1] \to [0,1]$ if we have $\beta_n = G(n/N)$ for all $n$. 
In this case, we refer to $G$ as a \emph{schedule generator}.
Let $\cbra{I_t^N}_{t \geq 0}$ be an index process for PT with $N+1$ chains 
and schedule $\mcP_N$
between a reference $\pi_0$ and target $\pi_1$ with $\Lambda(\pi_0, \pi_1) = \Lambda$. 
We define a non-negative and non-decreasing time dilation function 
$M:[0,\infty) \to [0,\infty)$ with $M(0) = 0$. 
We introduce such functions for the purposes of scaling the NRPT and RPT index 
processes appropriately to converge to their corresponding limits.
Finally,
$Z^N(t) = (W^N(t), \eps^N(t))$ with $W^N(t) = I_{M(t)}/N$ and $\eps^N(t) = \eps_{M(t)}$.

\cref{thm:index_weak_convergence} below characterizes the weak convergence of the 
NRPT and RPT index processes with an appropriate scaling of time through the 
time dilation function $M$. 
Our bounds on the rate of ergodicity of NRPT and RPT are based on the hitting times 
of the corresponding index processes and so we are interested in distribution of the 
supremum of these stochastic processes.
The discrete-time index processes we consider are embedded 
into continuous time and, as a result, are piecewise constant. 
A natural space for assessing their weak convergence is given by the space of 
c\`{a}dl\`{a}g functions $D[0,\infty)$ equipped with the Skorokhod metric 
\cite{billingsley1999convergence}. We refer readers to Chapter 3 of \cite{billingsley1999convergence} 
for more information on the Skorokhod space and how to interpret and establish weak convergence
in this space. 
When we refer to weak convergence of a sequence of stochastic processes $H_n(t)$ to $H(t)$, 
where each $t \mapsto H_n(t, \omega)$ and $t \mapsto H(t, \omega) \in D[0,\infty)$, we specifically mean that 
for all bounded and continuous functions $f$---with respect to the Skorokhod topology---that 
\[
  \EE[f(H_n)] \to \EE[f(H)], \qquad n \to \infty.  
\]

\bthm
\label{thm:index_weak_convergence}
Let $\pi_0$ and $\pi_1$ be distributions that satisfy \cref{assump:third_moment}. 
Suppose that 
\cref{assump:base_erg_collection}.\ref{assump:ELE} holds for all $N$.
Then, under the schedules $\mcP_N$ generated by $G$ given by $H^{-1}$ in \cref{eq:schedule_generator}, 
we have:
\begin{enumerate}
  \item For NRPT, if $Z^N(0)$ converges weakly to $Z_\Lambda(0)$
   and $M(t) = \floor{tN}$, then $Z^N$ converges weakly to $Z_\Lambda$ 
   with initial condition $Z_\Lambda(0)$.
  \item For RPT, if $W^N(0)$ converges weakly to $W(0)$
  and $M(t) = \floor{tN^2}$, then $W^N$ converges weakly to $W$ 
  with initial condition $W(0)$.
\end{enumerate}
In both cases, weak convergence occurs in the space of c\`{a}dl\`{a}g functions 
$D[0,\infty)$ equipped with the Skorokhod metric.
\ethm

In \cite{syed2021nrpt}, the authors present a result analogous to \cref{thm:index_weak_convergence} 
except that changes in the index process occur at event times distributed according 
to a Poisson process. 
There, versions of \cref{assump:third_moment,assump:base_erg_collection} 
are invoked.
One reason for doing so is that the resulting Markov process 
is time homogeneous and a generator-based proof technique can be used. 
For our purposes we require statements about the index process 
where events occur at fixed times. 
Further, it is also possible to embed the index process in $C[0,\infty)$ by means of a 
linear interpolation instead of piecewise-constant interpolation. However, this approach 
could result in a non-Markovian process. 

\bprfof{\cref{thm:index_weak_convergence}}

We proceed by proving the result for NRPT and RPT separately. In both cases we use 
the result of Theorem 4 in \cite{syed2021nrpt}, which shows that the weak convergence 
holds for a version of the index process in which the time is scaled according to a 
Poisson process, denoted by $\tilde{Z}^N$. We then show that the difference between 
$Z^N$ and $\tilde{Z}^N$ at finite collections of time points converges to zero in 
probability, establishing the weak convergence of the finite-dimensional 
distributions. We then establish tightness of the processes $Z^N$, concluding weak 
convergence of the entire process. 
That is, for every $\eps > 0$ there exists a compact subset $K \subset D[0,\infty)$ 
such that $\Pr_\text{NRPT}(Z^N \in K) > 1-\eps$.
In what follows, let $\cbra{\tilde{M}(\cdot)}$ be a homogeneous Poisson process with mean 
$m_N$ and define $\tilde{Z}^N(t) = (\tilde{W}^N(t), \tilde\eps^N(t))$, with 
$\tilde{W}^N(t) = I_{\tilde{M}(t)}/N$ and $\tilde\eps^N(t) = \eps_{\tilde M(t)}$.
Here, the process $\tilde Z^N$ is defined on the same probability space as $Z^N$ 
and in such a way that for any given $\omega \in \Omega$ 
the sample paths $t \mapsto \tilde Z^N(t)(\omega)$ 
and $t \mapsto Z^N(t)(\omega)$ are identical up to transformations in the time domain.

For tightness, we take note of the observation that 
we can use arguments for $C[0, t]$ equipped with the uniform norm because 
the limiting processes $Z_\Lambda(s)$ and $W(s)$ are almost-surely continuous in $s$.
That is, because the limiting processes are continuous, 
convergence in $D[0, \infty)$ holds if and only if convergence holds in $D[0, t]$ 
for all $t$ (Theorem 16.7 in \cite{billingsley1999convergence}). 
Furthermore, for the same reason, by the Corollary immediately after Theorem 13.4 
in \cite{billingsley1999convergence}, it suffices to check (7.6) and (7.7) 
in \cite{billingsley1999convergence}---which are conditions in $C[0,t]$---and the weak convergence of the finite-dimensional 
distributions to conclude weak convergence of the processes in $D[0,t]$ for all $t$ 
and hence also $D[0, \infty)$.

\bigskip \noindent \textbf{Case 1: NRPT} \\
For NRPT we take $m_N = N$. If $\tilde{Z}^N(0) = Z^N(0)$ converges weakly to $Z_\Lambda(0)$, 
it is established in Theorem 4 of \cite{syed2021nrpt} that $\tilde{Z}^N$ converges 
weakly to $Z_\Lambda$ with the initial condition $Z_\Lambda(0)$ in $D[0,\infty)$. 
It remains to be shown that all finite-dimensional marginals of $\tilde{Z}^N(t) - Z^N(t)$ 
converge to zero in probability.
First note that for any fixed $t \in [0, \infty)$, and any $\alpha \in (0.5, 1)$ we have 
\[
  \Pr_\text{NRPT}\rbra{\abs{\tilde{Z}^N(t) - Z^N(t)} > \eps}
  &= (S1) + (S2),
\]
where 
\[ 
  (S1) &= \Pr_\text{NRPT}\rbra{\abs{\tilde{M}(t) - \floor{Nt}} \leq N^\alpha} \cdot
    \Pr\rbra{\abs{\tilde{Z}^N(t) - Z^N(t)} > \eps \mid \abs{\tilde{M}(t) - \floor{Nt}} \leq N^\alpha} \\
  (S2) &= \Pr_\text{NRPT}\rbra{\abs{\tilde{M}(t) - \floor{Nt}} > N^\alpha} \cdot
    \Pr\rbra{\abs{\tilde{Z}^N(t) - Z^N(t)} > \eps \mid \abs{\tilde{M}(t) - \floor{Nt}} > N^\alpha}, 
\]
as a consequence of the law of total probability, conditioning on  
$\cbra{\abs{\tilde{M}(t) - \floor{Nt}} \leq N^\alpha}$ and 
$\cbra{\abs{\tilde{M}(t) - \floor{Nt}} > N^\alpha}$.
Because $\alpha \in (0.5, 1)$, we have $(S2) \to 0$ as $N \to \infty$ by an 
appropriate bound on the tails of a Poisson distribution.
That is, because $\alpha > 0.5$, $N^\alpha$ is asymptotically approximately 
in the tails of a shifted Poisson distribution with standard deviation proportional to $N^{1/2}$. 
More formally, 
\[
  \Pr_\text{NRPT}(|\tilde M(t) - \floor{N t}| > N^\alpha) 
  &\leq \Pr_\text{NRPT}\left(\abs{\tilde M(t) - Nt} > N^\alpha-1\right) \\  
  &\leq \frac{\EE[\abs{\tilde M(t) - Nt}^2]}{(N^\alpha-1)^2} \\ 
  &= \frac{Nt}{(N^\alpha-1)^2} \\
  &\to 0,
\]
as $N \to \infty$.
For $(S1)$, we note that the scaled index process for NRPT can change at a rate of at 
most one, and so the largest deviation of $(I_{\tilde M(t)} - I_{\floor{Nt}})/N$ is 
bounded by $(\tilde M(t) - \floor{Nt})/N$. Therefore,
\[
  &\Pr_\text{NRPT}\rbra{\abs{\tilde{Z}^N(t) - Z^N(t)} > \eps \mid \abs{\tilde{M}(t) - \floor{Nt}} \leq N^\alpha} \\
  &{\quad}\leq \Pr_\text{NRPT}\rbra{\abs{\tilde{M}(t) - \floor{Nt}}/N > \eps \mid \abs{\tilde{M}(t) - \floor{Nt}} \leq N^\alpha} \\ 
  &{\quad}\leq \Pr_\text{NRPT}(N^{\alpha-1} > \eps) \\
  &{\quad}\to 0,
\]
as $N \to \infty$. 
By a union bound this result also holds for any finite collection of times $t$ and hence 
we establish the convergence of all finite-dimensional marginals of the stochastic process.

We now establish tightness of the process $Z^N$.
Without loss of generality we prove tightness in $D[0,1]$ and the result also holds for $D[0,t]$.
Define the modulus of continuity for any $0 < \delta \leq 1$, and any process $z(t)$,
\[
  w_z(\delta) = \sup_{\abs{s-t} \leq \delta} \abs{z(s) - z(t)}. 
\] 
Then, by the Corollary immediately after Theorem 13.4 in \cite{billingsley1999convergence},
to establish that the sequence of probability measures $\mcL(Z^N)$ is 
tight in $D[0,1]$, we check the two conditions: 
\renewcommand{\theenumi}{(\roman{enumi})}
\begin{enumerate}
  \item For every $\eta > 0$ there exists an $a$ and $N_0$ such that for all $N \geq N_0$, 
    we have 
    \[
      \Pr_\text{NRPT}(\abs{Z^N(0)} \geq a) \leq \eta.
    \]
  \item For every $\eta, \eps > 0$, there is a $0 < \delta < 1$ and $N_0$ such that 
    for all $N \geq N_0$ we have 
    \[
      \Pr_\text{NRPT}(w_{Z^N}(\delta) \geq \eps) \leq \eta.  
    \]
\end{enumerate}
Because $Z^N(0)$ converges weakly to $Z_\Lambda(0)$ in $\reals$, we have that 
(i) holds from tightness of $Z^N(0)$.
To establish (ii), note that 
\[
  w_{Z^N}(\delta) 
  = \sup_{\abs{s-t} \leq \delta} \abs{Z^N(t) - Z^N(s)}
  \leq \sup_{\abs{s-t} \leq \delta} \frac{1}{N} m_N \abs{t-s}
  \leq \delta.
\]
Therefore, (ii) also holds for sufficiently small $\delta$, 
and we have that $\cbra{\mcL(Z^N)}_{N=1}^\infty$ is tight in $D[0,1]$.
We then apply the same arguments to conclude tightness in $D[0,t]$.
By the remark at the beginning of the proof, due to the continuity of the limit 
process, we establish weak convergence in the Skorokhod space $D[0,\infty)$.

\bigskip \noindent \textbf{Case 2: RPT} \\
In the case of RPT, we follow a similar argument as above. 
When $m_N = N^2$, Theorem 4 of \cite{syed2021nrpt} establishes that if 
$\tilde{W}^N(0)$ converges weakly to $W(0)$, 
then $\tilde{W}^N$ converges weakly to $W$ with the initial condition $W(0)$. 
We begin by showing that all finite-dimensional marginals of $\tilde{W}^N - W^N$ 
converge to zero in probability. 
Again, for any fixed $t \in [0, \infty)$, and any $\alpha \in (1, 2)$, we have 
\[
  \Pr_\text{RPT}\rbra{\abs{\tilde{W}^N(t) - W^N(t)} > \eps}
  &= (T1) + (T2),
\]
where
{\small 
\[ 
  (T1) &= \Pr_\text{RPT}\rbra{\abs{\tilde{M}(t) - \floor{N^2 t}} \leq N^\alpha} \cdot
    \Pr_\text{RPT}\rbra{\abs{\tilde{W}^N(t) - W^N(t)} > \eps \mid \abs{\tilde{M}(t) - \floor{N^2 t}} \leq N^\alpha} \\
  (T2) &= \Pr_\text{RPT}\rbra{\abs{\tilde{M}(t) - \floor{N^2 t}} > N^\alpha} \cdot
    \Pr_\text{RPT}\rbra{\abs{\tilde{W}^N(t) - W^N(t)} > \eps \mid \abs{\tilde{M}(t) - \floor{N^2 t}} > N^\alpha}. 
\]
}
Because $\alpha \in (1, 2)$, we have $(T2) \to 0$ by an appropriate bound on the 
tails of a Poisson distribution. 
For $(T1)$, first define $R^N$ to be a simple random walk on $\cbra{0, 1, 2, \ldots, N}$
with initial value $R^N(0) = N \cdot W^N(0)$. 
In this simple random walk we move to an adjacent state with equal probability of $0.5$, 
except at the boundaries $0$ and $N$ where we reflect with probability one. 
Because of this construction, we can define for any non-negative integer $t \geq 0$ 
a random variable $T(t) \leq t$ such that $I_t^N = R_{T(t)}^N$ almost surely. 
Furthermore, we note that for any $t \leq t'$ we have $T(t) \leq T(t')$ 
and $\abs{T(t') - T(t)} \leq \abs{t' - t}$. 
We additionally define an unbounded random walk $R$ with $R(0) = N \cdot W^N(0)$
that is coupled with $R^N$ in a similar way. Then, 
\[
  (T1) 
  &\leq \Pr_\text{RPT}\rbra{\abs{\tilde{W}^N(t) - W^N(t)} > \eps \mid 
    \abs{\tilde{M}(t) - \floor{N^2 t}} \leq N^\alpha} \\ 
  &= \Pr_\text{RPT}\rbra{\abs{I_{\tilde{M}(t)}^N - I^N_{\floor{N^2 t}}} > N \eps \mid 
    \abs{\tilde{M}(t) - \floor{N^2 t}} \leq N^\alpha} \\
  &= \Pr_\text{RPT}\rbra{\abs{R_{T(\tilde{M}(t))}^N - R^N_{T(\floor{N^2 t})}} > N \eps \mid 
    \abs{\tilde{M}(t) - \floor{N^2 t}} \leq N^\alpha} \\
  &\leq \Pr_\text{RPT}\rbra{\abs{R_{T(\tilde{M}(t))} - R_{T(\floor{N^2 t})}} > N \eps \mid 
    \abs{\tilde{M}(t) - \floor{N^2 t}} \leq N^\alpha} \quad \text{(Reflection princ.)} \\
  &= \Pr_\text{RPT}\rbra{\abs{R_{0} - R_{\abs{T(\floor{N^2 t})- T(\tilde{M}(t))}}} > N \eps \mid 
    \abs{\tilde{M}(t) - \floor{N^2 t}} \leq N^\alpha} \\
  &\leq \Pr_\text{RPT}\rbra{\abs{R_{0}^N - R^N_{\ceil{N^\alpha}}} > N \eps} \\
  &\leq N^{-1} \eps^{-1} \cdot \EE\sbra{\abs{R_{0}^N - R^N_{\ceil{N^\alpha}}}} 
    \quad \text{(Markov)} \\
  &= N^{-1+\alpha/2} \frac{\EE\sbra{\abs{R_{0}^N - R^N_{\ceil{N^\alpha}}}}}{\eps N^{\alpha/2}} \\
  &\to 0,
\]
as $N \to \infty$, because $\alpha \in (1,2)$.
By the same argument as for NRPT, we then argue that the finite-dimensional 
marginals converge in probability to zero. 

To establish tightness, we use a similar approach as for NRPT, replacing 
$\Pr_\text{NRPT}$ with $\Pr_\text{RPT}$ in the statements. 
Property (i) immediately follows. 
To establish (ii), in what follows fix $\eta, \eps > 0$. Note that 
\[
  w_{W^N}(\delta) 
  &= \sup_{\abs{s-t} \leq \delta} \abs{W^N(s) - W^N(t)} \\
  &= \sup_{\abs{s-t} \leq \delta} N^{-1} \abs{I^N_{\floor{s N^2}} - I^N_{\floor{t N^2}}} \\
  &= \sup_{\abs{s-t} \leq \delta} N^{-1} \abs{R^N_{T(\floor{s N^2})} - R^N_{T(\floor{t N^2})}} \\
  &\leq \sup_{\abs{s-t} \leq \delta} N^{-1} \abs{R^N_{\floor{s N^2}} - R^N_{\floor{t N^2}}} \\
  &\leq \sup_{\abs{s-t} \leq \delta} \abs{R_{\floor{s N^2}}/N - R_{\floor{t N^2}}/N} \\
  &= w_{B^N}(\delta),
\]
where we define $B^N(t) = R_{\floor{t N^2}}/N$.
Therefore,
\[
  \Pr_\text{RPT}(w_{W^N}(\delta) \geq \eps) 
  &\leq \Pr_\text{RPT}(w_{B^N}(\delta) \geq \eps).
\]
However, because we know that $B^N$ converges weakly to Brownian motion on $\reals$, 
it necessarily follows that we have tightness of the sequence of processes 
$\cbra{B^N}_{N=1}^\infty$ in $D[0, \infty)$ and hence also in $D[0, 1]$.
Define the modulus of continuity for $z \in D[0,1]$ as 
\[
  w'_z(\delta) = \inf_{\cbra{t_i}} \max_{1 \leq i \leq v} 
    \sup_{s, t \in [t_{i-1}, t_i)} \abs{z(s) - z(t)}, 
\]
with the infimum being taken over all $\delta$-sparse sets $\cbra{t_i}$
(i.e., $\min_{1 \leq i \leq v} (t_i - t_{i-1}) > \delta$).
By Theorem 13.2 in \cite{billingsley1999convergence}, tightness in $D[0,1]$ 
implies that for all $\eps > 0$,
\[
  \lim_{\delta \to 0} \limsup_{N} \Pr_\text{RPT}(w'_{B^N}(\delta) \geq \eps) = 0.
\] 
But, by the bound in (12.9) of \cite{billingsley1999convergence}, we also have that
\[
  w_{B^N}(\delta) \leq 2 w'_{B^N}(\delta) + \frac{1}{N}.  
\]
We therefore conclude that 
\[
  \lim_{\delta \to 0} \limsup_{N} \Pr_\text{RPT}(w_{B^N}(\delta) \geq \eps) 
  = \lim_{\delta \to 0} \limsup_{N} \Pr_\text{RPT}(w_{W^N}(\delta) \geq \eps) 
  = 0.
\]
Hence, we establish that property (ii) also holds for $W^N$. 
This establishes tightness and hence weak convergence in the 
Skorokhod space $D[0,1]$. The result for $D[0,t]$ is analogous and 
hence we can also conclude weak convergence in $D[0, \infty)$, 
because the limiting process is almost surely continuous. 
\eprfof

\subsection{Non-reversible parallel tempering}
\bprfof{\cref{thm:erg_NRPT}}
Fix $t \geq 1$. From \cref{thm:erg_NRPT_finite},
\[
  \mathrm{TV}_N (\bar{\mu} \bar{K}_{tN}^\text{NRPT}, \bar{\pi})
  &\leq \Pr_\text{NRPT}(\tau_N > tN-1) \\
  &= \Pr_\text{NRPT}\rbra{\sup_{0 \leq s \leq tN-1} I_{s,N} < N},
\]
where $I_{s,N}$ is an index process with $N$ chains initialized at $(0, +1)$.
We define another index process, $\tilde{I}_{s,N}$, which is also 
initialized at $(0, +1)$, but with $2N+1$ chains instead of $N+1$.
This artificial construction is used to obtain continuity which is needed for 
the continuous mapping theorem below.  
If the rejection rates between chains 
in $\tilde{I}_{s,N}$ are $r_{n,N}$ for $n \in \cbra{0, 1, \ldots, N}$, 
then we construct $\tilde{I}_{s,N}$ so that the $2N$ rejection rates are equal to 
$r_{0,N}, \ldots, r_{N-1,N}, r_{0,N}, \ldots, r_{N-1,N}$. 
Additionally, define the processes $\tilde{Z}_\Lambda$ and $\tilde{W}$ analogous to 
$Z_\Lambda$ and $W$ except that reflections occur at boundaries $\cbra{0, 2}$.
A similar result as in \cref{thm:index_weak_convergence} holds for 
$\tilde{I}_{\floor{sN},N}/N$ and $\tilde{I}_{\floor{sN^2},N}/N$, wherein the 
processes converge to either $\tilde{Z}_\Lambda$ or $\tilde{W}$ 
under the DEO and SEO communication schemes, respectively.

We apply \cref{thm:index_weak_convergence} and its mentioned extension.
Note that the supremum of a process in $D[0,\infty)$ over an interval is a 
continuous function (with respect to the Skorokhod topology over that interval) 
and so by the continuous mapping theorem, under $\Pr_\text{NRPT}$,
\[
  \sup_{\substack{0 \leq s \leq t-1 \\ s \in \reals}} \tilde{I}_{\floor{sN},N}/N 
  \xrightarrow{d} \sup_{0 \leq s \leq t-1} \tilde{Z}_\Lambda(s).
\]
Also note that $\Pr_\text{NRPT}\rbra{\sup_{0 \leq s \leq t-1} \tilde{Z}_\Lambda(s) = 1} = 0$.
Combining these, we have
\[
  \limsup_{N \to \infty} \mathrm{TV}_N(\bar{\mu} \bar{K}_{tN}^\text{NRPT}, \bar{\pi}) 
  &\leq \limsup_{N \to \infty} \Pr_\text{NRPT}\rbra{\sup_{\substack{0 \leq s \leq tN-1 \\ s \in \nats}} 
    I_{s,N}/N < 1} \\
  &= \limsup_{N \to \infty} \Pr_\text{NRPT}\rbra{\sup_{\substack{0 \leq s \leq tN-1 \\ s \in \reals}} 
    I_{\floor{s},N}/N < 1} \\
  &= \limsup_{N \to \infty} \Pr_\text{NRPT}\rbra{\sup_{\substack{0 \leq s \leq t-1/N \\ s \in \reals}} 
    I_{\floor{sN},N}/N < 1} \\
  &\leq \limsup_{N \to \infty} \Pr_\text{NRPT}\rbra{\sup_{\substack{0 \leq s \leq t-1 \\ s \in \reals}}
    I_{\floor{sN},N}/N < 1} \\
  &= \limsup_{N \to \infty} \Pr_\text{NRPT}\rbra{\sup_{\substack{0 \leq s \leq t-1 \\ s \in \reals}}
    \tilde{I}_{\floor{sN},N}/N < 1} \\
  &= \Pr_\text{NRPT}\rbra{\sup_{0 \leq s \leq t-1} \tilde{Z}_\Lambda(s) < 1} 
    \quad \text{(Portmanteau lemma)} \\
  &= \Pr_\text{NRPT}\rbra{\sup_{0 \leq s \leq t-1} Z_\Lambda(s) < 1} \\
  &= \Pr_\text{NRPT}(\tau_\infty > t-1).
\]
\eprfof

\bprfof{\cref{prop:PDMP_hitting_time_loose}}
Our proof can be formalized but here we only offer a brief intuition. 
Consider the PDMP $Z_\Lambda(t) = (W_\Lambda(t), \eps_\Lambda(t))$ 
that starts with value $(0, 1)$ at time $t=0$
and changes direction at exponentially distributed times with rate $\Lambda$ or at 
the boundaries $\cbra{0, 1}$.
Regardless of its position at time $t$, the process $Z_\Lambda$ can reach 
the position $(1, 1)$ within 2 units of time with probability at least $e^{-2\Lambda}$. 
Following the same argument as in the proof of \cref{prop:persistent_walk_bound},
we obtain the desired result.
\eprfof

\bprfof{\cref{thm:PDMP_hitting_time}}
We separate the proof into several steps.
We first write an integral equation for the hitting time of the process $Z_\Lambda$ 
(\cref{lem:integral_eq}).
We then take the Laplace transform of this integral equation (\cref{lem:Laplace_integral_eq}),  
and obtain the final solution by using a Bromwich integral inversion of the Laplace transform 
(\cref{lem:inverse_Laplace}). Combining these intermediate steps yields the desired result.
\eprfof

\subsubsection{Intermediate lemmas for proof of \cref{thm:PDMP_hitting_time}}
Let $\tau_{x \bullet}$ be the random time that it takes for $Z_\Lambda$  
to reach $(1,1)$ starting from position $x \in [0,1]$ 
and moving initially in the direction $\bullet \in \cbra{\uparrow, \downarrow}$.
We first develop an integral equation for the hitting time CDF.

\blem
\label{lem:integral_eq}
$\Pr_\text{NRPT}(\tau_{x\uparrow} > t-x+1)$ is the unique solution of
\[
  \label{eq:integral_eq}
  \begin{aligned}
  f(x, t)
    = &e^{-\Lambda(1-x)} \ind[t< 0] \\
      &+ \frac{1}{2}\Lambda \int^{t-2x}_{t-2} e^{-\Lambda (t-s - x)} f(0, s)\, \dee s \\
      &+ \frac{1}{2}\Lambda^2 \int_x^1 \int^{t}_{t+2(y-1)} e^{-\Lambda (t-s+y - x)} f(y, s)\, \dee s\, \dee y \\
      &+ \frac{1}{2}\Lambda^2 \int_0^x \int^{t+2(y-x)}_{t+2(y-1)} e^{-\Lambda (t-s+y - x)} f(y, s)\, \dee s\, \dee y
  \end{aligned}
\]
in the space of bounded, measurable functions $[0,1]\times \reals \to \reals$ equipped with the supremum norm.
Furthermore, the solution satisfies the following properties:
\bitems
\item For all $x\in[0,1]$, $t < 0$, $f(x,t) = 1$.
\item For all $x\in[0,1]$, $t\geq 0$, $f(x,t)$ is continuous.
\item For all $x\in[0,1]$, $f(x, 0) = 1-e^{-\Lambda(1-x)}$.
\item For all $t\geq 0$, $f(1,t) = 0$.
\eitems
\elem

\bprfof{\cref{lem:integral_eq}}
We use the rule of total probability, conditioning on the time $T$ of the first change in 
direction for the process $Z_\Lambda$; $T$ is distributed according 
to an exponential distribution with rate $\Lambda$ (not including reflections at boundaries). 
For any $0 \leq x \leq 1$ and $t \geq 0$, 
\[
  &\Pr_\text{NRPT}(\tau_{x \uparrow} \leq t) \\ 
  &= e^{-\Lambda(1-x)}\Pr_\text{NRPT}(\tau_{x \uparrow} \leq t \mid T > 1-x) 
    + \int_0^{1-x} \Lambda e^{-\Lambda v} \cdot 
    \Pr_\text{NRPT}(\tau_{x \uparrow} \leq t \mid T = v) \, \dee v.
\]
If a bounce does not occur before $1-x$ time units elapse (i.e., $T > 1-x$), then $\tau_{x\uparrow} = 1-x$.
If a bounce occurs at $T=v < 1-x$ time units, the process moving upwards from $x$ is located at $x+v$
and begins moving downwards. Therefore
\[
\Pr_\text{NRPT}(\tau_{x \uparrow} \leq t \mid T > 1-x) &= \ind[1-x \leq t]\\
\Pr_\text{NRPT}(\tau_{x \uparrow} \leq t \mid T = v) &=\Pr_\text{NRPT}(\tau_{(x+v) \downarrow} \leq t-v),
\]
and using the change of variables $v \to u = x+v$,
\[
  &= e^{-\Lambda(1-x)} \ind[1-x \leq t]
    + \int_x^1 \Lambda e^{-\Lambda (u - x)} \cdot 
    \Pr_\text{NRPT}(\tau_{u\downarrow} \leq t - u + x) \, \dee u.
\] 
Using similar arguments, for any $u\in[0,1]$, $t \geq 0$,
\[
  \Pr_\text{NRPT}(\tau_{u \downarrow} \leq t) 
  &= e^{-\Lambda u} \cdot \Pr_\text{NRPT}(\tau_{0\uparrow} \leq t - u) 
    + \int_0^u \Lambda e^{-\Lambda(u-y)} \cdot 
    \Pr_\text{NRPT}(\tau_{y\uparrow} \leq t - u + y) \, \dee y.  
\]
Substituting the equation for $\Pr_\text{NRPT}(\tau_{u \downarrow} \leq t)$ into the equation for 
$\Pr_\text{NRPT}(\tau_{x \uparrow} \leq t)$ shows that 
$\Pr_\text{NRPT}(\tau_{x \uparrow} \leq t)$ is a bounded function $g:[0,1]\times\reals \to \reals$ that satisfies
\[
g(x, t)
  = &e^{-\Lambda(1-x)} \ind[t+x-1\geq 0] \\
    &+ \Lambda \int_x^1 e^{-\Lambda (2u - x)} g(0, t+x-2u) \, \dee u\\
    &+ \Lambda^2 \int_x^1 \int_0^u e^{-\Lambda (2u - y - x)} g(y, t+x-2u+y) \, \dee y \, \dee u.
\]
Let $f : [0,1]\times \reals \to \reals$ be defined via $g(x,t) = 1-f(x,t+x-1)$ where $g$ is a solution to \cref{eq:integral_eq}; thus
$f(x,t) = 1-g(x, t-x+1)$ and therefore $f$ and $g$ uniquely determine each other. We rewrite the integral equation \cref{eq:integral_eq} in terms of $f$,
and change the coordinates $(x, t) \to (x, t+x-1)$ to find that
$\Pr_\text{NRPT}(\tau_{x\uparrow} > t-x+1)$ satisfies
\[
1-f(x, t)
  = &e^{-\Lambda(1-x)} \ind[t\geq 0] \\
    &+ \Lambda \int_x^1 e^{-\Lambda (2u - x)} (1-f(0, t-2u))\, \dee u\\
    &+ \Lambda^2 \int_x^1 \int_0^u e^{-\Lambda (2u - y - x)} (1-f(y, t-2u+2y))\, \dee y \, \dee u.
\]
Evaluating the constant integrands, followed by the transformation of variables $u\to s = t-2u$ in the first integral and $u\to s = t-2u+2y$ in the second integral,
yields \cref{eq:integral_eq}.

To show that this equation has a unique solution,
note that we can rewrite it in the form $f = A(f)$ for an operator $A$ on the space of bounded, measurable, real-valued functions
$[0,1]\times\reals \to \reals$ equipped with the uniform norm. But for any two functions $f, g$ in this space, 
\[
  &\|A(f) - A(g)\|_\infty\\
  &\leq \Biggl|\!\Biggl| e^{\Lambda(x-t)} 
  \Bigl(\frac{\!\Lambda}{2} \int^{t-2x}_{t-2} e^{\Lambda s} \dee s \\
    &+ \frac{\Lambda^2}{2} \! \sbra{\int_x^1 \! \int^{t}_{t+2(y-1)} e^{-\Lambda (y-s)} \dee s \dee y
    + \int_0^x \! \int^{t+2(y-x)}_{t+2(y-1)} e^{-\Lambda (y-s)} \dee s \dee y}
  \Bigr)\Biggr|\!\Biggr|_\infty \cdot\|f-g\|_\infty\\
  &= \left\|1 - e^{\Lambda(x-1)}\right\|_\infty\|f-g\|_\infty\\
  &= (1-e^{-\Lambda})\|f-g\|_\infty.
\]
Therefore, $A$ is a contraction, and the fixed point solution is unique by the Banach fixed point theorem.
The Banach fixed point theorem also guarantees that initializing $f_0$ and iterating $f_{n+1} = A(f_n)$ will 
yield a sequence $f_0, f_1, \dots$ that converges uniformly to
the solution $f = A(f)$. If we initialize $f_0(x,t) = 1$, then by inspection of \cref{eq:integral_eq}, for 
all $n\in \nats$, $f_n(x, t) = 1$ for $t < 0$, and hence the solution satisfies this as well.
Similarly, by inspection, if $f_n$ has a single jump discontinuity on the line $t = 0$, but is continuous otherwise, then $f_{n+1}$
has the same structure. 
Therefore $f_n$ is continuous for $t\geq 0$ and converges uniformly to $f$ on that set,
and hence $f$ is continuous for $t\geq 0$. 
The last two statements about $f(x,0)$ and $f(1,t)$ arise by inspection of \cref{eq:integral_eq}.
\eprfof 

By (formally) differentiating the integral equation in \cref{lem:integral_eq}, we can obtain a PDE system,
\[
\begin{aligned}
  f_{xx} + 2 f_{xt} - 2\Lambda f_t &= 0, \quad t \geq 0\\
  f(x, 0) &= 1 - e^{-\Lambda(1-x)}, \quad x \in [0,1] \\
  f(1, t) &= 0, \quad t \geq 0 \\
  f_x(0, t) &= 0, \quad t \geq 0\\
  f(x, t) &= 1, \quad t < 0.
  \end{aligned} \label{eq:transformed_diff_eq} 
\] 
This equation can be shown to have a unique solution by considering the difference $u = f_1 - f_2$ of two solutions $f_1, f_2$, 
and showing that the energy integral 
\[
E(t) = \int_0^1 (\Lambda u(x,t) - u_x(x,t))^2 \dee x
\]
satisfies $E(0) = 0$, $E(t) \geq 0$,
and $\der{E}{t} \leq 0$. One can (again, formally) solve this PDE in $x,t$ by taking a Laplace transform along $t \in [0,\infty)$, 
which yields a 2nd order linear PDE in $x$ that can be solved using standard techniques.
However, it is challenging to derive this PDE and prove uniqueness of solutions rigorously 
due to technical difficulties related to interchange of derivatives and integrals in the integral equation \cref{eq:integral_eq}.

Instead, we take the Laplace transform of the integral equation in
\cref{lem:integral_eq} itself, and solve for the Laplace transform.  This
approach yields the same solution as found formally via the PDE in
\cref{eq:transformed_diff_eq}, but is easier to make rigorous.  For $x \in
[0,1]$ and $z \in \comps$, define the Laplace transform of
$\Pr_{\text{NRPT}}(\tau_{x\uparrow} > t-x+1)$ as
\[
  \label{eq:joint_Laplace}
  F(x,z) = \int_0^\infty e^{-zt} \cdot \Pr_{\text{NRPT}}(\tau_{x\uparrow} > t-x+1)  \, \dee t,
\]
wherever the integral is defined and finite. 

\blem 
\label{lem:Laplace_integral_eq} 
For $\Lambda \geq 1$, on the region 
$\cbra{x\in[0,1], \, z\in \comps : z\neq 0, \, \text{Re}(z) > -\frac{1}{\Lambda+\sqrt{2}}}$, 
we have that $F(x, \cdot)$ has an analytic continuation, which has the form
\[
  \label{eq:F_Laplace_integral_soln}
  F(x,z) &= \frac{D(1, z) - e^{z(1-x)}D(x,z)}{z D(1, z)},
\]
where
\[
  D(x, z) &= \cosh(x r(z)) + \frac{z}{r(z)}\sinh(x r(z)), \qquad 
  r(z) = \sqrt{(z+\Lambda)^2-\Lambda^2}.
\]
Furthermore, for all $x\in[0,1]$, the rightmost pole of $F$ 
on the negative real axis is at a point $z^\star \leq -\frac{1}{\Lambda+\sqrt{2}}$.
\elem 

\bprfof{\cref{lem:Laplace_integral_eq}} 
First, note that for $\text{Re}(z) \geq 0$,
\[
\left|F(x,z)\right| \leq \int_0^\infty \Pr_{\text{NRPT}}(\tau_{x\uparrow} > t-x+1)  \, \dee t
\leq \int_0^\infty \Pr_{\text{NRPT}}(\tau_{0\uparrow} > t)  \, \dee t = \EE_{\text{NRPT}}[\tau_{0\uparrow}] < \infty,
\]
where the finiteness of the final term follows by \cref{prop:PDMP_hitting_time_loose}.
Therefore, $F$ is uniformly bounded on the region $[0,1]\times\{z \in \comps : \text{Re}(z) \geq 0\}$.
Integrating \cref{eq:integral_eq}, and using the shorthand $f(x,t)$ for $\Pr_{\text{NRPT}}\left(\tau_{x\uparrow} > t-x+1\right)$,
we obtain
\[
  \begin{aligned}
  F(x,z)
  = &\int_0^\infty e^{-\Lambda(1-x)} \ind[t< 0] e^{-zt} \, \dee t\\
    &+ \frac{1}{2}\Lambda \int_0^\infty \int^{t-2x}_{t-2} e^{-\Lambda (t-s - x)} 
      f(0, s) e^{-zt} \, \dee s \, \dee t\\
    &+ \frac{1}{2}\Lambda^2 \int_0^\infty \int_x^1 \int^{t}_{t+2(y-1)} 
      e^{-\Lambda (t-s+y - x)} f(y, s) e^{-zt} \, \dee s\, \dee y \, \dee t\\
    &+ \frac{1}{2}\Lambda^2 \int_0^\infty \int_0^x \int^{t+2(y-x)}_{t+2(y-1)} 
      e^{-\Lambda (t-s+y - x)} f(y, s) e^{-zt} \, \dee s\, \dee y \, \dee t.
  \end{aligned}
\]
The first term vanishes. For the second, by the monotonicity of $f$ in its second argument
and the fact that $f(x,t) = 1$ for $t<0$,
\[
  \int_0^\infty \! \int^{t-2x}_{t-2} \left|e^{-\Lambda (t-s - x)}  f(0, s) e^{-zt}\right| \, \dee s \, \dee t
  &\leq \int_0^\infty \! f(0, t-2) e^{-\text{Re}(z)t} \int^{t-2x}_{t-2} e^{-\Lambda (t-s - x)} \, \dee s \, \dee t \\
  &\leq c \int_0^\infty f(0, t-2) \, \dee t < \infty,
\]
for some $0 \leq c < \infty$ independent of $x$ and $z$.
The result is therefore finite and uniformly bounded for all $x \in [0,1]$ and $z \in \comps$ with $\text{Re}(z) \geq 0$.
The remaining two terms can be shown to satisfy the same property using the same technique.
Therefore, interchange of integrals is justified on this region via Fubini's theorem. By appropriate
transformations of variables, integral interchange, and noting that $f(x,t) = 1$ for $t<0$, we find
that for all $x\in[0,1]$, and on $\cbra{z\neq0, \text{Re}(z) \geq 0}$, we have
$F(x,z)$ satisfies the integral equation
\[
  \label{eq:Laplace_integral_eq}
  F(x,z) = \frac{1}{2} \sbra{\Lambda (T1) + \Lambda^2 (T2) + \Lambda^2 (T3)}, \qquad x \in [0,1],
\]
where 
\[
  (T1) &= F(0,z) \int_{-2}^{-2x} e^{\Lambda(u+x) + uz} \, \dee u 
    - \frac{1}{z} \int_{-2}^{-2x} e^{\Lambda(u+x) + uz} (1-e^{-uz}) \, \dee u  \\
  (T2) &= \int_x^1 \int_{2(y-1)}^0 F(y,z) e^{\Lambda(x-y+u)+uz} \, \dee u \, \dee y \\ 
          &{\qquad}- \frac{1}{z} \int_x^1 \int_{2(y-1)}^0 e^{\Lambda(x-y+u)+uz} (1-e^{-uz}) \, \dee u \, \dee y \\
  (T3) &= \int_0^x \int_{2(y-1)}^{2(y-x)} F(y,z) e^{\Lambda(x-y+u)+uz} \, \dee u \, \dee y \\
          &{\qquad}- \frac{1}{z} \int_0^x \int_{2(y-1)}^{2(y-x)} 
          e^{\Lambda(x-y+u)+uz} (1-e^{-uz}) \, \dee u \, \dee y.
\]
Fix $z \in \comps$, $z\neq 0, \text{Re}(z) \geq 0$; \cref{eq:Laplace_integral_eq} can be viewed
as an integral equation for measurable functions $[0,1]\to\comps$.
We can then rewrite \cref{eq:Laplace_integral_eq} as a fixed-point equation $\hf = A\hf$ for some mapping
on the space of such functions $\hf$. Further, for any two such functions $\hf, \hg$,
\[
  &\|A \hf(\cdot, z) - A \hg(\cdot, z)\|_\infty \\
  &\leq \frac{\Lambda e^{x\Lambda}}{2} \Biggl|\!\Biggl|
    \int_{-2}^{-2x} \abs{e^{u\Lambda + uz}} \, \dee u 
    + \Lambda\int_x^1 \int_{2(y-1)}^{0} \abs{e^{u\Lambda + uz - \Lambda y}}\, \dee u \, \dee y \\
    &{\qquad}+ \Lambda\int_0^x \int_{2(y-1)}^{2(y-x)} \abs{e^{u\Lambda + uz - \Lambda y}} \, \dee u \, \dee y 
    \Biggr|\!\Biggr|_\infty \cdot \|\hf(\cdot, z) - \hg(\cdot, z)\|_\infty.
\]
The norms in the first term depend only on $\text{Re}(z)$, and since $u \leq 0$ and $\text{Re}(z) \geq 0$, 
we can maximize them by setting $z=0$. We then evaluate the integrals to find that
\[
  \|A \hf(\cdot, z) - A \hg(\cdot, z)\|_\infty 
  &\leq (1-e^{-\Lambda(1-x)}) \|(\hf-\hg)(\cdot,z)\|_\infty
\leq (1-e^{-\Lambda})\|(\hf-\hg)(\cdot,z)\|_\infty,
\] 
and hence $A$ is a contraction. Therefore, by the Banach fixed point theorem, for each 
$z \in \comps$ with $z\neq 0, \text{Re}(z) \geq 0$, we have 
$F(\cdot,z)$ is the unique bounded function $[0,1]\to\comps$ that solves \cref{eq:Laplace_integral_eq}.
By substitution into \cref{eq:Laplace_integral_eq}, we find that for all $z\in\comps$ such that $z\neq 0, \text{Re}(z) \geq0$,
the function
\[
  \tdF(\cdot, z) : [0,1] \to \comps, \quad \tdF(x, z)  = \frac{D(1, z) - e^{z(1-x)}D(x,z)}{z D(1, z)}
\]
solves \cref{eq:Laplace_integral_eq}. Therefore, if we can show 
$\tdF(\cdot, z)$ is bounded for each $z \neq 0, \text{Re}(z) \geq 0$,
then $F(x,z) = \tdF(x,z)$ on this set. We instead prove a stronger property: that $\tdF(x,z)$ in \cref{eq:F_Laplace_integral_soln} is 
analytic on $[0,1]\times \{z \in \comps : z\neq 0, \text{Re}(z) > -1/(\Lambda+\sqrt{2})\}$, and that $\tdF$ has a pole at $z^\star \leq -1/(\Lambda+\sqrt{2})$
on the negative real line. ($\tilde F$ being analytic implies boundedness once we fix $z$ and consider 
the function over the compact set for $x \in [0,1]$.)
Note that points of non-differentiability of $\tdF(x,z)$ can occur only where there are branch 
cuts due to the square root in $r(z)$, or where $D(1,z) = 0$. 
For the former, note that a Taylor series expansion of $\sinh(xr(z))$ and $\cosh(xr(z))$ in $D(x,z)$ yields 
only even powers of $r(z)$, eliminating branch cuts. For the latter,
note that we can rewrite $D$ as
\[
  D(x,z) = \frac{1}{2} e^{xr(z)}\left(1 + \frac{z}{r(z)}\right) + 
    \frac{1}{2}e^{-xr(z)}\left(1 - \frac{z}{r(z)}\right).
\]
For $z\in\mcR$, where 
$\mcR = \{z\in\comps : \text{Re}(z) > -1/(\Lambda+\sqrt{2}), z\notin (-1/(\Lambda+\sqrt{2}), 0]\}$,
we have that $\text{Re}(z/r(z)) > 0$, and hence  
\[
  \abs{\frac{1}{2} e^{xr(z)}\left(1 + \frac{z}{r(z)}\right)} 
  > \abs{\frac{1}{2}e^{-xr(z)}\left(1 - \frac{z}{r(z)}\right)}.
\]
Furthermore, each of the expressions
\[
  \frac{1}{2} e^{xr(z)}\left(1 + \frac{z}{r(z)}\right), \qquad 
  \frac{1}{2}e^{-xr(z)}\left(1 - \frac{z}{r(z)}\right)
\]
are analytic for $z\in\mcR$. Therefore
by Rouch\'e's theorem, $D$ has the same number of zeros
as the function
\[
 e^{xr(z)} \left(1+\frac{z}{r(z)}\right)
\]
on each of a sequence of connected regions $\mcR_n \subsetneq \mcR$ with 
$\cup_{n=1}^\infty \mcR_n = \mcR$. 
For $\Lambda \geq 1$, the zeros can only occur where $z=-r(z)$, i.e., $z=0$, 
which is not contained in any $\mcR_n$. Therefore, $D(1,z)$ has no zeros on $\mcR$.

It remains to show that $D(1,z)$ has no zeros when $z \in (-1/(\Lambda+\sqrt{2}), 0)$, and that 
it has a zero at some $z \leq -1/(\Lambda+\sqrt{2})$.
To prove this, note that for $-2\Lambda < z < 0$, 
\[
  r(z) = i \cdot \sqrt{\abs{z^2 + 2 \Lambda z}},
\]
and so
\[
  \cosh(r(z)) &= \cos(\sqrt{\abs{z^2 + 2 \Lambda z}}) \\
  \sinh(r(z)) &= i \cdot \sin(\sqrt{\abs{z^2 + 2 \Lambda z}}) \\
  \frac{z}{r(z)} &= \frac{i}{\sqrt{\abs{1 + 2\Lambda/z}}}.
\]
Therefore, $D(1, z) = 0$ on $-\Lambda \leq z < 0$ if and only if
\[
  \cosh(r(z)) + \frac{z}{r(z)} \sinh(r(z)) &= 0 \\
  \iff \tan(\sqrt{\abs{z^2+2\Lambda z}}) &= \sqrt{\abs{1 + 2\Lambda/z}} \\
  \iff  \exists \, n \in \mathbb{Z} \text{ s.t. } \sqrt{\abs{z^2+2\Lambda z}} + \pi \cdot n 
    &= \arctan(\sqrt{\abs{1 + 2\Lambda/z}}).
\]
For the case $n=0$, we solve
\[
  \sqrt{\abs{z^2+2\Lambda z}}  = \arctan(\sqrt{\abs{1 + 2\Lambda/z}}).
\]
The left side of this equation is monotonically decreasing from $\Lambda$ to 0 for $-\Lambda \leq z < 0$,
while the right side is monotonically increasing from $\arctan(1)$ to $\pi/2$ for $-\Lambda \leq z < 0$.
Since $\arctan(1) < 1 \leq \Lambda$, there is precisely one solution to the equation on $-\Lambda \leq z < 0$.
By substitution, we see that at $z=-1/(\Lambda+\sqrt{2})$, the left hand side is bounded above by the right, and
hence the solution must occur at $z\leq -1/(\Lambda+\sqrt{2})$.

Finally, there can be no solution when $n>0$ because the left side of the equation is bounded below by $\pi$,
while the right side is bounded above by $\pi/2$, and there can similarly be no solution for $z > -1/(\Lambda+\sqrt{2})$
when $n < 0$ because we already know the left side is bounded above by the right on that region.
Therefore, $D(1,z)$ has no zeros for $z\in(-1/(\Lambda+\sqrt{2}), 0)$, and hence $\tdF$ is analytic on 
$[0,1]\times \{z\in\comps: \text{Re}(z) > -1/(\Lambda+\sqrt{2}), z\neq 0\}$, and the proof is complete.
\eprfof

\blem
\label{lem:inverse_Laplace}
For all $\Lambda \geq 1$, $0 < \gamma < \frac{1}{\Lambda+\sqrt{2}}$, and $t > 0$,
\[
  \Pr_\text{NRPT}(\tau_\infty > t+1) \leq C(\gamma, \Lambda) \exp(-\gamma t),
\]
where
\[
  C(\gamma, \Lambda) &= \frac{1}{\pi} \sup_{t\geq 0} \lim_{R \to \infty} \int_{0}^R \text{Re}\left(e^{ixt} F(-\gamma + ix) \right)\dee x.
\]
\elem

\begin{figure}
  \includegraphics[width=0.5\textwidth]{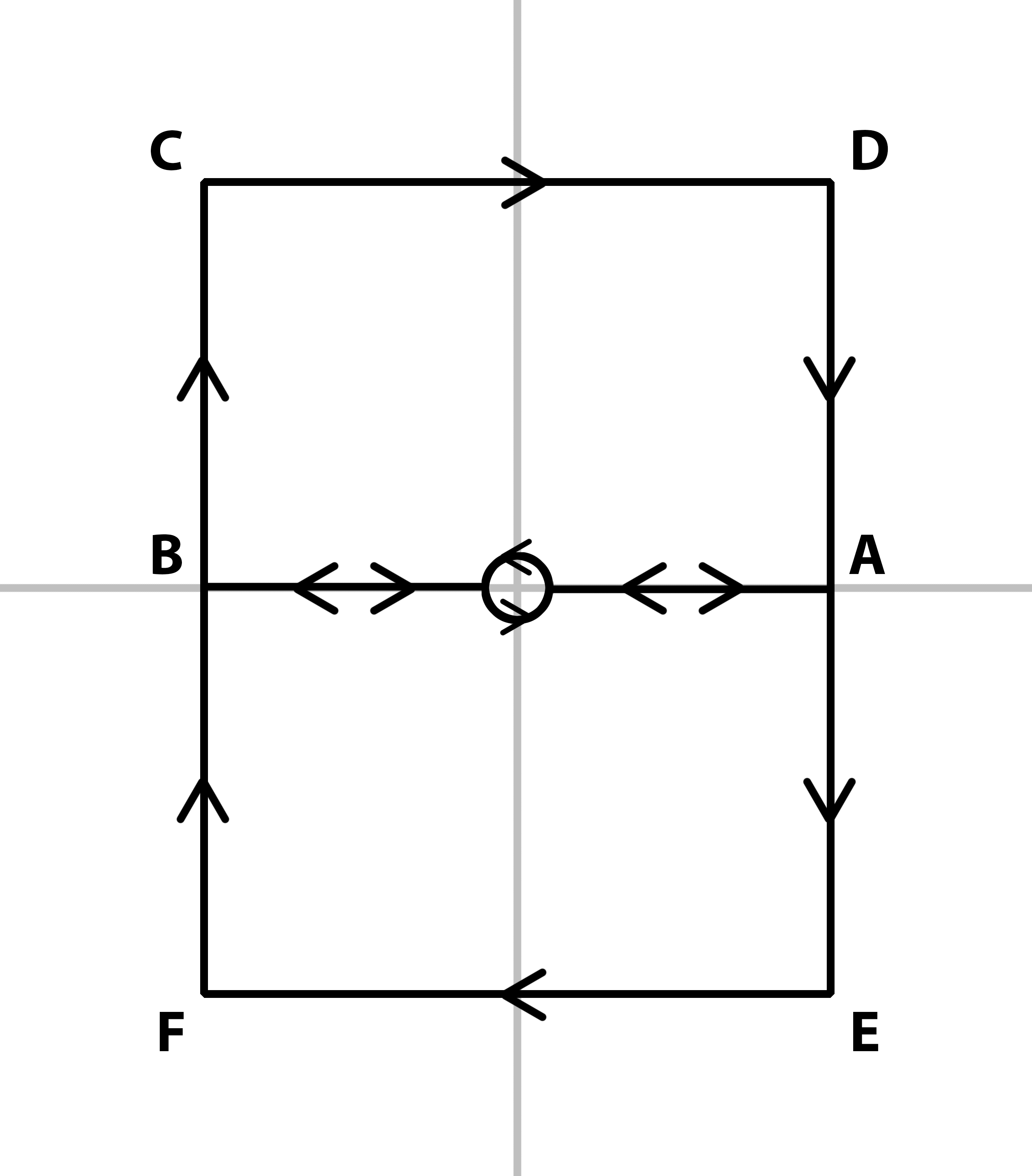}
  \caption{The contour of integration used to move the integration line to the left 
  for the Bromwich inversion formula. 
  Around $z = 0$ we integrate around a ball with radius $1/R$.
  The length $AD = R$ and $A = \gamma$, while $B = -\gamma$.}
  \label{fig:contour}
\end{figure}

\bprfof{\cref{lem:inverse_Laplace}}
Fix $\Lambda \geq 1$. By substituting $x=0$ into the result of \cref{lem:Laplace_integral_eq},
we find that the analytic continuation of the Laplace transform of $\Pr_{\text{NRPT}}(\tau_\infty > t+1)$ is
\[
  F(z) = F(0,z) = \frac{D(1,z) - e^{z}}{z D(1,z)}
\]
for $z\in\comps$ such that $z\neq 0$, $\text{Re}(z) > -\frac{1}{\Lambda+\sqrt{2}}$.
$F$ is analytic on this region, and its rightmost pole occurs on the negative real 
axis at $z \leq -\frac{1}{\Lambda+\sqrt{2}}$.
One can verify the conditions in \cref{lem:Bromwich_inversion} to apply the 
Bromwich inversion formula for any arbitrary $0 < \gamma < \frac{1}{\Lambda+\sqrt{2}}$:
\[
  \Pr_{\text{NRPT}}(\tau_\infty > t+1) &= \lim_{R\to\infty} \frac{1}{2\pi i} \int_{\gamma-iR}^{\gamma+iR} e^{z t} F(z) \dee z.\label{eq:brom1}
\]
The next step is to move the line of integration to have real component $-\gamma$ instead of $\gamma$, which
provides the means to bound the rate of decay of $\Pr_{\text{NRPT}}$. For this purpose we use
the Cauchy integral theorem (see \cref{fig:contour}). In particular, 
using the fact that $F$ is analytic when $\text{Re}(z) > -1/(\Lambda+\sqrt{2})$, $z\neq 0$,
we note that the contour integral over the upper and lower rectangles are both 0:
\[
  \int_{\text{DA}} + \int_{\text{AB}} + \int_{\text{BC}} + \int_{\text{CD}} = 0, \qquad  
  \int_{\text{AE}} + \int_{\text{EF}} + \int_{\text{FB}} + \int_{\text{BA}} = 0.
\]
Adding these two equations together and rearranging yields
\[
  \int_{\text{ED}} &= \int_{\text{FC}} + \int_{\text{CD}}+ \int_{\text{EF}} + \oint_{B_{1/R}(0)}.
\]
The integral along $\text{ED}$ is precisely the Bromwich inversion integral in \cref{eq:brom1} for fixed $R$,
and the integral along $\text{FC}$ is the same but shifted horizontally to lie along the real component $-\gamma$.
The integral on the contour around $z=0$ with radius $1/R$ vanishes by continuity.
We show that the remaining two components converge to 0 as $R\to\infty$. Note that for sufficiently large $|\text{Im}(z)|$,
by \cref{prop:Fnormbound}, there exists $c<\infty$ such that
\[
  |F(z)| \leq \frac{c}{|z|}.
\]
Therefore,
\[
  \max\left\{\left|\int_{\text{EF}}\right|, \left|\int_{\text{CD}}\right|\right\} \leq 2\gamma c/R,
\]
and so taking the limit as $R\to\infty$ yields
\[
  \Pr_{\text{NRPT}}(\tau_\infty > t+1) &= \lim_{R\to\infty} \frac{1}{2\pi i} \int_{-\gamma-iR}^{-\gamma+iR} e^{z t} F(z) \dee z.
\]
Now, using a transformation of variables $z = -\gamma + ix$ to convert the integral to one on the real line,
\[
  \Pr_{\text{NRPT}}(\tau_\infty > t+1) &= \lim_{R\to\infty} \frac{1}{2\pi} \int_{-R}^{R} e^{(-\gamma+ix) t} F(-\gamma+ix) \dee x\\
  &= e^{-\gamma t} \lim_{R\to\infty} \frac{1}{2\pi} \left(\int_{-R}^{0} e^{ixt} F(-\gamma+ix) \dee x+\int_{0}^{R} e^{ixt} F(-\gamma+ix) \dee x\right)\\
  &= e^{-\gamma t}\lim_{R\to\infty} \frac{1}{2\pi} \left(\int_{0}^{R} e^{-ixt} F(-\gamma-ix) \dee x+\int_{0}^{R} e^{ixt} F(-\gamma+ix) \dee x\right)\\
  &= e^{-\gamma t}\lim_{R\to\infty} \frac{1}{\pi} \int_{0}^{R} \text{Re}\left(e^{ixt} F(-\gamma+ix)\right) \dee x\\
  &\leq e^{-\gamma t}\sup_{t\geq 0}\lim_{R\to\infty} \frac{1}{\pi} \int_{0}^{R} \text{Re}\left(e^{ix t} F(-\gamma+ix)\right) \dee x\\
  &= C(\gamma,\Lambda) e^{-\gamma t}.
\]
\eprfof

We state a sufficient set of conditions for the Bromwich inversion formula to hold.
This is a well-known result in complex analysis, but we provide a proof here in terms of Lebesgue 
integrals for completeness.

\blem
\label{lem:Bromwich_inversion}
Let $f:[0, \infty) \to \reals$ be continuously differentiable 
with $\lim_{t \to \infty} f(t) = 0$.
Consider any $\gamma \in \reals$ such that 
\[
  \max \cbra{\int_0^\infty e^{-\gamma t} \abs{f(t)} \, \dee t, 
    \int_0^\infty e^{-\gamma t} \abs{f'(t)} \, \dee t} < \infty,
\]
with $\lim_{t \to \infty} e^{-\gamma t} f(t) = 0$.
Define
\[
  F(s) = \int_0^\infty e^{-st} f(t) \, \dee t,   
\]
for $s \in \comps$ where the integrand above is Lebesgue integrable. 
Then, for all $t > 0$,
\[
  f(t) 
  = \lim_{R \to \infty} 
    \frac{1}{2\pi i} \int_{\gamma - iR}^{\gamma + iR} e^{st} F(s) \, \dee s.
\]
\elem 

\bprfof{\cref{lem:Bromwich_inversion}}
Define 
\[
  S(x) = \frac{1}{2} + \frac{1}{\pi} \int_0^x \frac{\sin y}{y} \, \dee y,
\]
which has the pointwise property for $x \in \reals$ that 
\[
  \lim_{R \to \infty} S(Rx)
  = \begin{cases}
      1, & x > 0 \\
      1/2, & x = 0 \\
      0, & x < 0
    \end{cases}.
\]
From our assumptions of Lebesgue integrability, we also have that 
\[
  \frac{\partial}{\partial u} f(u) e^{-\gamma (u-t)}
  = f'(u) e^{-\gamma (u-t)} - \gamma f(u) e^{-\gamma (u-t)}
\]
is also Lebesgue integrable with respect to $u$.

Then,
\[
  \frac{1}{2\pi i} \int_{\gamma - iR}^{\gamma + iR} e^{st} F(s) \, \dee s 
  &= \frac{1}{2\pi} \int_{-R}^R e^{(\gamma + ix) t} F(\gamma + ix) \, \dee x \\
  &= \frac{1}{2\pi} \int_{-R}^R  
    \int_0^\infty f(u) e^{-(\gamma + ix) u} e^{(\gamma + ix) t} \, \dee u \, \dee x \\
  &= \int_0^\infty f(u) e^{-(\gamma + ix) u} 
    \rbra{ \frac{1}{2\pi} \int_{-R}^R  e^{(\gamma + ix) t} \, \dee x } \, \dee u
    \quad \text{(Fubini)} \\
  &= \int_0^\infty f(u) e^{-\gamma (u-t)} 
    \rbra{ \frac{1}{2\pi} \int_{-R}^R  e^{(t-u) ix} \, \dee x } \, \dee u \\
  &= \int_0^\infty f(u) e^{-\gamma (u-t)} \cdot \frac{\sin(R(u-t))}{\pi (u-t)} \, \dee u \\
  &= \left[f(u) e^{-\gamma (u-t)} \cdot S(R(u-t))\right]_{u=0}^{u=\infty} \\
    &{\quad}- \int_0^\infty \frac{\partial}{\partial u} \rbra{ f(u) e^{-\gamma (u-t)} }
      \cdot S(R(u-t)) \, \dee u \\
  &= f(0) e^{\gamma t} \cdot S(-Rt)
    - \int_0^\infty \frac{\partial}{\partial u} \rbra{ f(u) e^{-\gamma (u-t)} }
      \cdot S(R(u-t)) \, \dee u.
\]
Therefore, 
\[
  \lim_{R \to \infty} \frac{1}{2\pi i} \int_{\gamma - iR}^{\gamma + iR} e^{st} F(s) \, \dee s
  &= \lim_{R \to \infty} 
    - \int_0^\infty \frac{\partial}{\partial u} \rbra{ f(u) e^{-\gamma (u-t)} }
      \cdot S(R(u-t)) \, \dee u \\
  &= - \int_t^\infty \frac{\partial}{\partial u} \rbra{ f(u) e^{-\gamma (u-t)} } \, \dee u
    \quad \text{(Dominated convergence)} \\
  &= -\left[f(u) e^{-\gamma (u-t)}\right]_{u=t}^{u=\infty} \\
  &= f(t).
\]
\eprfof

\newpage 

\newpage 
\subsubsection{Bounds for the uniform constant in \cref{thm:erg_NRPT}}
In this subsection we detail the steps taken to prove the bounds on $C(\Lambda)$. 
We analyze the overall $C(\gamma,\Lambda)$ term, where for $0 < \gamma < 1/(\Lambda+\sqrt{2})$ and $\Lambda \geq 1$,
\[
  C(\gamma, \Lambda) =  
  \frac{1}{\pi} \sup_{t \geq 0} \lim_{R \to \infty} \int_0^R \text{Re}\left(e^{ix t} F(-\gamma+ix)\right) \dee x.
\]

\bprop
\label{prop:C_Lambda_bound}
For $\Lambda \geq 1$, $1/(4\Lambda) \leq \gamma < 1/(\Lambda+\sqrt{2})$, $B \geq \sqrt{6}(\Lambda+1)$, and 
any $0 < \eps \leq 1/(136\Lambda)$, we have $C(\gamma, \Lambda) \geq 0$ and 
\[
  C(\gamma, \Lambda) 
  &\leq \frac{(1+e^{-\Lambda})\left(2+\pi\right)}{\pi}
    + \frac{15 \eps \cdot e^{-\gamma}}{\pi\gamma} + 
    \frac{765 \Lambda e^{-\frac{3\Lambda}{4}}}{\pi B} \\
    &{\quad}+ \frac{4 e^{-\gamma}}{\pi k} \sbra{-\sqrt{k} \log(1-e^{-\sqrt{k}}) + 2 e^{-\sqrt{k}} } \\
    &{\quad}+ \frac{4 e^{-\gamma}}{\pi \gamma k} \sbra{-\sqrt{\eps k} \log(1-e^{-\sqrt{\eps k}}) + 2e^{-\sqrt{\eps k}}},
\]
where 
\[
  k &= (\Lambda-\gamma) \rbra{\sqrt{1+\frac{B^2}{4\Lambda^2}} - \frac{B}{2\Lambda}}.
\]
In particular, for the case where $B = \sqrt{6}(\Lambda+1)$, $\eps = 1/(136\Lambda)$, 
$1/(\Lambda+2) \leq \gamma < 1/(\Lambda+\sqrt{2})$, and $\Lambda \geq 1$, we have 
\[
  C(\gamma, \Lambda) 
  &\leq \frac{1}{\pi} \left[(1+e^{-\Lambda})\left(2+\pi\right)
    + 251 + \frac{30}{136\Lambda} 
    + \frac{765 \Lambda e^{-\frac{3\Lambda}{4}}}{\sqrt{6}(\Lambda+1)} \right] \\ 
  &\leq 106.
\]
\eprop

\bprfof{\cref{prop:C_Lambda_bound}}
We write $\int_0^\infty$ as a shorthand for $\lim_{R \to \infty} \int_0^R$, 
even though the quantities may not necessarily be absolutely (Lebesgue) integrable over $[0, \infty)$. 

For any $B \geq 0$,
\[
  &\int_{0}^{\infty} \text{Re}\left(e^{ix t} F(-\gamma+ix)\right) \dee x\\
  &= \int_{0}^{B} \text{Re}\left(\frac{e^{ix t}}{-\gamma+ix} \left((-\gamma+ix) F(-\gamma+ix) - 1+1\right)\right) \dee x\\
    &{\quad}+\int_{B}^{\infty} \text{Re}\left(\frac{e^{ix t}}{-\gamma+ix} \left((-\gamma+ix) F(-\gamma+ix) - (1-e^{-\Lambda}) + (1-e^{-\Lambda})\right)\right) \dee x\\
  &= (T1) + (T2) + (T3), 
\] 
where 
\[
  (T1) &= \int_{0}^{B} \text{Re}\left(\frac{e^{ix t}}{-\gamma+ix}\right) \dee x + (1-e^{-\Lambda})\int_{B}^{\infty} \text{Re}\left(\frac{e^{ix t}}{-\gamma+ix}\right) \dee x \\
  (T2) &= \int_{0}^{B} \text{Re}\left(\frac{e^{ix t}}{-\gamma+ix} \left((-\gamma+ix) F(-\gamma+ix) - 1\right)\right) \dee x\\
  (T3) &= \int_{B}^{\infty} \text{Re}\left(\frac{e^{ix t}}{-\gamma+ix} \left((-\gamma+ix) F(-\gamma+ix) - (1-e^{-\Lambda})\right)\right) \dee x.
\]
We claim that, under the conditions of the stated result, 
\[
  (T1) &\leq (1+e^{-\Lambda})\left(2+\pi\right) \\ 
  (T2) &\leq \frac{15 \eps \cdot e^{-\gamma}}{\gamma} 
    + \frac{4 e^{-\gamma}}{k} \sbra{-\sqrt{k} \log(1-e^{-\sqrt{k}}) + 2 e^{-\sqrt{k}} } \\
    &{\qquad}+ \frac{4 e^{-\gamma}}{\gamma k} \sbra{-\sqrt{\eps k} \log(1-e^{-\sqrt{\eps k}}) + 2e^{-\sqrt{\eps k}}} \\
  (T3) &\leq \frac{765 \Lambda e^{-\frac{3\Lambda}{4}}}{B}.
\]

For $(T1)$, note that
\[
  \text{Re}\left(\frac{e^{ix t}}{-\gamma+ix}\right) &= \frac{x \sin(tx)}{\gamma^2+x^2} - \frac{\gamma \cos(tx)}{\gamma^2+x^2}
\]
Therefore, for $0 \leq a \leq b$,
\[
  \int_a^b \text{Re}\left(\frac{e^{ix t}}{-\gamma+ix}\right)\dee x
  &= \int_a^b \frac{x \sin(tx)}{\gamma^2+x^2}\dee x - \int_a^b\frac{\gamma \cos(tx)}{\gamma^2+x^2} \dee x\\
  &= \int_{ta}^{tb} \frac{u^2}{t^2 \gamma^2+u^2}\frac{\sin u}{u}\dee u - \int_a^b \frac{\gamma \cos(tx)}{\gamma^2+x^2} \dee x.
\]
We can find local maxima and minima of the integral for $x\geq0$
\[
  0 &= \der{}{x}\int_0^x \frac{u^2}{t^2 \gamma^2+u^2}\frac{\sin u}{u}\dee u\\
  &=\frac{x^2}{t^2 \gamma^2+x^2}\frac{\sin x}{x}\\
  \implies x &= k\pi, \,\, k\in\{0,1,\dots\}
\]
at $x=\pi$ it is a maximum, at $x=2\pi$ it is a minimum, and so on.
The local maxima decrease in value and the local minima increase in value.
Therefore, the largest deviation is on the first interval, and so
\[
  \int_{ta}^{tb} \frac{u^2}{t^2\gamma^2+u^2}\frac{\sin u}{u}\dee u
  &\leq \int_0^\pi \frac{u^2}{t^2\gamma^2+u^2}\frac{\sin u}{u}\dee u\\
  &\leq \frac{\pi^2}{t^2\gamma^2+\pi^2}\int_0^\pi \frac{\sin u}{u}\dee u\\
  &\leq \frac{2\pi^2}{t^2\gamma^2+\pi^2}.
\]
For the second term,
\[
  -\int_a^b \frac{\gamma \cos(tx)}{\gamma^2+x^2} \dee x
  \leq \gamma\int_a^b \frac{1}{\gamma^2+x^2} \dee x
  &= \tan^{-1}\left(\frac{b}{\gamma}\right) - \tan^{-1}\left(\frac{a}{\gamma}\right) 
  \leq \pi.
\]
Combining these, we get for $(T1)$ that for $t \geq 0$, 
\[
  (T1) 
  &=\int_{0}^{B} \text{Re}\left(\frac{e^{ix t}}{-\gamma+ix}\right) \dee x 
    + (1-e^{-\Lambda})\int_{B}^{\infty} \text{Re}\left(\frac{e^{ix t}}{-\gamma+ix}\right) \dee x\\
  &=\int_{0}^{\infty} \text{Re}\left(\frac{e^{ix t}}{-\gamma+ix}\right) \dee x  
    - e^{-\Lambda}\int_{B}^{\infty} \text{Re}\left(\frac{e^{ix t}}{-\gamma+ix}\right) \dee x\\
  &\leq (1+e^{-\Lambda})\left(\frac{2\pi^2}{t^2\gamma^2+\pi^2}+ \pi\right) \\ 
  &\leq (1+e^{-\Lambda})\left(2+\pi\right),
\]
which is independent of $\gamma$.

For $(T2)$, fix $0 < \eps \leq B$, and note that 
\[
  (T2) 
  &= \int_{0}^{B} \text{Re}\left(\frac{e^{ix t}}{-\gamma+ix} \left((-\gamma+ix) F(-\gamma+ix) - 1\right)\right) \dee x \\ 
  &= -\int_0^B\text{Re}\left(\frac{e^{ixt-\gamma+ix}}{(-\gamma+ix)D(1,-\gamma+ix)}\right) \dee x \\ 
  &\leq e^{-\gamma} \int_0^B\frac{1}{\sqrt{\gamma^2+x^2}|D(1,-\gamma+ix)|} \dee x \\ 
  &= e^{-\gamma} \sbra{\int_0^\eps \frac{1}{\sqrt{\gamma^2+x^2}|D(1,-\gamma+ix)|} \dee x 
    + \int_\eps^B \frac{1}{\sqrt{\gamma^2+x^2}|D(1,-\gamma+ix)|} \dee x}.
\] 
Now, on the interval $[0, \eps]$, provided that $\eps \leq 136 \Lambda$, 
we have by \cref{prop:C_Lambda_compact_2} that 
$\abs{D(1,-\gamma+ix)} \geq 0.07$, and so for the first term we have
\[
  e^{-\gamma} \int_0^\eps \frac{1}{\sqrt{\gamma^2+x^2}|D(1,-\gamma+ix)|} \dee x 
  &\leq \frac{15 \eps \cdot e^{-\gamma}}{\gamma}. 
\] 
For the second term, note that by \cref{lem:nonnegrz}, for $z = -\gamma+ix$ we have  
\[
  \abs{D(1,z)}  
  &= \frac{1}{2} \abs{e^{r(z)}\rbra{1+\frac{z}{r(z)}} + e^{-r(z)}\rbra{1 - \frac{z}{r(z)}}} 
  \geq \sinh(\text{Re}(r(z))).
\]
Then, by \cref{lem:rerz}, setting 
\[
  k &= (\Lambda-\gamma) \rbra{\sqrt{1+\frac{B^2}{4\Lambda^2}} - \frac{B}{2\Lambda}},
\]
we have
\[
  |D(1,-\gamma+ix)| 
  &\geq \sinh\rbra{\sqrt{x(-\gamma+\Lambda)} \sqrt{\sqrt{1+\frac{B^2}{4\Lambda^2}} - \frac{B}{2\Lambda}}} \\ 
  &\geq \frac{\exp\left(\sqrt{x(-\gamma+\Lambda)} \sqrt{\sqrt{1+\frac{B^2}{4\Lambda^2}} - \frac{B}{2\Lambda}}\right)-1}{2}.
\]
Therefore, 
\[
  &e^{-\gamma} \int_\eps^B \frac{1}{\sqrt{\gamma^2+x^2}|D(1,-\gamma+ix)|} \dee x \\
  &\leq e^{-\gamma} \int_\eps^B \frac{2}{\sqrt{\gamma^2+x^2}\left[\exp\left(\sqrt{x(-\gamma+\Lambda)} 
    \sqrt{\sqrt{1+\frac{B^2}{4\Lambda^2}} - \frac{B}{2\Lambda}}\right)-1\right]} \dee x.
\]
We now split this term into two parts, integrating over $[\eps, 1]$ and $[B,1]$,
noting that $\eps \leq 1/(136\Lambda) < 1$. Define the polylogarithm function of order 2,
\[
  \text{Li}_2(z) = \sum_{n=1}^\infty \frac{z^n}{n^2}, \qquad \abs{z} < 1.
\]
Note that for real $z \geq 0$, $\text{Li}_2(z)$ is increasing and $\lim_{z \to 0} \text{Li}_2(z) = 0$.
Furthermore, $0 \leq \text{Li}_2(z) \leq 2z$ for $z \geq 0$, because 
\[
  \text{Li}_2(z) - 2z 
  &= -z + \sum_{n=2}^\infty \frac{z^n}{n^2} 
  = z \rbra{-1 + \sum_{n=2}^\infty \frac{z^{n-1}}{n^2}}
  \leq z \rbra{\frac{\pi^2}{6}-2} 
  \leq 0. 
\] 
Then, over the domain $[1,B]$,
\[
  &e^{-\gamma} \int_1^B \frac{2}{\sqrt{\gamma^2+x^2} \left[\exp\left(\sqrt{x k} \right)-1\right]} \dee x \\
  &\leq e^{-\gamma} \int_1^B \frac{2}{\left[\exp\left(\sqrt{x k} \right)-1\right]} \dee x \\
  &= \frac{4 e^{-\gamma}}{k} 
    \sbra{-\sqrt{k}\log(1-e^{-\sqrt{k}}) + \sqrt{Bk} \log(1-e^{-\sqrt{Bk}}) 
    + \text{Li}_2(e^{-\sqrt{k}}) - \text{Li}_2(e^{-\sqrt{Bk}})} \\ 
  &\leq \frac{4 e^{-\gamma}}{k} \sbra{-\sqrt{k} \log(1-e^{-\sqrt{k}}) + \text{Li}_2(e^{-\sqrt{k}}) } \\
  &\leq \frac{4 e^{-\gamma}}{k} \sbra{-\sqrt{k} \log(1-e^{-\sqrt{k}}) + 2 e^{-\sqrt{k}} }
\]
Similarly, over $[\eps, 1]$, we get 
\[
  &e^{-\gamma} \int_\eps^1 \frac{2}{\sqrt{\gamma^2+x^2} \left[\exp\left(\sqrt{x k} \right)-1\right]} \dee x \\
  &\leq \frac{2 e^{-\gamma}}{\gamma} \int_\eps^1 \frac{1}{\left[\exp\left(\sqrt{x k} \right)-1\right]} \dee x \\
  &= \frac{4 e^{-\gamma}}{\gamma k} 
    \sbra{-\sqrt{\eps k}\log(1-e^{-\sqrt{\eps k}}) + \sqrt{k} \log(1-e^{-\sqrt{k}}) 
    + \text{Li}_2(e^{-\sqrt{\eps k}}) - \text{Li}_2(e^{-\sqrt{k}})} \\ 
  &\leq \frac{4 e^{-\gamma}}{\gamma k} \sbra{-\sqrt{\eps k} \log(1-e^{-\sqrt{\eps k}}) + 2e^{-\sqrt{\eps k}}}.
\]
Combining all of the bounds, we get 
\[
  (T2) 
  &\leq \frac{15 \eps \cdot e^{-\gamma}}{\gamma} 
    + \frac{4 e^{-\gamma}}{k} \sbra{-\sqrt{k} \log(1-e^{-\sqrt{k}}) + 2e^{-\sqrt{k}} } \\ 
    &{\quad}+ \frac{4 e^{-\gamma}}{\gamma k} \sbra{-\sqrt{\eps k} \log(1-e^{-\sqrt{\eps k}}) + 2e^{-\sqrt{\eps k}}}.
\]
For the case where $B = \sqrt{6}(\Lambda+1)$, $\eps = 1/(136\Lambda)$,
and $1/(\Lambda+2) \leq \gamma < 1/(\Lambda+\sqrt{2})$, note that
\[
  k &= (\Lambda-\gamma) \rbra{\sqrt{1+\frac{B^2}{4\Lambda^2}} - \frac{B}{2\Lambda}}.
\]
\[
  k 
  \geq \rbra{\Lambda-\frac{1}{\Lambda+\sqrt{2}}} 
    \rbra{\sqrt{1+\frac{6(\Lambda+1)^2}{4\Lambda^2}} - \frac{\sqrt{6}(\Lambda+1)}{2\Lambda}}
  \geq 0.11, 
\]
and 
\[
  k \leq 0.36 \Lambda.
\]
Similarly, 
\[
  \eps k 
  \geq \frac{1}{136} \rbra{1-\frac{1}{\Lambda(\Lambda+\sqrt{2})}} 
    \rbra{\sqrt{1+\frac{6(\Lambda+1)^2}{4\Lambda^2}} - \frac{\sqrt{6}(\Lambda+1)}{2\Lambda}}
  \geq 0.0008, 
\]
and 
\[
  \eps k \leq 0.003.  
\]
Therefore,
\[
  (T2)
  &\leq   
  \frac{15 (\Lambda+2)}{136 \Lambda} 
    + \frac{4}{k} \sbra{-\sqrt{k} \log(1-e^{-\sqrt{k}}) + 2e^{-\sqrt{k}} } \\ 
    &{\quad}+ \frac{4}{\gamma k} \sbra{-\sqrt{\eps k} \log(1-e^{-\sqrt{\eps k}}) + 2e^{-\sqrt{\eps k}}} \\ 
  &\leq \frac{15 (\Lambda+2)}{136 \Lambda} 
    + \frac{4(1.27)}{\sqrt{k}} + \frac{8(0.72)}{k} + \frac{4(3.58) \eps}{\gamma \sqrt{\eps k}} + 
    \frac{8(0.98) \eps}{\gamma \eps k} \\ 
  &\leq \frac{15}{136} + \frac{30}{136 \Lambda} 
    + \frac{4(1.27)}{\sqrt{0.11}} + \frac{8(0.72)}{0.11} + 
    \frac{1+\sqrt{2}}{136} \rbra{\frac{4(3.58)}{\sqrt{0.0008}} + \frac{8(0.98)}{0.0008}} \\ 
  &< 251 + \frac{30}{136\Lambda}.
\]

For $(T3)$, note that by \cref{prop:Fnormbound}, as long as $B \geq \sqrt{6}(\Lambda+1)$,
\[
  (T3)
  &= \int_B^\infty \text{Re}\rbra{\frac{e^{ix t}}{-\gamma+ix}\left((-\gamma+ix) F(-\gamma+ix)-(1-e^{-\Lambda})\right)} \dee x \\
  &\leq \int_B^\infty \frac{1}{\sqrt{\gamma^2+x^2}}\left|(-\gamma+ix) F(-\gamma+ix)-(1-e^{-\Lambda})\right| \dee x \\
  &\leq 765 \Lambda e^{-\frac{3\Lambda}{4}} 
    \int_B^\infty \frac{1}{\sqrt{\gamma^2+x^2}}\frac{1}{x} \dee x \\
  &\leq \frac{765 \Lambda e^{-\frac{3\Lambda}{4}}}{B}.
\]
When $B = \sqrt{6}(\Lambda+1)$, we get 
\[
  (T3) 
  &\leq \frac{765 \Lambda e^{-\frac{3\Lambda}{4}}}{\sqrt{6}(\Lambda+1)}.
\]
\eprfof

\subsubsection{Supplementary results for the proof of \cref{prop:C_Lambda_bound}}

First we show that $\text{Re}(z/r(z)) \geq 0$ on some useful region.

\blem
\label{lem:nonnegrz}
Let $z=a+ib$ for $a,b\in\reals$, $\Lambda \geq 1$, and $-\frac{1}{\Lambda+\sqrt{2}} < a$,
and $z \neq 0$.
Then,
\[
  \text{Re}(z/r(z)) \geq 0.
\]
\elem

\bprfof{\cref{lem:nonnegrz}}
If $z=a+ib$ for $a,b\in\reals$, then by definition of complex division,
\[
  \text{Re}\left(\frac{z}{r(z)}\right) &= \frac{a\text{Re}(r(z)) + b\text{Im}(r(z))}{|r(z)|^2}.
\]
Therefore, provided $z \neq 0$, 
\[
  \text{Re}\left(\frac{z}{r(z)}\right) \geq 0 \iff a \text{Re}(r(z)) + b\text{Im}(r(z)) \geq 0.
\]
Using the definition of the complex square root and the fact that $a+\Lambda > 0$,
\[
  \text{Re}(r(z)) &= \sqrt{\frac{1}{2}\left(c + d\right)}\\
  \text{Im}(r(z)) &= \sgn(b) \sqrt{\frac{1}{2}\left(c - d\right)},
\]
where we define
\[
  c &= \sqrt{(a^2-b^2+2a\Lambda)^2 + (2b(a+\Lambda))^2}\\
  d &= a^2-b^2+2a\Lambda \\ 
  \sgn(b) &= \begin{cases}
    1,  & b \geq 0 \\ 
    -1, & b < 0.
  \end{cases}
\]
(Note that $\sgn(b) \neq 0$ when $b = 0$.)
Therefore, if $a\geq 0$, the result trivially holds by inspection. 
On the other hand, when $a < 0$, we have $a \text{Re}(r(z)) \leq 0$ and $b\text{Im}(r(z)) \geq 0$, so
\[
  \text{Re}\left(\frac{z}{r(z)}\right) \geq 0 &\iff  b^2\text{Im}(r(z))^2 \geq a^2 \text{Re}(r(z))^2\\
  &\iff b^2(c - d) \geq a^2 (c + d)\\
  &\iff (b^2-a^2)c  \geq (a^2+b^2)d.
\]
Note that $c\geq 0$, and since $-\frac{1}{\Lambda+\sqrt{2}} \leq a < 0$, 
$d \leq a(a+2\Lambda) < 0$. So if $b^2 \geq a^2$, the result again holds trivially by inspection.
On the other hand, if $b^2 < a^2$, then
\[
  \text{Re}\left(\frac{z}{r(z)}\right) \geq 0 &\iff  (b^2-a^2)^2c^2  \leq (a^2+b^2)^2d^2\\
  &\iff (b^2-a^2)^2\left((a^2-b^2+2a\Lambda)^2 + (2b(a+\Lambda))^2\right) \\ 
    &{\quad}\leq (a^2+b^2)^2(a^2-b^2+2a\Lambda)^2\\
  &\iff (b^2-a^2)^24b^2(a+\Lambda)^2 \leq ((a^2+b^2)^2-(b^2-a^2)^2)(a^2-b^2+2a\Lambda)^2\\
  &\iff (b^2-a^2)^24b^2(a+\Lambda)^2 \leq 4a^2b^2(a^2-b^2+2a\Lambda)^2.
\]
If $b^2 = 0$, the result holds again by inspection. If instead $b^2 \neq 0$, we continue the derivation:
\[
  \text{Re}\left(\frac{z}{r(z)}\right) \geq 0 
  &\iff (b^2-a^2)^2(a+\Lambda)^2 \leq a^2(a^2-b^2+2a\Lambda)^2\\
  &\iff (b^2-a^2)^2(a+\Lambda)^2 \leq a^2((a^2-b^2)^2 + 4a^2\Lambda^2 + 4(a^2-b^2)a\Lambda)\\
  &\iff 0 \leq (a^2-b^2)^2(-\Lambda^2 + 2a\Lambda) + 4a^4\Lambda^2 + 4(a^2-b^2)a^3\Lambda\\
  &\iff 0 \leq -\Lambda a^4 -\Lambda b^4 + 2\Lambda a^2b^2 + 2a^5 \\ 
    &{\qquad \qquad} + 2ab^4 - 4a^3b^2 + 4a^4\Lambda + 4a^5 - 4b^2a^3\\
  &\iff 0 \leq 3\Lambda a^4 - 4a^3b^2 + 2a^2b^2(\Lambda-2a) + (2a-\Lambda)b^4 + 6a^5\\
  &\iff 0 \leq 3\Lambda a^4 - 4a^3b^2 + a^2b^2(\Lambda-2a) + b^2(a^2-b^2)(\Lambda-2a) + 6a^5\\
  &\iff 0 \leq 3a^4(\Lambda +2a) - 4a^3b^2 + a^2b^2(\Lambda-2a) + b^2(a^2-b^2)(\Lambda-2a).
\]
and since $a < 0$ and $b^2 < a^2$, all of these terms are positive as long as
\[
\Lambda + 2a \geq 0 \iff a \geq -\Lambda/2.
\]
Since $\Lambda \geq 1$, we have that
\[
a \geq -1/(\Lambda + \sqrt{2}) \geq -1/(1+\sqrt{2}) \geq -1/2 \geq -\Lambda/2.
\]
\eprfof

Next, we analyze the limits of $r(z)$ and $z/r(z)$ when $|z|$ is large.

\blem
\label{lem:asymprz}
Let $z = a+ib$ for $a,b\in\reals$, $\Lambda \geq 1$, $a+\Lambda  > 0$, and $0 \leq \epsilon < 1$.
\bitems
\item If $\frac{b^2\Lambda^2}{(a^2+b^2+2a\Lambda)^2} \leq \frac{\epsilon}{(1-\epsilon)^2}$, and $a^2+b^2+2a\Lambda > 0$, then
\[
\left|\text{Im}(r(z)) -b\right| &\leq\sqrt{\epsilon}\frac{\Lambda(1+\epsilon)}{2(1-\epsilon)}.
\]
\item If $\frac{\Lambda(\Lambda+a)}{(a^2+b^2)} \leq \frac{\epsilon}{(1-\epsilon)^2}$, and $a^2+b^2 > 0$, then
\[
\left|\text{Re}(r(z)) - (a+\Lambda)\right| &\leq \epsilon\Lambda.
\]
\item If both of the above sets of conditions hold,
\[
\left|\text{Re}\left(\frac{z}{r(z)}\right) - 1\right|
&\leq
\frac{2\epsilon}{1+\epsilon^2}
+ \frac{|a|\Lambda(1+\epsilon)}{(a^2+b^2)}
+ \frac{\sqrt{\epsilon}|b|\Lambda(1+\epsilon)}{2(a^2+b^2)(1-\epsilon)}\\
\left|\text{Im}\left(\frac{z}{r(z)}\right)\right| &\leq 
\frac{|b|\Lambda(1+\epsilon)}{(a^2+b^2)}
+\frac{\sqrt{\epsilon}|a|\Lambda(1+\epsilon)}{2(a^2+b^2)(1-\epsilon)}.
\]
\eitems
\elem

\bprfof{\cref{lem:asymprz}}
If $z=a+ib$ for $a,b\in\reals$, then by definition of the complex square root and the fact that $a+\Lambda > 0$,
\[
  \text{Re}(r(z)) &= \sqrt{\frac{1}{2}\left(\sqrt{(a^2-b^2+2a\Lambda)^2 + (2b(a+\Lambda))^2} + a^2-b^2+2a\Lambda \right)}\\
  \text{Im}(r(z)) &= \sgn(b) \sqrt{\frac{1}{2}\left(\sqrt{(a^2-b^2+2a\Lambda)^2 + (2b(a+\Lambda))^2} - a^2+b^2-2a\Lambda\right)},
\]
where $\sgn(b) = 1$ if $b=0$.
The inner square root can be simplified in various ways:
\[
  &(a^2-b^2+2a\Lambda)^2 + (2b(a+\Lambda))^2 \\
  &= a^4 + b^4 + 4a^2\Lambda^2 - 2a^2b^2 - 4b^2a\Lambda + 4a^3\Lambda + 4b^2(a^2 + \Lambda^2 + 2a\Lambda)\\
  &= a^4 + b^4 + 4a^2\Lambda^2 + 2a^2b^2 + 4b^2a\Lambda + 4a^3\Lambda + 4b^2\Lambda^2 \\
  &= (a^2+b^2+2a\Lambda)^2 + 4b^2\Lambda^2 \label{eq:reformsqrt1}\\ 
  &= (a^2+b^2)^2 + 4\Lambda(\Lambda+a)(a^2+b^2). \label{eq:reformsqrt2}
\]
The last two equalities are useful forms of the simplification.
We use the first equality for the imaginary component and the second for the real component.
\[
  \text{Re}(r(z)) &= \sqrt{\frac{1}{2}\left(\sqrt{(a^2+b^2)^2 + 4\Lambda(\Lambda+a)(a^2+b^2)} + a^2-b^2+2a\Lambda \right)}\\
  \text{Im}(r(z)) &= \sgn(b) \sqrt{\frac{1}{2}\left(\sqrt{(a^2+b^2+2a\Lambda)^2 + 4b^2\Lambda^2} - a^2+b^2-2a\Lambda\right)}.
\]
Therefore, when $a^2+b^2 > 0$,
\[
  \text{Re}(r(z)) &= \sqrt{\frac{1}{2}\left((a^2+b^2)\sqrt{1 + \frac{4\Lambda(\Lambda+a)}{a^2+b^2}} + a^2-b^2+2a\Lambda \right)},
\]
and when $a^2+b^2+2a\Lambda > 0$,
\[
\text{Im}(r(z)) &= \sgn(b) \sqrt{\frac{1}{2}\left((a^2+b^2+2a\Lambda)\sqrt{1 + \frac{4b^2\Lambda^2}{(a^2+b^2+2a\Lambda)^2}} - a^2+b^2-2a\Lambda\right)}.
\]
Next, we use the square root bounds:
\[
  \forall \, 0\leq \epsilon < 1, \quad 0 \leq x \leq \frac{4\epsilon}{(1-\epsilon)^2},  \qquad 1 + \frac{1-\epsilon}{2}x \leq \sqrt{1+x} &\leq 1+\frac{1+\epsilon}{2}x.
\]
If $0\leq \epsilon < 1$, $4b^2\Lambda^2/(a^2+b^2+2a\Lambda)^2 \leq \frac{4\epsilon}{(1-\epsilon)^2}$, and $a^2+b^2+2a\Lambda > 0$, 
we can apply the bounds to the inner square root in the imaginary term to find that
\[
\left|\text{Im}(r(z)) -b\right| &\leq |b|\max \left\{\left|\sqrt{1 + \frac{(1+\epsilon)\Lambda^2}{(a^2+b^2+2a\Lambda)}} - 1\right|,\left|\sqrt{1 + \frac{(1-\epsilon)\Lambda^2}{(a^2+b^2+2a\Lambda)}} - 1\right|\right\}\\
&\leq|b|\left|\sqrt{1 + \frac{(1+\epsilon)\Lambda\sqrt{\epsilon}}{|b|(1-\epsilon)}} - 1\right|\\
&\leq\sqrt{\epsilon}\frac{\Lambda(1+\epsilon)}{2(1-\epsilon)}.
\]
If $0\leq \epsilon < 1$,  $4\Lambda(\Lambda+a)/(a^2+b^2) \leq \frac{4\epsilon}{(1-\epsilon)^2}$, and $a^2+b^2 > 0$, 
we can apply the square root bounds to the inner square root in the real term to find that
\[
\left|\text{Re}(r(z)) - (a+\Lambda)\right| &\leq (a+\Lambda)\max\left\{\left|\sqrt{1+\epsilon\frac{\Lambda}{\Lambda+a}} - 1\right|,\left|\sqrt{1-\epsilon\frac{\Lambda}{\Lambda+a}} - 1\right| \right\}\\
&\leq(a+\Lambda)\max\left\{\epsilon\frac{\Lambda}{2(\Lambda+a)},\epsilon\frac{\Lambda}{\Lambda+a} \right\}\\
&\leq\epsilon \Lambda.
\]

Using the reformulation of the square root term in \cref{eq:reformsqrt1,eq:reformsqrt2}, we get two equalities for the norm:
\[
|r(z)|^2 &= \sqrt{(a^2+b^2)^2 + 4\Lambda(\Lambda+a)(a^2+b^2)} = \sqrt{(a^2+b^2+2a\Lambda)^2 + 4b^2\Lambda^2}.
\]
We now analyze the real component of $z/r(z)$ under the stipulated conditions on $a,b$: 
\[
  &\text{Re}\left(\frac{z}{r(z)}\right) \\ 
  &= \frac{a\text{Re}(r(z)) + b\text{Im}(r(z))}{|r(z)|^2}\\
  &= \frac{a(\text{Re}(r(z))-(a+\Lambda)) + a(a+\Lambda) + b(\text{Im}(r(z)) - b) + b^2}{(a^2+b^2)\sqrt{1 + \frac{4\Lambda(\Lambda+a)}{(a^2+b^2)^2}}}\\
  &= \frac{a(a+\Lambda) + b^2}{(a^2+b^2)\sqrt{1 + \frac{4\Lambda(\Lambda+a)}{(a^2+b^2)^2}}}
    + \frac{a(\text{Re}(r(z))-(a+\Lambda)) + b(\text{Im}(r(z)) - b)}{(a^2+b^2)\sqrt{1 + \frac{4\Lambda(\Lambda+a)}{(a^2+b^2)^2}}}\\
  &= \frac{1}{\sqrt{1 + \frac{4\Lambda(\Lambda+a)}{(a^2+b^2)^2}}} + \frac{a\Lambda}{(a^2+b^2)\sqrt{1 + \frac{4\Lambda(\Lambda+a)}{(a^2+b^2)^2}}} \\
    &{\quad} + \frac{a(\text{Re}(r(z))-(a+\Lambda)) + b(\text{Im}(r(z)) - b)}{(a^2+b^2)\sqrt{1 + \frac{4\Lambda(\Lambda+a)}{(a^2+b^2)^2}}}.
\]
Since $4\Lambda(\Lambda+a)/(a^2+b^2) \leq \frac{4\epsilon}{(1-\epsilon)^2}$,
\[
  &\left|\text{Re}\left(\frac{z}{r(z)}\right) - 1\right| \\ 
  &\leq 1-\frac{1}{\sqrt{1 + \frac{4\epsilon}{(1-\epsilon)^2}}}
    + \frac{|a|\Lambda}{(a^2+b^2)\sqrt{1 + \frac{4\Lambda(\Lambda+a)}{(a^2+b^2)^2}}} \\
    &{\qquad} + \frac{|a|\left|\text{Re}(r(z))-(a+\Lambda)\right|+ |b|\left|\text{Im}(r(z)) - b\right|}{(a^2+b^2)\sqrt{1 + \frac{4\Lambda(\Lambda+a)}{(a^2+b^2)^2}}}\\
  &\leq 1-\frac{1}{\sqrt{1 + \frac{4\epsilon}{(1-\epsilon)^2}}}
    + \frac{|a|\Lambda}{(a^2+b^2)\sqrt{1 + \frac{4\Lambda(\Lambda+a)}{(a^2+b^2)^2}}}
    + \frac{|a|\epsilon\Lambda+ |b|\sqrt{\epsilon}\frac{\Lambda(1+\epsilon)}{2(1-\epsilon)}}{(a^2+b^2)\sqrt{1 + \frac{4\Lambda(\Lambda+a)}{(a^2+b^2)^2}}}.
\]
Using the bound $\sqrt{1+x} \leq 1+\frac{1}{2}x$ in the second term
and bounding $\sqrt{1+\dots} \geq 1$ in the other denominators,
\[
\left|\text{Re}\left(\frac{z}{r(z)}\right) - 1\right|
&\leq
\frac{2\epsilon}{1+\epsilon^2}
+ \frac{|a|\Lambda(1+\epsilon)}{(a^2+b^2)}
+ \frac{\sqrt{\epsilon}|b|\Lambda(1+\epsilon)}{2(a^2+b^2)(1-\epsilon)}.
\]
For the imaginary component,
\[
  \text{Im}\left(\frac{z}{r(z)}\right)
  &= \frac{b\text{Re}(r(z)) - a\text{Im}(r(z))}{|r(z)|^2}\\
  &= \frac{b(\text{Re}(r(z))-(a+\Lambda)) + b(a+\Lambda) - a(\text{Im}(r(z)) - b) - ab}{|r(z)|^2}\\
  &= \frac{b\Lambda}{(a^2+b^2)\sqrt{1 + \frac{4\Lambda(\Lambda+a)}{(a^2+b^2)^2}}}
  +\frac{b(\text{Re}(r(z))-(a+\Lambda)) - a(\text{Im}(r(z)) - b)}{(a^2+b^2)\sqrt{1 + \frac{4\Lambda(\Lambda+a)}{(a^2+b^2)^2}}}.
\]
Since $4\Lambda(\Lambda+a)/(a^2+b^2) \leq \frac{4\epsilon}{(1-\epsilon)^2}$, and $\sqrt{1+\dots} \geq 1$,
\[
\left|\text{Im}\left(\frac{z}{r(z)}\right)\right| &\leq 
\frac{|b|\Lambda(1+\epsilon)}{(a^2+b^2)}
+\frac{\sqrt{\epsilon}|a|\Lambda(1+\epsilon)}{2(a^2+b^2)(1-\epsilon)}.
\]
\eprfof

d

The following result allows us to shift the Bromwich integral line of integration to the left, 
which we use to bound $C(\gamma,\Lambda)$.

\bprop
\label{prop:Fnormbound}
Let $z=a+ib$ for $a,b\in\reals$, and let $\Lambda\geq 1$.
If $b^2 \geq 6(\Lambda+1)^2$ and $|a| < \frac{1}{\Lambda+\sqrt{2}}$, then
\[
|zF(z) - (1-e^{-\Lambda})| &\leq \frac{765\Lambda}{|b|} e^{-\frac{3\Lambda}{4}}.
\]
\eprop

\bprf
Since $b^2 \geq 6(\Lambda+1)^2$, and $\Lambda \geq 1$,
we have that
\[
b^2 > \max\left\{ (\Lambda+1)^2,  \frac{2\Lambda}{1+\sqrt{2}},  \frac{\Lambda^2\left(\Lambda+\frac{1}{\Lambda+\sqrt{2}}\right)}{\Lambda - \frac{1}{\Lambda+\sqrt{2}}}\right\}.
\]
Using the first bound,
\[
b^2 > (\Lambda+1)^2 &\implies |b| > \Lambda + 1 \\
&\implies |b| > \frac{\Lambda \pm \sqrt{\Lambda^2 + \frac{8\Lambda}{\Lambda+\sqrt{2}}}}{2}\\
&\implies |b|^2 - |b|\Lambda - \frac{2\Lambda}{\Lambda+\sqrt{2}} > 0\\
&\implies a^2+|b|^2+2a\Lambda - |b|\Lambda  > 0\\
&\implies \frac{b^2\Lambda^2}{(a^2+b^2+2a\Lambda)^2} < 1.
\]
Using the second bound,
\[
b^2 \geq \frac{2\Lambda}{\Lambda+\sqrt{2}} > 2|a|\Lambda \geq -2a\Lambda \geq -a^2 - 2a\Lambda \implies a^2+b^2+2a\Lambda > 0.
\]
And the final bound,
\[
  &b^2 \geq \frac{\Lambda^2\left(\Lambda+\frac{1}{\Lambda+\sqrt{2}}\right)}{\Lambda - \frac{1}{\Lambda+\sqrt{2}}}
  > \frac{\Lambda^2(\Lambda+|a|)}{\Lambda-|a|}
  \geq \frac{\Lambda^2(\Lambda+a)}{\min(\Lambda, \Lambda+a)} - a^2 \\ 
  &\implies 
  \frac{\Lambda(\Lambda+a)}{a^2+b^2} < \min\left(1, \frac{a+\Lambda}{\Lambda}\right).
\]
Note also that
\[
  \frac{b^2\Lambda^2}{(a^2+b^2+2a\Lambda)^2} 
  &> \frac{b^2\Lambda(\Lambda+a)}{(a^2+b^2+2a\Lambda)^2} \\ 
  &= \frac{b^2\Lambda(\Lambda+a)}{(a^2+b^2+2a\Lambda)(a^2+b^2+2a\Lambda)} \\
  &> \frac{b^2\Lambda(\Lambda+a)}{(b^2)(a^2+b^2)} > \frac{\Lambda(\Lambda+a)}{a^2+b^2}.
\]
Therefore, all conditions of \cref{lem:asymprz} hold with
\[
  \epsilon = \frac{b^2\Lambda^2}{(a^2+b^2+2a\Lambda)^2} &= \frac{b^2\Lambda^2}{b^4 + 2b^2(a^2+2a\Lambda) + (a^2+2a\Lambda)^2} \\
  &\leq \frac{\Lambda^2}{b^2\left(1 + \frac{2a(a+2\Lambda)}{b^2}\right)}\\
  &\leq \frac{\Lambda^2}{b^2\left(1 + \frac{4\Lambda a}{b^2}\right)}\\
  &\leq \frac{\Lambda^2}{b^2\left(1 - \frac{4\Lambda }{6(\Lambda+1)^2}\right)}\\
  &\leq \frac{\Lambda^2}{b^2\left(1 - \frac{4 }{6(\Lambda+1)}\right)}\\
  &\leq \frac{\Lambda^2}{b^2\left(2/3\right)}\\
  &= \frac{3\Lambda^2}{2b^2}.
\]

By definition of $F(z)$,
\[
|zF(z)- (1-e^{-\Lambda})| &= e^{-\Lambda}\frac{\left| D(1,z) - e^{z+\Lambda}\right|}{|D(1,z)|}.
\]
By the reverse triangle inequality,
\[
|D(1,z)| &= \left|\frac{1}{2} e^{r(z)}\left(1 + \frac{z}{r(z)}\right) + 
    \frac{1}{2}e^{-r(z)}\left(1 - \frac{z}{r(z)}\right)\right|\\
    &\geq \frac{1}{2} \left|e^{r(z)}\left(1 + \frac{z}{r(z)}\right)\right| - 
    \frac{1}{2}\left|e^{-r(z)}\left(1 - \frac{z}{r(z)}\right)\right|\\
&= \frac{1}{2} e^{\text{Re}(r(z))}\left|1 + \frac{z}{r(z)}\right| - 
    \frac{1}{2}e^{-\text{Re}(r(z))}\left|1 - \frac{z}{r(z)}\right|.
\]
By \cref{lem:nonnegrz}, $|1+z/r(z)| \geq |1-z/r(z)|$ and $|1+z/r(z)| \geq |z/r(z)|$, so
\[
|D(1,z)| &\geq \left|1+\frac{z}{r(z)}\right| \sinh(\text{Re}(r(z))) \geq \left|\frac{z}{r(z)}\right| \sinh(\text{Re}(r(z))).
\]
Applying \cref{lem:asymprz},
\[
\left|\text{Re}(r(z)) - (a+\Lambda)\right| \leq \epsilon\Lambda \leq \frac{3\Lambda^3}{2b^2}.
\]
Further, 
\[
\left|\frac{z}{r(z)}\right| &= \frac{|z|}{|r(z)|}
= \left(1 + \frac{4\Lambda(\Lambda+a)}{a^2+b^2}\right)^{-1/4}
\geq (1+4\epsilon)^{-1/4}
\geq (1+6\Lambda^2/b^2)^{-1/4}.
\]
Therefore,
\[
  |D(1,z)| &\geq (1+6\Lambda^2/b^2)^{-1/4}\sinh\left(a+\left(1-\frac{3\Lambda^2}{2b^2}\right)\Lambda\right).
\]
Next, for the numerator,
\[
  \left|D(1,z) - e^{z+\Lambda}\right|
  &= \left|\frac{1}{2} e^{r(z)}\left(1 + \frac{z}{r(z)}\right)-e^{z+\Lambda} +  \frac{1}{2}e^{-r(z)}\left(1 - \frac{z}{r(z)}\right)\right|\\
  &\leq \abs{e^{z+\Lambda}} \left|e^{r(z)-z-\Lambda}\frac{1}{2}\left(1 + \frac{z}{r(z)}\right)-1\right| +  \frac{1}{2}e^{-\text{Re}(r(z))}\left|1 - \frac{z}{r(z)}\right|\\
  &\leq \abs{e^{z+\Lambda}} \left|e^{r(z)-z-\Lambda}\right| \left|\frac{1}{2}\left(1 + \frac{z}{r(z)}\right)-1\right| \\ 
    &{\quad} + e^{a+\Lambda}\left|e^{r(z)-z-\Lambda}-1\right| +  \frac{1}{2}e^{-\text{Re}(r(z))}\left|1 - \frac{z}{r(z)}\right|\\
  &= e^{a+\Lambda}\left|e^{r(z)-z-\Lambda}-1\right| +  \cosh(\text{Re}(r(z)))\left|1 - \frac{z}{r(z)}\right|,
\]
and
\[
  \left|e^{r(z)-z-\Lambda}-1\right| 
  &\leq \left|\text{Re}\left(e^{r(z)-z-\Lambda}-1\right)\right| + \left|\text{Im}\left(e^{r(z)-z-\Lambda}-1\right)\right|\\
  &= \left|e^{\text{Re}(r(z)) -a-\Lambda}\cos(\text{Im}(r(z))-b)-1\right| \\ 
    &{\quad} + e^{\text{Re}(r(z)) -a-\Lambda}\left|\sin(\text{Im}(r(z))- b)\right|\\
  &\leq e^{\text{Re}(r(z)) -a-\Lambda} (1- \cos(\text{Im}(r(z))-b))
    +e^{\left|\text{Re}(r(z)) -a-\Lambda\right|}-1 \\
    &{\quad} + e^{\text{Re}(r(z)) -a-\Lambda}\left|\text{Im}(r(z))- b\right|\\
  &\leq e^{\epsilon\Lambda} (\text{Im}(r(z))-b)^2
    +e^{\epsilon\Lambda}-1
    + e^{\epsilon\Lambda}\left|\text{Im}(r(z))- b\right|\\
  &\leq e^{\epsilon\Lambda}\left(\left|\text{Im}(r(z))-b\right| + 1\right)^2 - 1\\
  &\leq e^{\frac{3\Lambda^3}{2b^2}}\left(\left|\text{Im}(r(z))-b\right| + 1\right)^2 - 1.
\]
Also, since $\epsilon \leq 3\Lambda^2/2b^2 \leq 3\Lambda^2/(12(\Lambda+1)^2) \leq 1/16$,
\[
  \left|\text{Im}(r(z))-b\right| &\leq \sqrt{\epsilon}\frac{\Lambda}{2}\frac{1+\epsilon}{1-\epsilon}
  \leq \sqrt{\frac{3\Lambda^2}{2b^2}}\frac{\Lambda}{2}\frac{17}{15}
  \leq \frac{7}{10}\frac{\Lambda^2}{|b|}
\]
and
\[
  \cosh(\text{Re}(r(z))) \leq \cosh\left(a+\left(1+\frac{3\Lambda^2}{2b^2}\right)\Lambda\right).
\]
Finally,
\[
  \left|1-\frac{z}{r(z)}\right| &\leq \left|\text{Im}\frac{z}{r(z)}\right| + \left|\text{Re}\frac{z}{r(z)} - 1\right|.
\]
So, since $\epsilon \leq 3\Lambda^2/2b^2 \leq 3\Lambda^2/(12(\Lambda+1)^2) \leq 1/16$,
\[
  \left|1-\frac{z}{r(z)}\right| \! &\leq \!
  \frac{2\epsilon}{1+\epsilon^2}
  + \frac{|a|\Lambda(1+\epsilon)}{(a^2+b^2)}
  + \frac{\sqrt{\epsilon}|b|\Lambda(1+\epsilon)}{2(a^2+b^2)(1-\epsilon)}
  +\frac{|b|\Lambda(1+\epsilon)}{(a^2+b^2)}
  +\frac{\sqrt{\epsilon}|a|\Lambda(1+\epsilon)}{2(a^2+b^2)(1-\epsilon)}\\
  &= \frac{2\epsilon}{1+\epsilon^2} + \frac{(|a|+|b|)\Lambda(1+\epsilon)}{(a^2+b^2)}\left(1 + \frac{\sqrt{\epsilon}}{2(1-\epsilon)}\right)\\
  &\leq 2\frac{3\Lambda^2}{2b^2} + \frac{2|b|\Lambda(1+1/16)}{b^2}\left(1 + \frac{\sqrt{1/16}}{2(1-1/16)}\right)\\
  &= \frac{3\Lambda^2}{b^2} + \frac{578\Lambda }{240 |b|}\\
  &\leq \frac{3\Lambda^2}{\sqrt{6}|b|(\Lambda+1)} + \frac{578\Lambda }{240 |b|}\\
  &\leq \frac{3\Lambda}{2\sqrt{6}|b|} + \frac{578\Lambda }{240 |b|}\\
  &\leq \frac{7\Lambda}{2|b|}.
\]
Putting these all together,
\[
  &|zF(z)- (1-e^{-\Lambda})| \\
  &\leq e^{-\Lambda}\frac{e^{a+\Lambda}\left(e^{\frac{3\Lambda^3}{2b^2}}\left(\frac{7}{10}\frac{\Lambda^2}{|b|} + 1\right)^2 - 1\right) + \frac{7\Lambda}{2|b|}\cosh\left(a+\left(1+\frac{3\Lambda^2}{2b^2}\right)\Lambda\right)}{(1+6\Lambda^2/b^2)^{-1/4}\sinh\left(a+\left(1-\frac{3\Lambda^2}{2b^2}\right)\Lambda\right)}\\
  &\leq e^{-\Lambda}\frac{e^{a+\Lambda}\left(e^{\frac{3\Lambda^3}{2b^2}}\left(\frac{7}{10}\frac{\Lambda^2}{|b|} + 1\right)^2 - 1\right) + \frac{7\Lambda}{2|b|}\cosh\left(a+\left(1+\frac{3\Lambda^2}{2b^2}\right)\Lambda\right)}{(1+1/4)^{-1/4}\sinh\left(a+\left(1-\frac{3\Lambda^2}{2b^2}\right)\Lambda\right)}\\
  &\leq e^{-\Lambda}\frac{\cosh\left(a+\left(1+\frac{3\Lambda^2}{2b^2}\right)\Lambda\right)}{(1+1/4)^{-1/4}\sinh\left(a+\left(1-\frac{3\Lambda^2}{2b^2}\right)\Lambda\right)}
\left(2 e^{\frac{3\Lambda^3}{2b^2}}\left(\frac{7}{10}\frac{\Lambda^2}{|b|} + 1\right)^2 - 2 + \frac{7\Lambda}{2|b|}\right),
\]
where the last line follows because $e^x \leq 2\cosh(x)$. 
Next, since $e^x \leq xe^x+1$, $\Lambda \geq 1$, and
$b^2 \geq 6(\Lambda+1)^2$, we have that $(1+1/4)^{1/4} \leq 1.1$,
\[
 \frac{3\Lambda^2}{2b^2} \leq \frac{1}{16}, \qquad e^{\frac{3\Lambda^3}{2b^2}} \leq \frac{3\Lambda^3}{2b^2} e^{\frac{3\Lambda^3}{2b^2}} + 1 \leq \frac{3\Lambda^3}{2b^2} e^{\frac{\Lambda}{16}} + 1.
\]
Additionally, for $0 \leq \eta < 1$,
\[
  \frac{\cosh(a+\Lambda + \eta \Lambda)}{\sinh(a+\Lambda-\eta\Lambda)} 
  \leq e^{2\eta \Lambda}\left(\frac{1+e^{2/(\Lambda+\sqrt{2})-2\Lambda}}{1-e^{2/(\Lambda+\sqrt{2})-15/8\Lambda}}\right).
\]
Therefore,
\[
  &|zF(z)- (1-e^{-\Lambda})| \\
  &\leq 1.1 e^{-\frac{7\Lambda}{8}}
    \left(\frac{1+e^{2/(\Lambda+\sqrt{2})-2\Lambda}}{1-e^{2/(\Lambda+\sqrt{2})-15/8\Lambda}}\right)
    \left(2\left(\frac{3\Lambda^3}{2b^2} e^{\frac{\Lambda}{16}}+1\right)\left(\frac{49\Lambda^4}{100b^2} + \frac{14\Lambda^2}{10|b|} + 1\right) - 2 + \frac{7\Lambda}{2|b|}\right)\\
  &= \frac{1.1\Lambda}{|b|} e^{-\frac{7\Lambda}{8}} 
    \left(\frac{1+e^{2/(\Lambda+\sqrt{2})-2\Lambda}}{1-e^{2/(\Lambda+\sqrt{2})-15/8\Lambda}}\right)
    \left(e^{\frac{\Lambda}{16}}\frac{3\Lambda^2}{|b|}\left(\frac{49\Lambda^4}{100 b^2} + \frac{7\Lambda^2}{5|b|} + 1\right) 
    +\frac{49\Lambda^3}{50|b|} + \frac{14\Lambda}{5} + \frac{7}{2}\right)\\
  &\leq \frac{1.1\Lambda}{|b|} e^{-\frac{7\Lambda}{8}} 
    \left(\frac{1+e^{2/(\Lambda+\sqrt{2})-2\Lambda}}{1-e^{2/(\Lambda+\sqrt{2})-15/8\Lambda}}\right)
    \left(e^{\frac{\Lambda}{16}}\frac{3\Lambda}{\sqrt{6}}\left(\frac{49\Lambda^2}{600} + \frac{7\Lambda}{5\sqrt{6}} + 1\right) 
    +\frac{49\Lambda^2}{50\sqrt{6}} + \frac{14\Lambda}{5} + \frac{7}{2}\right)\\
  &= \frac{1.1\Lambda}{|b|} e^{-\frac{13\Lambda}{16}} 
    \left(\frac{1+e^{2/(\Lambda+\sqrt{2})-2\Lambda}}{1-e^{2/(\Lambda+\sqrt{2})-15/8\Lambda}}\right) \cdot \\ 
    &{\quad} 
    \left(\Lambda^3 \frac{49}{200\sqrt{6}} + \Lambda^2 \rbra{\frac{21}{30} + e^{-\frac{\Lambda}{16}} \frac{49}{50\sqrt{6}}} + 
    \Lambda\rbra{\frac{3}{\sqrt{6}} + e^{-\frac{\Lambda}{16}} \frac{14}{5}} + \frac{7}{2} e^{-\frac{\Lambda}{16}}\right)\\
  &\leq \frac{1.1\Lambda}{|b|} e^{-\frac{13\Lambda}{16}}
    \left(\frac{1+e^{2/(\Lambda+\sqrt{2})-2\Lambda}}{1-e^{2/(\Lambda+\sqrt{2})-15/8\Lambda}}\right) 
    \left( 0.101\Lambda^3 + 1.076\Lambda^2 + 3.855\Lambda + 3.288\right)\\
  &\leq \frac{765\Lambda}{|b|} e^{-\frac{3\Lambda}{4}}.
\]
\eprf

Next, we analyze the behaviour of $\text{Re}(r(z))$ on a compact set. 

\blem
\label{lem:rerz}
Let $z = a+ib$ and suppose that $a+\Lambda >0$, $a^2-b^2+2a\Lambda < 0$, and 
$0 \leq b \leq B$ for some $B \geq 0$. Then,
\[
  \text{Re}(r(z)) 
  \geq \sqrt{b(a+\Lambda)} \sqrt{\sqrt{1+\frac{B^2}{4\Lambda^2}} - \frac{B}{2\Lambda}}.
\]
\elem

\bprfof{\cref{lem:rerz}}
By the definition of the complex square root and because $a+\Lambda>0$,
\[
  \text{Re}(r(z)) 
  &= \sqrt{\frac{1}{2}\left(\sqrt{(a^2-b^2+2a\Lambda)^2 + (2b(a+\Lambda))^2} + a^2-b^2+2a\Lambda \right)} \\
  &= \sqrt{\frac{1}{2}\left(2b(a+\Lambda)\sqrt{1 + \frac{(a^2-b^2+2a\Lambda)^2}{(2b(a+\Lambda))^2}} 
    + a^2-b^2+2a\Lambda \right)} \\
  &= \sqrt{b(a+\Lambda)} \sqrt{\sqrt{1 + \frac{(a^2-b^2+2a\Lambda)^2}{(2b(a+\Lambda))^2}} 
    + \frac{a^2-b^2+2a\Lambda}{2b(a+\Lambda)}}.
\]
Now, because $a^2-b^2+2a\Lambda < 0$,
\[
  \abs{\frac{a^2-b^2+2a\Lambda}{2b(a+\Lambda)}} 
  &= \frac{b^2-a^2-2a\Lambda}{2b(a+\Lambda)} 
  \leq \frac{B}{2\Lambda}.
\]
Finally, since $\sqrt{1+x^2} - x/2$ is a decreasing function of $x$, we have 
\[
  \text{Re}(r(z))  
  &\geq \sqrt{b(a+\Lambda)} \sqrt{\sqrt{1+\frac{B^2}{4\Lambda^2}} - \frac{B}{2\Lambda}}.
\]
\eprfof

To bound $\abs{D(1,z)}$ we make use of the following result. 

\blem 
\label{lem:r_re_im_increasing}
For $\Lambda \geq 1$, $0 < \gamma < 1/(\Lambda + \sqrt{2})$, and $x > 0$, both 
$\text{Re}(r(-\gamma + ix))$ and $\text{Im}(r(-\gamma + ix))$ are strictly increasing functions of $x$.
\elem

\bprfof{\cref{lem:r_re_im_increasing}} 
To see this, set $z = -\gamma + ix$ and recall that for $x > 0$,
\[
  \text{Re}(r(z)) &= \sqrt{\frac{1}{2}\left(c(-\gamma,x) + d(-\gamma,x)\right)}, \qquad 
  \text{Im}(r(z)) = \sqrt{\frac{1}{2}\left(c(-\gamma) - d(-\gamma, x)\right)},
\]
where 
\[
  c(-\gamma,x) &= \sqrt{(\gamma^2-x^2-2\gamma\Lambda)^2 + (2x (-\gamma+\Lambda))^2}, \qquad
  d(-\gamma,x) = \gamma^2-x^2-2\gamma\Lambda.
\]
We can establish that for $x > 0$,
\[
  \frac{\partial}{\partial x} c(-\gamma, x) + d(-\gamma, x) > 0,
\]
provided that $\Lambda > \gamma$. 
This condition holds for $\Lambda \geq 1$  and $0 < \gamma < 1/(\Lambda + \sqrt{2})$.
Similarly, 
\[
  \frac{\partial}{\partial x} c(-\gamma, x) - d(-\gamma, x) > 0.
\]
Because $\sqrt{x}$ is strictly increasing on $x \geq 0$, we obtain the desired result.
\eprfof

Finally, we provide a useful bound on $\abs{D(1,-\gamma+ix)}$ when $x$ is small. 

\bprop
\label{prop:C_Lambda_compact_2}
For $\Lambda \geq 1$, $1/(4\Lambda) \leq \gamma < 1/(\Lambda+\sqrt{2})$, and $0 < \eps \leq 1/(136\Lambda)$,
\[
  \inf_{x \in [0,\eps]} \abs{D(1,-\gamma+ix)} \geq 0.07.  
\]
\eprop

\bprfof{\cref{prop:C_Lambda_compact_2}}
Setting $z = -\gamma+ix$, we turn to the representation 
\[
  \abs{D(1,z)} 
  &= \abs{\cosh(r(z)) + \frac{z}{r(z)} \sinh(r(z))} \\ 
  &\geq \abs{\cosh(r(z))} - \abs{\frac{z}{r(z)}} \abs{\sinh(r(z))}.
\]
We can establish that, uniformly over $0 < \gamma < -1/(\Lambda+\sqrt{2})$ and $\Lambda \geq 1$, 
\[
  \abs{D(1,-\gamma)} 
  \geq \abs{\cosh(r(-\gamma))} - \abs{\frac{-\gamma}{r(-\gamma)}} \abs{\sinh(r(-\gamma))} 
  \geq 0.14.
\]
To see this, note that 
\[
  D(1,-\gamma) 
  = \cos(\sqrt{\gamma(-\gamma+2\Lambda)}) + 
    \frac{-\gamma}{\sqrt{\gamma (-\gamma+2\Lambda)}} \sin(\sqrt{\gamma(-\gamma+2\Lambda)}).
\]
The absolute value of the first term, $\abs{\cos(\sqrt{\gamma(-\gamma+2\Lambda)})}$, 
is a decreasing function of $\gamma$ for $\Lambda \geq 1$ and $0 < \gamma < 1/(\Lambda+\sqrt{2})$ 
because $\sqrt{\gamma(-\gamma+2\Lambda)} \leq \pi/2$ for $\Lambda \geq 1$. 
Therefore, for $\Lambda \geq 1$, 
\[
  \abs{\cos(\sqrt{\gamma(-\gamma+2\Lambda)})} 
  &\geq \cos\rbra{\sqrt{\frac{1}{\Lambda+\sqrt{2}}\rbra{2\Lambda-\frac{1}{\Lambda+\sqrt{2}}}}}.
\]
Similarly, both $\abs{-\gamma/(\sqrt{\gamma(-\gamma+2\Lambda)})}$ and $\abs{\sin(\sqrt{\gamma(-\gamma+2\Lambda)})}$ 
are increasing functions of $\gamma$ for $\Lambda \geq 1$ and $0 < \gamma < 1/(\Lambda+\sqrt{2})$. 
Hence, 
\[
  &\abs{\frac{-\gamma}{\sqrt{\gamma (-\gamma+2\Lambda)}}} \abs{\sin(\sqrt{\gamma(-\gamma+2\Lambda)})} \\
  &{\quad} \leq \frac{1/(\Lambda+\sqrt{2})}{\sqrt{1/(\Lambda+\sqrt{2})(2\Lambda-1/(\Lambda+\sqrt{2}))}} \cdot 
    \sin\rbra{\sqrt{1/(\Lambda+\sqrt{2})(2\Lambda-1/(\Lambda+\sqrt{2}))}} \\ 
  &{\quad} \leq 1/(\Lambda+\sqrt{2}). 
\]
The difference between these two bounds in terms of $\Lambda$ is a decreasing function of $\Lambda \geq 1$ 
over $\Lambda \in [1, 100]$. That is, setting 
\[
  h(\Lambda) = \cos\rbra{\sqrt{\frac{1}{\Lambda+\sqrt{2}}\rbra{2\Lambda-\frac{1}{\Lambda+\sqrt{2}}}}} 
    - \frac{1}{\Lambda + \sqrt{2}},
\]
we have $\inf_{\Lambda \in [1, 100]} h(\Lambda) \geq 0.155$. Also, 
\[
  \inf_{\Lambda \in [100, \infty)} h(\Lambda) 
  \geq \lim_{\Lambda \to \infty} h(\Lambda) - \frac{1}{100+\sqrt{2}} \geq 0.14,
\] 
and hence $\abs{D(1,-\gamma)} \geq 0.14$. 

From here on, set $w = r(-\gamma+ix) - r(-\gamma)$, which implicitly depends on $x$.
For a given $w \in \comps$, there exists a $\xi \in N_{\abs{w}}(r(-\gamma))$ such that 
\[
  \abs{\cosh(r(-\gamma)+w) - \cosh(r(-\gamma))} 
  &= \abs{w} \abs{\sinh(\xi)} 
  \leq \abs{w} e^{\abs{w}}.
\]
To obtain the last inequality, note that $\xi = r(-\gamma) + \tilde w$, where 
$\abs{\tilde w} \leq \abs{w}$, and so 
\[
  \abs{\sinh(\xi)} 
  &\leq \abs{\sinh(\text{Re}(\xi))} \abs{\cos(\text{Im}(\xi))} 
    + \abs{\cosh(\text{Re}(\xi))} \abs{\sin(\text{Im}(\xi))} \\
  &\leq \abs{\sinh(\text{Re}(\tilde w))} + \abs{\cosh(\text{Re}(\tilde w))} \\ 
  &\leq \sinh(\abs{w}) + \cosh(\abs{w}) \\ 
  &= e^{\abs{w}}.
\]
Using similar arguments, we also have that 
\[
  \abs{\sinh(r(-\gamma)+w) - \sinh(r(-\gamma))} \leq \abs{w} e^{\abs{w}}.    
\]
Note that, by applying \cref{lem:r_re_im_increasing}, we get 
\[
  \abs{\frac{-\gamma+ix}{r(a)+w} - \frac{-\gamma}{r(-\gamma)}}    
  \leq \frac{\abs{r(-\gamma)} x + \gamma \abs{w}}{\abs{r(-\gamma)}^2}.
\]

Now, observe that if $\abs{a - a'} \leq \eps_1$ and $\abs{b - b'} \leq \eps_2$, then 
$\abs{ab - a'b'} \leq \eps_1 \eps_2 + \eps_1 \abs{b} + \eps_2 \abs{a}$. 
We can then combine all bounds above to obtain 
\[
  \abs{D(1,-\gamma+ix) - D(1,-\gamma)} 
  \leq (1+\gamma) \abs{w} e^{\abs{w}} + (x + \abs{w}e^{\abs{w}}) 
    \frac{\abs{r(-\gamma)} x + \gamma \abs{w}}{\abs{r(-\gamma)}^2}.
\]
Finally, we have 
\[
  \abs{w} 
  = \abs{r(-\gamma + ix) - r(-\gamma)} 
  \leq \frac{x (\gamma + x + \Lambda)}{\sqrt{\gamma(2\Lambda - \gamma)}}. 
\]
This can be obtained by means of another Taylor series expansion and application of 
\cref{lem:r_re_im_increasing}. Namely, for some $\xi \in [0, x]$,
\[
  \abs{r(-\gamma+ix) - r(-\gamma)} 
  &= \abs{x} \abs{r'(-\gamma+i\xi)} \\
  &\leq \frac{x (\gamma + x + \Lambda)}{\abs{\sqrt{(-\gamma+i\xi)^2+2\Lambda(-\gamma+ix)}}} \\ 
  &\leq \frac{x (\gamma + x + \Lambda)}{\sqrt{\gamma(2\Lambda-\gamma)}}.
\]

For the following, note that for $1/(4\Lambda) \leq \gamma < 1/(\Lambda+\sqrt{2})$, we have
$0.5 \leq \abs{r(-\gamma)} \leq \sqrt{2}$.
Suppose $x \leq k/\Lambda$ for some $k \geq 0$. Then, we can obtain the bound $\abs{w} \leq 6k$.
We want 
\[
  \abs{D(1,-\gamma+ix) - D(1,-\gamma)} \leq 0.14/2,  
\]
and we can check that by setting $k \leq 1/136$, we can satisfy this inequality.
That is, using $x \leq 1/(136\Lambda)$ suffices.
\eprfof

\newpage
\subsection{Reversible parallel tempering}
\bprfof{\cref{thm:erg_reversible}}
The proof is almost identical to the proof of \cref{thm:erg_NRPT} except that 
$\Pr_\text{NRPT}$, $tN$, and $Z_\Lambda$ are replaced with $\Pr_\text{RPT}$, $tN^2$, 
and $W$, respectively.
\eprfof

\bprfof{\cref{thm:Brownian_hitting_time}}
Let $\tilde{W}(t)$ be Brownian motion on $[-1,1]$ with reflections at $\{-1,1\}$ 
for $t \geq 0$ with $\tilde{W}(0) = 0$.
By the reflection principle,
\[
  \Pr_\text{RPT}(\tau_\infty > t) 
  = \Pr_\text{RPT}\rbra{ \sup_{0 \leq s \leq t} |\tilde{W}(s)| < 1 }.
\]
From here, define 
$y(x,t) = \Pr_\text{RPT}(\sup_{0 \leq s \leq t} |\tilde{W}(s)| < 1 \mid \tilde{W}(0) = x)$.
When $\tilde{W}(0) = x \in [-1,1]$, by appealing to the infinitesimal generator of 
Brownian motion, we have that $y$ is the solution to the PDE 
\[
  \begin{aligned} 
  \label{eq:Brownian_PDE}
  \frac{1}{2} y_{xx} &= y_t \\
  y(x, 0) &= 1, \quad x \in (-1, 1) \\
  y(1, t) = y(-1, t) &= 0, \quad t \geq 0.
  \end{aligned}
\]
We now proceed to solve this partial differential equation. 
If we assume that 
\[
  y(x,t) = \sum_{k=1}^\infty a_k(t) \cdot \sin\rbra{\frac{k \pi (x+1)}{2}},  
\]
for some coefficients $a_k(t)$, then the PDE and initial condition yield
\[
  a_k'(t) &= -\frac{k^2 \pi^2}{8} a_k(t) \\
  a_k(0) &= \frac{2-2\cos(k\pi)}{k\pi}.
\]
Solving this ordinary differential equation, we find that 
\[
  a_k(t) = \frac{2}{k\pi}(1-(-1)^k) \exp(-k^2 \pi^2 t /8),  \quad k \geq 1,  
\]
so that 
\[
  \Pr_\text{RPT}(\tau_\infty > t)
  = \sum_{k=1}^\infty \frac{2}{k\pi}(1-(-1)^k) \sin(k\pi/2) \exp(-k^2 \pi^2 t /8).
\]
Note that for $t \geq 1$ we can write
\[
  \Pr_\text{RPT}(\tau_\infty > t)
  &\leq \frac{4}{\pi} \exp(-\pi^2 t/8) \sum_{k=0}^\infty \exp(-k \pi^2 t /8) \\
  &= \frac{4}{\pi} \exp(-\pi^2 t/8) \frac{1}{1 - \exp(-\pi^2 t/8)} \\
  &\leq 2 \exp(-\pi^2 t/8).
\]
\eprfof

\newpage
\section{Experiments}
\label{sec:appendix_experiments} 

\subsection{Additional pairwise correlation plots and traceplots of energies}

\begin{figure}[t]
  \centering
  \begin{subfigure}{0.245\textwidth}
      \centering
      \includegraphics[width=\textwidth]{img/simple-mix/pairwise_corr_chain_0.png}
  \end{subfigure}
  \begin{subfigure}{0.245\textwidth}
      \centering
      \includegraphics[width=\textwidth]{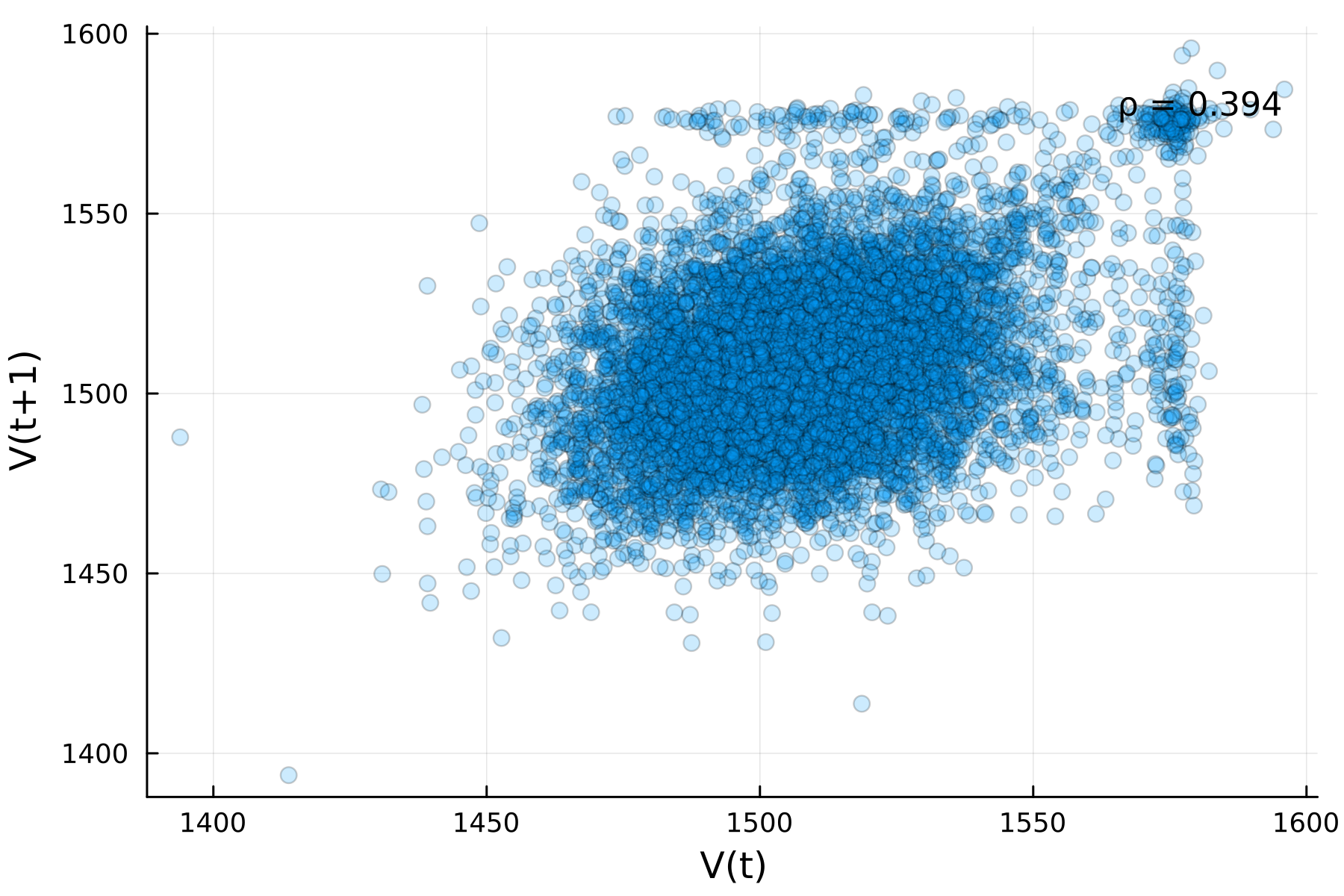}
  \end{subfigure}
  \begin{subfigure}{0.245\textwidth}
      \centering
      \includegraphics[width=\textwidth]{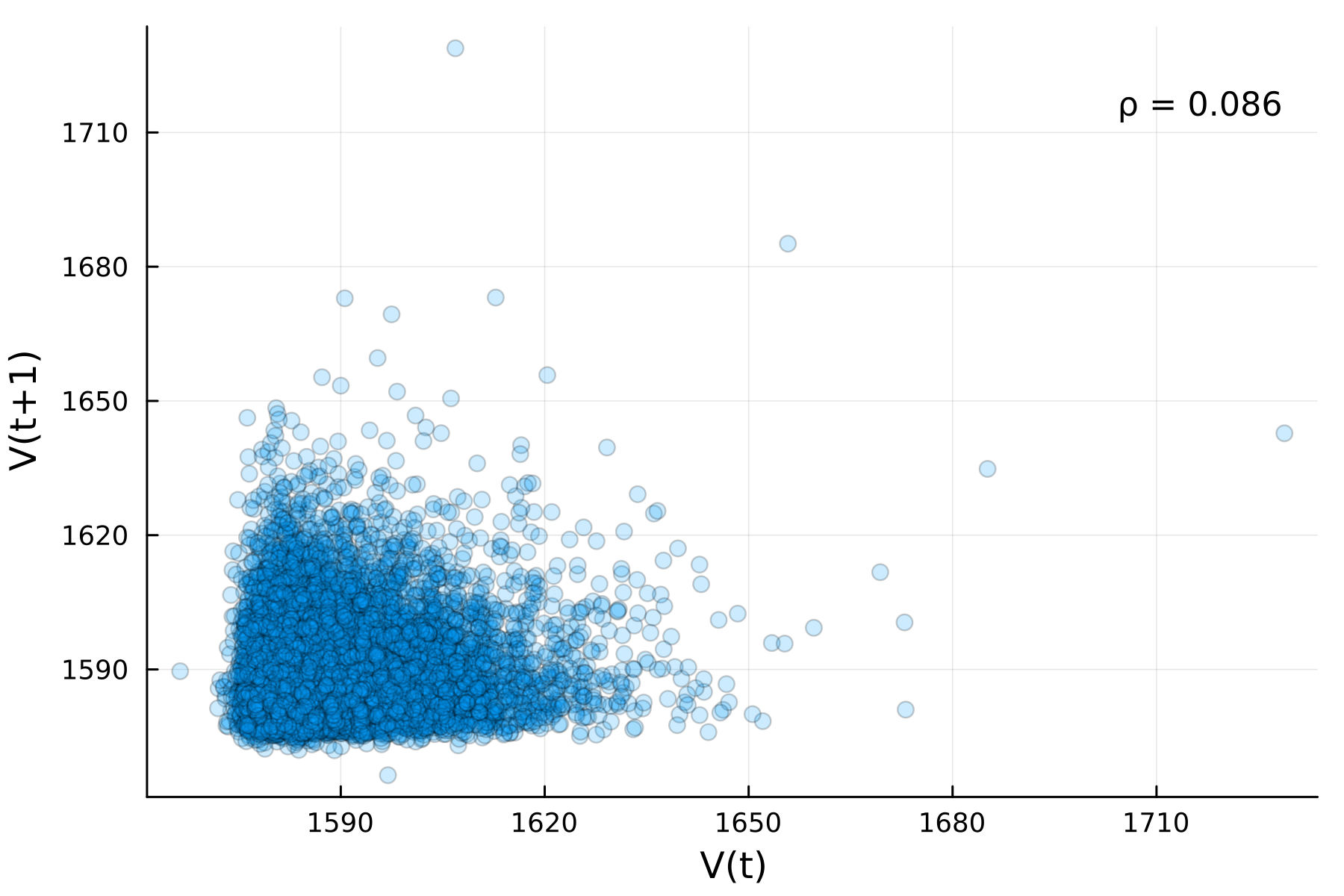}
  \end{subfigure}
  \begin{subfigure}{0.245\textwidth}
      \centering
      \includegraphics[width=\textwidth]{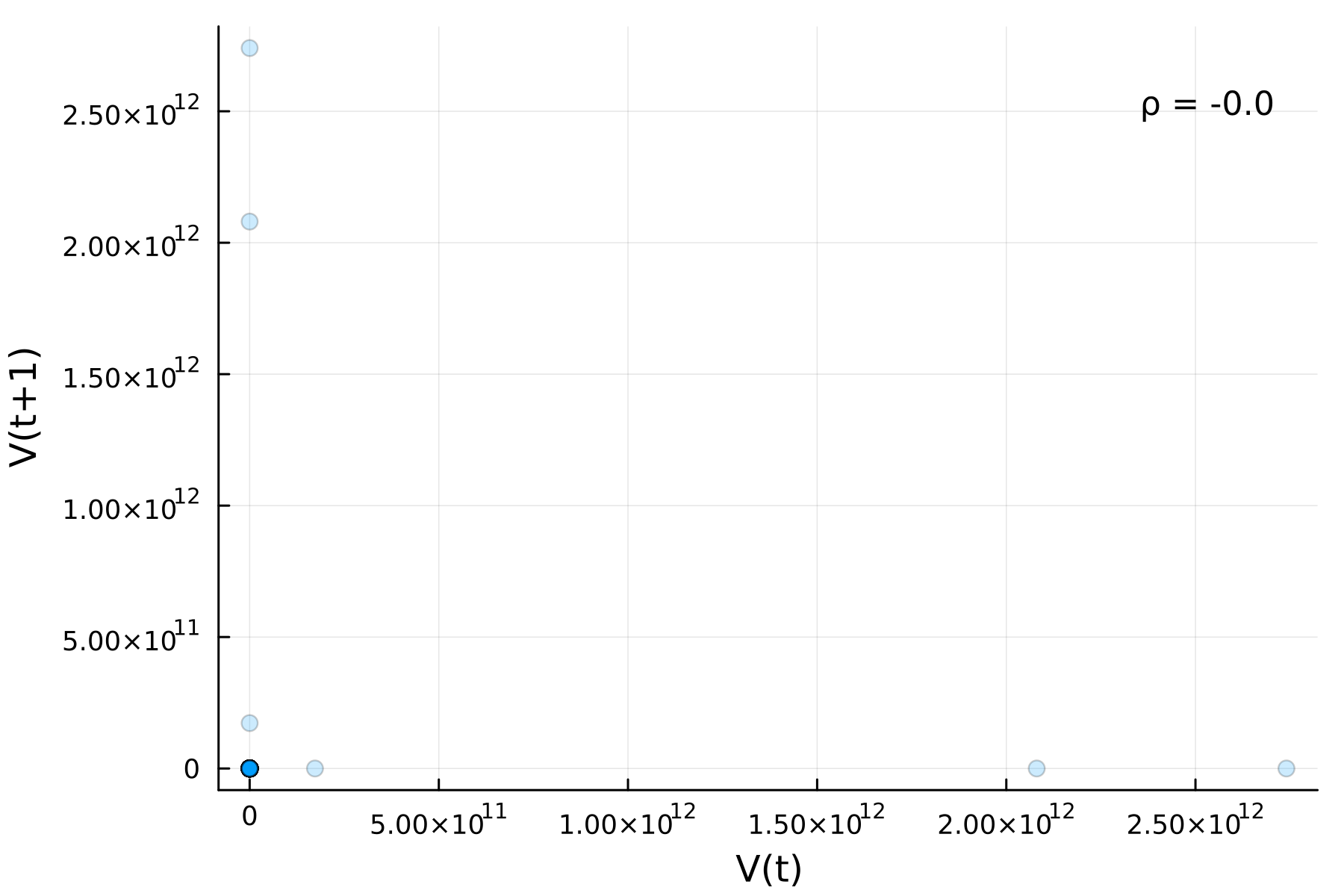}
  \end{subfigure}
  \begin{subfigure}{0.245\textwidth}
      \centering
      \includegraphics[width=\textwidth]{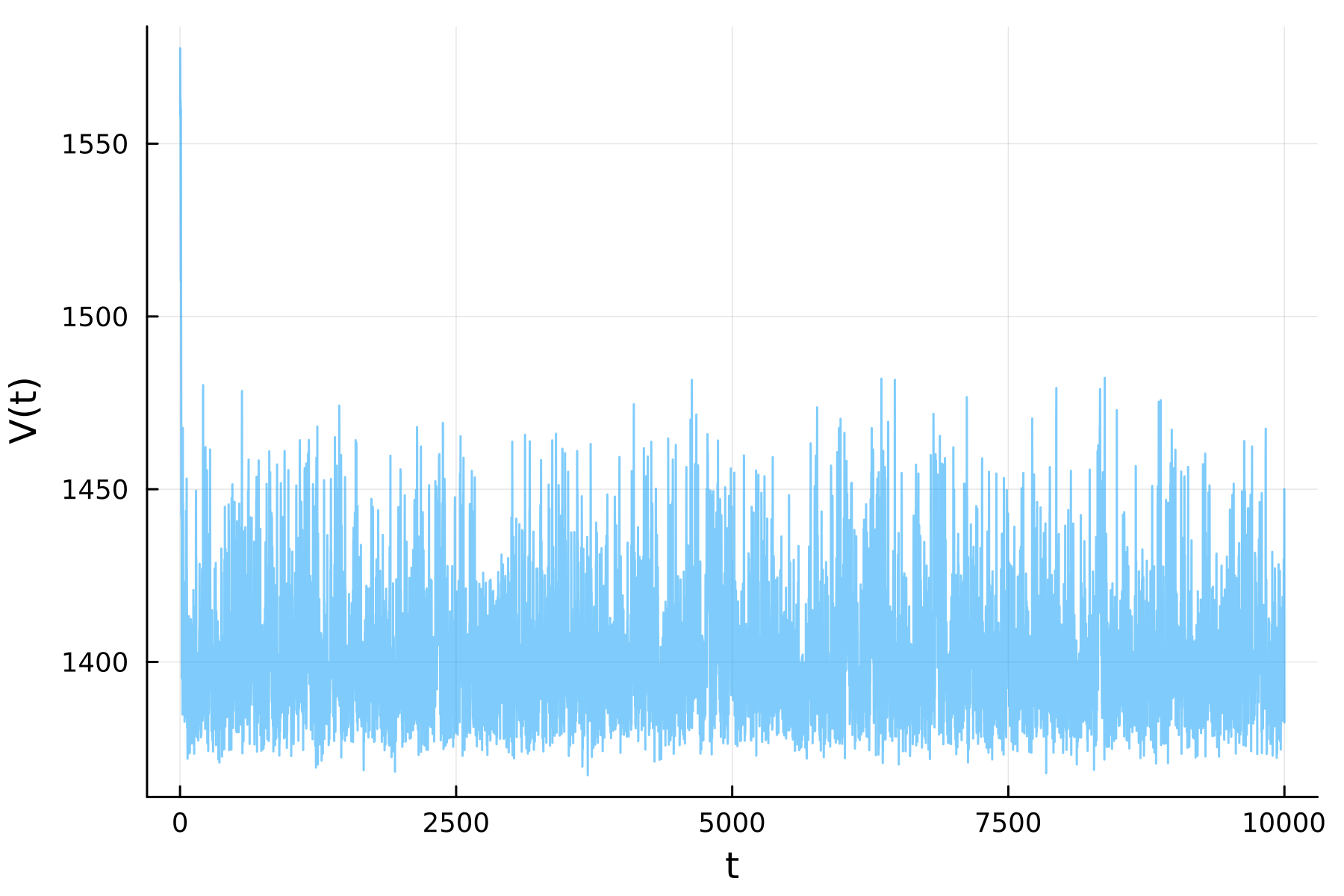}
  \end{subfigure}
  \begin{subfigure}{0.245\textwidth}
      \centering
      \includegraphics[width=\textwidth]{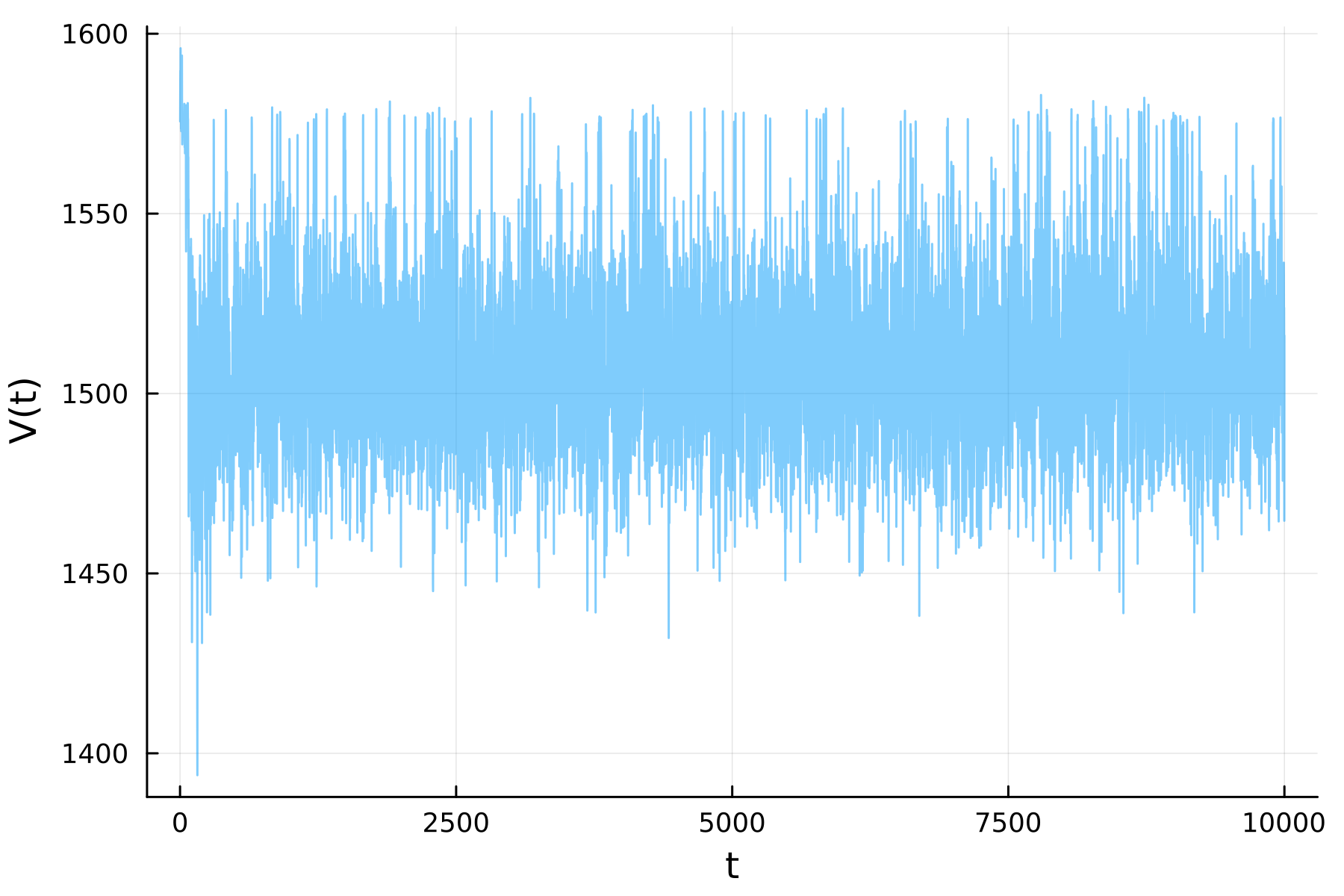}
  \end{subfigure}
  \begin{subfigure}{0.245\textwidth}
      \centering
      \includegraphics[width=\textwidth]{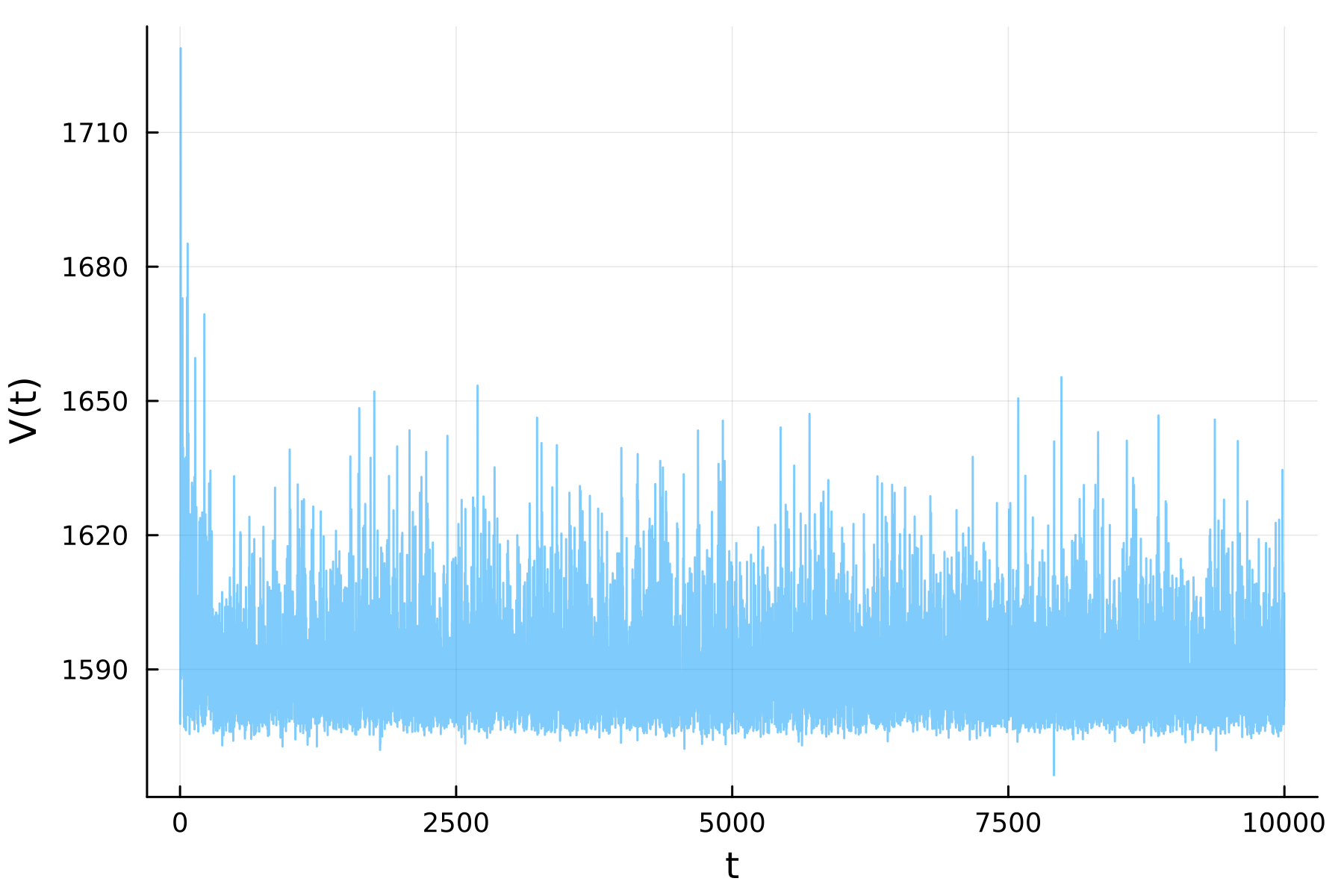}
  \end{subfigure}
  \begin{subfigure}{0.245\textwidth}
      \centering
      \includegraphics[width=\textwidth]{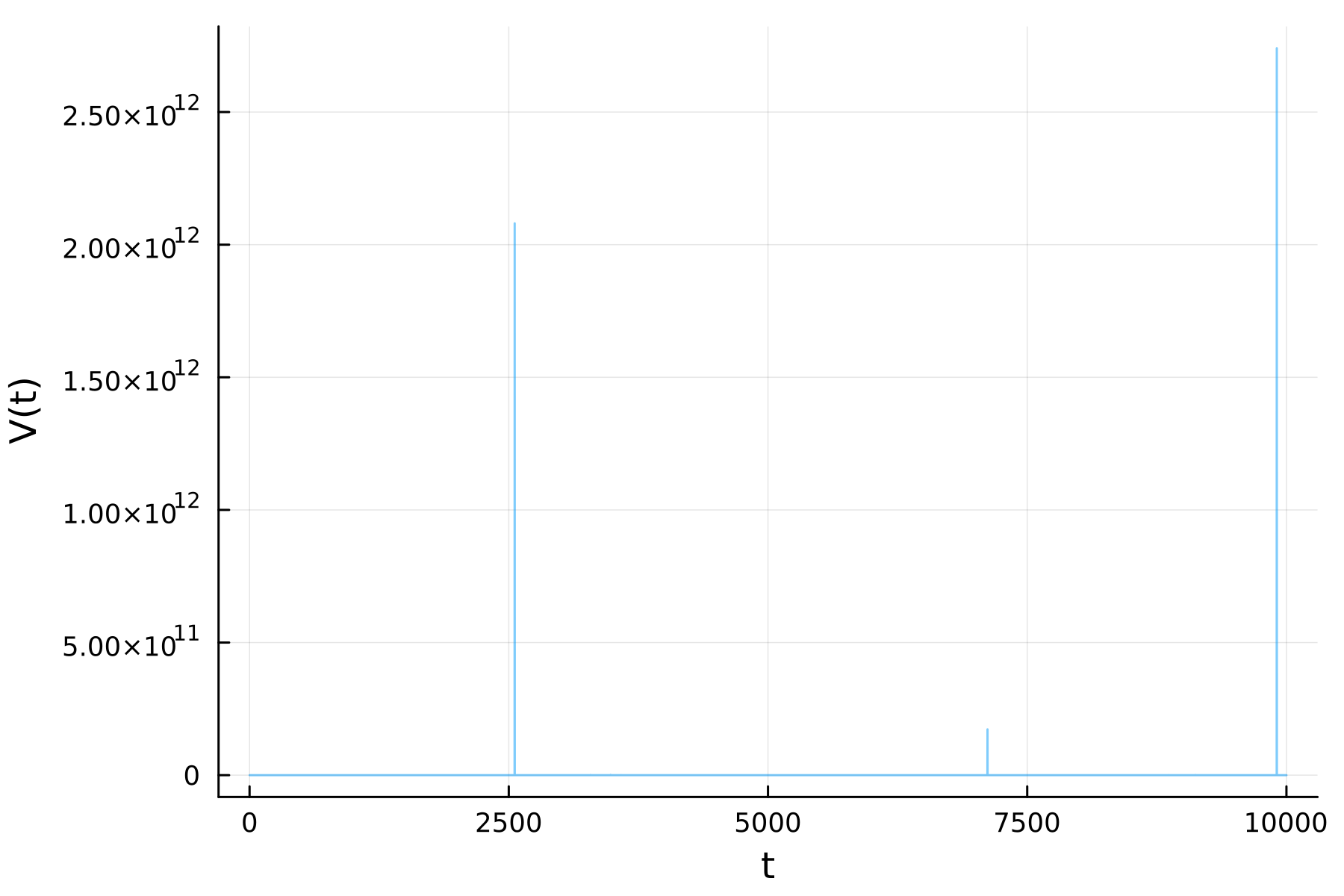}
  \end{subfigure}

  \begin{subfigure}{0.245\textwidth}
      \centering
      \includegraphics[width=\textwidth]{img/transfection/pairwise_corr_chain_0.png}
  \end{subfigure}
  \begin{subfigure}{0.245\textwidth}
      \centering
      \includegraphics[width=\textwidth]{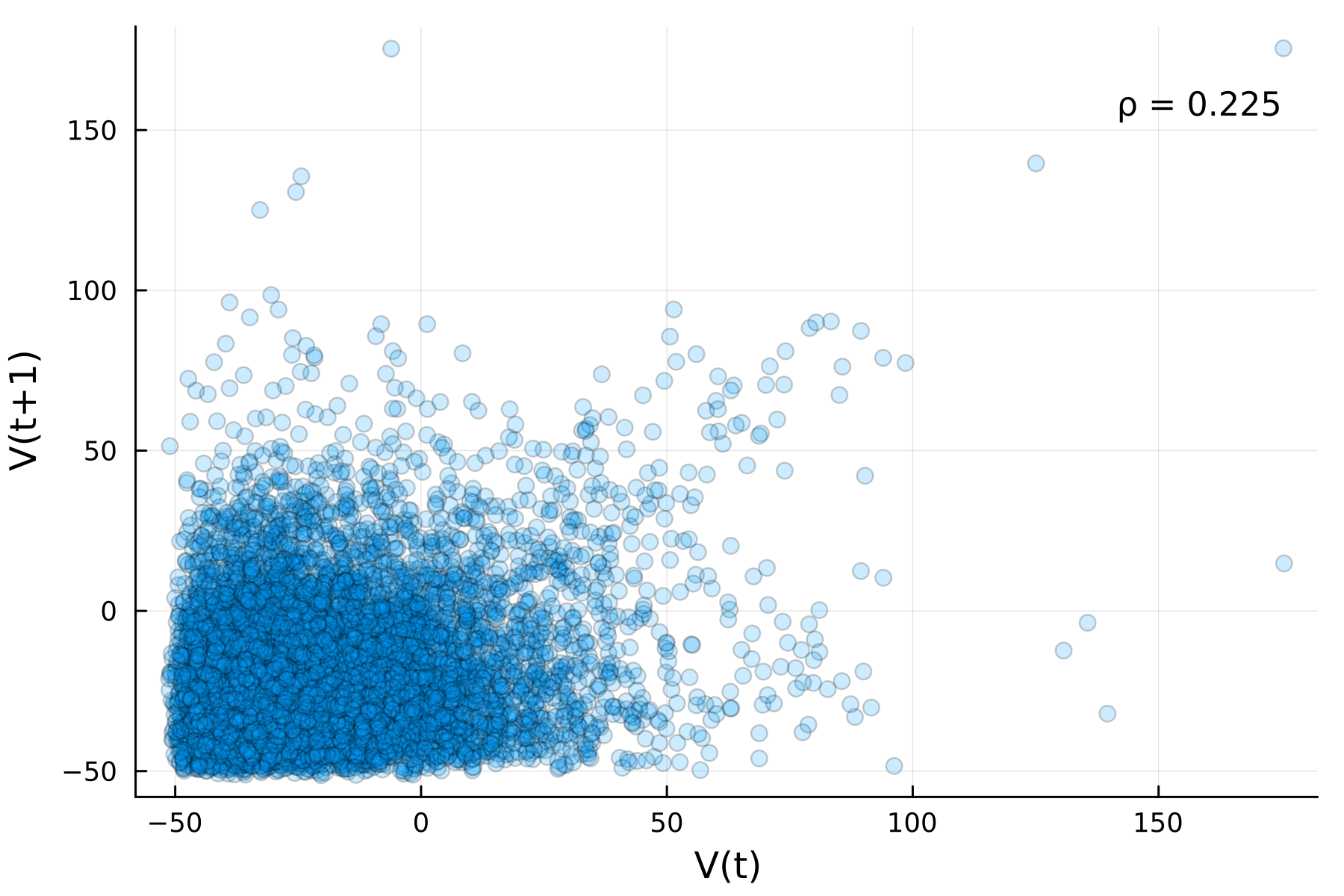}
  \end{subfigure}
  \begin{subfigure}{0.245\textwidth}
      \centering
      \includegraphics[width=\textwidth]{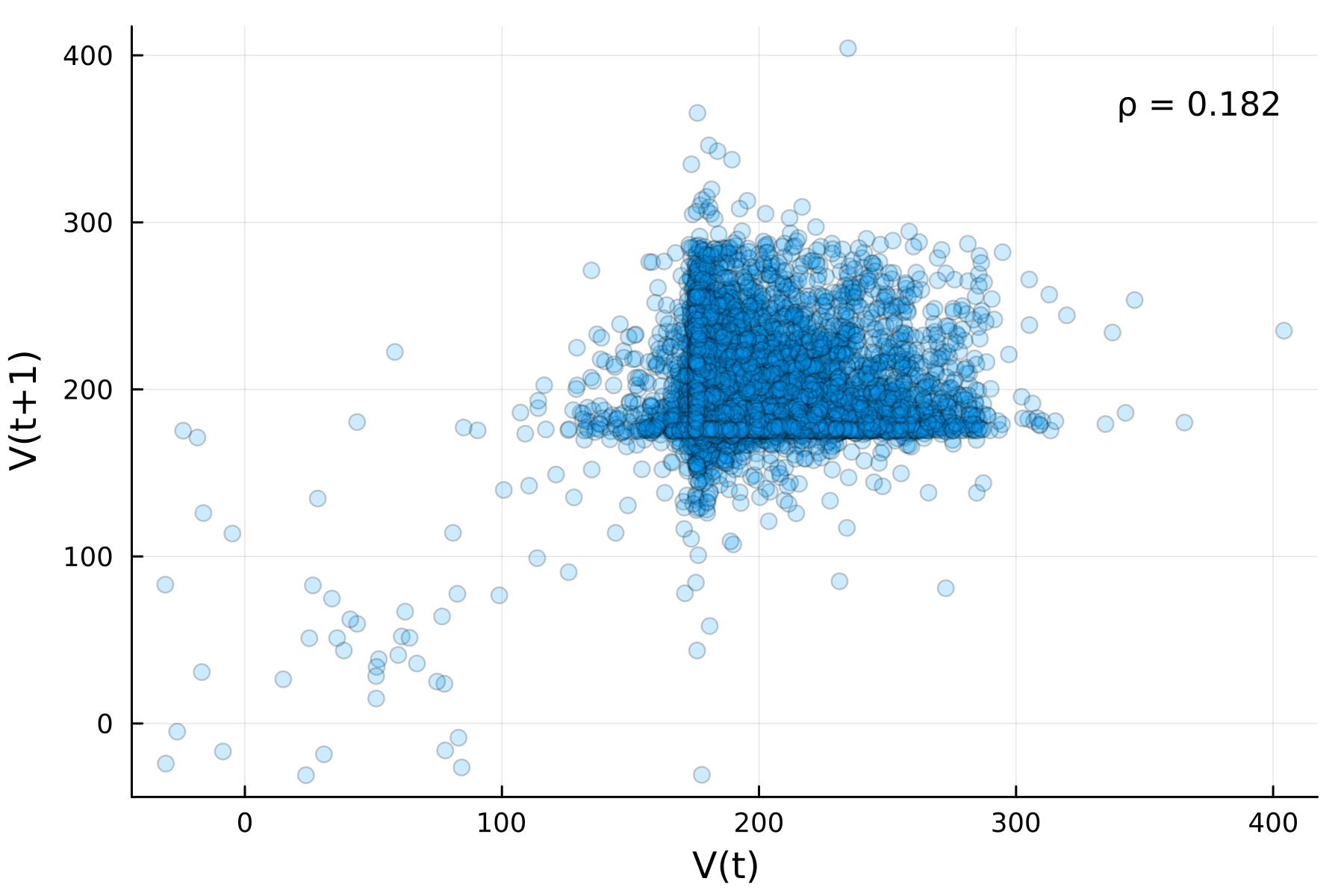}
  \end{subfigure}
  \begin{subfigure}{0.245\textwidth}
      \centering
      \includegraphics[width=\textwidth]{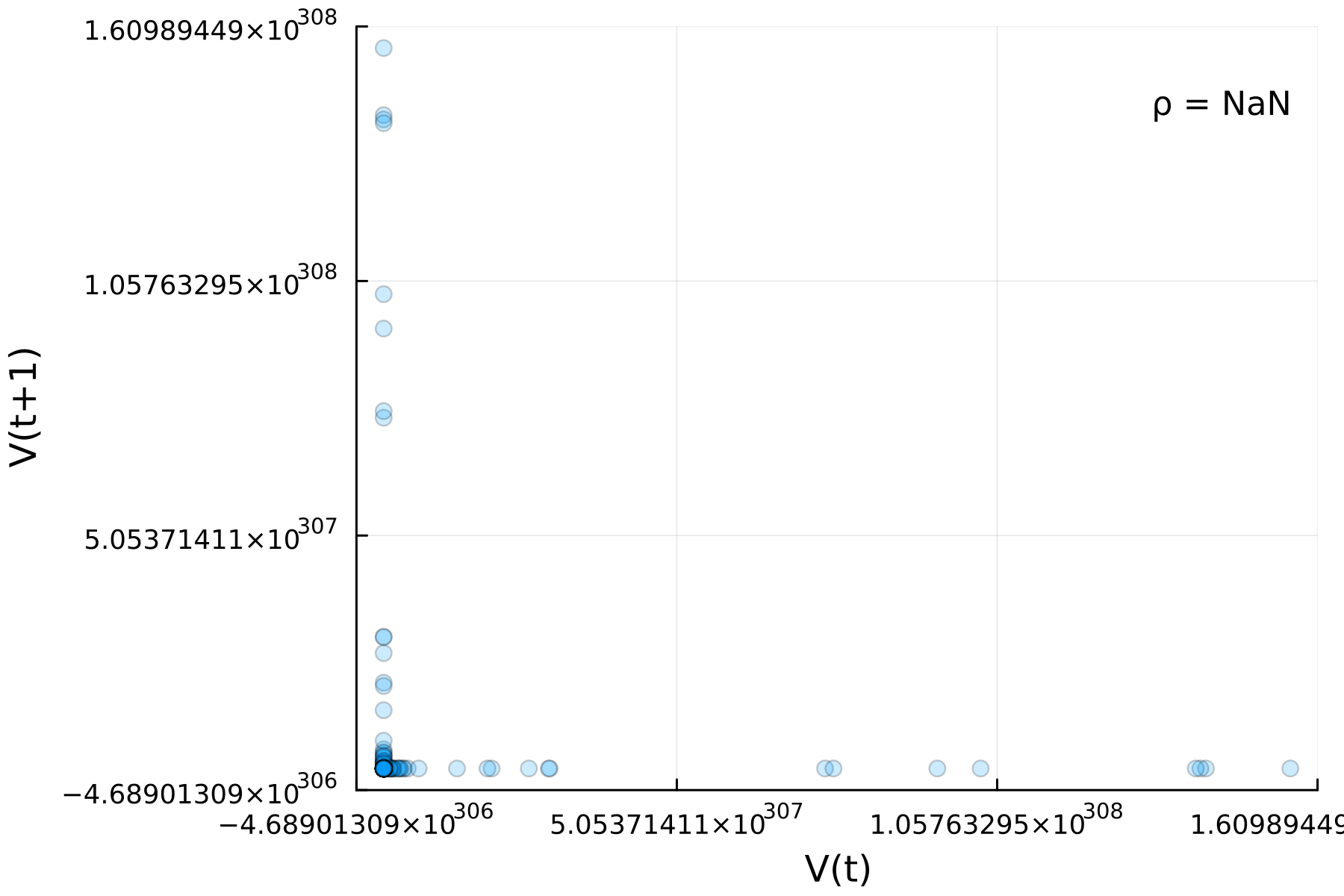}
  \end{subfigure}
  \begin{subfigure}{0.245\textwidth}
      \centering
      \includegraphics[width=\textwidth]{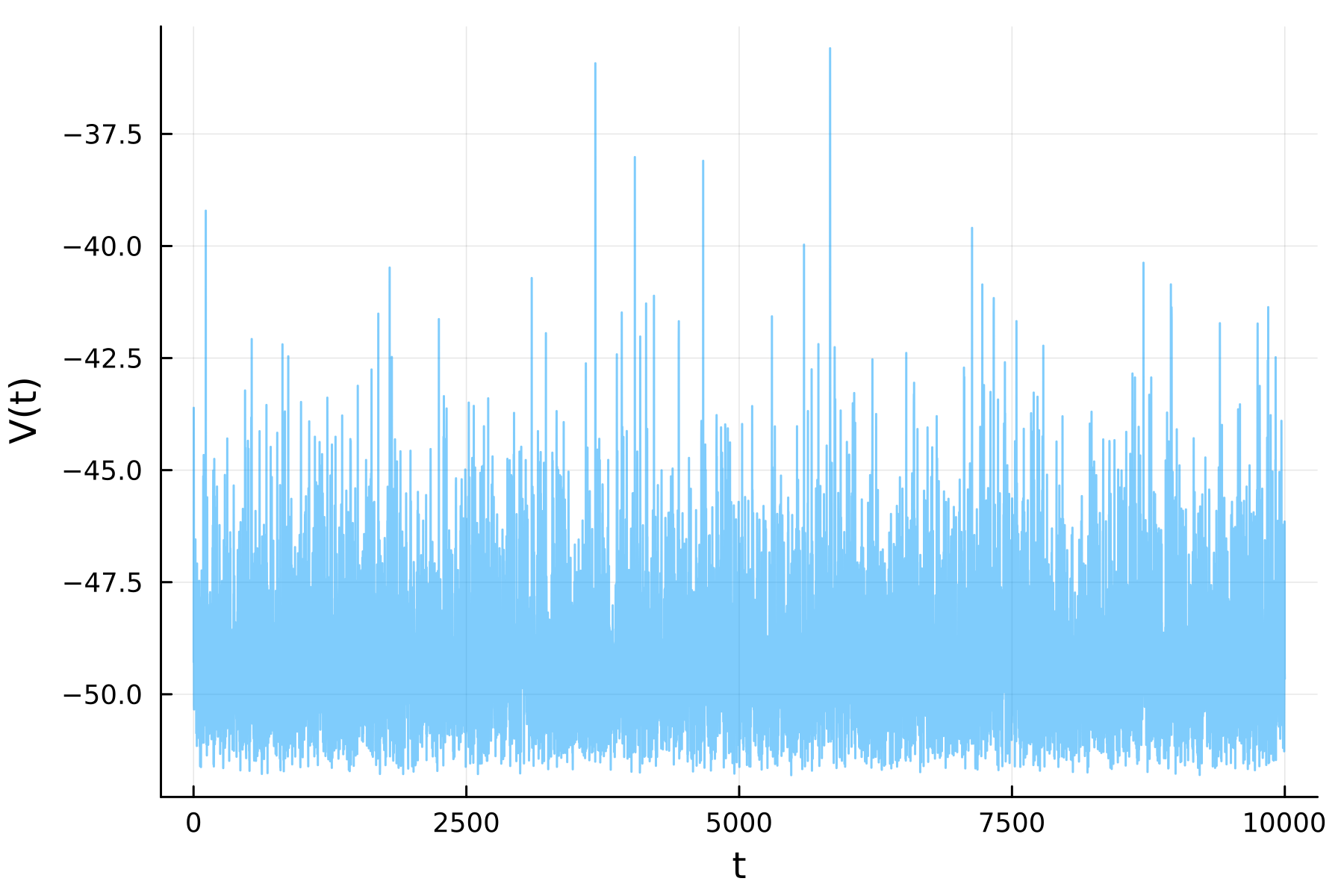}
  \end{subfigure}
  \begin{subfigure}{0.245\textwidth}
      \centering
      \includegraphics[width=\textwidth]{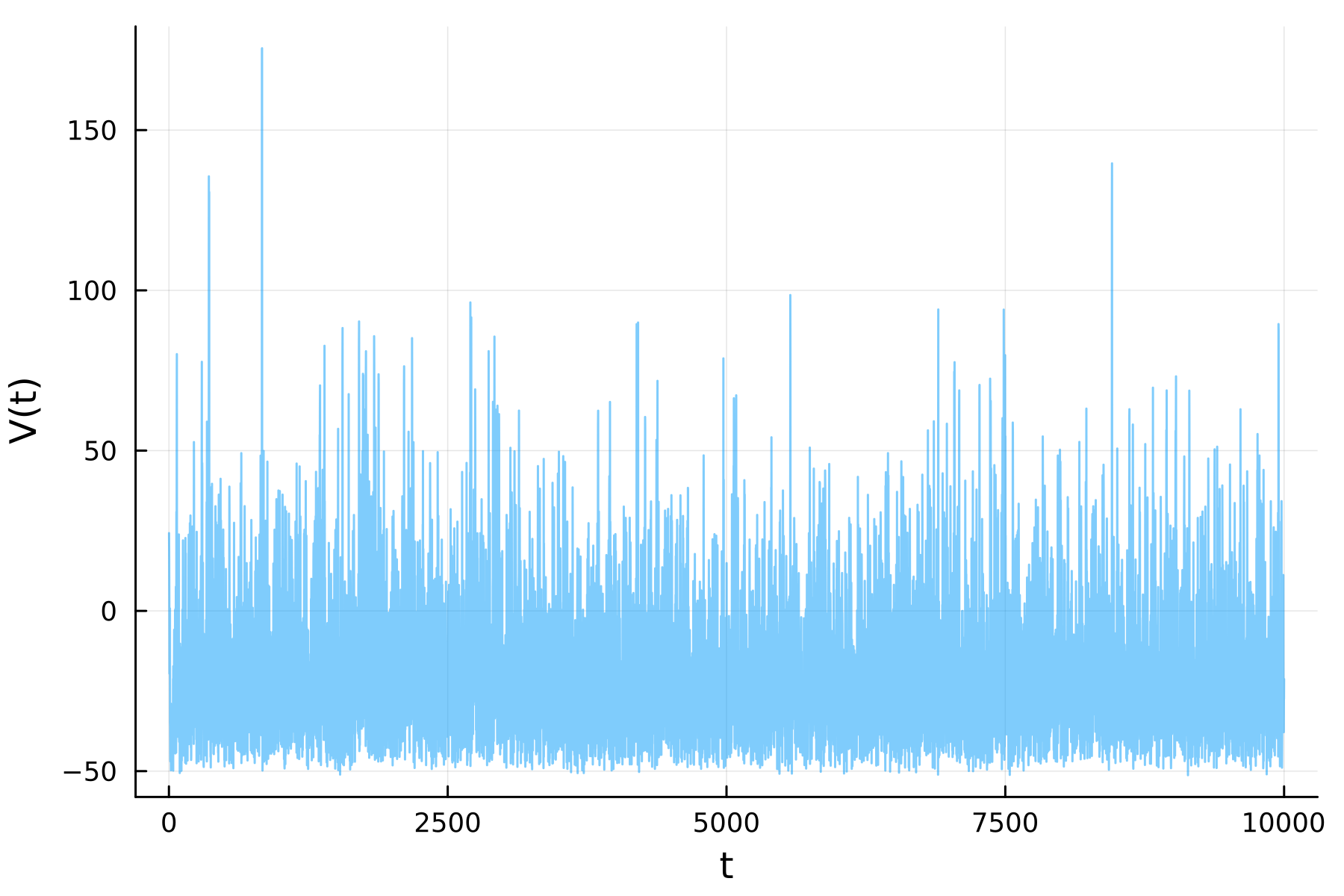}
  \end{subfigure}
  \begin{subfigure}{0.245\textwidth}
      \centering
      \includegraphics[width=\textwidth]{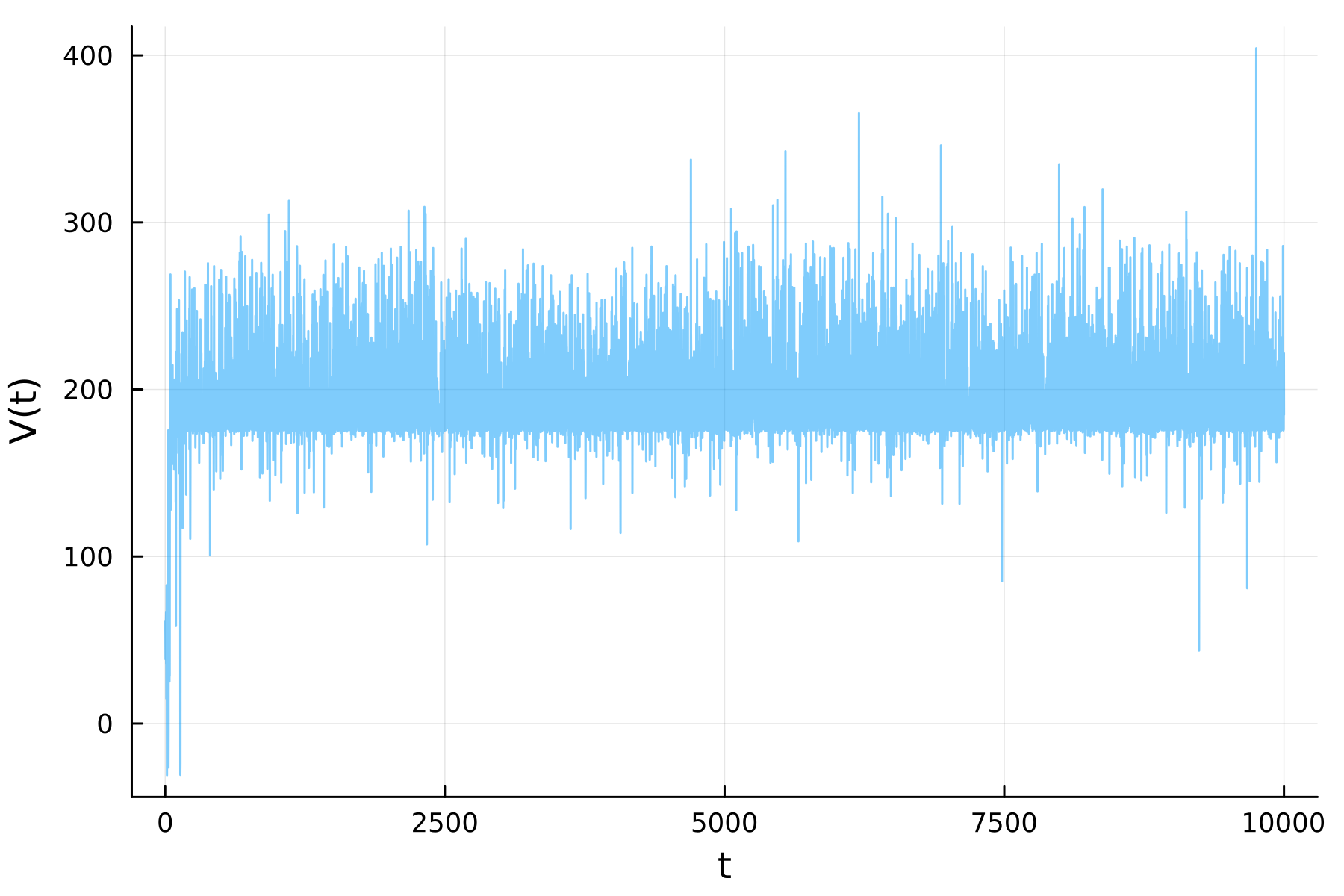}
  \end{subfigure}
  \begin{subfigure}{0.245\textwidth}
      \centering
      \includegraphics[width=\textwidth]{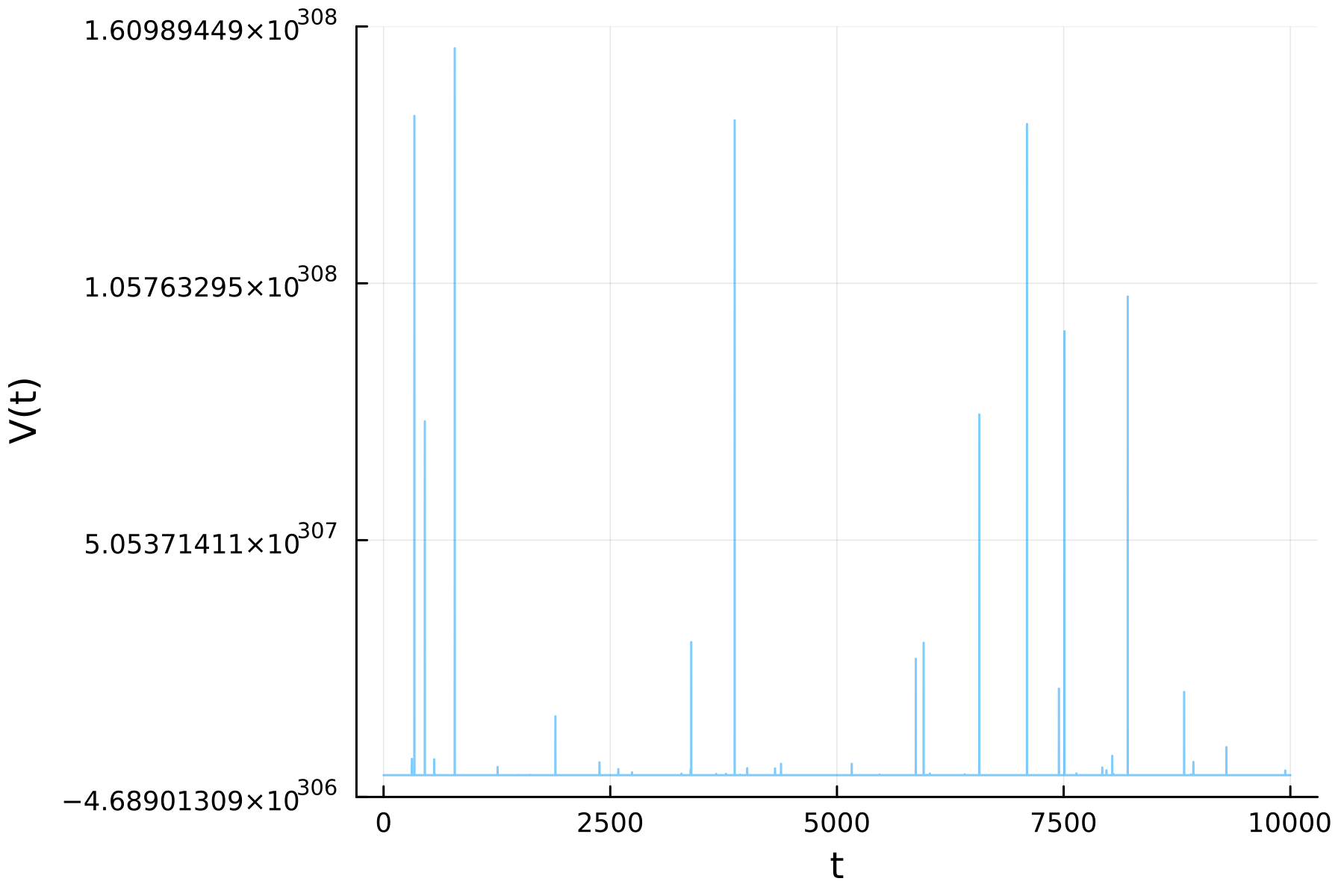}
  \end{subfigure}

  \begin{subfigure}{0.245\textwidth}
      \centering
      \includegraphics[width=\textwidth]{img/Ising-critical/pairwise_corr_chain_0.png}
  \end{subfigure}
  \begin{subfigure}{0.245\textwidth}
      \centering
      \includegraphics[width=\textwidth]{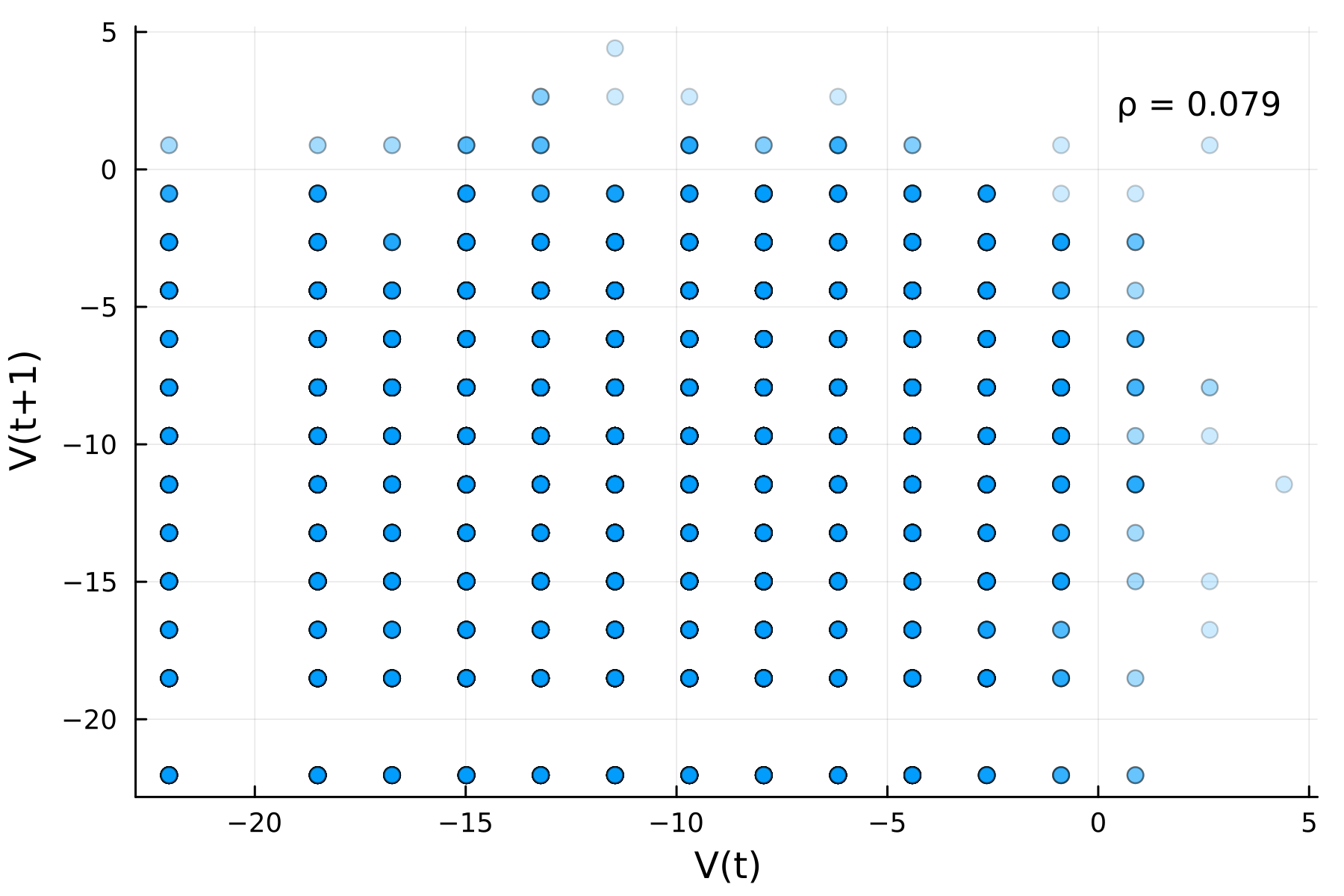}
  \end{subfigure}
  \begin{subfigure}{0.245\textwidth}
      \centering
      \includegraphics[width=\textwidth]{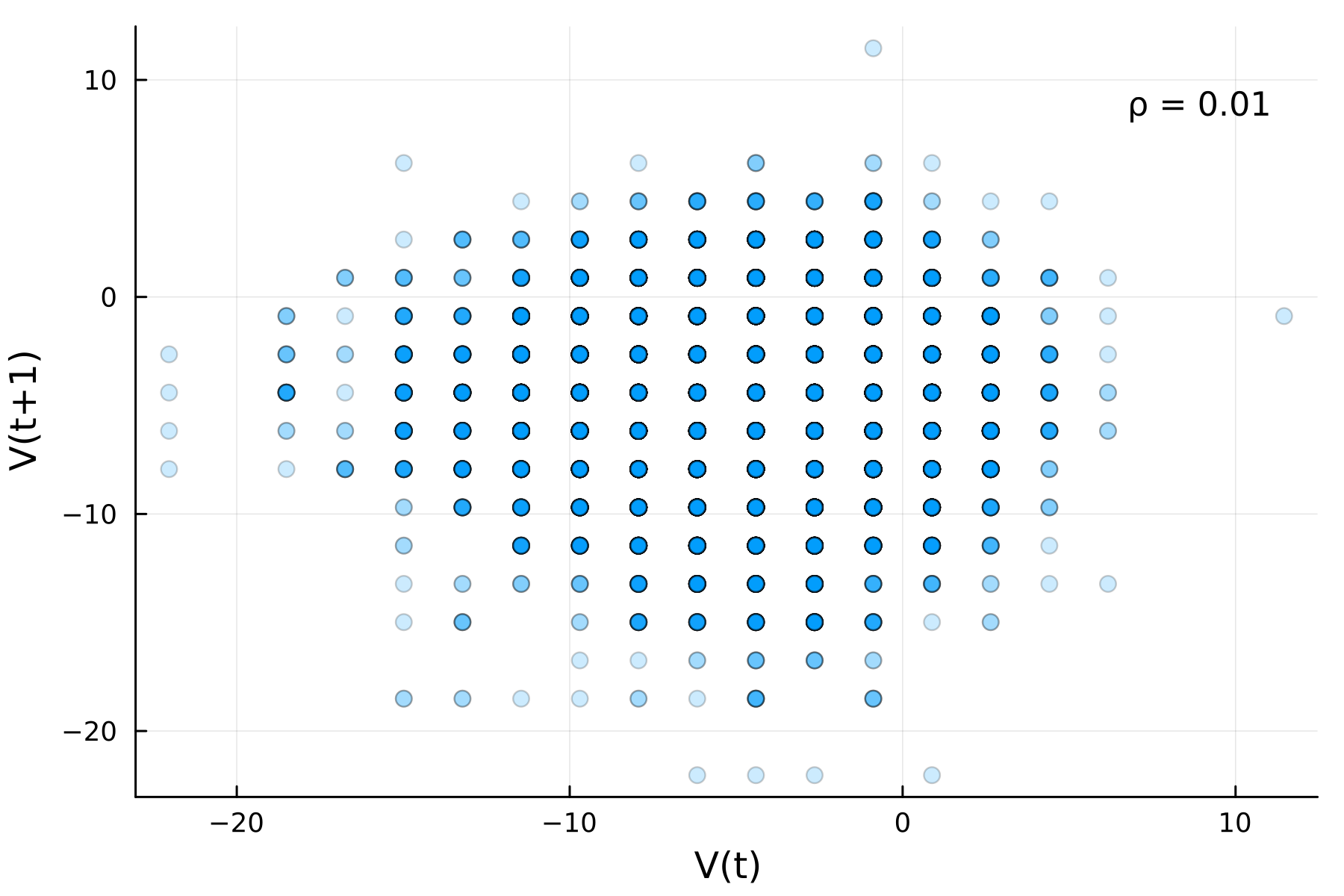}
  \end{subfigure}
  \begin{subfigure}{0.245\textwidth}
      \centering
      \includegraphics[width=\textwidth]{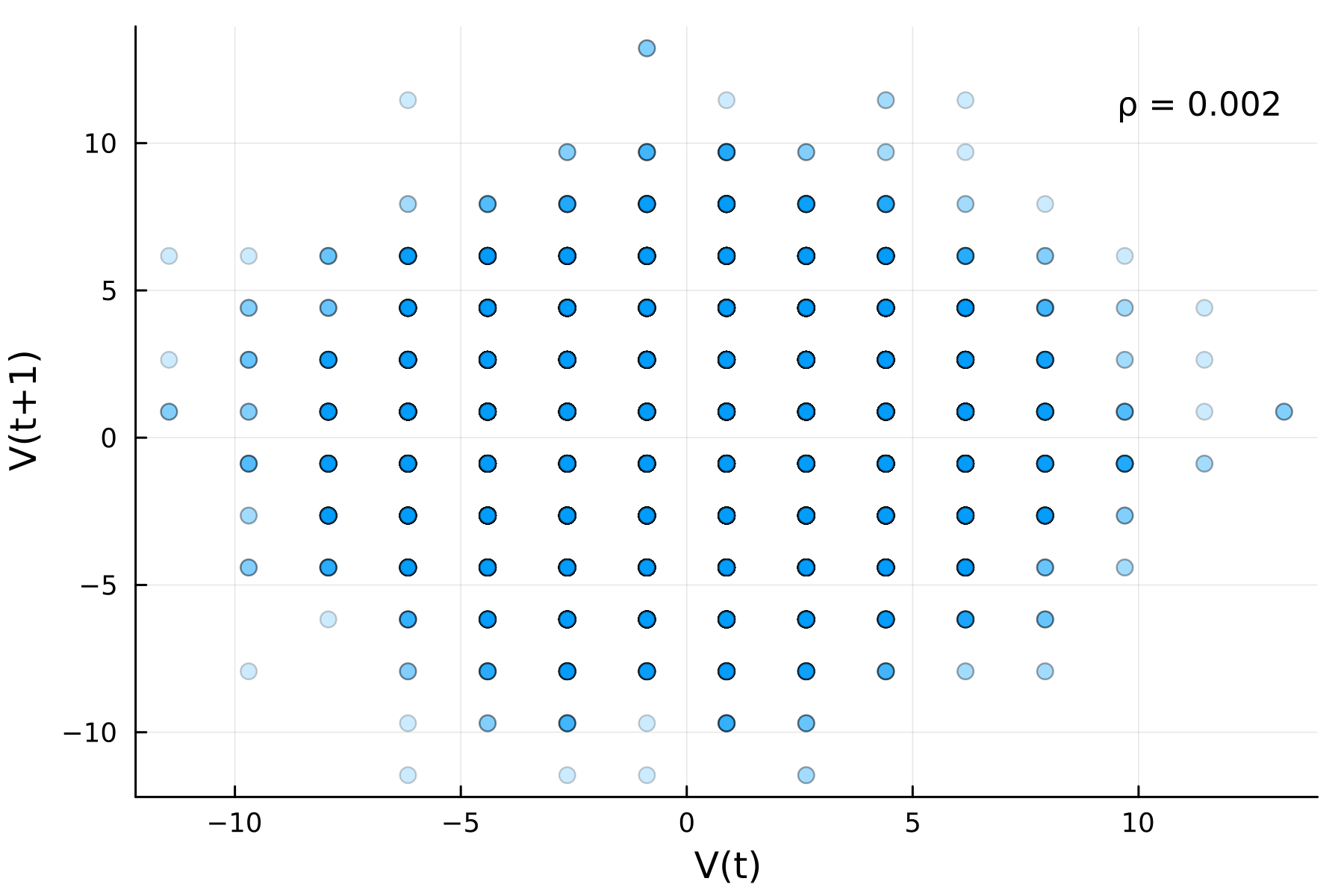}
  \end{subfigure}
  \begin{subfigure}{0.245\textwidth}
      \centering
      \includegraphics[width=\textwidth]{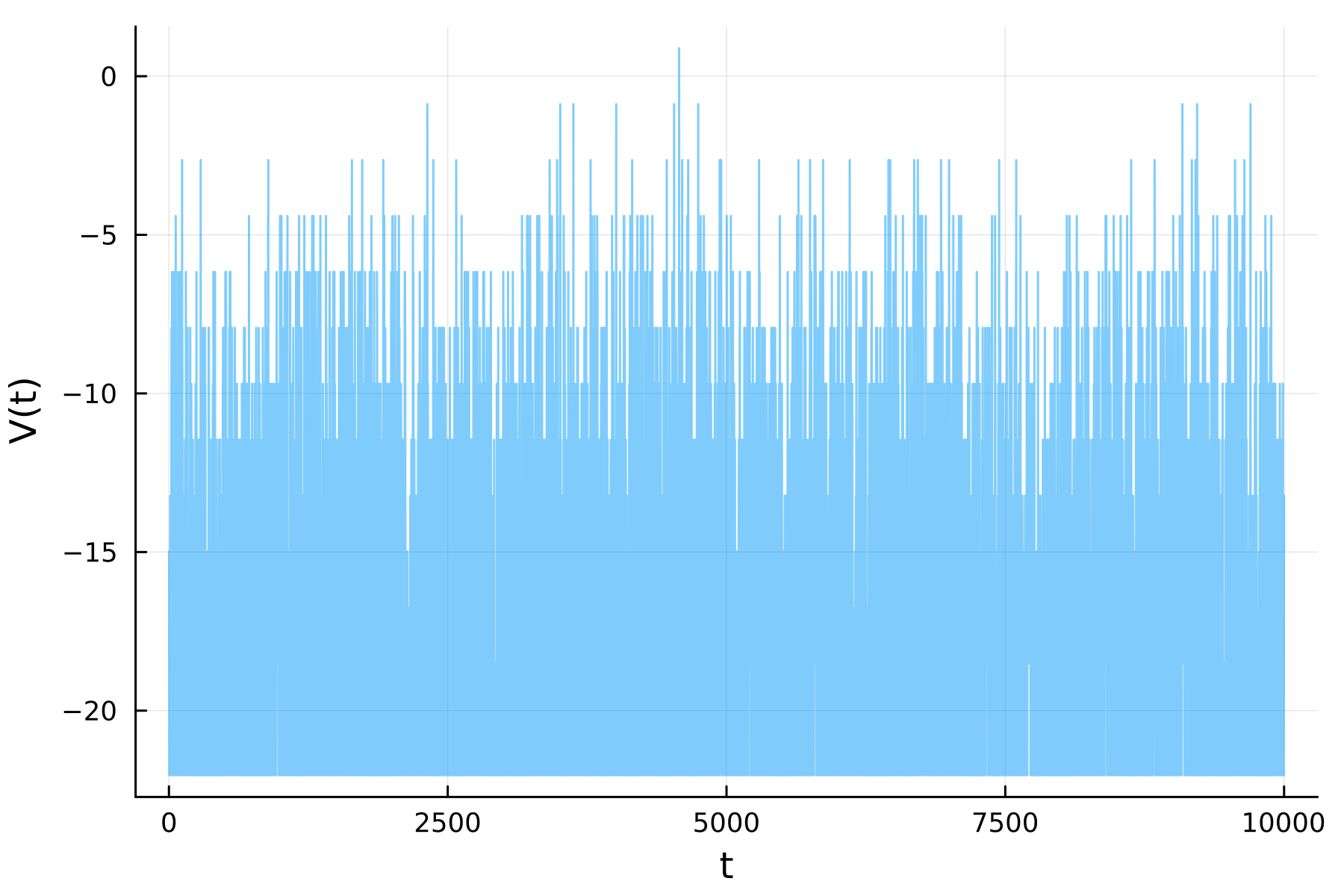}
  \end{subfigure}
  \begin{subfigure}{0.245\textwidth}
      \centering
      \includegraphics[width=\textwidth]{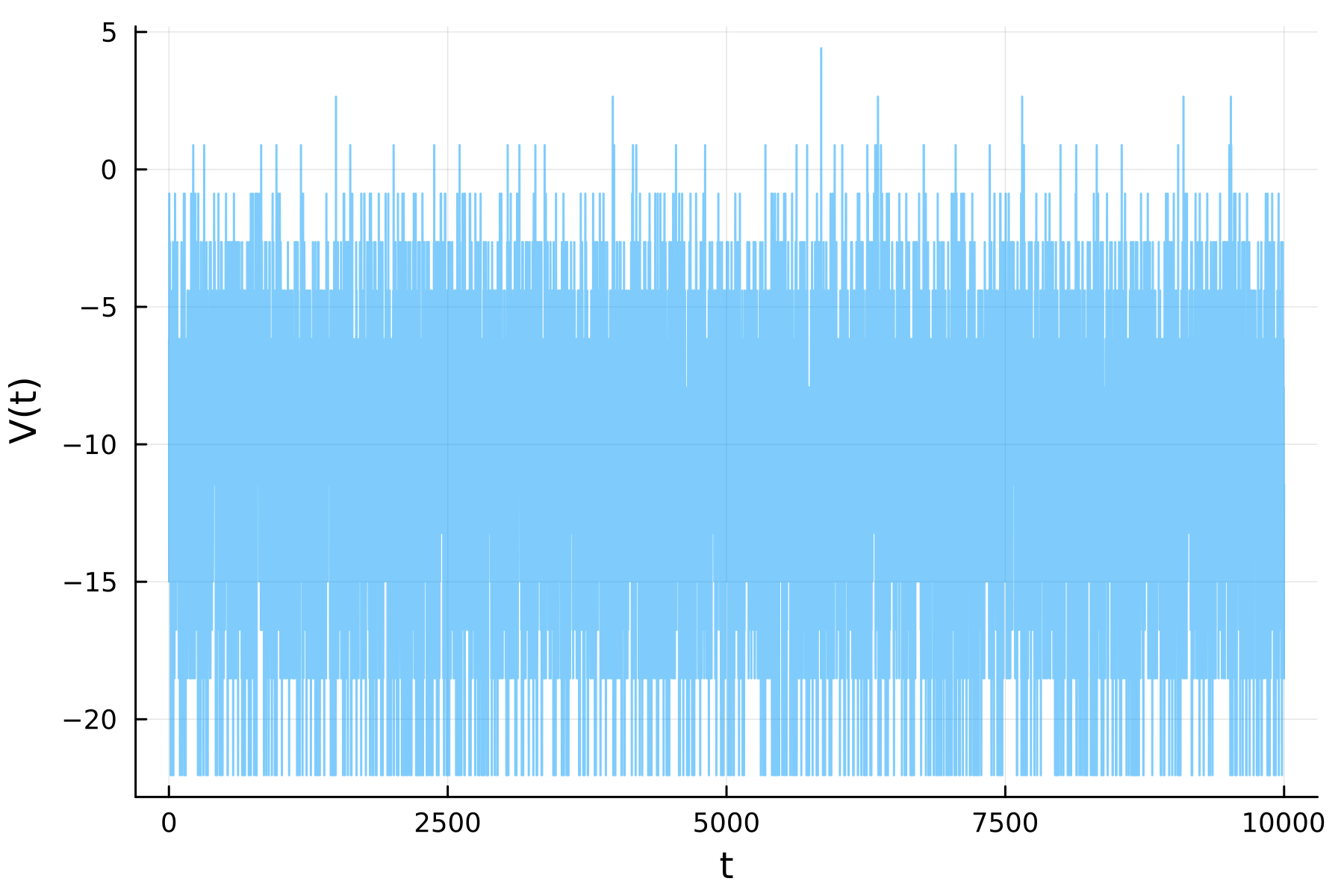}
  \end{subfigure}
  \begin{subfigure}{0.245\textwidth}
      \centering
      \includegraphics[width=\textwidth]{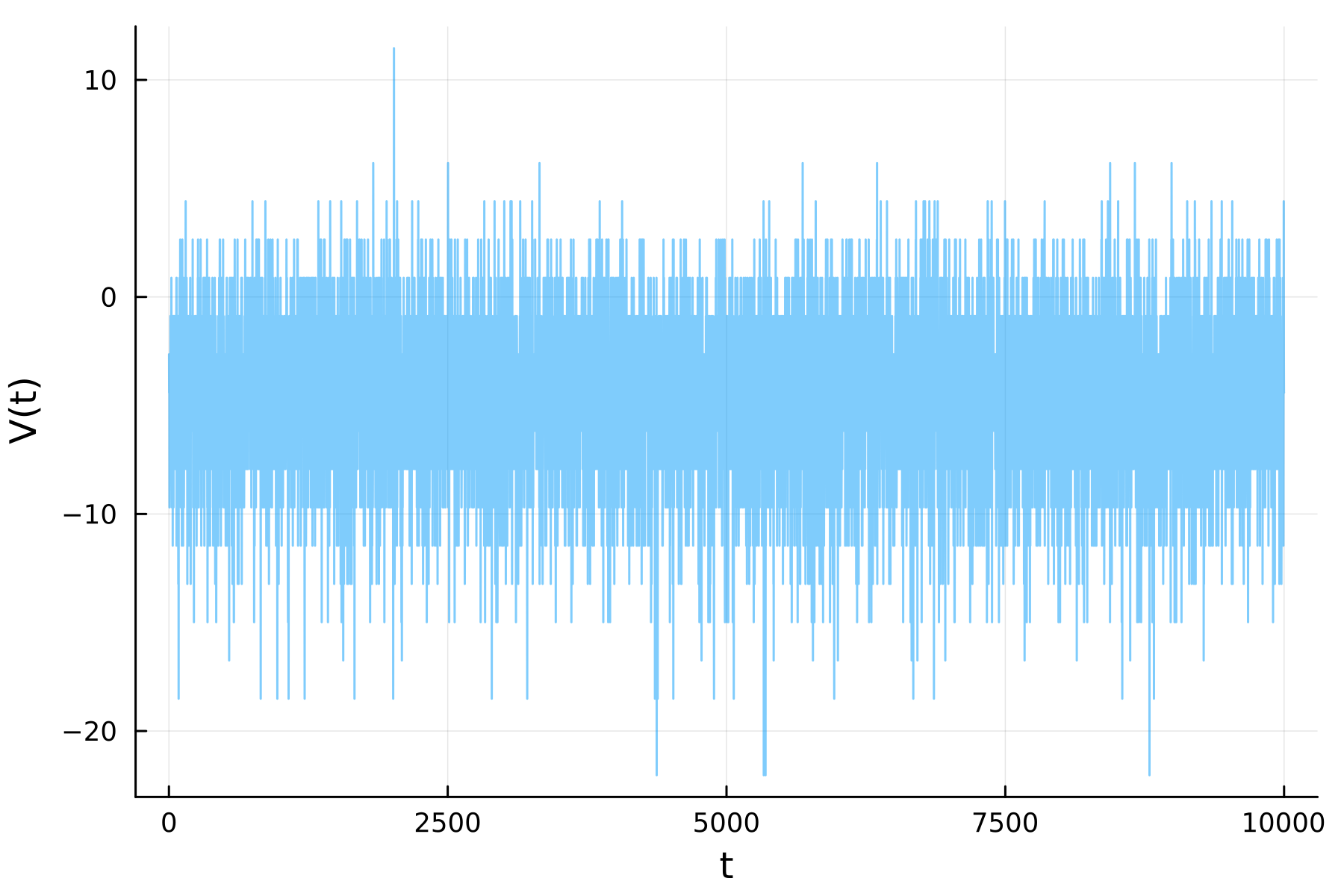}
  \end{subfigure}
  \begin{subfigure}{0.245\textwidth}
      \centering
      \includegraphics[width=\textwidth]{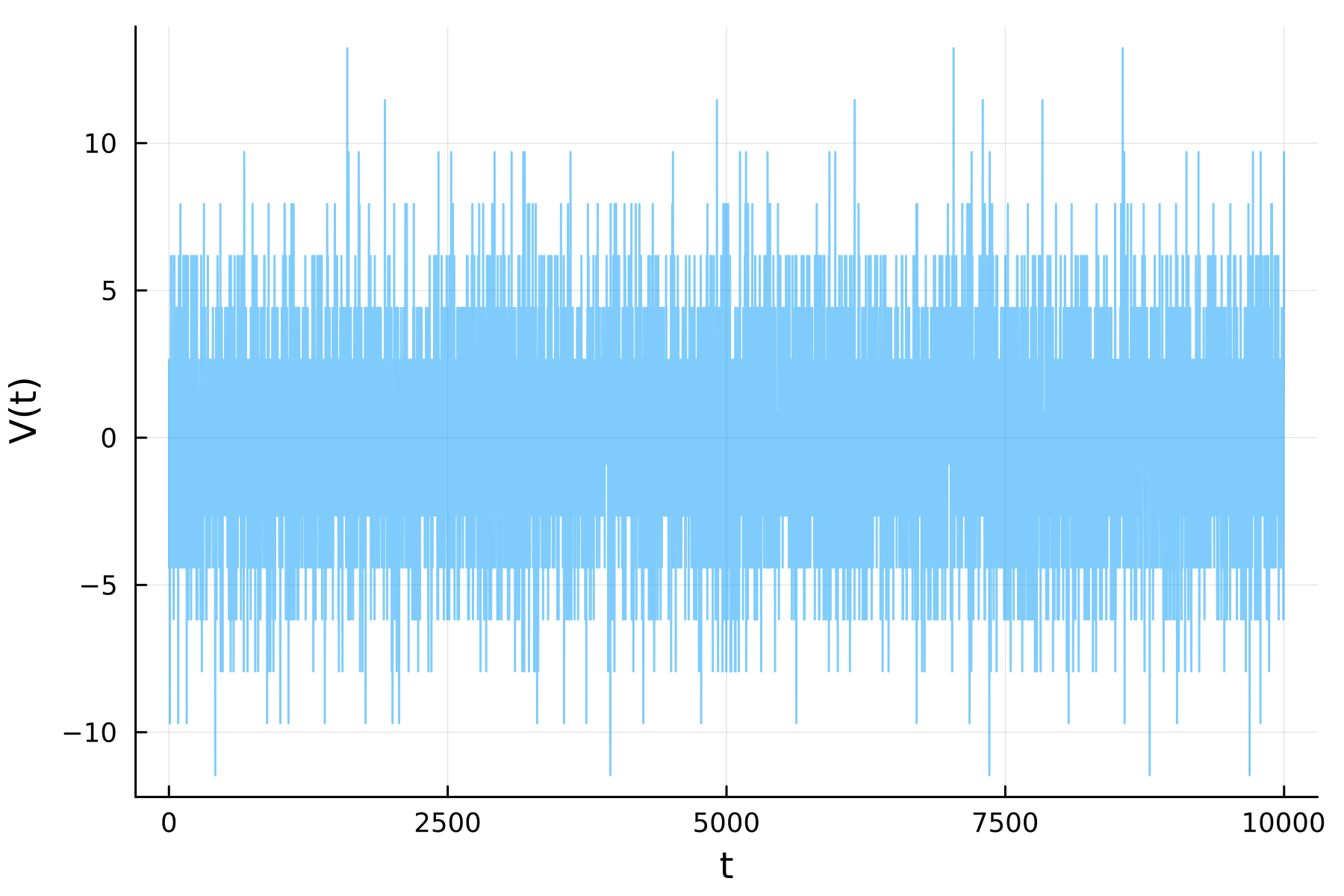}
  \end{subfigure}

  \begin{subfigure}{0.245\textwidth}
      \centering
      \includegraphics[width=\textwidth]{img/chromobreak/pairwise_corr_chain_0.png}
  \end{subfigure}
  \begin{subfigure}{0.245\textwidth}
      \centering
      \includegraphics[width=\textwidth]{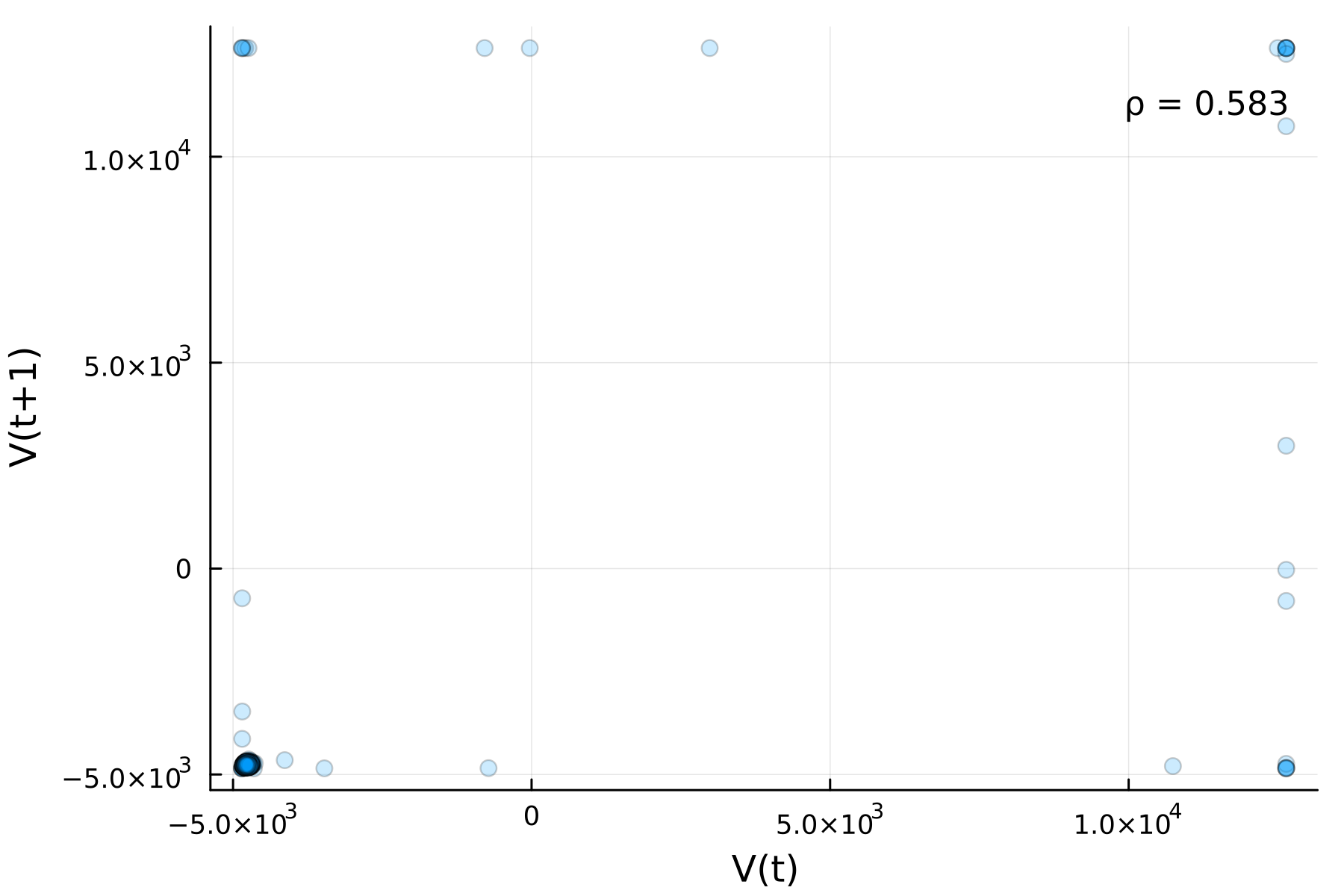}
  \end{subfigure}
  \begin{subfigure}{0.245\textwidth}
      \centering
      \includegraphics[width=\textwidth]{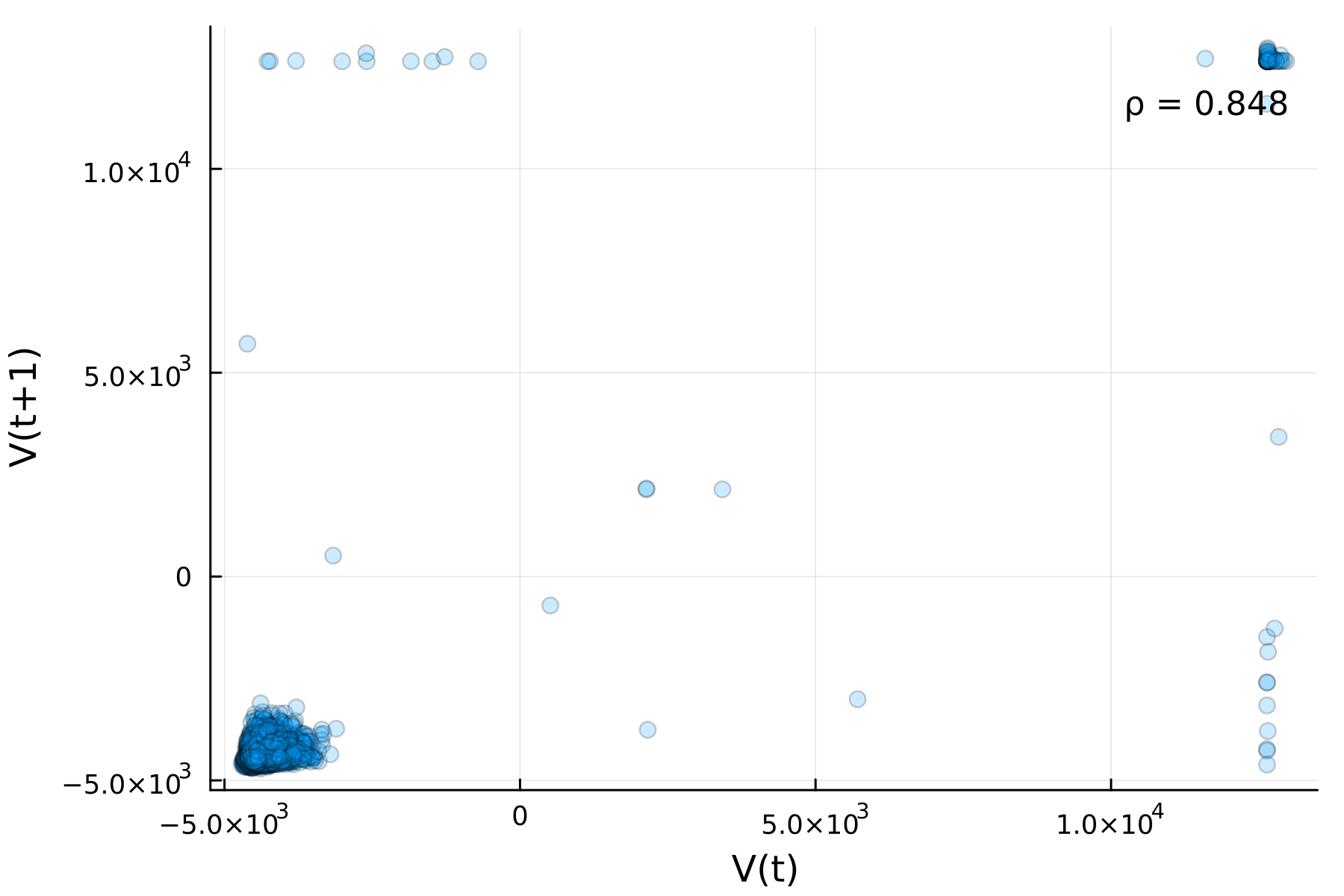}
  \end{subfigure}
  \begin{subfigure}{0.245\textwidth}
      \centering
      \includegraphics[width=\textwidth]{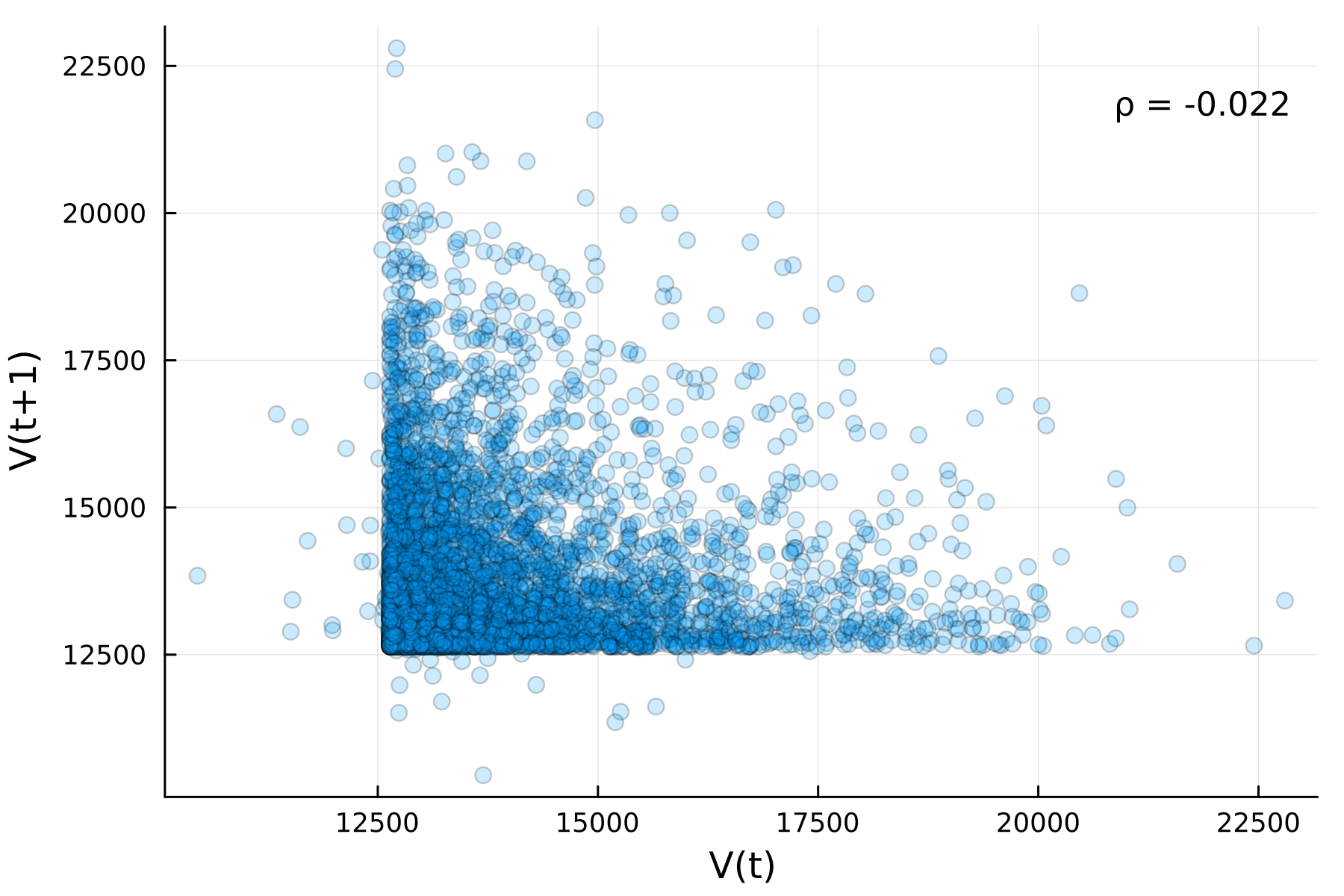}
  \end{subfigure}
  \begin{subfigure}{0.245\textwidth}
      \centering
      \includegraphics[width=\textwidth]{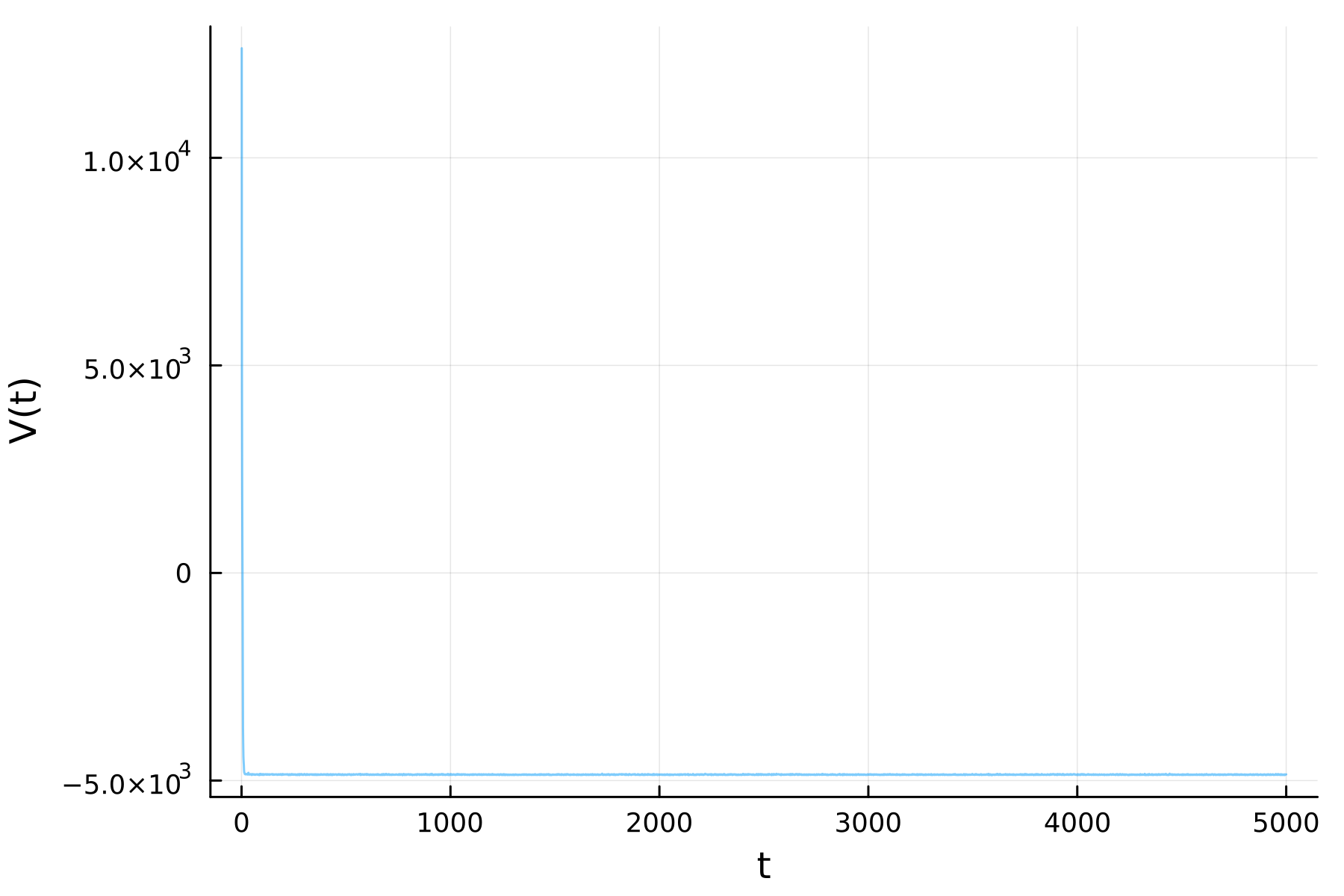}
  \end{subfigure}
  \begin{subfigure}{0.245\textwidth}
      \centering
      \includegraphics[width=\textwidth]{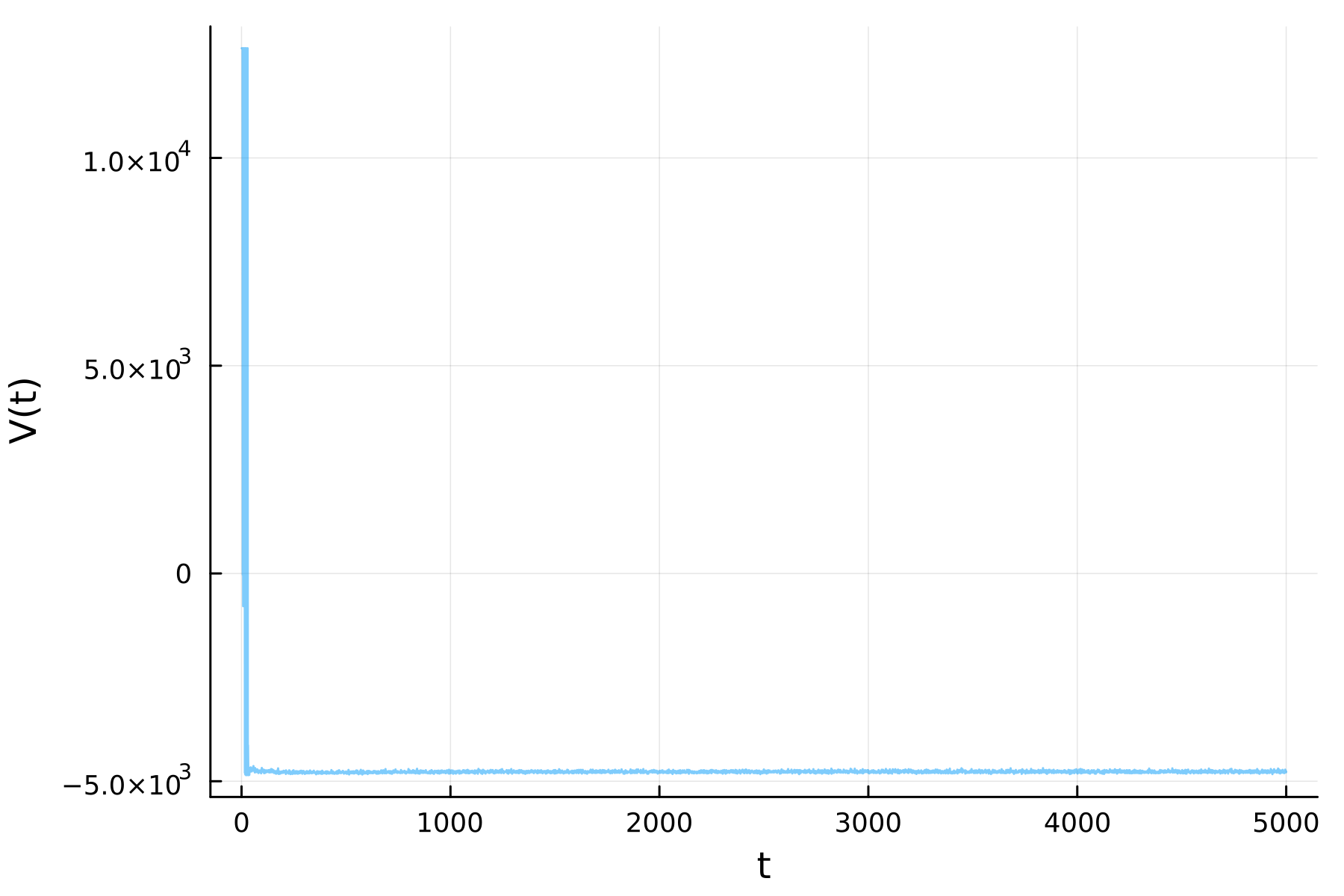}
  \end{subfigure}
  \begin{subfigure}{0.245\textwidth}
      \centering
      \includegraphics[width=\textwidth]{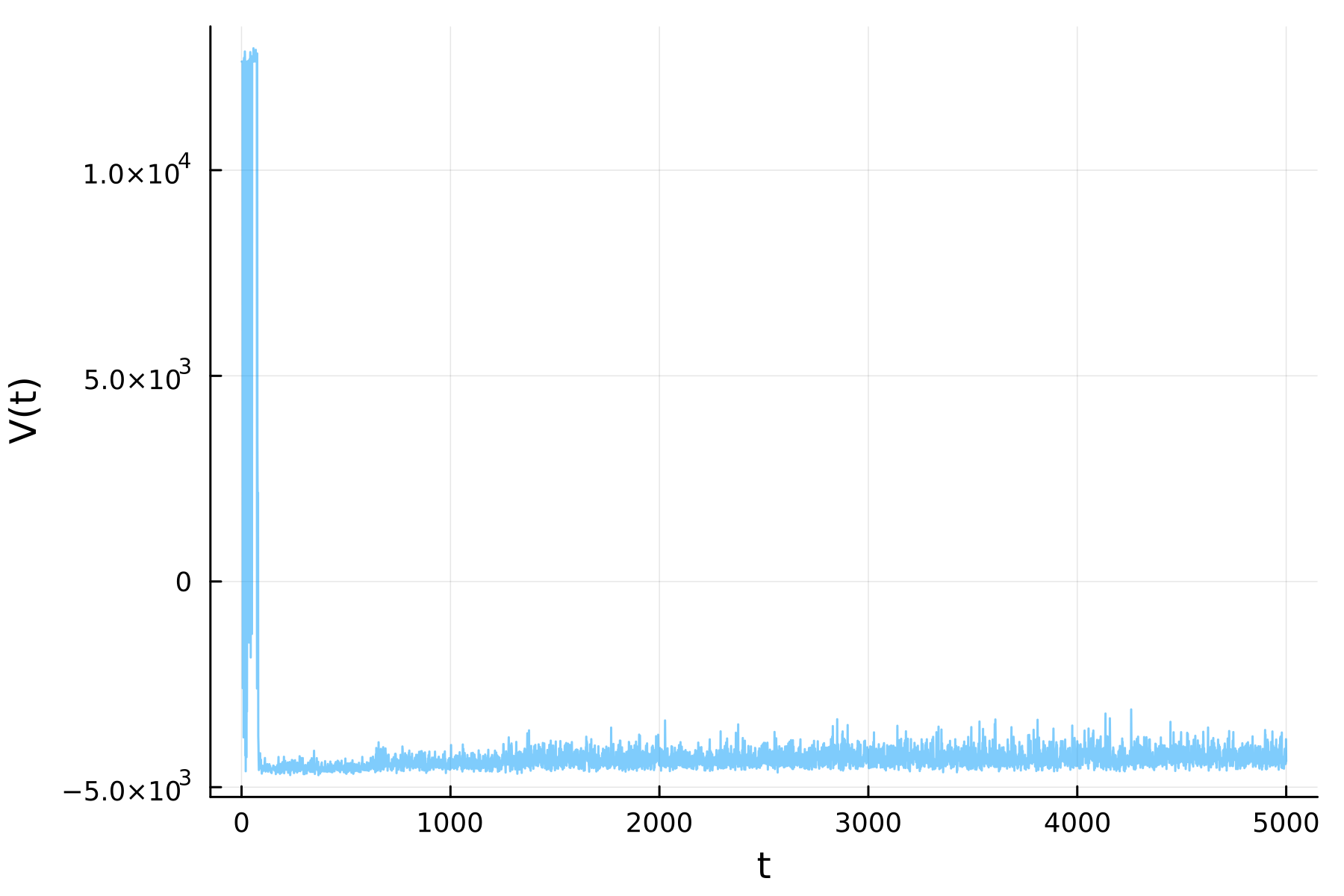}
  \end{subfigure}
  \begin{subfigure}{0.245\textwidth}
      \centering
      \includegraphics[width=\textwidth]{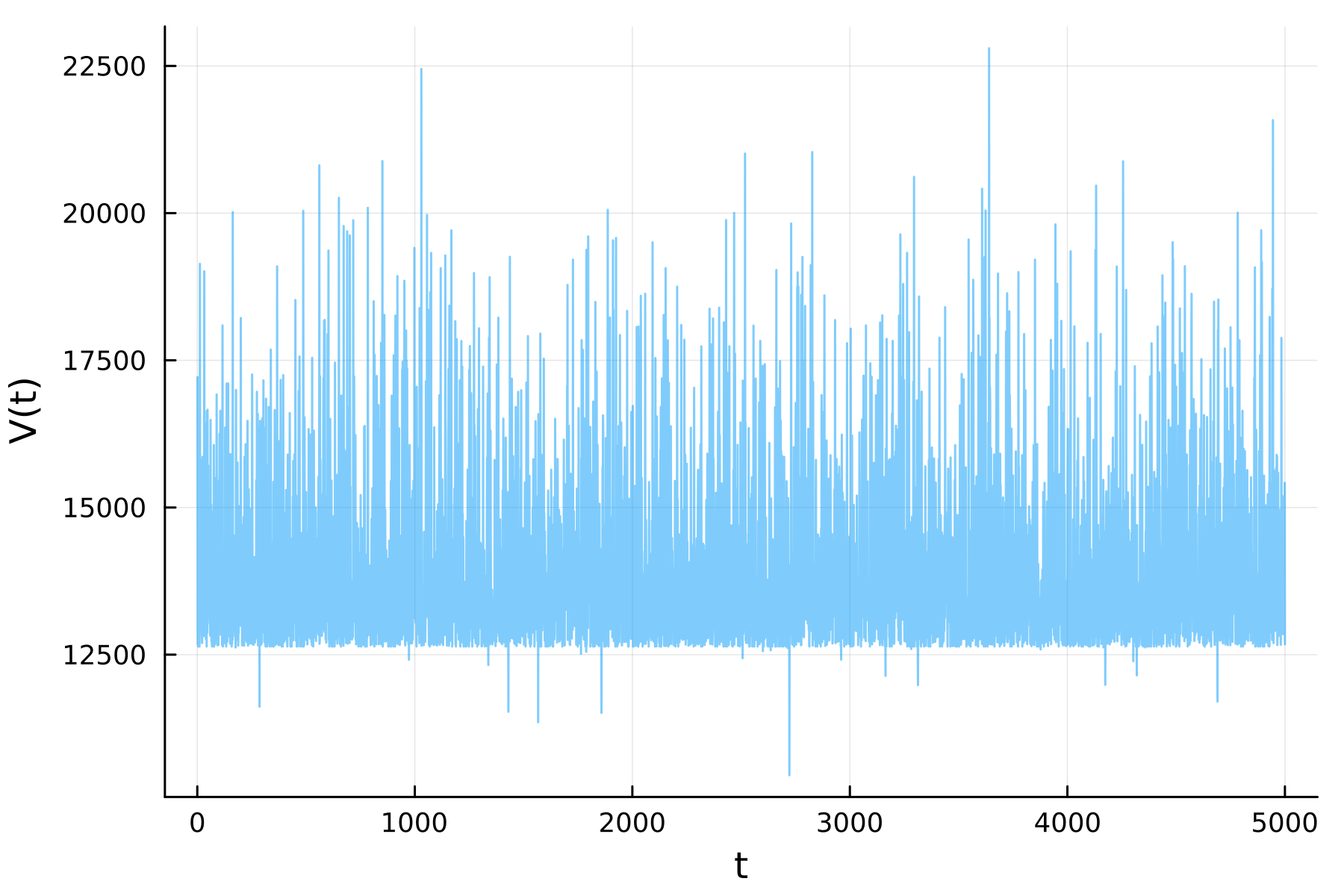}
  \end{subfigure}
  \caption{Pairwise correlation plots for the energy pairs $\{(V_t, V_{t+1})\}_{t=0}^{T-1}$
  from various chains along the annealing path, along with corresponding traceplots 
  of the energy values. 
  \textbf{Top to bottom:} Bayesian mixture model, ODE parameter estimation, Ising model, 
  copy number inference. 
  For the Bayesian mixture and Ising models, the chains are 29 (target), 20, 10, and 0 (reference).
  For the ODE and copy number inference models, the chains are 49 (target), 33, 17, and 0 (reference).}
  \label{fig:additional_pairwise_corr_traceplot}
\end{figure}

Additional pairwise correlation plots and traceplots of energies are presented in 
\cref{fig:additional_pairwise_corr_traceplot}.

\subsection{Limit of finite-chain bounds}
\begin{figure}[t]
  \centering 
  \begin{subfigure}{0.49\textwidth}
      \centering
      \includegraphics[width=\textwidth]{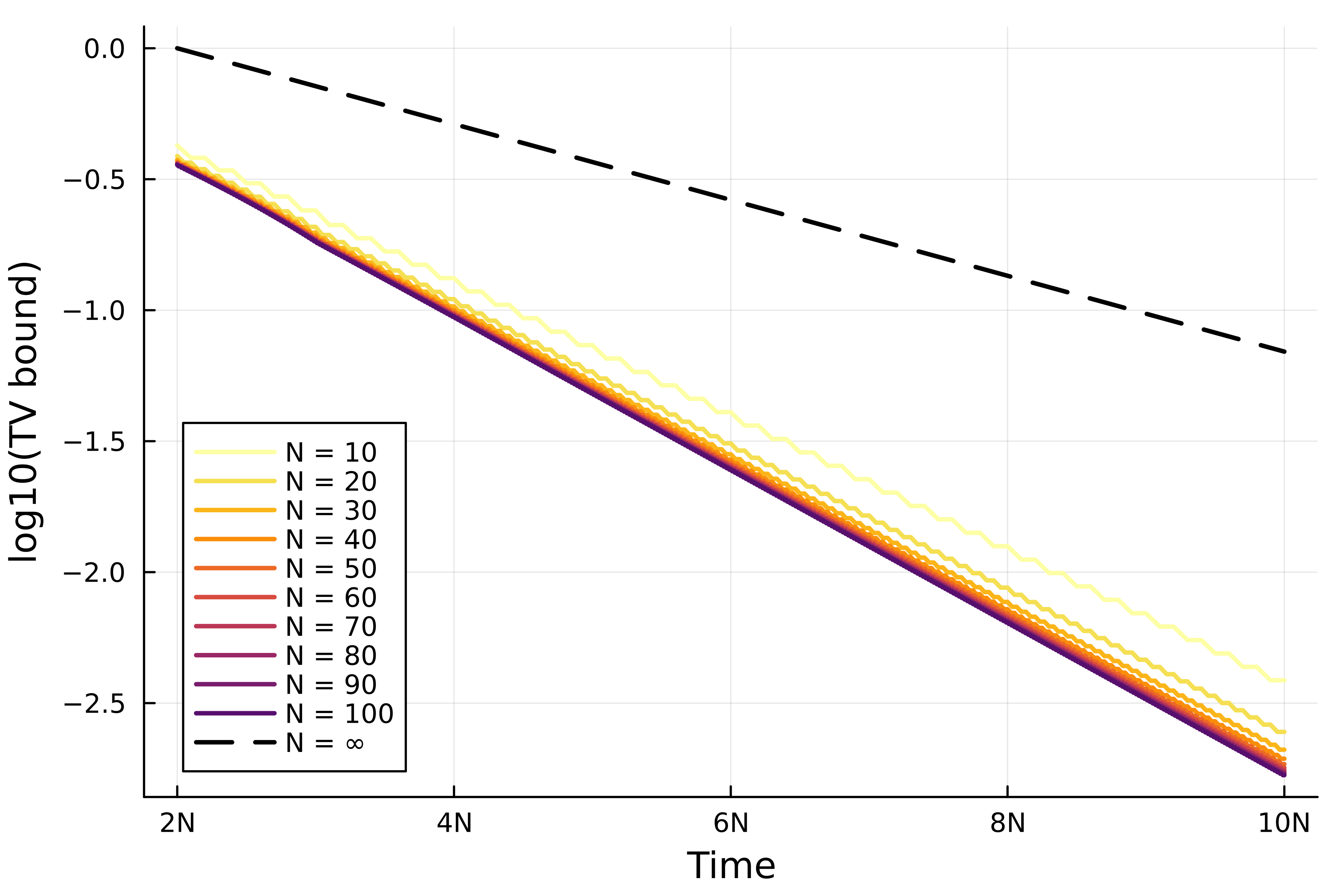}
  \end{subfigure}
  \begin{subfigure}{0.49\textwidth}
      \centering
      \includegraphics[width=\textwidth]{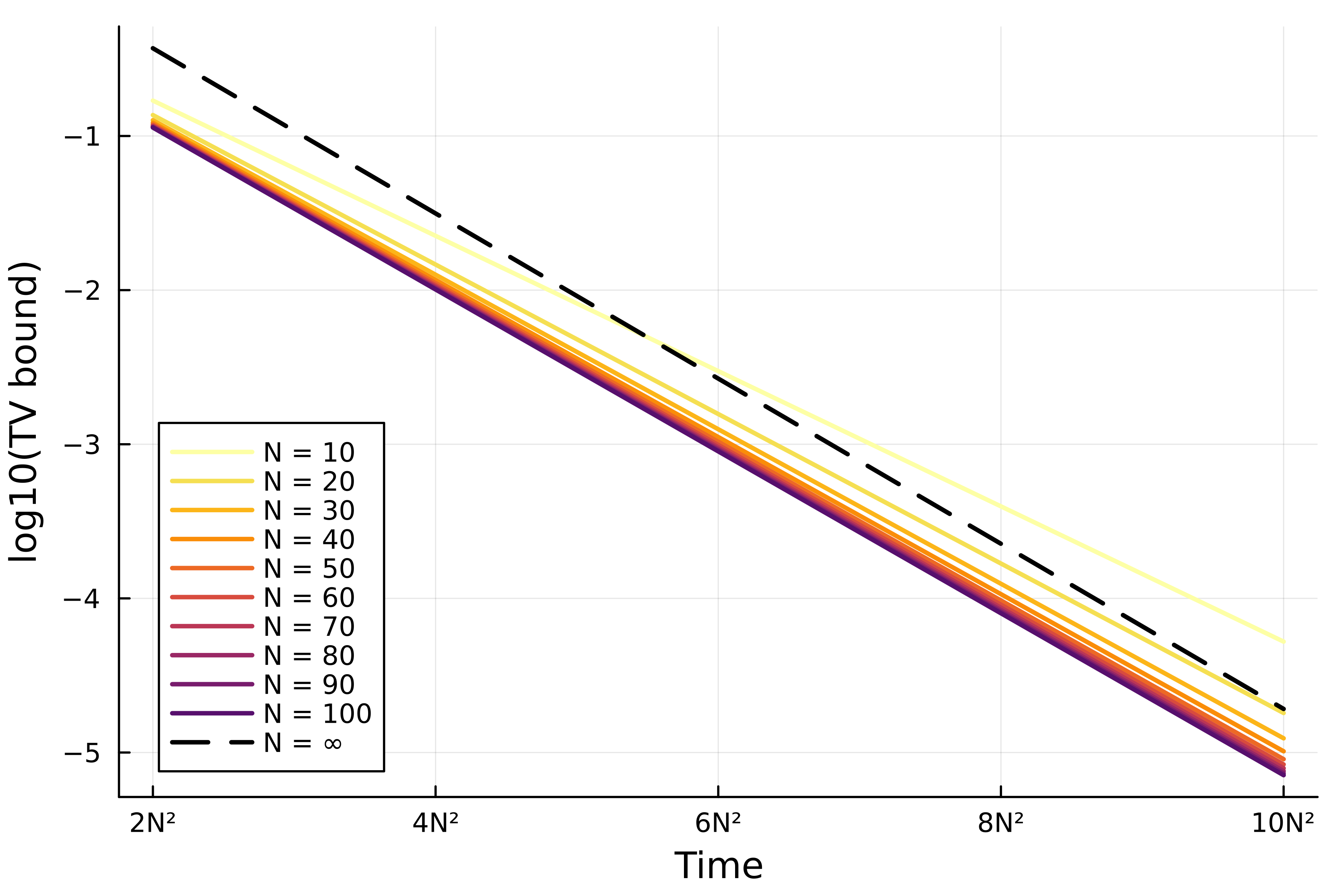}
  \end{subfigure}
  \begin{subfigure}{0.49\textwidth}
      \centering
      \includegraphics[width=\textwidth]{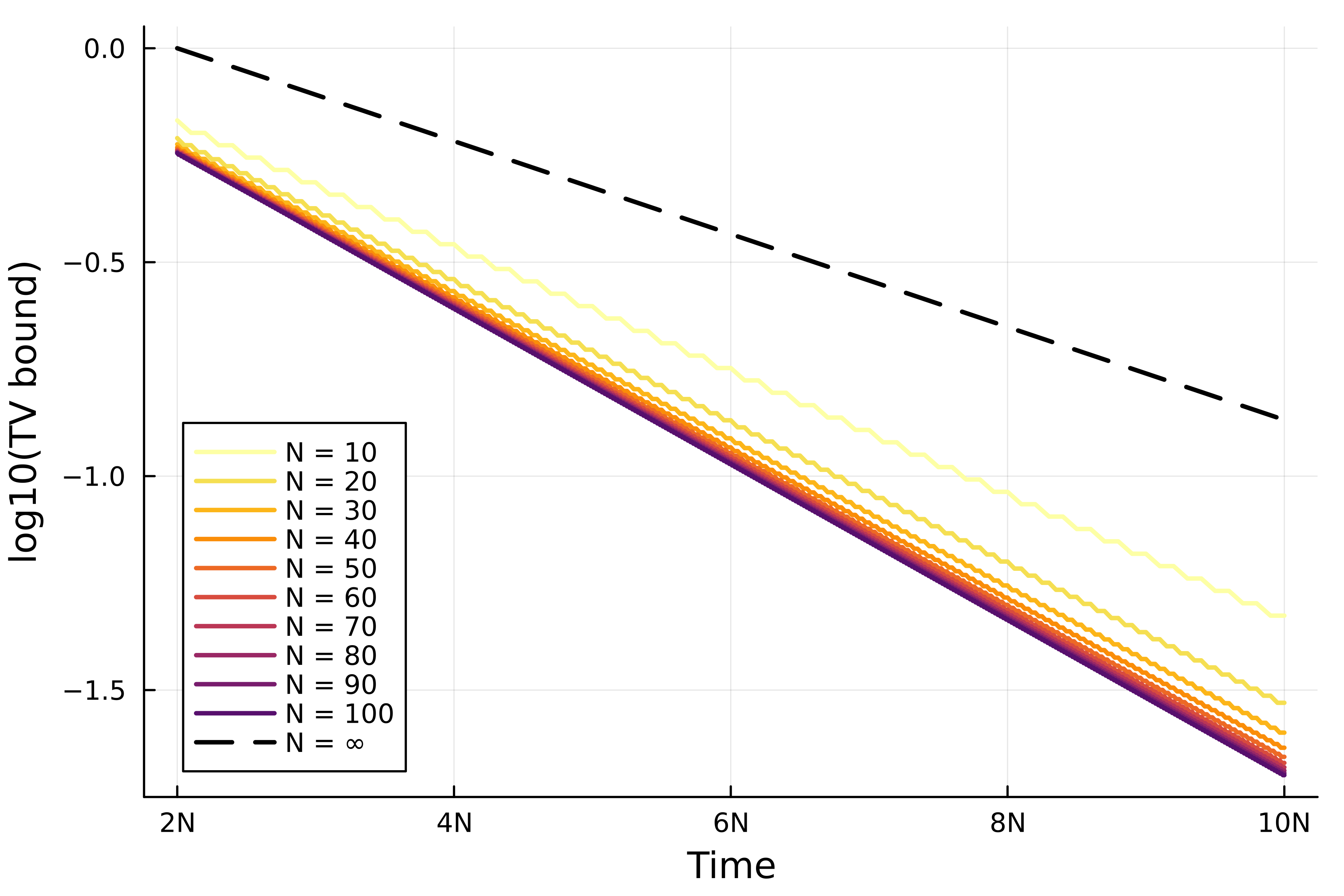}
  \end{subfigure}
  \begin{subfigure}{0.49\textwidth}
      \centering
      \includegraphics[width=\textwidth]{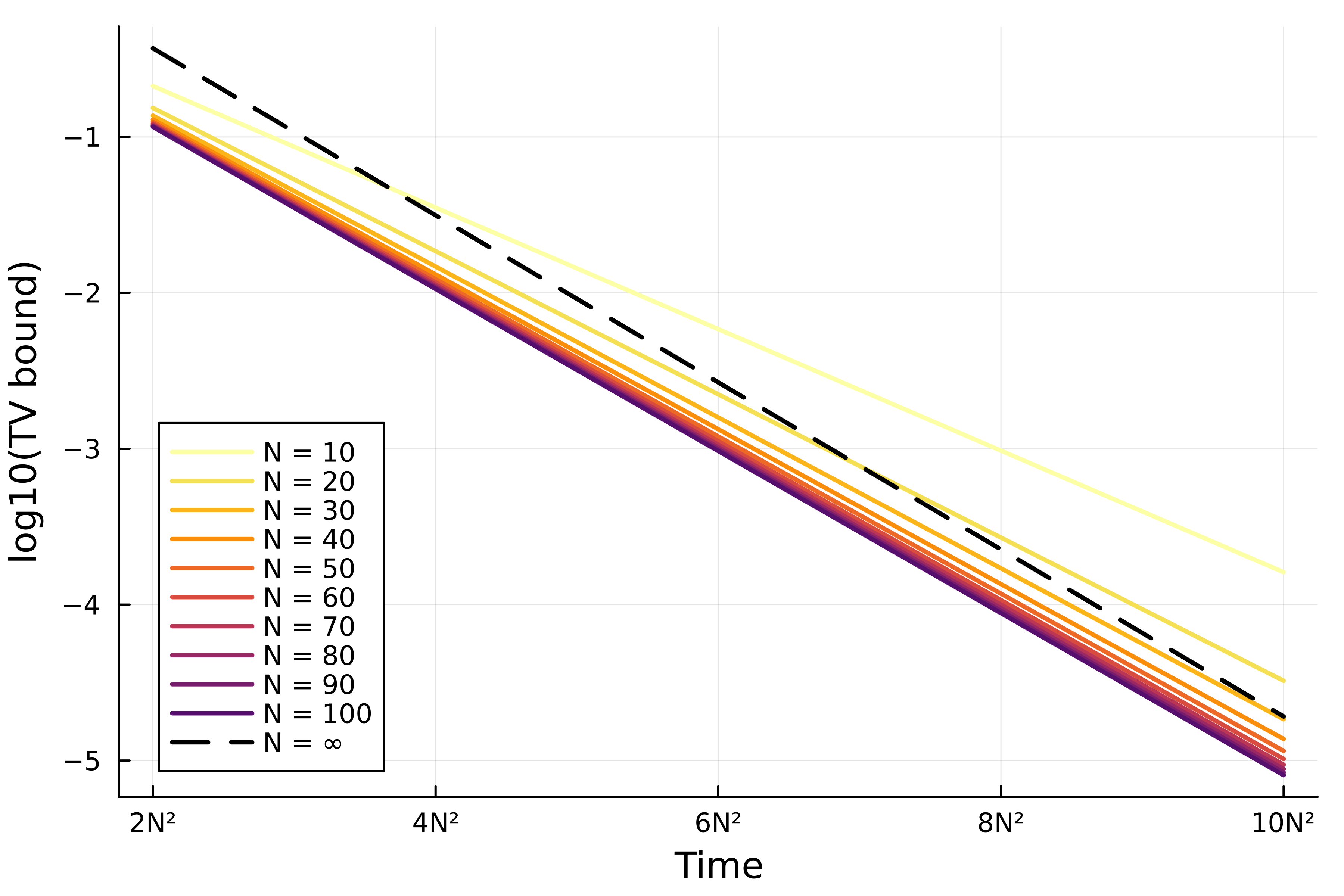}
  \end{subfigure}
  \begin{subfigure}{0.49\textwidth}
      \centering
      \includegraphics[width=\textwidth]{img/survival_curves_asymptotic_Lambda_4.0_DEO.png}
  \end{subfigure}
  \begin{subfigure}{0.49\textwidth}
      \centering
      \includegraphics[width=\textwidth]{img/survival_curves_asymptotic_Lambda_4.0_SEO.png}
  \end{subfigure}
  \begin{subfigure}{0.49\textwidth}
      \centering
      \includegraphics[width=\textwidth]{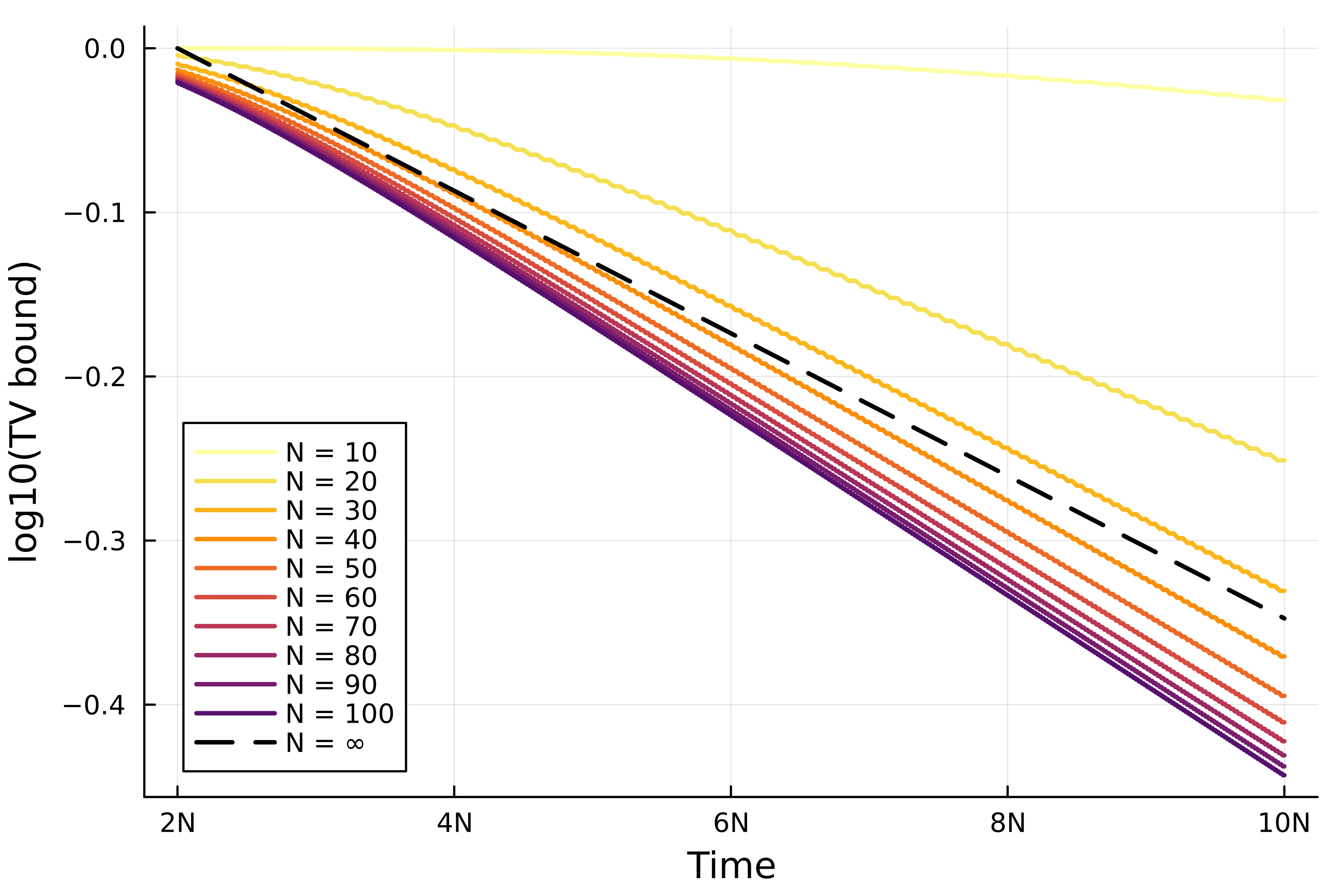}
  \end{subfigure}
  \begin{subfigure}{0.49\textwidth}
      \centering
      \includegraphics[width=\textwidth]{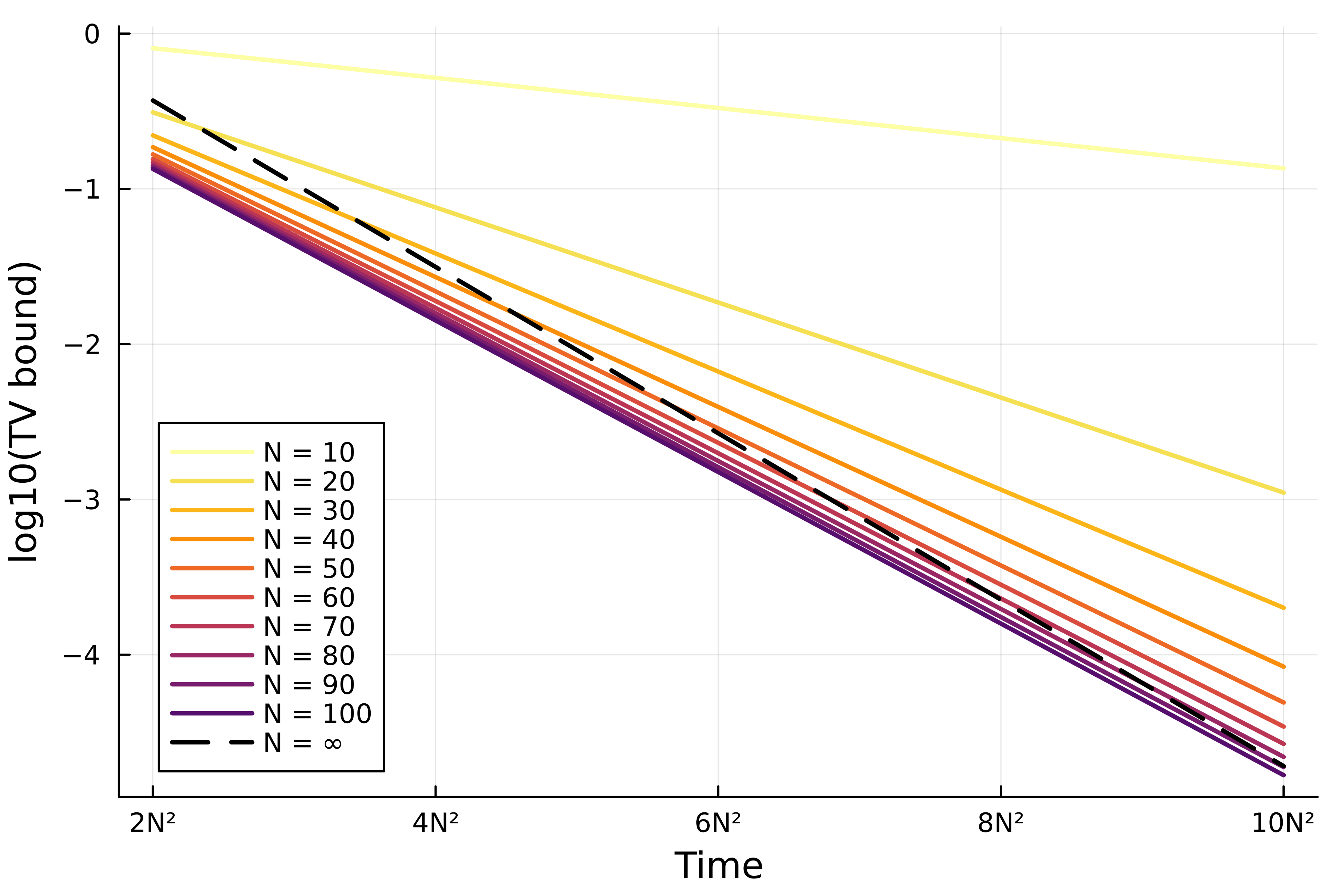}
  \end{subfigure}
  \caption{Finite-chain and infinite-chain TV distance bounds for varying $\Lambda$ and $N$. 
    Within each figure, $N$ varies in the set $N \in \cbra{10, 20, \ldots, 100}$.
    \textbf{Rows:} $\Lambda \in \cbra{1, 2, 4, 8}$. 
    \textbf{Columns:} NRPT and RPT communication schemes.
    The horizontal axis is scaled by a factor of $N$ and $N^2$ for NRPT and RPT, respectively.}
  \label{fig:additional_finite_infinite_chain}
\end{figure}

Several plots assessing the convergence of the finite-chain rates to those 
obtained in the infinite-chain limit are presented in \cref{fig:additional_finite_infinite_chain}.

\subsection{Plots of the Laplace transform $F(z)$} 
\begin{figure}[t]
  \centering
  \begin{subfigure}{0.32\textwidth}
      \centering
      \includegraphics[width=\textwidth]{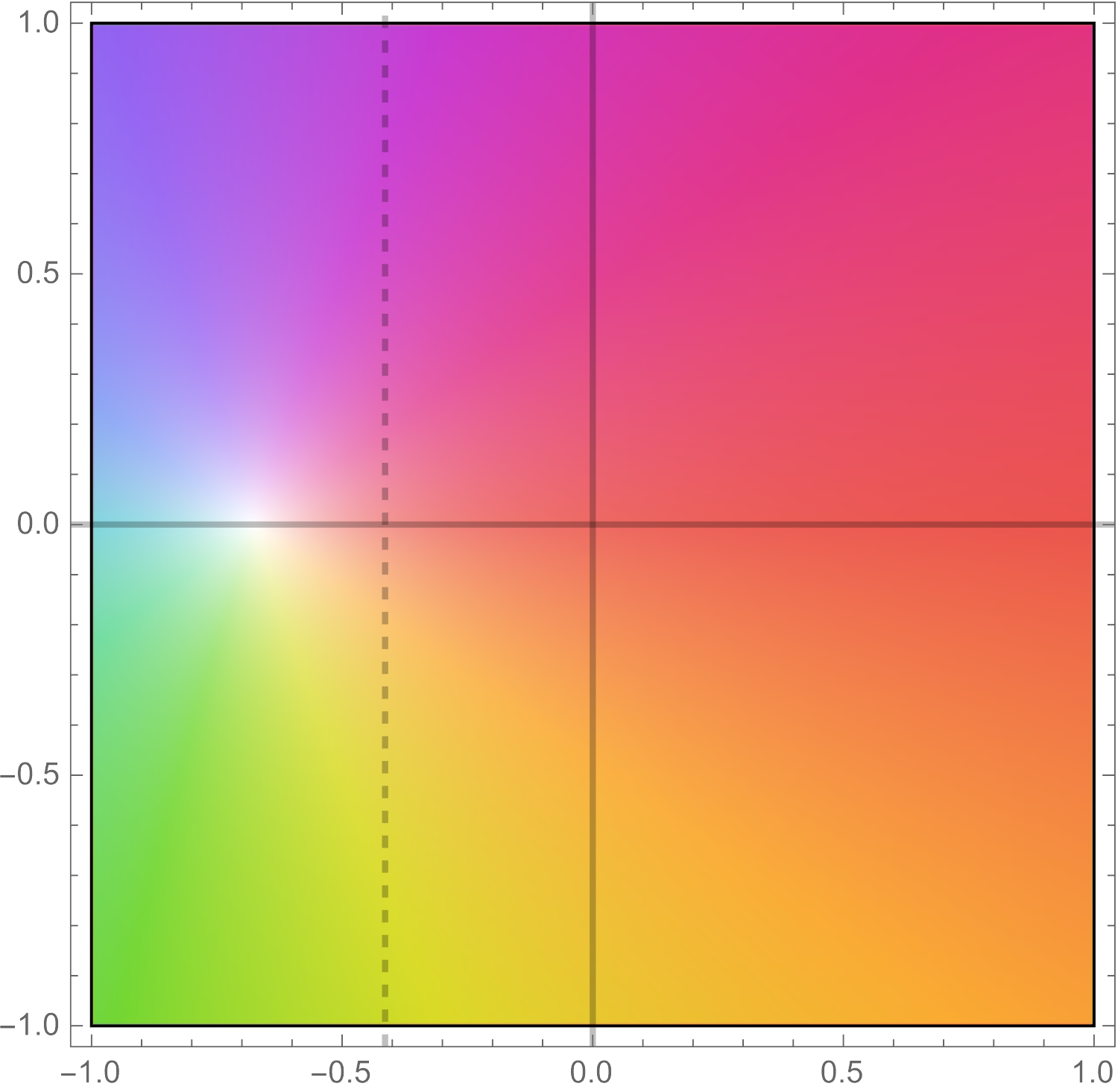}
  \end{subfigure}
  \begin{subfigure}{0.32\textwidth}
      \centering
      \includegraphics[width=\textwidth]{img/pole_search_Lambda_1_wide.png}
  \end{subfigure}
  \begin{subfigure}{0.32\textwidth}
      \centering
      \includegraphics[width=\textwidth]{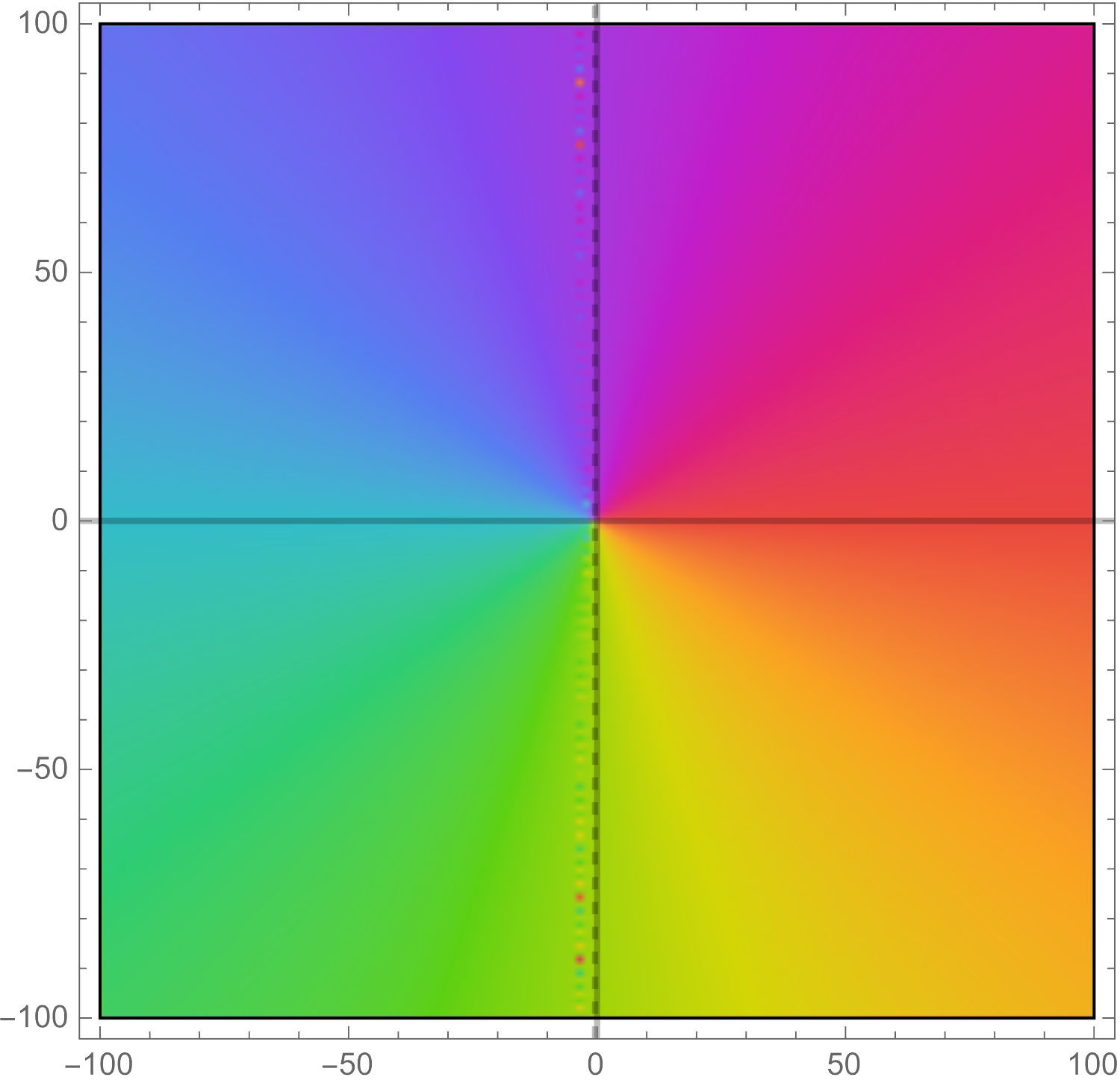}
  \end{subfigure}

  \begin{subfigure}{0.32\textwidth}
      \centering
      \includegraphics[width=\textwidth]{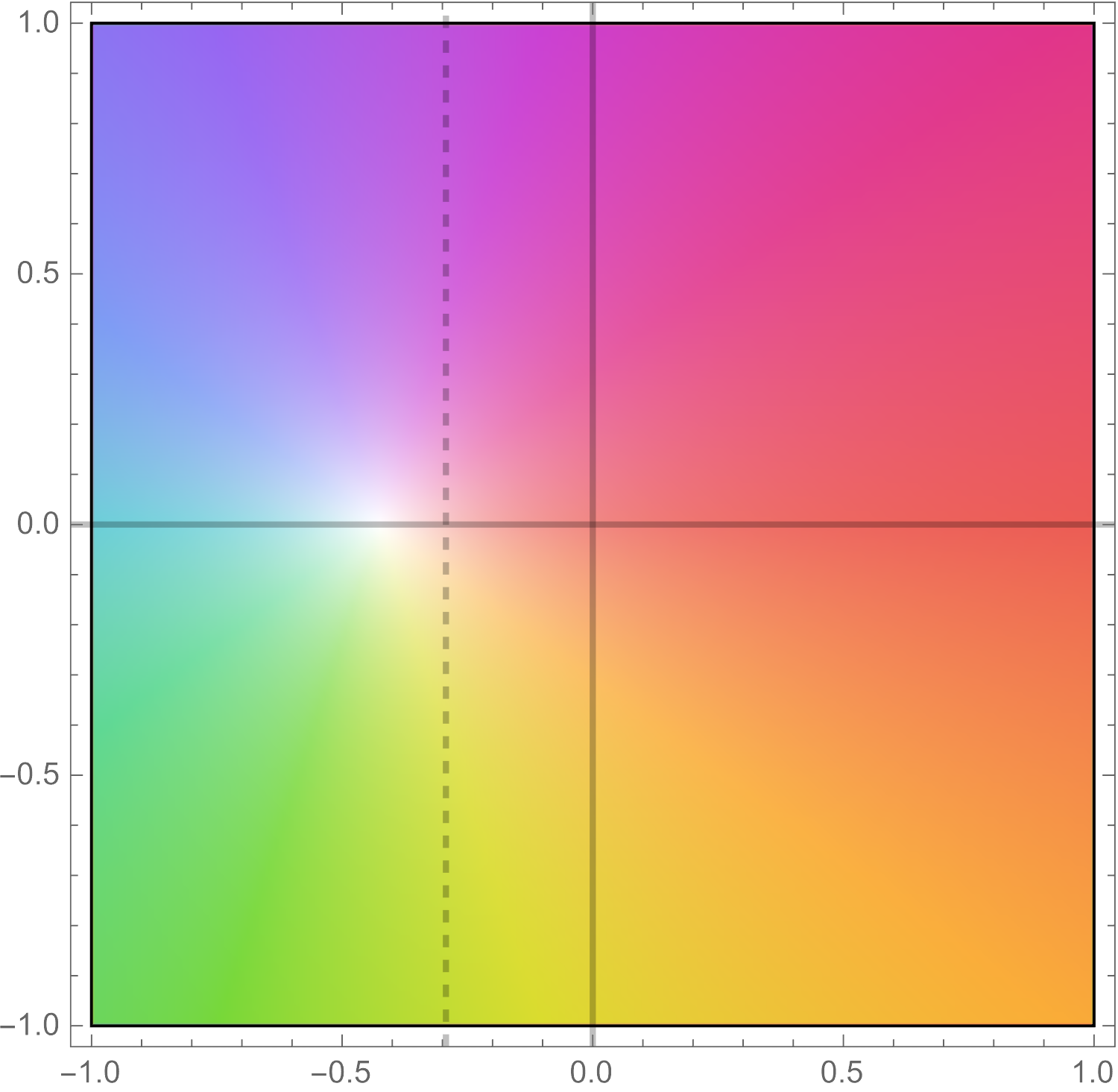}
  \end{subfigure}
  \begin{subfigure}{0.32\textwidth}
      \centering
      \includegraphics[width=\textwidth]{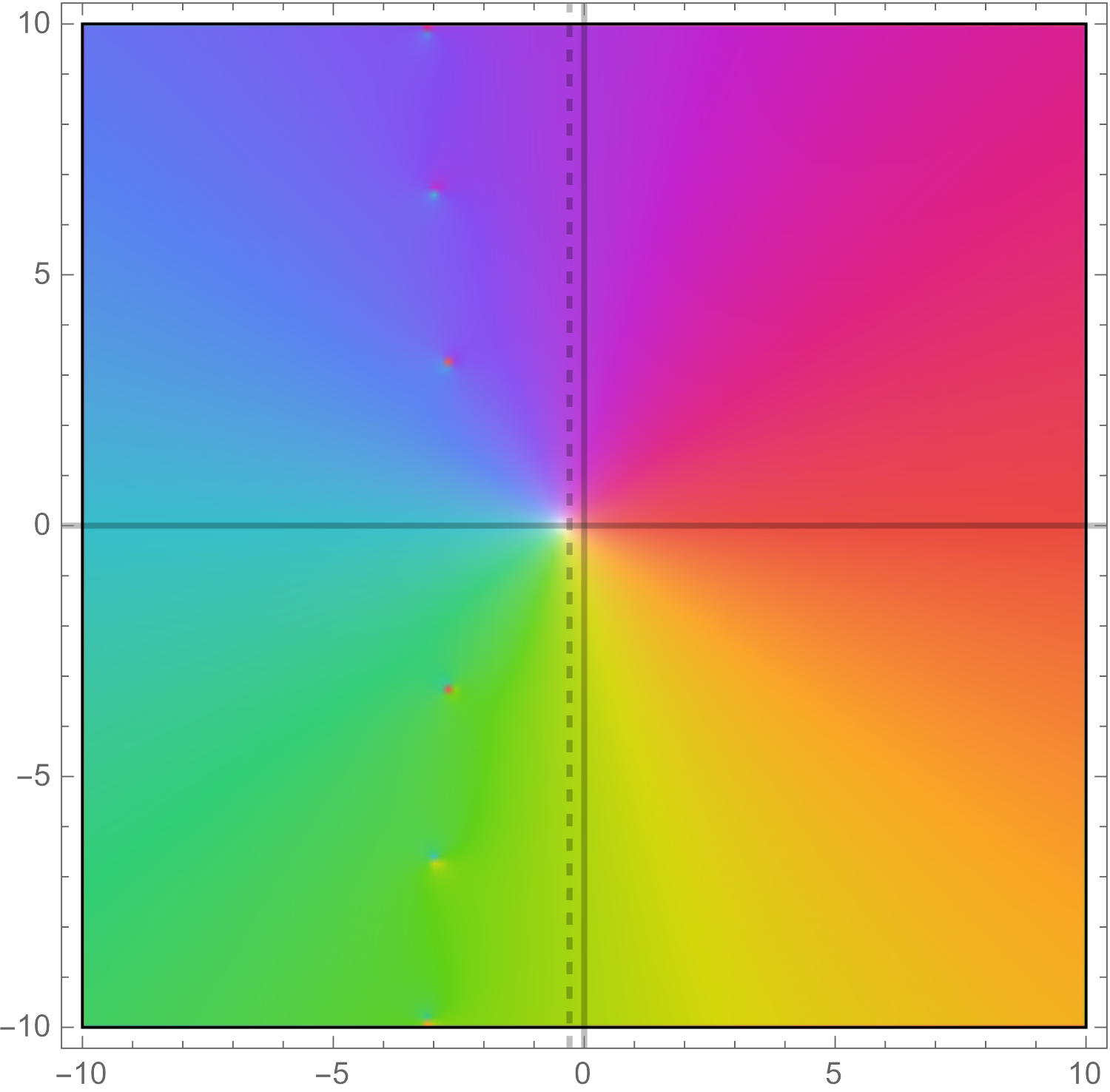}
  \end{subfigure}
  \begin{subfigure}{0.32\textwidth}
      \centering
      \includegraphics[width=\textwidth]{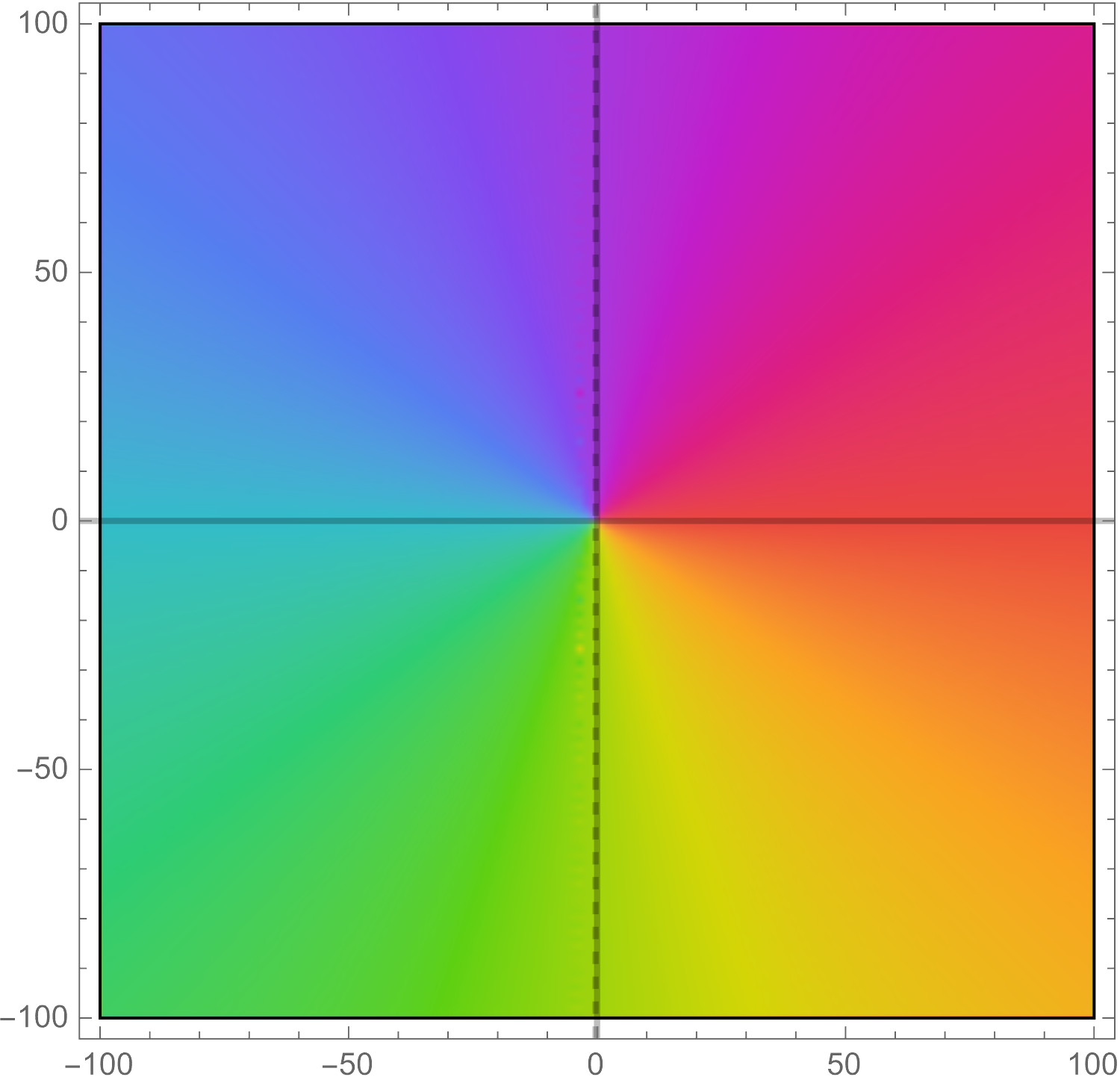}
  \end{subfigure}

  \begin{subfigure}{0.32\textwidth}
      \centering
      \includegraphics[width=\textwidth]{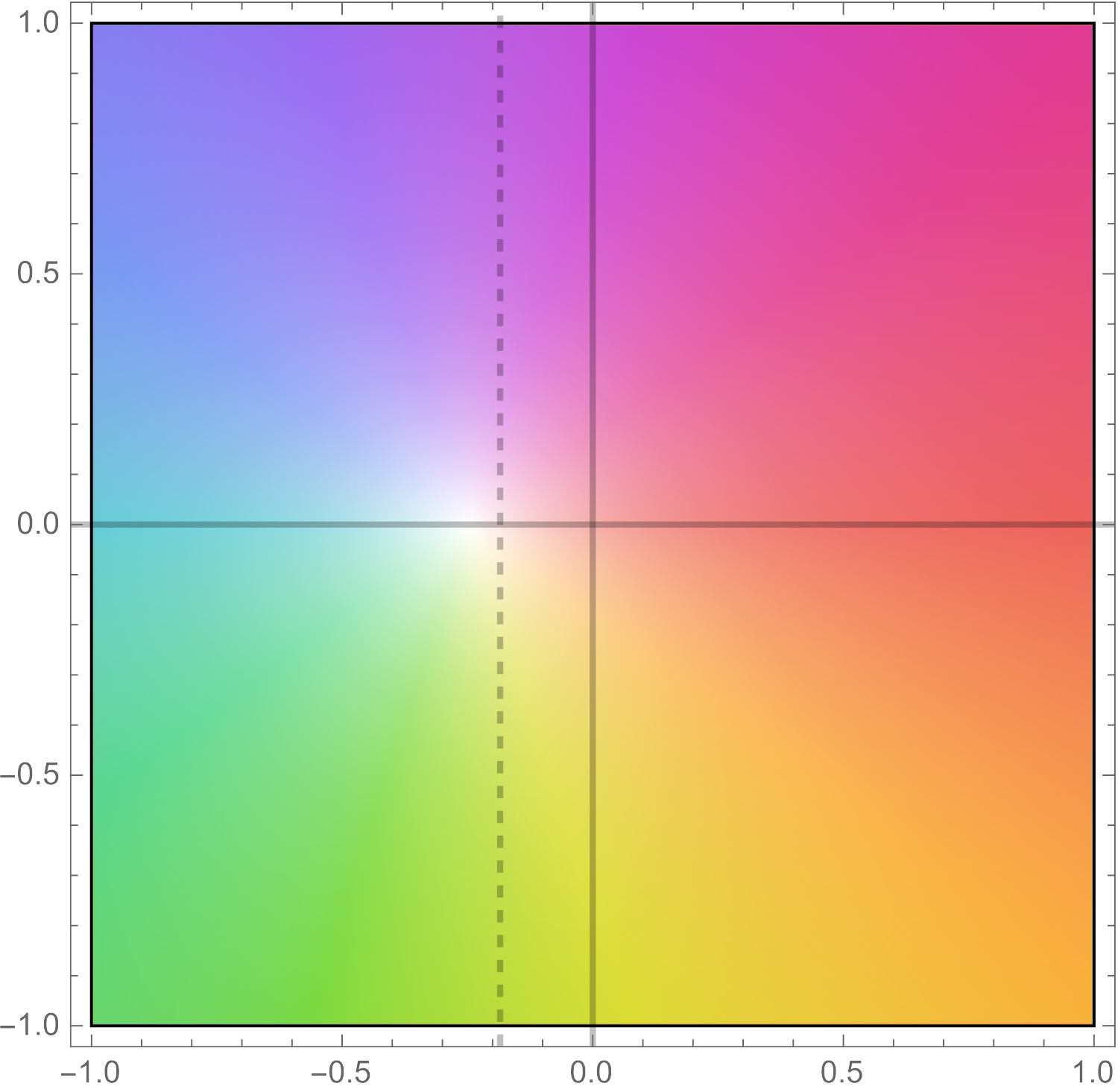}
  \end{subfigure}
  \begin{subfigure}{0.32\textwidth}
      \centering
      \includegraphics[width=\textwidth]{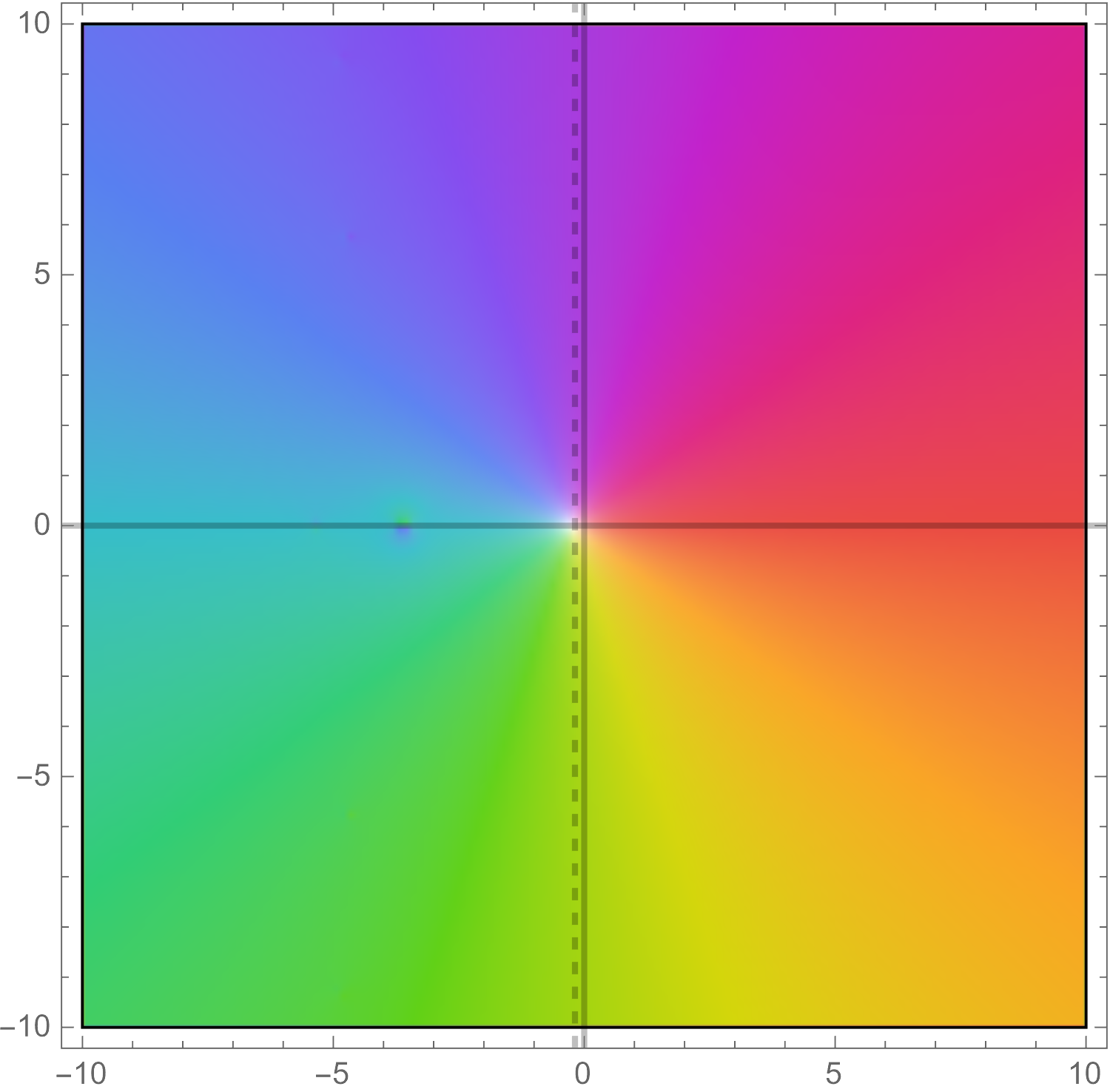}
  \end{subfigure}
  \begin{subfigure}{0.32\textwidth}
      \centering
      \includegraphics[width=\textwidth]{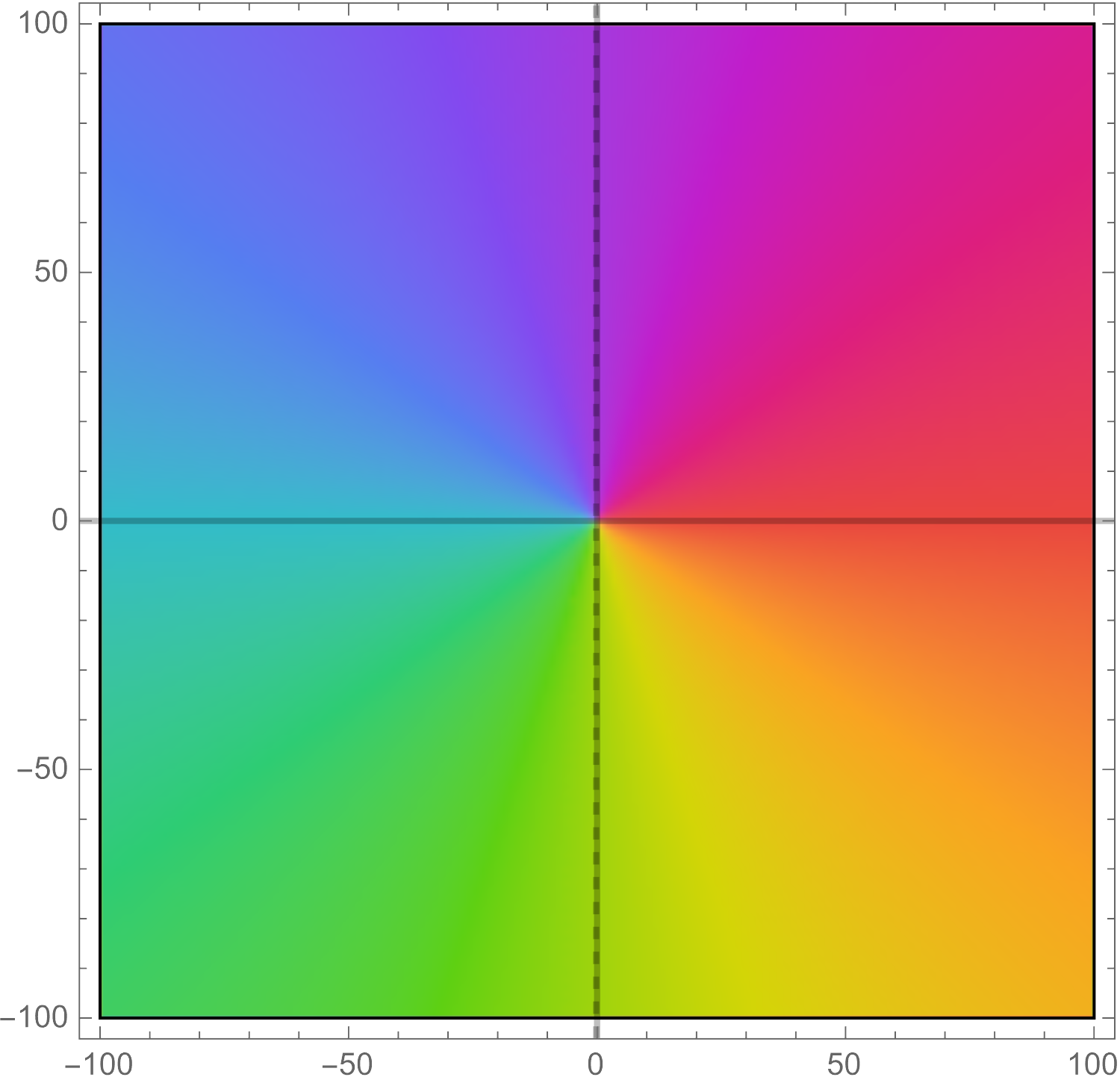}
  \end{subfigure}

  \begin{subfigure}{0.32\textwidth}
      \centering
      \includegraphics[width=\textwidth]{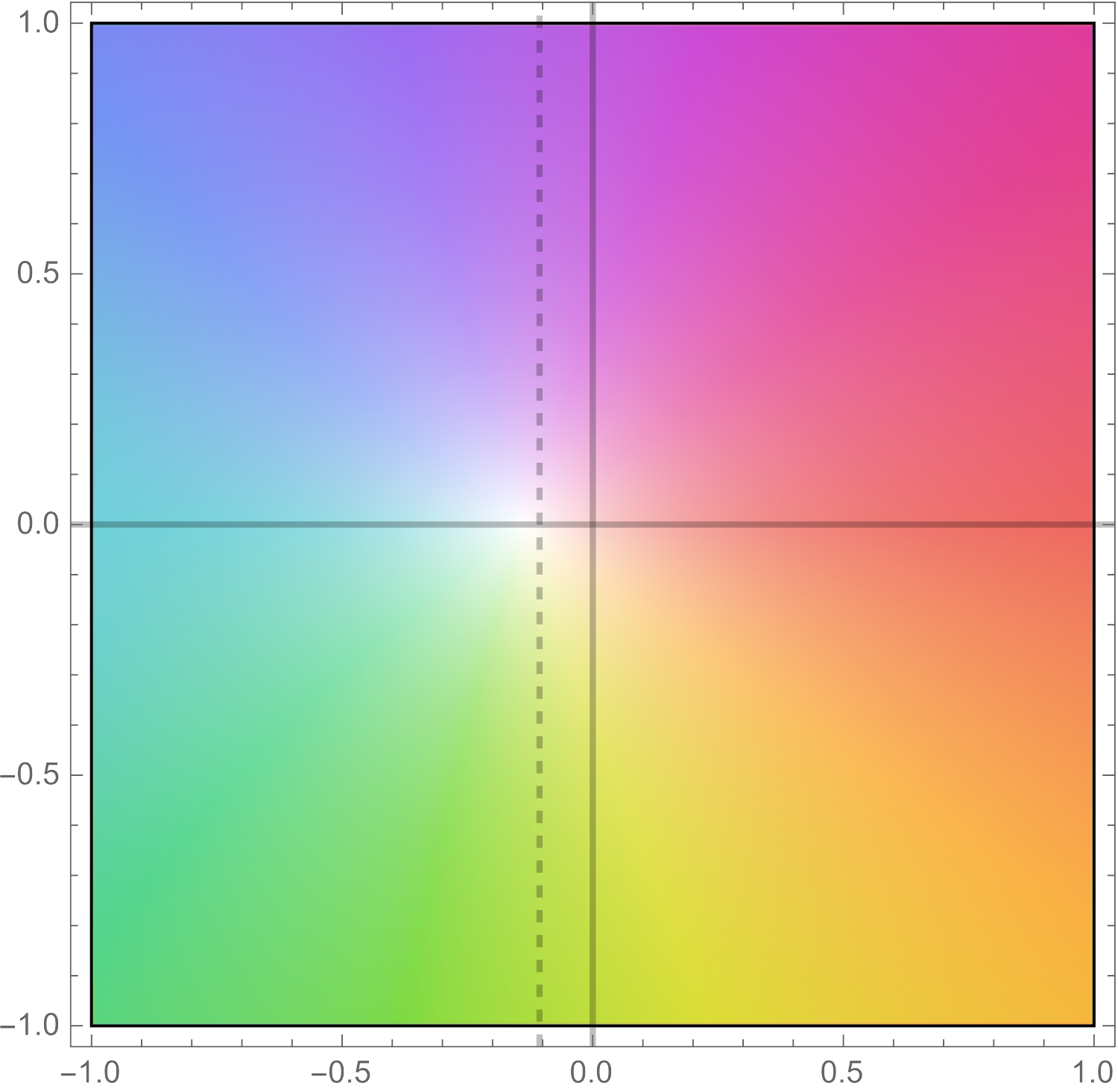}
  \end{subfigure}
  \begin{subfigure}{0.32\textwidth}
      \centering
      \includegraphics[width=\textwidth]{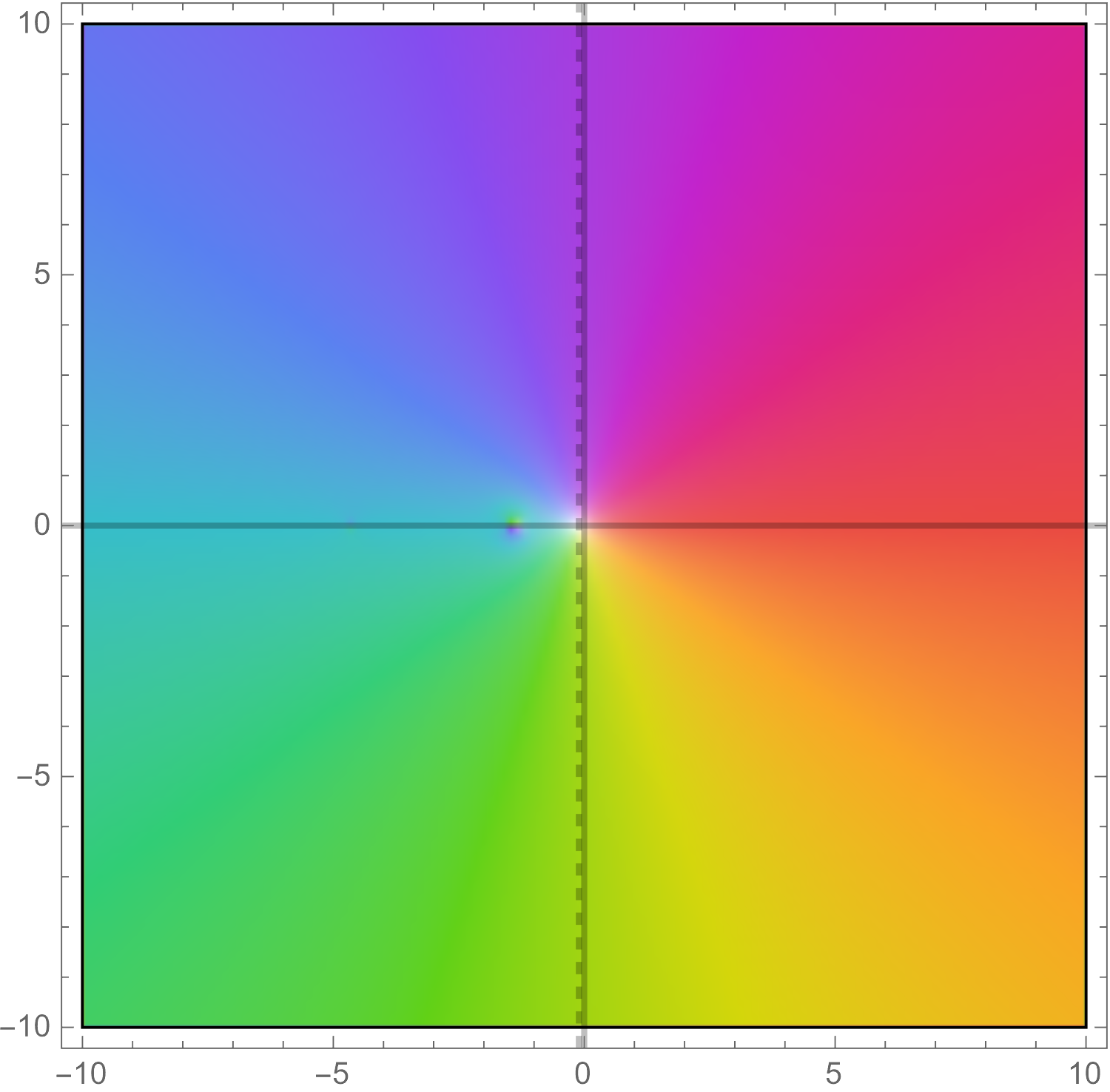}
  \end{subfigure}
  \begin{subfigure}{0.32\textwidth}
      \centering
      \includegraphics[width=\textwidth]{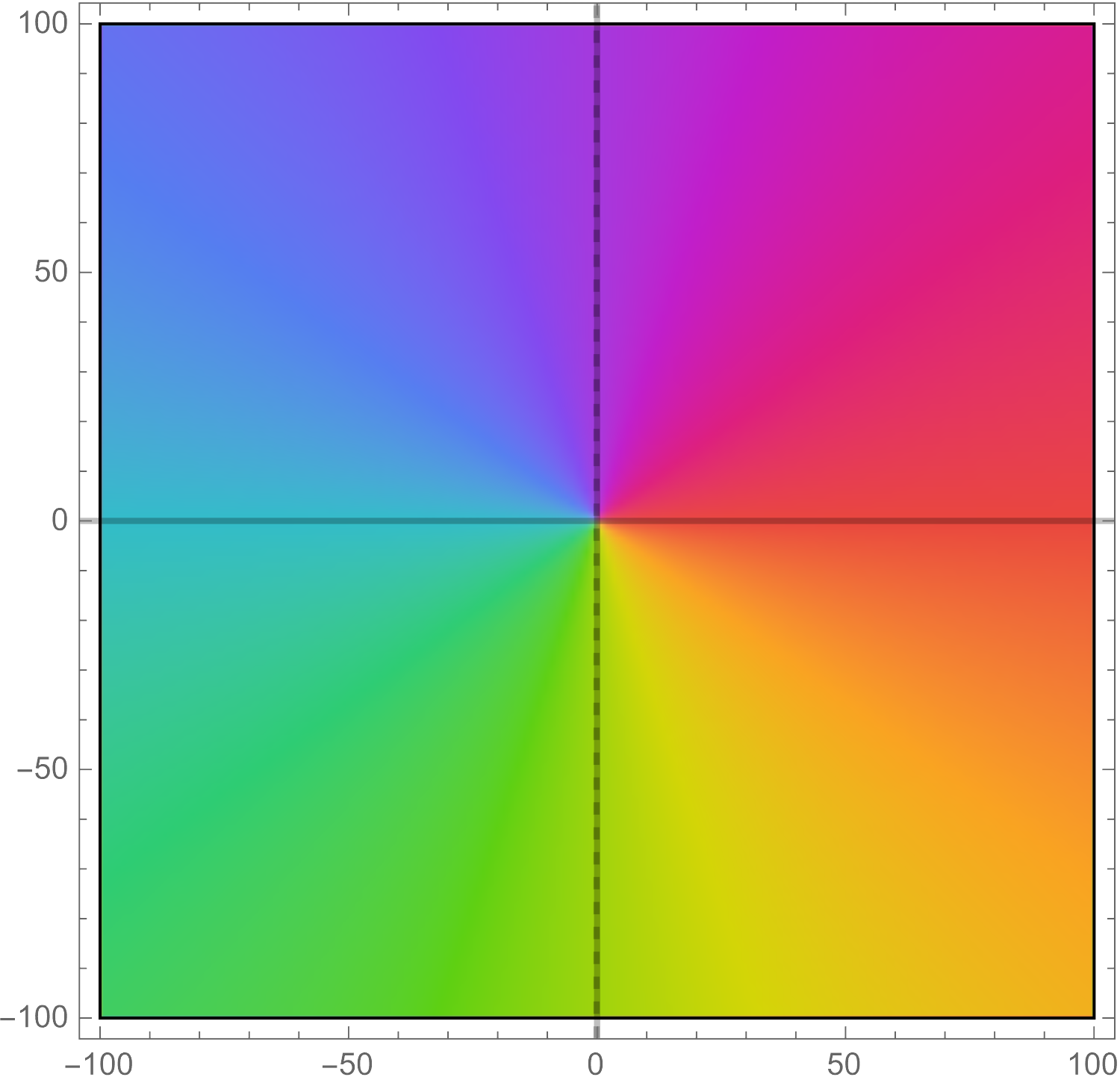}
  \end{subfigure}

  \begin{subfigure}{0.32\textwidth}
      \centering
      \includegraphics[width=\textwidth]{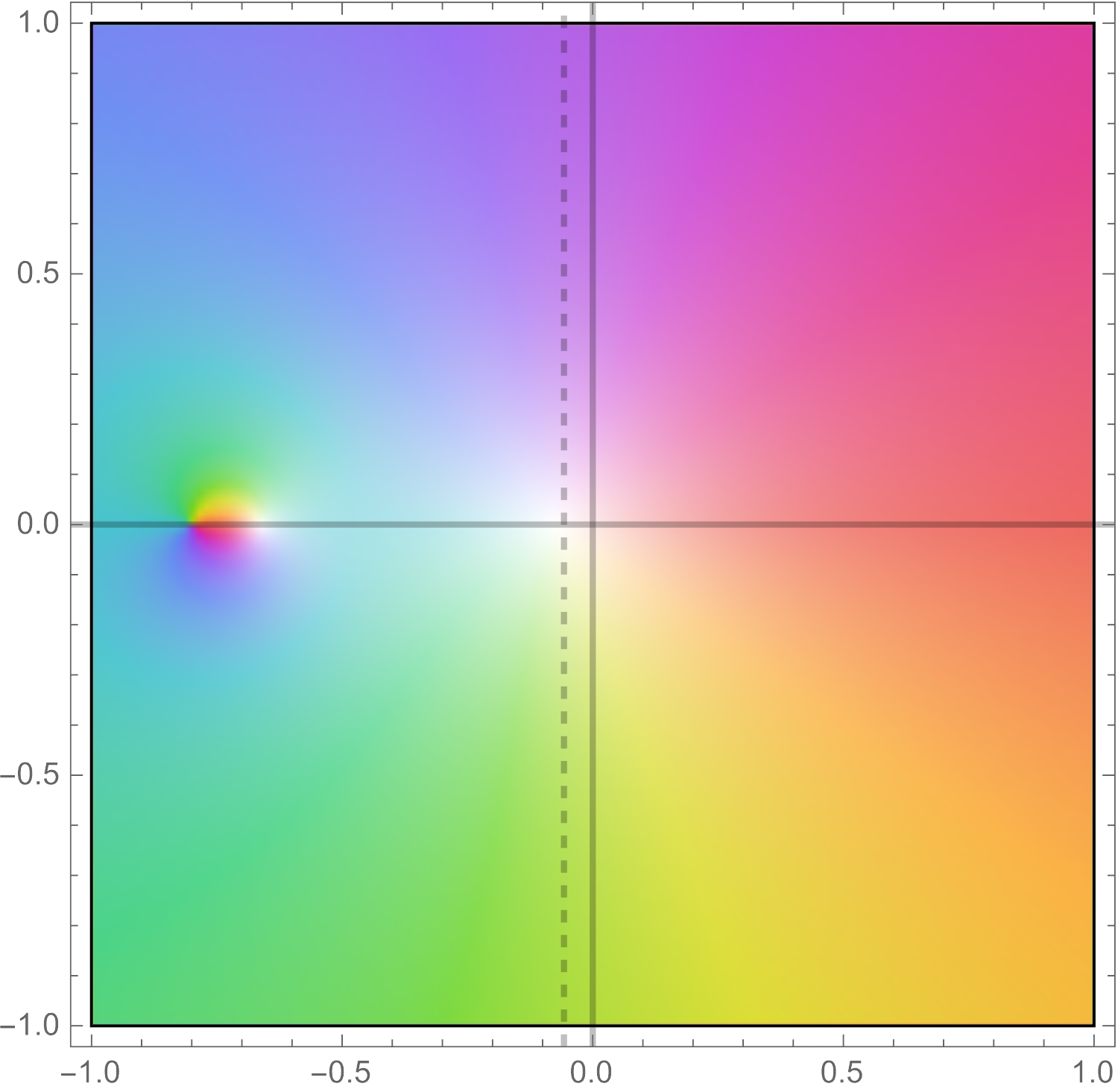}
  \end{subfigure}
  \begin{subfigure}{0.32\textwidth}
      \centering
      \includegraphics[width=\textwidth]{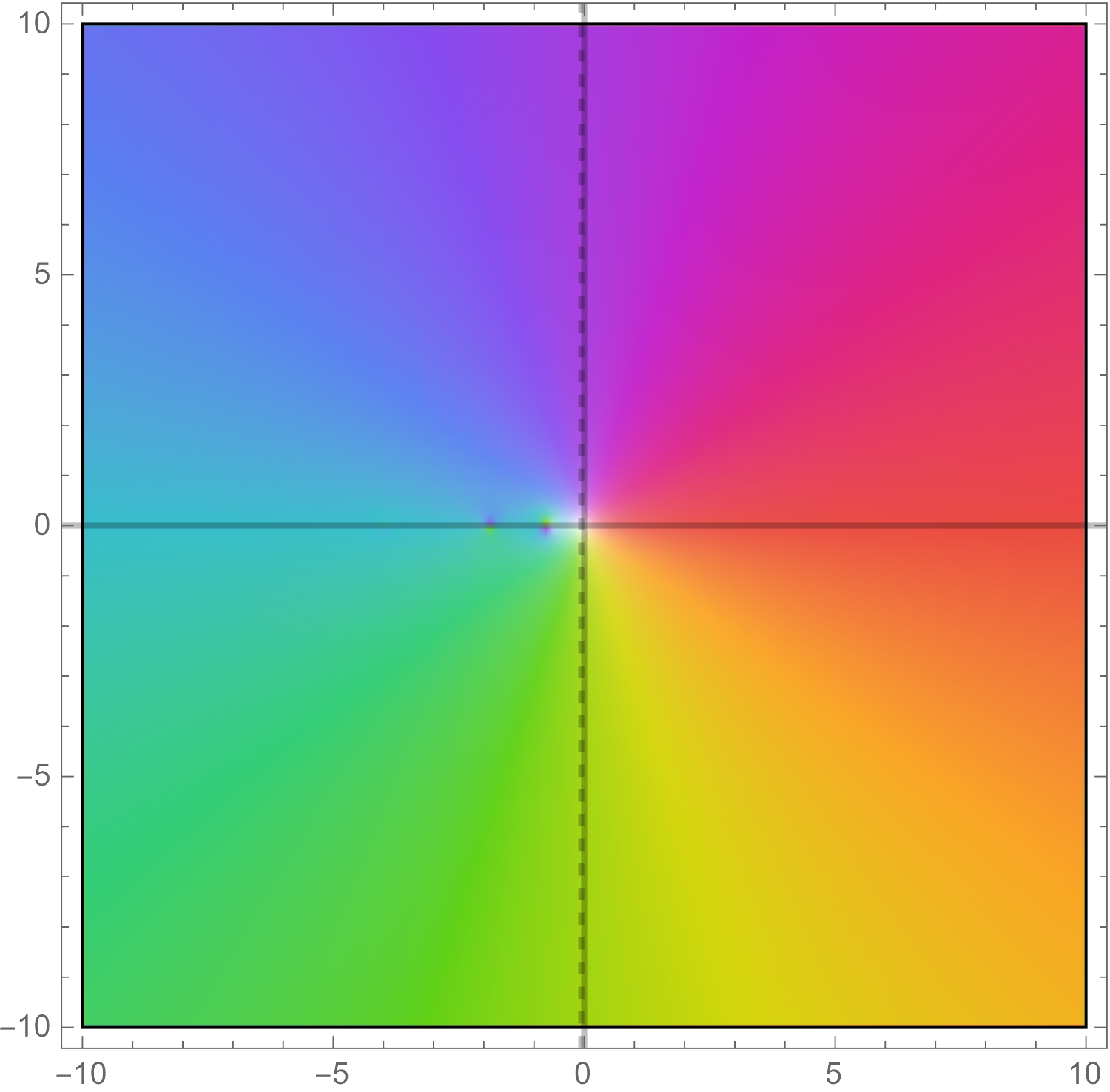}
  \end{subfigure}
  \begin{subfigure}{0.32\textwidth}
      \centering
      \includegraphics[width=\textwidth]{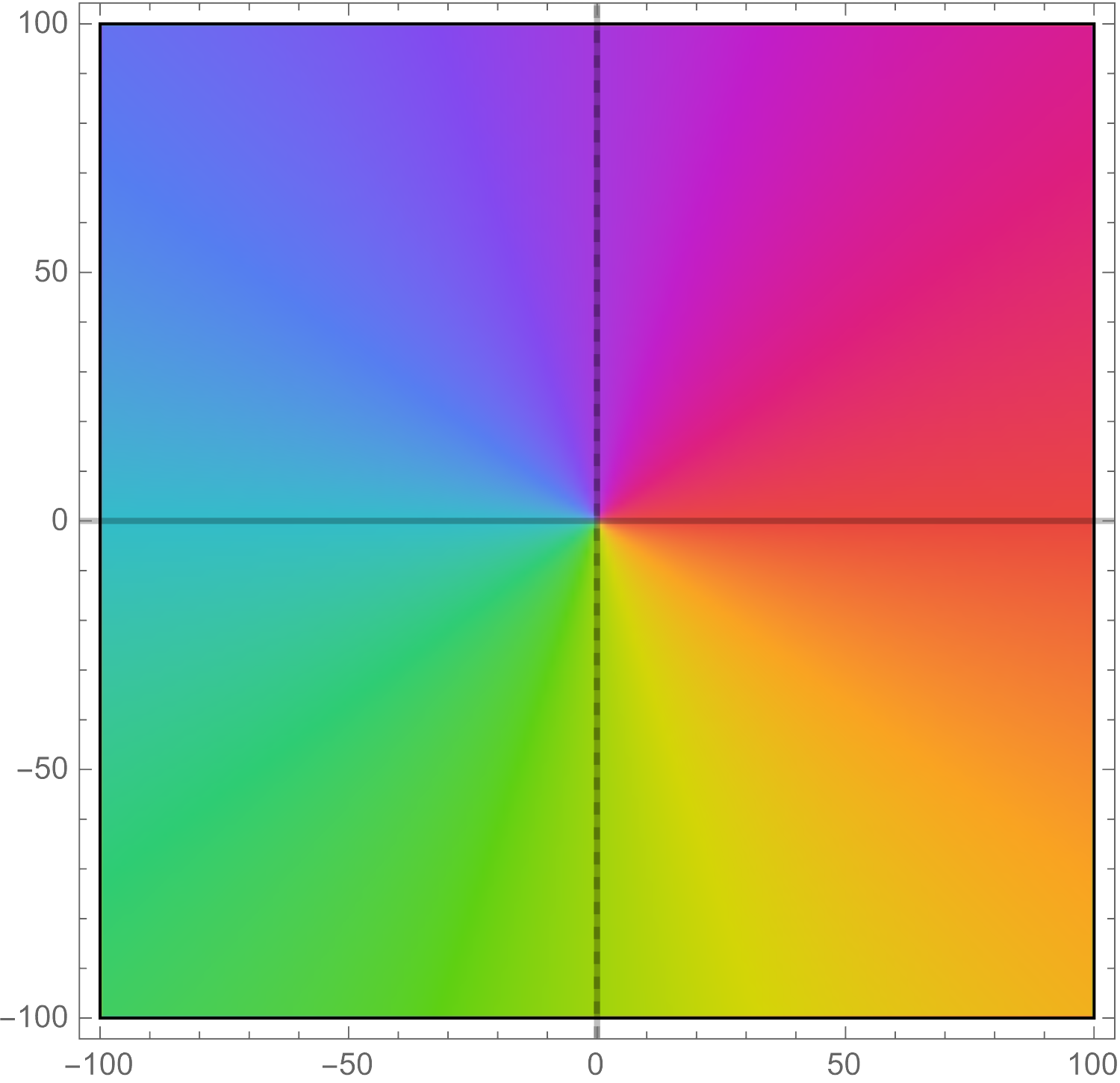}
  \end{subfigure}
  \caption{Plots of $F(z)$. Brightness indicates the magnitude of $\abs{F(z)}$ 
    and colours indicate the argument of $F(z)$. 
    The dashed vertical line in each plot represents our bound on the location 
    of the largest pole, given by $-1/(\Lambda+\sqrt{2})$.
    \textbf{Rows:} $\Lambda \in \cbra{1,2,4,8,16}$. 
    \textbf{Columns:} Plots over $\cbra{z = x+iy : -r \leq x, y \leq r}$ for 
    $r \in \cbra{1, 10, 100}$.}
  \label{fig:pole_search_1}
\end{figure}

\begin{figure}[t]
  \centering
  \begin{subfigure}{0.32\textwidth}
      \centering
      \includegraphics[width=\textwidth]{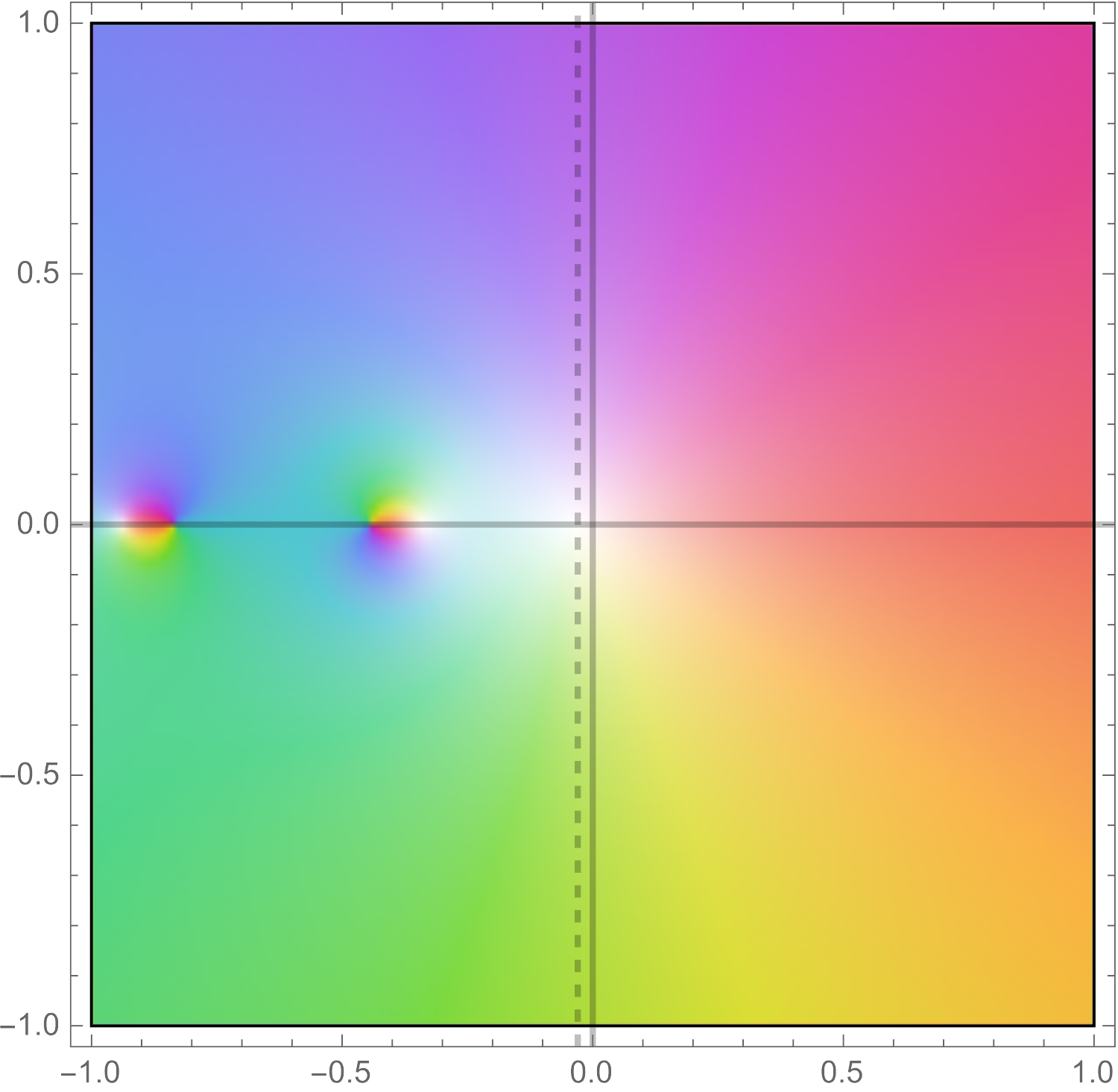}
  \end{subfigure}
  \begin{subfigure}{0.32\textwidth}
      \centering
      \includegraphics[width=\textwidth]{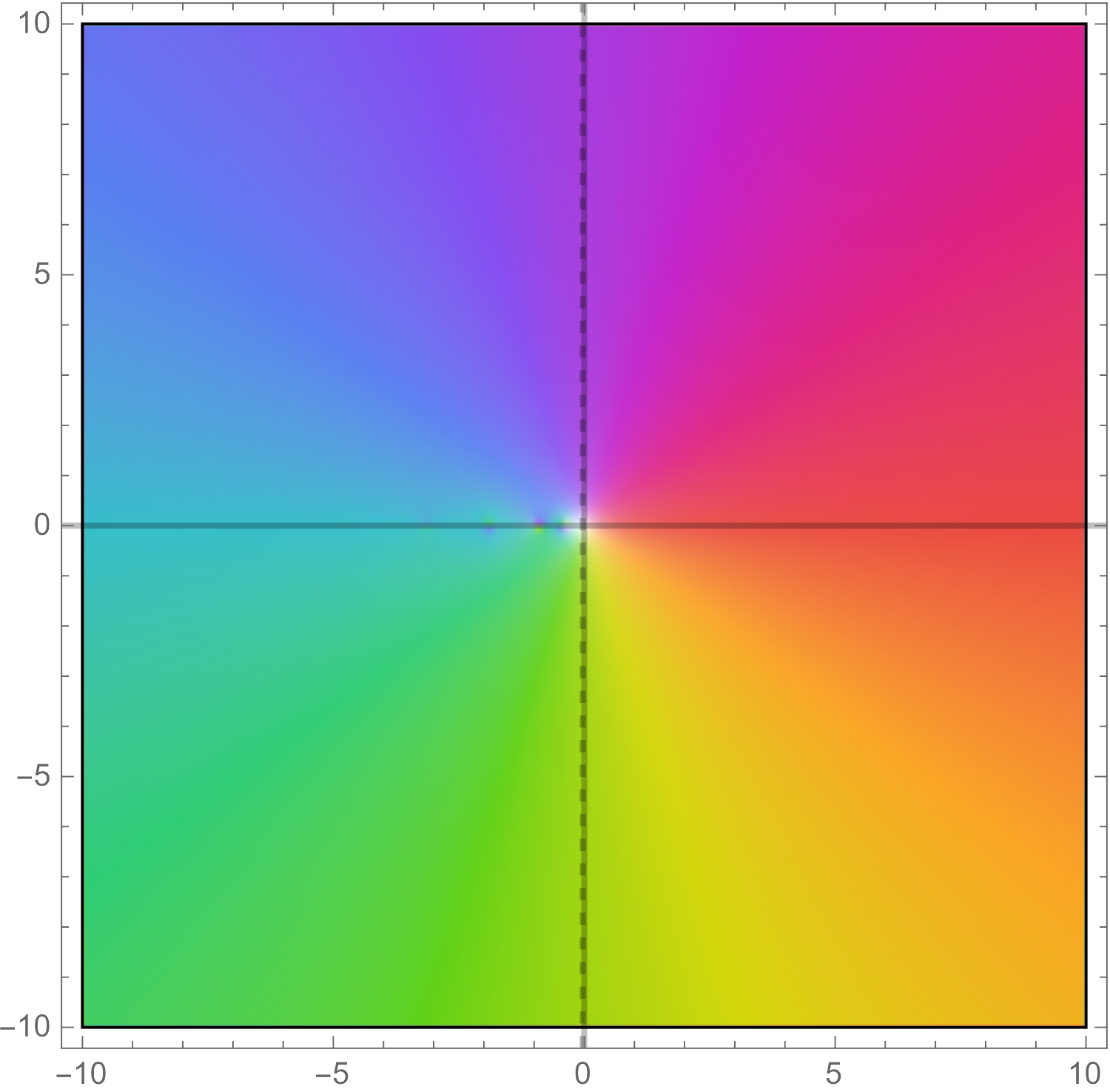}
  \end{subfigure}
  \begin{subfigure}{0.32\textwidth}
      \centering
      \includegraphics[width=\textwidth]{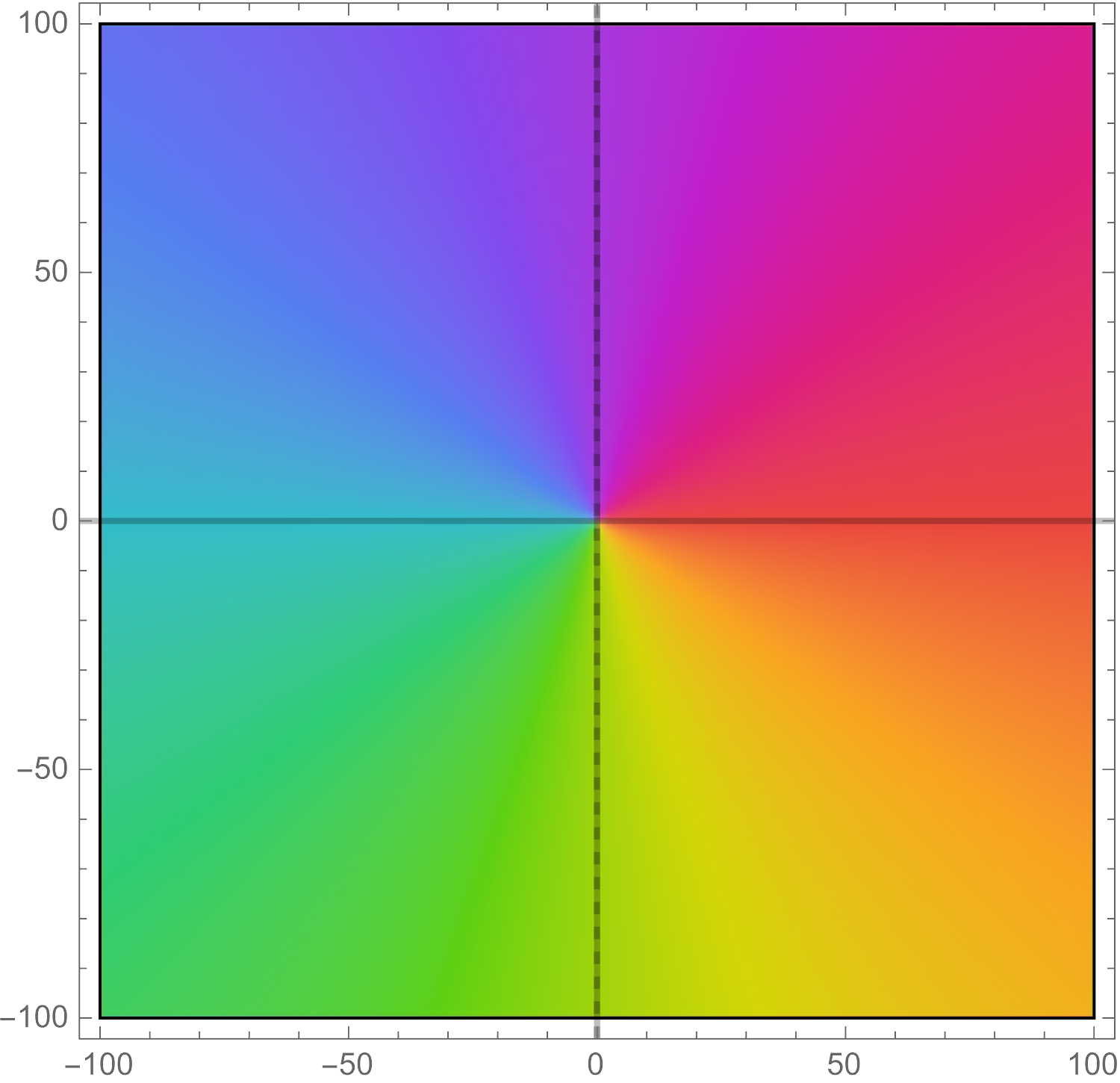}
  \end{subfigure}

  \begin{subfigure}{0.32\textwidth}
      \centering
      \includegraphics[width=\textwidth]{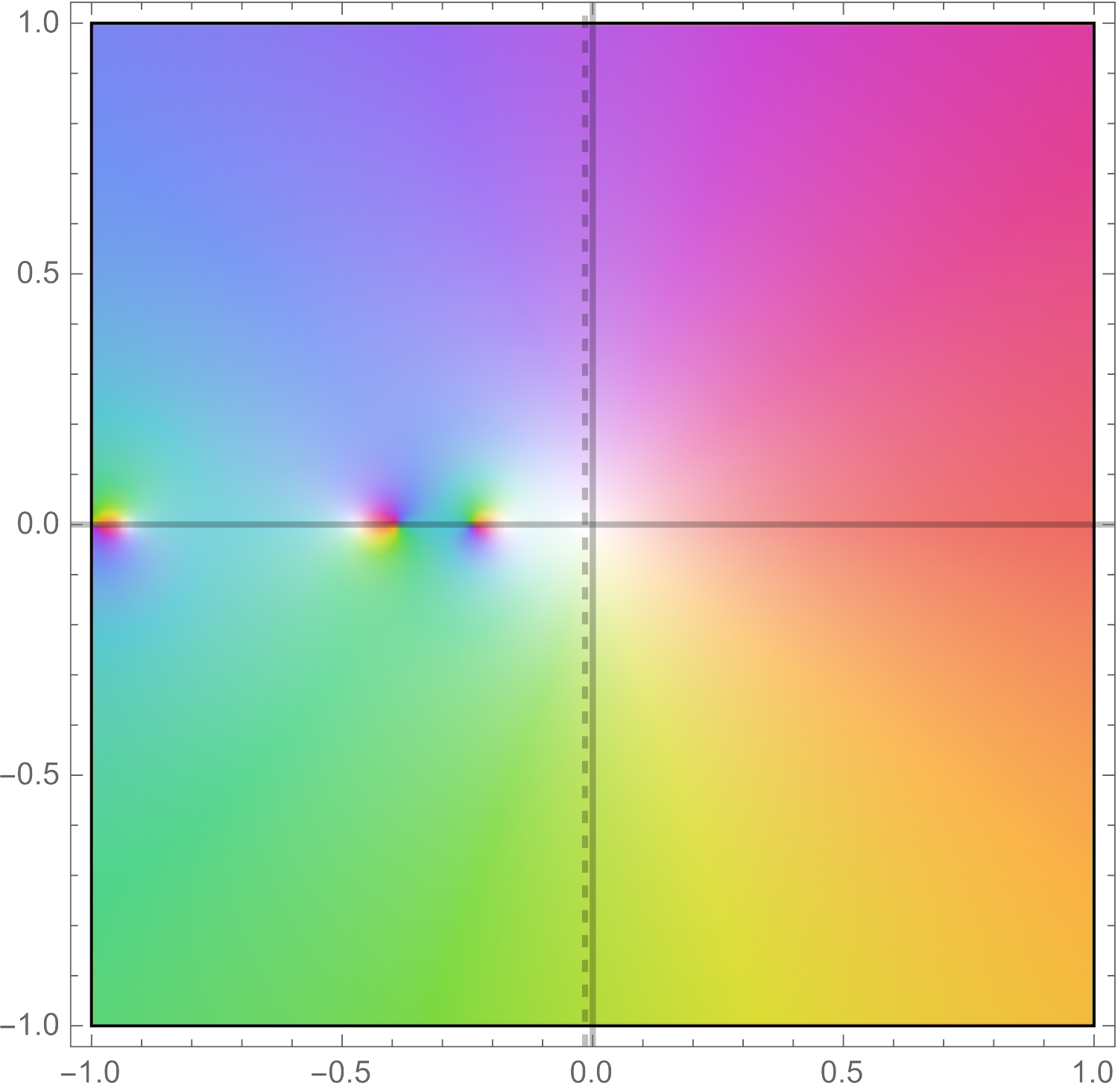}
  \end{subfigure}
  \begin{subfigure}{0.32\textwidth}
      \centering
      \includegraphics[width=\textwidth]{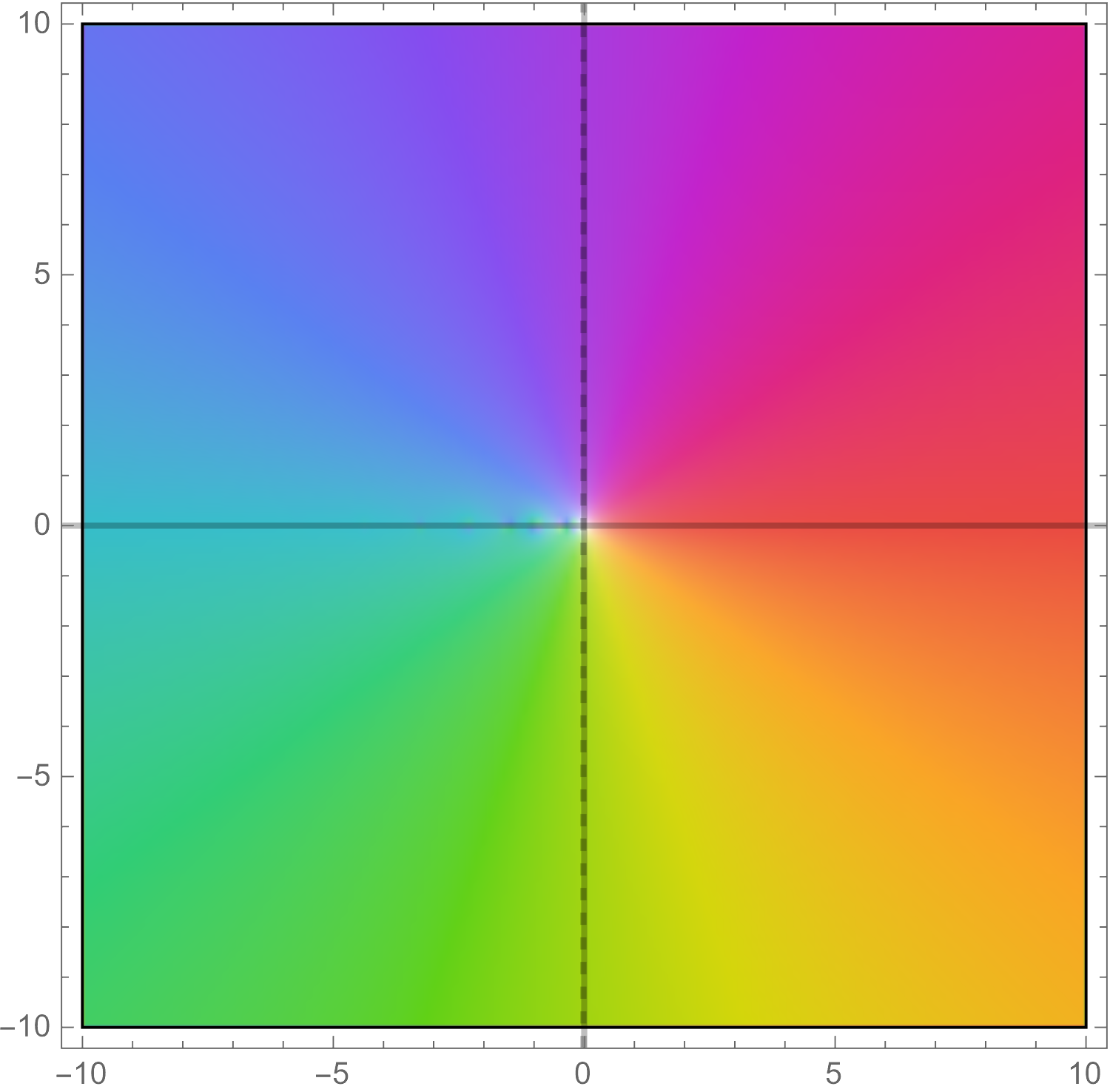}
  \end{subfigure}
  \begin{subfigure}{0.32\textwidth}
      \centering
      \includegraphics[width=\textwidth]{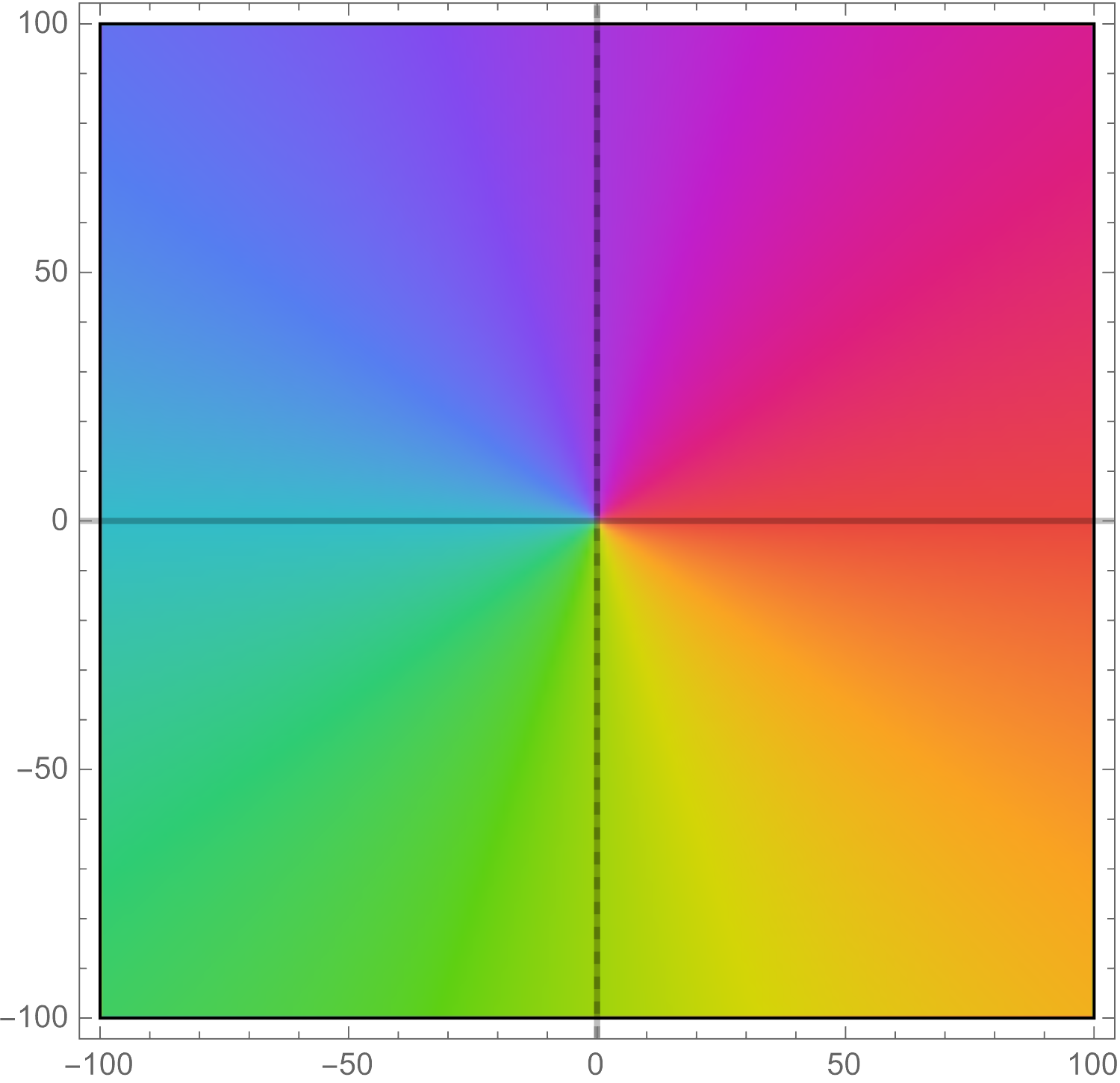}
  \end{subfigure}

  \begin{subfigure}{0.32\textwidth}
      \centering
      \includegraphics[width=\textwidth]{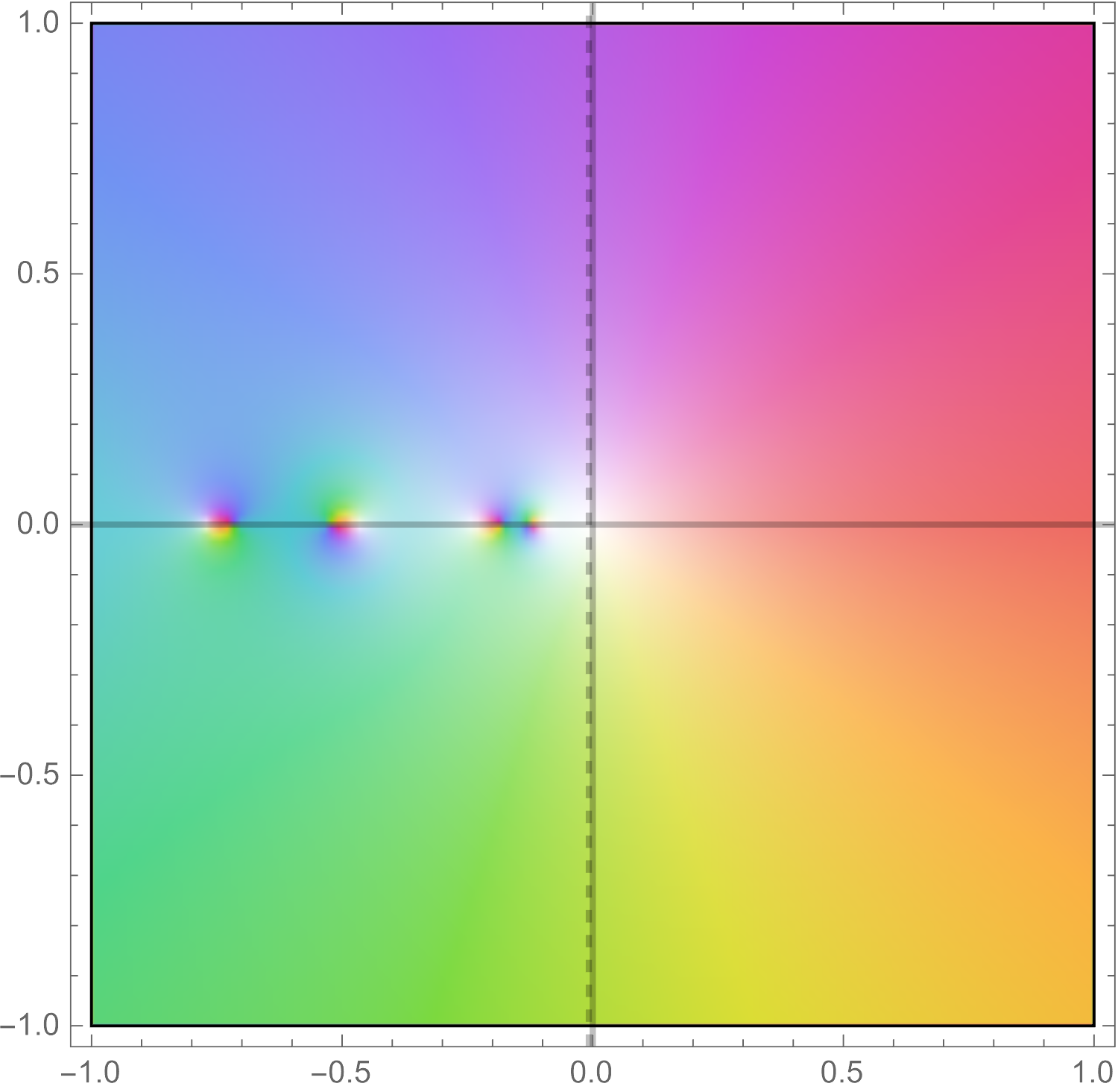}
  \end{subfigure}
  \begin{subfigure}{0.32\textwidth}
      \centering
      \includegraphics[width=\textwidth]{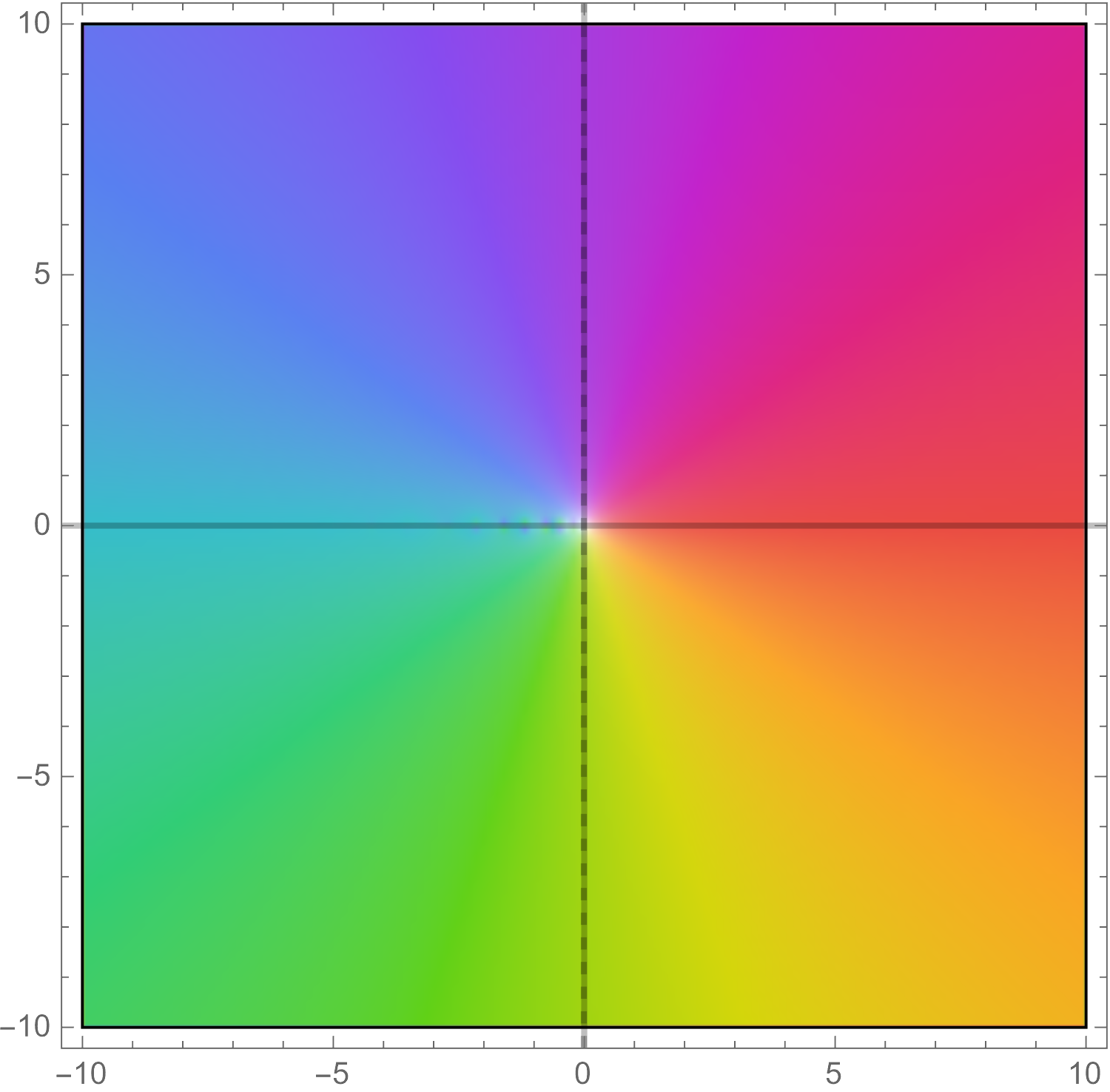}
  \end{subfigure}
  \begin{subfigure}{0.32\textwidth}
      \centering
      \includegraphics[width=\textwidth]{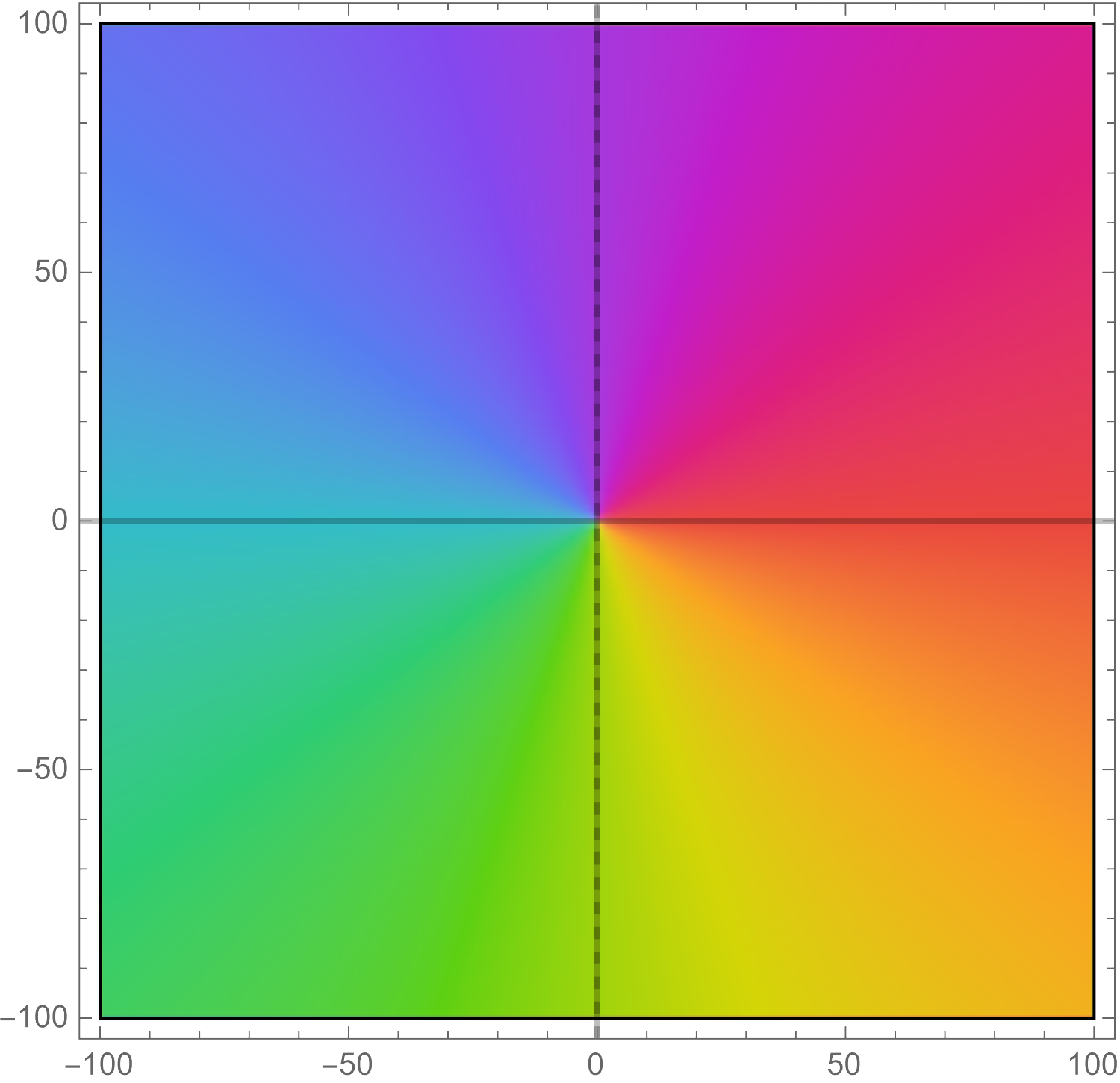}
  \end{subfigure}

  \begin{subfigure}{0.32\textwidth}
      \centering
      \includegraphics[width=\textwidth]{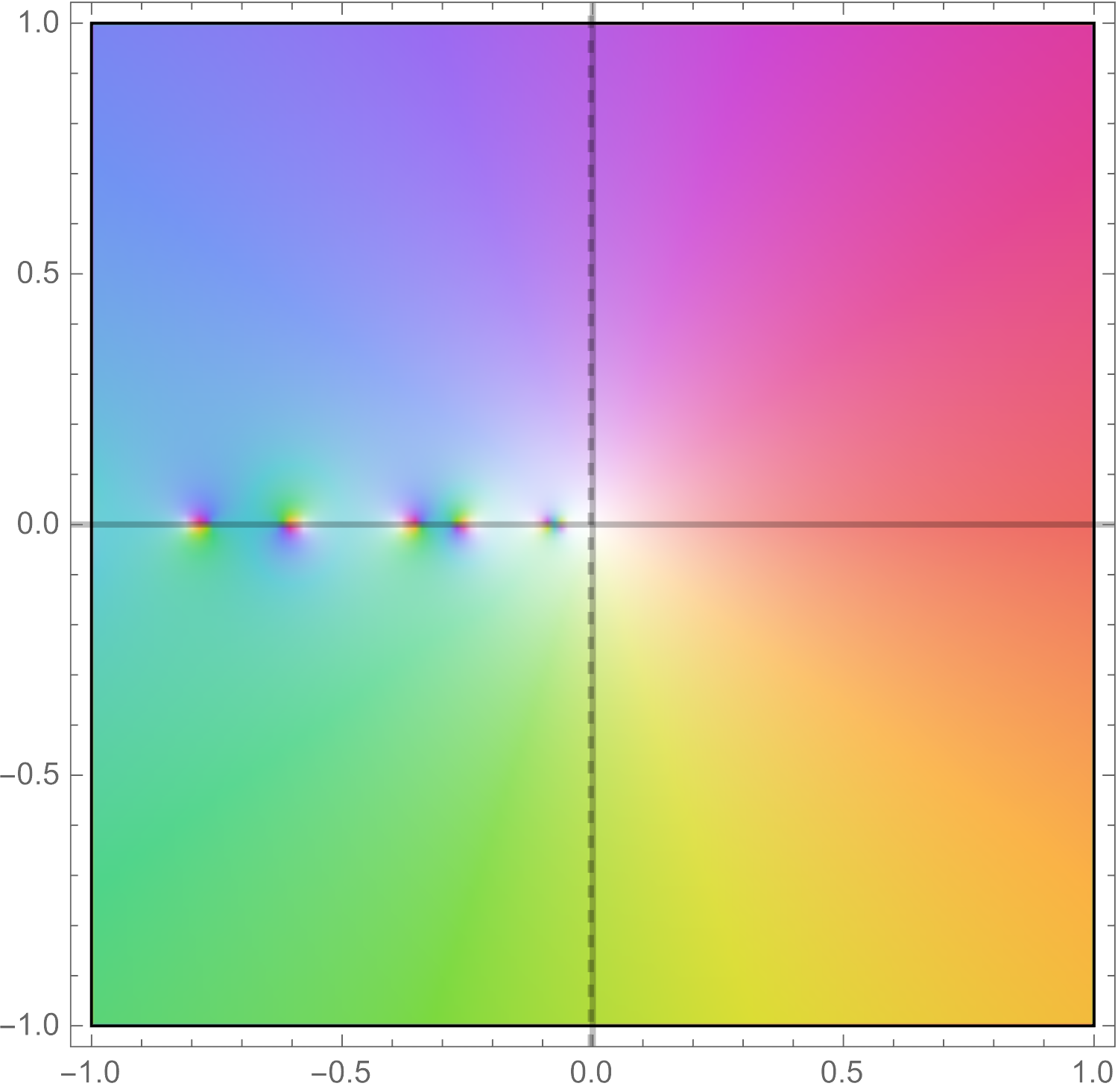}
  \end{subfigure}
  \begin{subfigure}{0.32\textwidth}
      \centering
      \includegraphics[width=\textwidth]{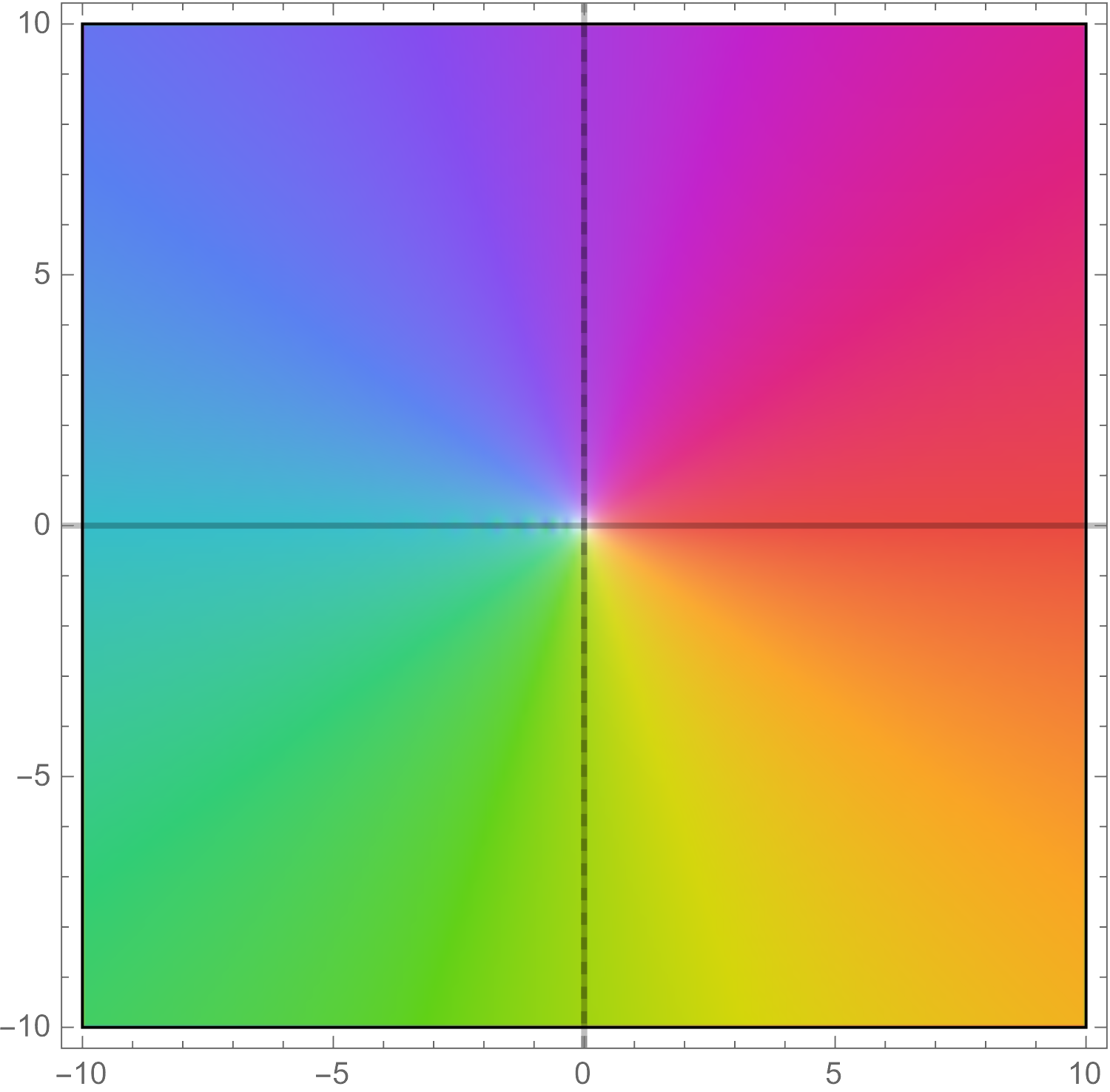}
  \end{subfigure}
  \begin{subfigure}{0.32\textwidth}
      \centering
      \includegraphics[width=\textwidth]{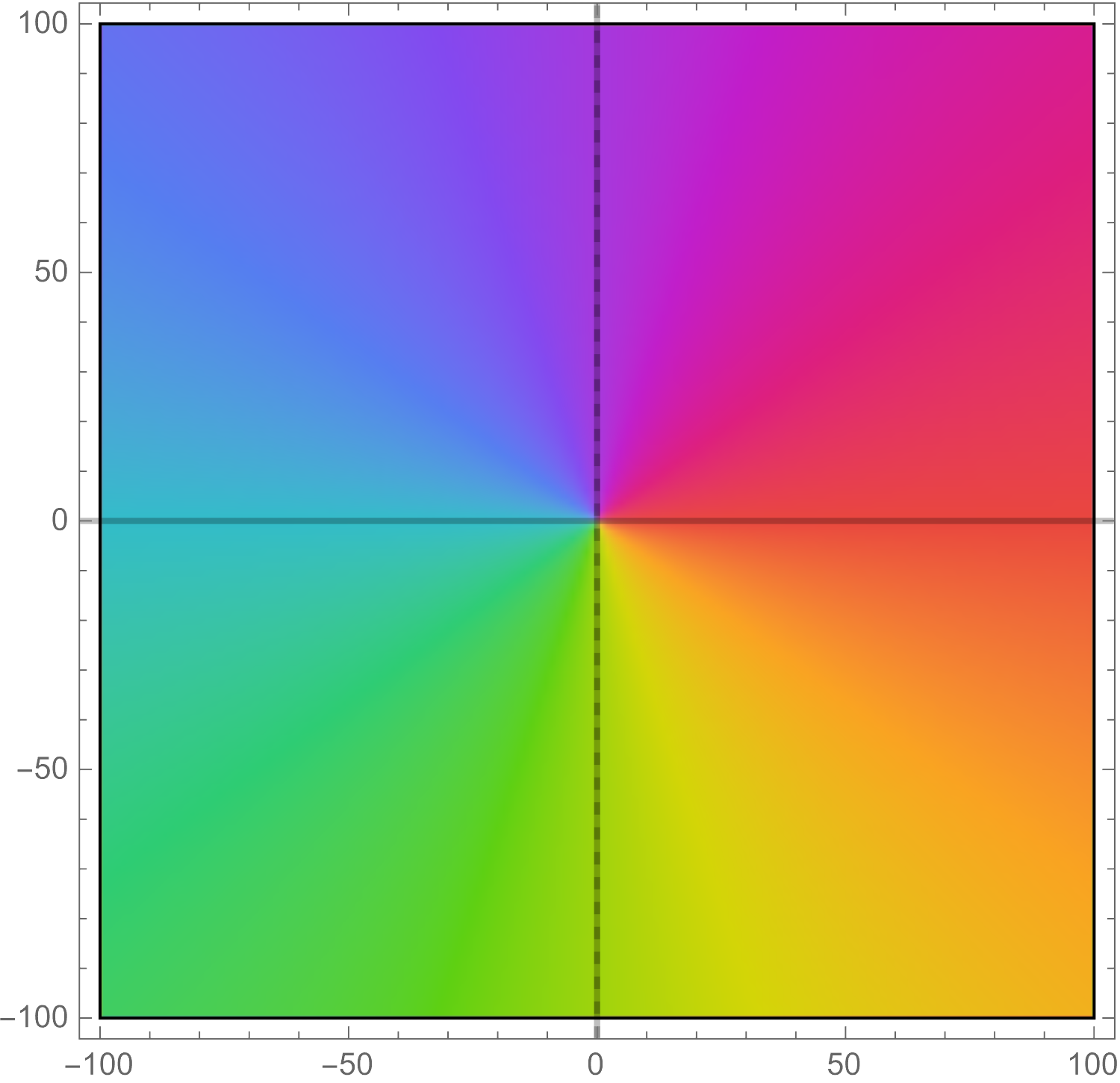}
  \end{subfigure}

  \begin{subfigure}{0.32\textwidth}
      \centering
      \includegraphics[width=\textwidth]{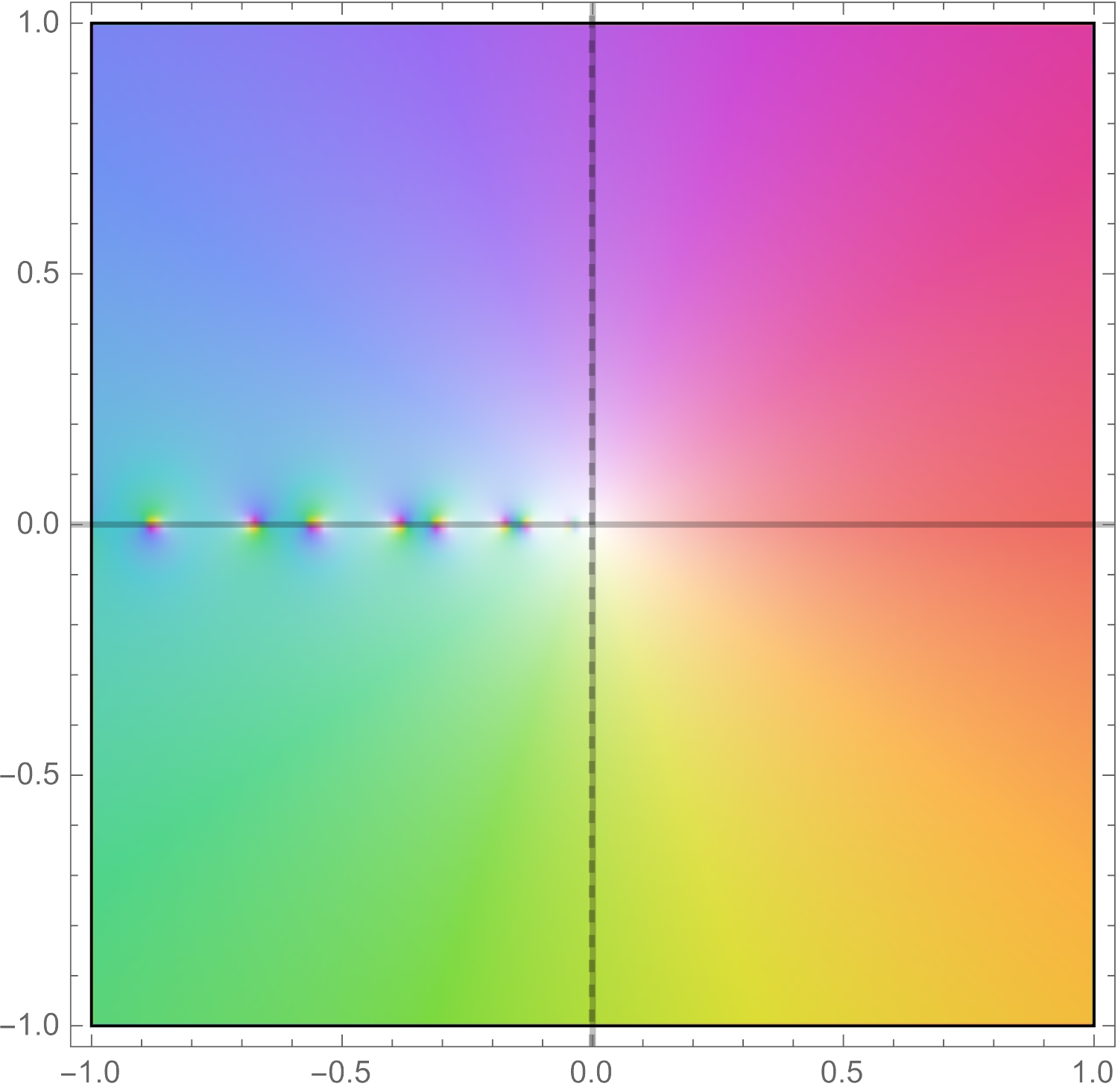}
  \end{subfigure}
  \begin{subfigure}{0.32\textwidth}
      \centering
      \includegraphics[width=\textwidth]{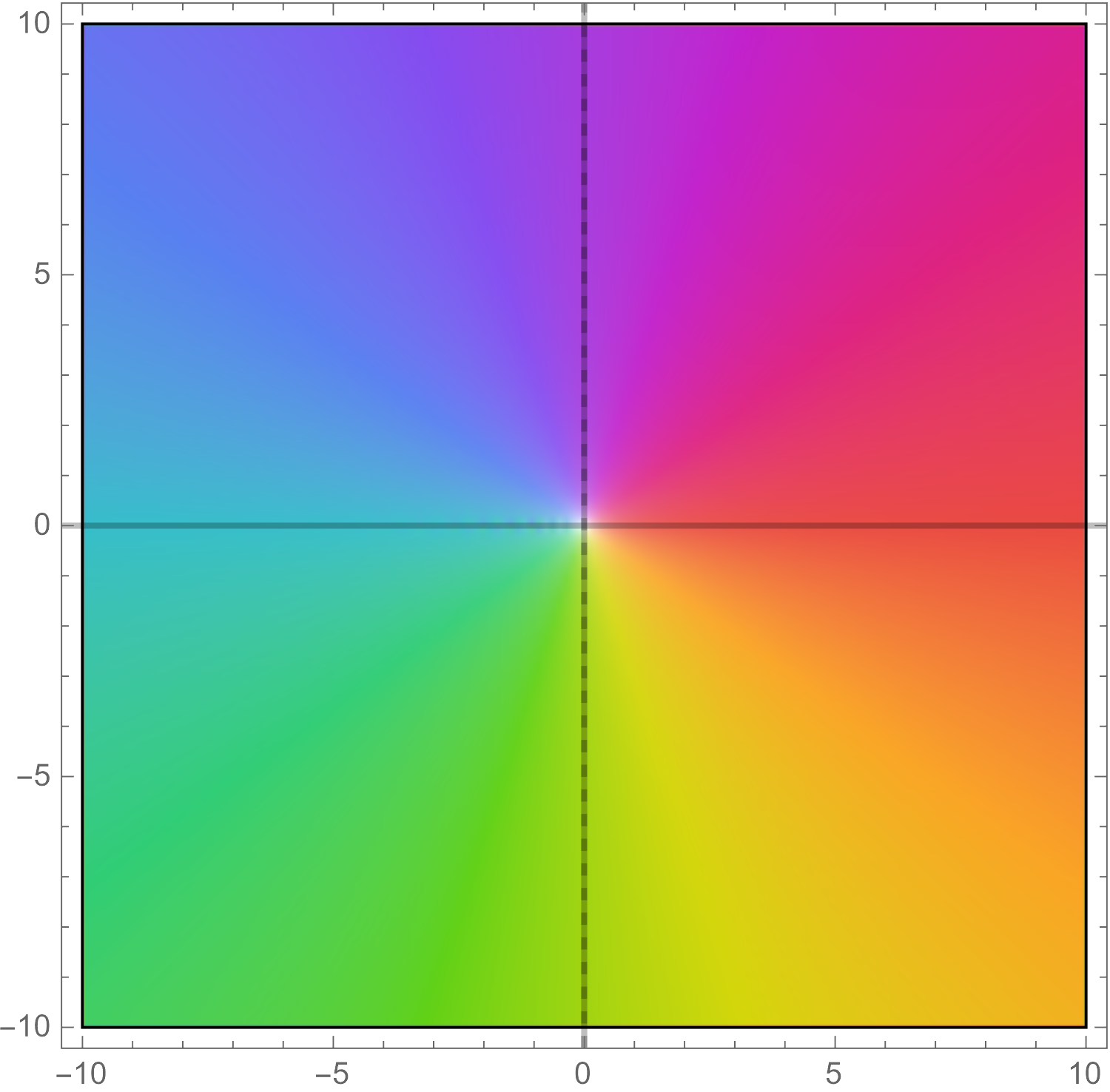}
  \end{subfigure}
  \begin{subfigure}{0.32\textwidth}
      \centering
      \includegraphics[width=\textwidth]{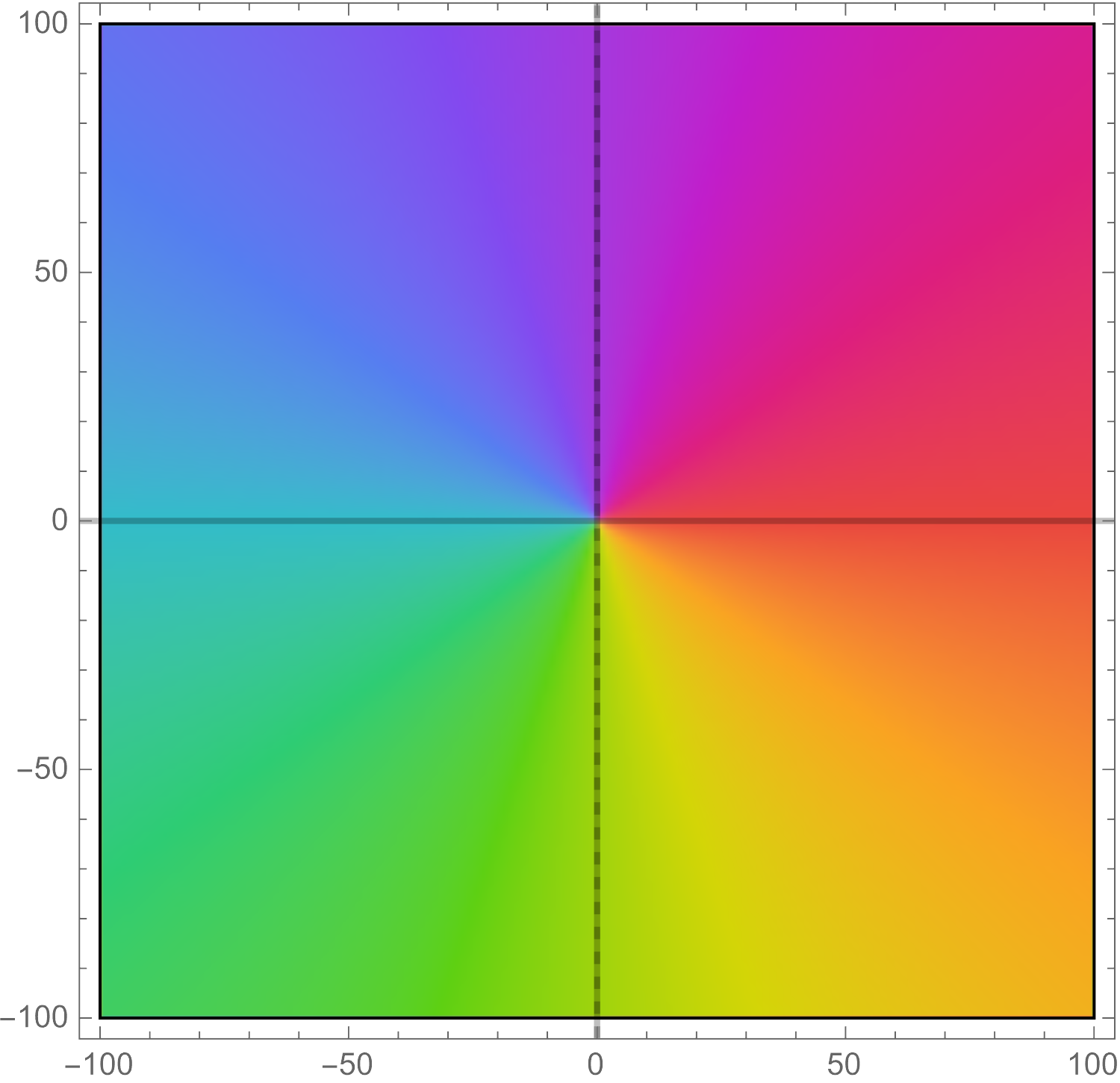}
  \end{subfigure}
  \caption{The same setting as \cref{fig:pole_search_1} with rows 
    corresponding to $\Lambda \in \cbra{32, 64, 128, 256, 512}$.}
  \label{fig:pole_search_2}
\end{figure}

Plots of $F(z)$ are presented in \cref{fig:pole_search_1,fig:pole_search_2}.

\subsection{Estimation of $C(\Lambda, t)$}
To estimate 
\[
 C(\Lambda, t) 
 = \lim_{R \to \infty} \frac{1}{\pi} \int_0^R \text{Re}\rbra{e^{ixt} 
    F\rbra{-\frac{1}{\Lambda+2} + ix}} \, \dee x,
\]
we use Mathematica's numerical inverse Laplace transform. Specifically, we evaluate
\texttt{InverseLaplaceTransform[F[z-1/(L+2)], z, t]} for a grid of values of $t$. 
In the definition above, \texttt{F[z-1/(L+2)]} evaluates to $F(z-1/(\Lambda+2))$.
Mathematica automatically selects an appropriate method for numerical inversion.

\end{appendix}

\begin{acks}[Acknowledgments]
We acknowledge use of the ARC Sockeye computing
platform from the University of British Columbia.
\end{acks}

\begin{funding}
NS acknowledges the support of a Vanier Canada Graduate Scholarship. 
SS acknowledges the support of a Florence Nightingale Bicentennial Fellowship.
ABC and TC acknowledge the support of an NSERC Discovery Grant. 
\end{funding}



\bibliographystyle{imsart-number} 
\bibliography{main.bib}       


\end{document}